%% file: mythesis.tex
\documentclass[11pt,expanded,copyright]{fsuthesis}

\usepackage{amsmath,amssymb,graphicx,multirow,tabularx,physics}
\usepackage{davitex}

\usepackage{float}     % To force figures on first pages
\usepackage{caption}   % Figure across multiple pages

\usepackage{amsthm}
\newtheorem{theorem}{Theorem}
\newtheorem{lemma}{Lemma}
\newtheorem{proposition}{Proposition}

\newcommand{\nconf}{N_\text{conf}}
\newcommand{\nsimil}{N_\text{sim}}

\usepackage{hyperref}
\hypersetup{
  colorlinks=true,
  linkcolor=black,
  citecolor=black,
  urlcolor=black,
}

\renewcommand*{\bibname}{References}
\renewcommand*{\appendixtocname}{Appendices}

\title{Scale Setting and Topological Observables in Pure SU(2) LGT}
\author{David A. Clarke}
\college{College of Arts and Sciences}
\department{Department of Physics}
\manuscripttype{Dissertation}
\degree{Doctor of Philosophy}                 
\degreeyear{2018}
\defensedate{September 14, 2018}

\subject{Particle physics theory}
\keywords{lattice field theory; scale; topology; continuum limit;%
finite size scaling; phase transition; gradient flow; cooling}

\committeeperson{Bernd Berg}{Professor Co-Directing Dissertation}
\committeeperson{Laura Reina}{Professor Co-Directing Dissertation}
\committeeperson{Thomas Albrecht-Schmitt}{University Representative}
\committeeperson{Rachel Yohay}{Committee Member}
\committeeperson{Peter Hoeflich}{Committee Member}

\begin{document}

\frontmatter
\maketitle
\makecommitteepage

\begin{dedication}
\centering
\vspace*{\fill}
This work is dedicated to my mother, who does everything in her power to 
support me, and to my father, whose curiosity and encouragement
kindled my interests in math and science.
\vspace*{\fill}
\end{dedication}

\begin{acknowledgments}
Many calculations used the FSU HEP theory cluster. In particular 
I would like to thank Joe Ryan for help with all aspects of the 
cluster. Some calculations in the topological charge project used the
computational resources of the FSU Astrophysics group, for which I would
like to thank Peter Hoeflich. Our calculations relied heavily
on the resources of the 
National Energy Research Scientific Computing Center (NERSC), a 
US Department of Energy (DOE) Office of Science User Facility 
supported by the DOE under Contract DE-AC02-05CH11231.

In part I was supported by the US Department of Energy under contract 
DE-SC0010102. I also received financial support from the J. W. Nelson
Endowment and the Dirac Endowed Fellowship.
My conference travel was subsidized by the FSU HEP group,
the Baugh Travel Scholarship, and the Congress of Graduate Students.  
Laura Reina made sure there was room in the budget for me to travel to
conferences. My graduate stipend was also funded because she agreed 
to be a co-chair.

I would like to thank the people who helped proofread this dissertation,
in particular Luis Mendoza, Samuel Glockner, and the members of the committee.

Most of all, I would like to thank my advisor, Bernd Berg, 
who played the leading role in my development as a physicist. 
None of this would have 
been possible without his patience, advocation, and support. 

\end{acknowledgments}

\tableofcontents
\listoftables
\listoffigures

\begin{listofsymbols}
\centering
\begin{tabular}{ll}
$\exists$ 	& There exists \\
$\forall$	& For all \\
$\in$		& Is a member of the set \\
$\equiv$	& Is defined as \\
$a$		& Lattice spacing \\
$\beta$		& $\SU(N_c)$ coupling constant; beta function \\
$\chi$		& Topological susceptibility; Polyakov loop susceptibility\\
$c$		& Speed of light\\
$\co_x$		& $\cos(x)$\\
$g$		& $\SU(N_c)$ bare coupling\\
$\hbar$		& Planck's constant \\
$k_B$		& Boltzmann's constant \\
$\log$		& Natural logarithm \\
$\mathbb{N}$	& Natural numbers \\
$N_c$		& Number of colors \\
$N_f$		& Number of fermion flavors \\
$N_s$		& Lattice extension in a spatial dimension \\
$N_\tau$	& Lattice extension in Euclidean time dimension;
                  temperature direction\\
$q$		& Gaussian or Student difference test; goodness-of-fit\\
$\sigma$	& String tension\\
$\sigma_i$	& A Pauli matrix\\
$\SU(N_c)$	& Special unitary group of degree $N_c$\\
$\s_x$		& $\sin(x)$\\
$T_c$		& Deconfining phase transition temperature\\
$\tauint$	& Integrated autocorrelation time\\
$\mathbb{Z}$	& Integers
\end{tabular}
\end{listofsymbols}

% make sure that abbreviations are only included if they are actually used
\begin{listofabbrevs}
\centering
\begin{tabular}{ll}
BC	&	Boundary condition \\
CDF	&	Cumulative distribution function \\
CLT	&	Central limit theorem\\
HEP	&	High energy physics\\
HB	&	Heat bath\\
LFT	&	Lattice field theory\\
LGT	&	Lattice gauge theory\\
LHS	&	Left hand side\\
LLN	&	Law of large numbers\\
MCMC	&	Markov chain Monte Carlo\\
MCOR	&	Monte Carlo plus over-relaxation \\
MPI	&	Message passing interface \\
NERSC	&	National Energy Research Scientific Computing Center \\
OR	&	Over-relaxation \\
PDF	&	Probability distribution function \\
QCD	&	Quantum chromodynamics \\
QFT	&	Quantum field theory \\
RG	&	Renormalization group \\
RHS	&	Right hand side \\
SM	&	Standard model \\
UV	&	Ultraviolet \\
\end{tabular}
\end{listofabbrevs}

\begin{abstract}
In this dissertation, we investigate the approach
of pure $\SU(2)$ lattice gauge theory to its continuum limit using
the deconfinement temperature, six gradient scales, and six cooling scales.
We find that cooling scales exhibit similarly good scaling behavior as gradient
scales, while being computationally more efficient.
In addition, we estimate systematic error in continuum limit extrapolations of
scale ratios by comparing standard scaling to asymptotic scaling.
Finally we study topological observables in pure $\SU(2)$ using
cooling to smooth the gauge fields, and investigate the sensitivity of cooling
scales to topological charge.
We find that large numbers of cooling
sweeps lead to metastable charge sectors, without destroying physical
instantons, provided the lattice spacing is fine enough and the volume
is large enough. Continuum limit estimates of the topological susceptibility
are obtained, of which we favor $\chi^{1/4}/T_c=0.643(12)$. Differences
between cooling scales in different topological sectors turn out to be 
too small to be detectable within our statistical error.
\end{abstract}

\mainmatter

\input chapter1 % introduction
\input chapter2 % theoretical preliminaries
\input chapter3 % MCMC simulations
\input chapter4 % comparison of scales
\input chapter5 % topological charge
\input chapter6 % conclusions 

\appendix
\input appendix6 % supplementary figures 
\input appendix5 % probability and statistics 
\input appendix2 % calculational details

\bibliographystyle{plain}
\bibliography{myrefs}

\begin{biosketch}
\section*{Education}
\centering
\begin{tabular}{ll}
  2015-2018 & Ph.D. Physics, Florida State University,
              Tallahassee, Florida.\vspace{1mm}\\
  2013-2015 & M.S. Physics, Florida State University,
              Tallahassee, Florida.\vspace{1mm}\\
  2008-2013 & B.S. Physics, Ohio State University, 
              Columbus, Ohio.\vspace{1mm}\\
  2008-2013 & B.S. Mathematics, Ohio State University, 
              Columbus, Ohio.
\end{tabular}

\section*{Publications}
\begin{enumerate}
\item B. A. Berg and D. A. Clarke, ``Topological charge and cooling scales
      in pure SU(2) lattice gauge theory", Phys. Rev. D, 97 (2018)
      {\footnotesize DOI}:10.1103/PhysRevD.97.054506.
\item B. A. Berg and D. A. Clarke, ``Estimates of scaling violations for pure
      SU(2) LGT", Eur. Phys. J., 175 (2018) 
      {\footnotesize DOI}:10.1051/epjconf/201817510007.
\item B. A. Berg and D. A. Clarke, ``Deconfinement, gradient, and cooling 
      scales for pure SU(2) lattice gauge theory", Phys. Rev. D, 95 (2017)
      {\footnotesize DOI}:10.1103/PhysRevD.95.094508.
\end{enumerate}

\section*{Presentations}
\begin{enumerate}
\item D. A. Clarke, ``Topological charge and cooling scales in pure
      SU(2) LGT", Presentation at American Physical Society
      April Meeting, Columbus OH, USA (2018).
\item D. A. Clarke, ``Topological charge in pure SU(2) LGT",
      Presentation at Florida State University High Energy Physics Seminar,
      Tallahassee FL, USA (2018).
\item D. A. Clarke, ``Estimates of scaling violations for pure SU(2) LGT", 
      Presentation at $35^{\rm th}$ International Symposium on 
      Lattice Field Theory, Granada, Spain (2017).
\item D. A. Clarke, ``A Comparison of scales in pure SU(2) LGT",
      Presentation at Florida State University High Energy Physics Seminar,
      Tallahassee FL, USA (2017).
\end{enumerate}

\section*{Awards}
\begin{enumerate}
  \item Dirac Endowed Fellowship, Florida State University, 2018.
  \item J. W. Nelson Endowment, Florida State University, 2017.
  \item Baugh Scholarship, Florida State University, 2017.
\end{enumerate}

\end{biosketch}

\end{document}

%% file: chapter1.tex
\chapter{Introduction}\label{ch:intro}

\begin{figure}
  \centering
  \includegraphics[width=0.95\linewidth]{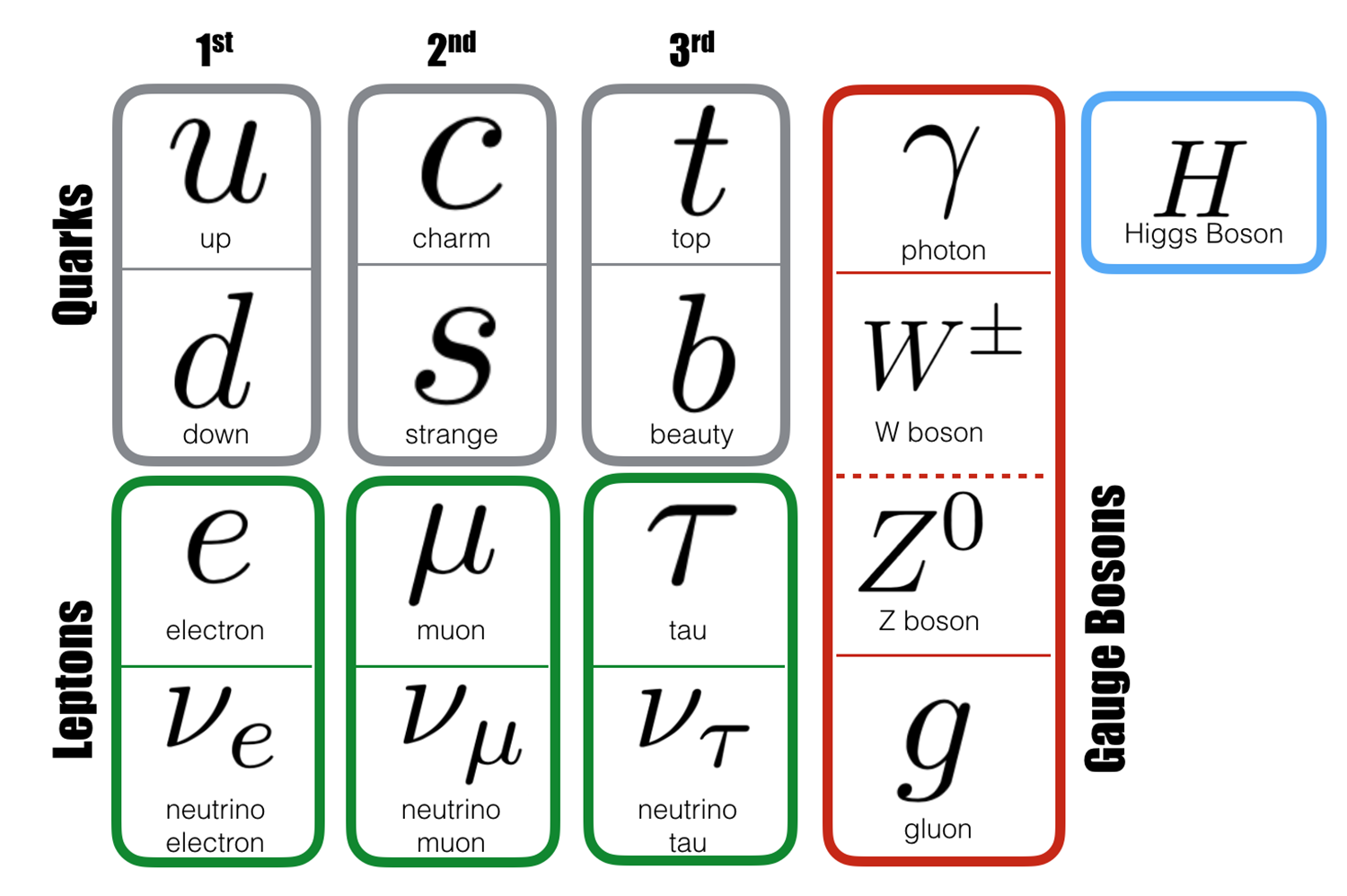}
  \caption{Summary of elementary SM particles. The first three columns give
           the three generations of matter particles. Image taken
           from the Physics Institute at University of Zurich~\cite{wiki_SM}.}
  \label{fig:SM}
\end{figure}

The Standard Model (SM) of particle physics classifies all known
{\it elementary particles}, i.e. particles with no known substructure,
and describes three fundamental forces: the electromagnetic,
weak, and strong forces. Elementary particles can be divided into
{\it matter particles} (quarks and leptons); {\it gauge bosons}, which mediate
the three aforementioned forces; and a {\it scalar boson}, the Higgs boson,
whose field interacts directly with elementary particles that thereby
acquire their mass. For each particle there exists a corresponding
antiparticle; sometimes a particle is its own antiparticle.
Figure~\ref{fig:SM} gives a schematic overview of the SM. 
The SM has a long history of experimental confirmations culminating 
in the 2012 discovery of the Higgs boson by the ATLAS and CMS 
experiments~\cite{aad_observation_2012,chatrchyan_observation_2012}.

The theoretical framework underlying the SM is Quantum Field Theory (QFT).
In QFT, the strength of an interaction is parameterized by some coupling
$g$, and in practice, one obtains analytic results in the small coupling
limit by Taylor expanding in $g$. This is known as a {\it perturbative}
calculation. Not all quantities lend themselves well to perturbative
methods. In particular physical observables $m$ with units of mass behave as
\begin{equation}
  m\sim e^{-1/g^2},
\end{equation} 
which is zero to all orders in perturbation theory. To calculate such
a quantity therefore requires an alternative, {\it non-perturbative}
method. Lattice Field Theory (LFT), which was introduced in 1974 by
Wilson~\cite{wilson_confinement_1974}, gives access to non-perturbative
quantities, supplementing perturbative calculations.  
One of the early successes of LFT came with the 1980 paper of 
Creutz~\cite{creutz_monte_1980}, which supported quark confinement.
Lattice calculations can also test the SM, for instance by calculating
baryon and meson spectra from first principles.  
Along this vein, lattice calculations can achieve
arbitrary precision in principle, provided enough computing
power is available. 

Lattice simulations of the full SM are not yet within our grasp, 
so for the time being, we are restricted to examinations of parts of the SM.
Nowadays one can study, for instance, quantum chromodynamics (QCD) with
$N_f=4$, which is a theory with gluons and four fermion flavors
~\cite{rae_ground_2016}.
Even with this restriction, useful information about 
the SM can still be gleaned.
Two of the simplest theories pure $\SU(3)$, which is a theory
of gluons only, and pure $\SU(2)$, which is a theory of gluon-like particles
only. When the particle content of an LFT includes only gauge bosons, it
is usually referred to as a {\it lattice gauge theory} (LGT).
Because of their relative computational simplicity, LGTs are often used as
a proving ground for new algorithms and techniques, allowing for
high precision calculations with modest computational resources.

Lattice calculations begin by discretizing space-time, where space-time
points are separated by a finite lattice spacing. The physical theory
is recovered in the infinite volume continuum limit, where one sends 
the volume to infinity and the lattice spacing to zero compared to
a physical length.
Using lattice regularization, one can
calculate dimensionless length ratios
\begin{equation}
  r_{ij}=\frac{\ell_i}{\ell_j}
\end{equation}
in the continuum limit, where each $\ell$ is some physical length;
for instance $\ell_i$ could be the characteristic wavelength of a pion.
Therefore if one wishes to extract $\ell_i$ from the lattice, one
must know $\ell_j$ precisely and accurately. Not all reference scales
$\ell_j$ are equally suited for this purpose; one reason is that different
reference scales may require different computational effort.
Choosing a reference scale is what we mean by {\it scale setting}.

Scale setting is an important source of error for the purpose
of extracting dimensionful quantities from the lattice, because the
precision of the reference scale propagates to the final result.
It is important to find a reference scale that can be computed
with high statistical precision, since modern lattice 
calculations, in particular those that compare against or supplement
experimental results, often aim at relative statistical error bars of 
1\% or smaller~\cite{thomas_meson_2017}. Additional details about scale
setting can be found in the review by Sommer~\cite{sommer_scale_2014}. 

Scale setting enjoyed renewed interest with the introduction
of L\"uscher's gradient flow~\cite{luscher_properties_2010},
from which a novel reference scale, the gradient scale, was defined.
The gradient flow also gained popularity as a technique for dampening
local UV fluctuations; such techniques are called {\it smoothing}
or {\it smearing}. In a pure $\SU(3)$ study, Bonati and 
D'Elia~\cite{bonati_comparison_2014} 
showed that for topological observables, smoothing using standard 
cooling, introduced originally by Berg~\cite{berg_dislocations_1981},
produces similar results as the gradient flow, while
progressing through flow time much faster. In the same
paper, they suggested that cooling could be used to define a
cooling scale in a similar manner as the gradient scale.
In this context, we decided to investigate cooling
scales in pure $\SU(2)$ LGT. Since high precision
results are computationally even less demanding for $\SU(2)$ than $\SU(3)$, we
were able to reach a greatly enhanced accuracy when compared with
Bonati and D'Elia.

The gauge bosons of the SM are thought to be excitations of underlying
{\it fields}, mathematical objects whose value depends on their space-time
location. Vacuum configurations of $\SU(N_c)$ gauge fields have 
intrinsic topologies classified by an integer topological charge. 
Configurations of the same charge can be continuously deformed into 
one another, i.e. they are topologically equivalent or {\it homeomorphic}.
The topology of gauge fields is relevant to physical quantities in our
world; in particular the mass of the $\eta'$ meson 
depends on the topological charge distribution
\cite{t_hooft_computation_1976,witten_current_1979,veneziano_u1_1979}.
On the lattice, configurations updated by Markov Chain Monte Carlo 
(MCMC) algorithms can get stuck on configurations of a particular charge, 
so that the distribution of configurations is not well-sampled. This 
{\it topological freezing} can lead to a bias in observables;
finding ways to circumvent this issue is an active area of 
research~\cite{luscher_stochastic_2018}.
Encouraged by the recent success of standard cooling as a 
smoothing algorithm for pure $\SU(3)$, we investigated the 
topology of pure $\SU(2)$ LGT and obtained an accurate estimate
of the $\SU(2)$ topological susceptibility.

The structure of this dissertation is as follows: In 
Chapter~\ref{ch:preliminaries}, the lattice formulation, along with
background theory for scale setting and topology on the lattice, is introduced.
Chapter~\ref{ch:MCMC} reviews MCMC along with details of how we
implemented computer simulations. Numerical results for our project are given
in Chapters~\ref{ch:comparison} and \ref{ch:top}. Conclusions are given in 
Chapter~\ref{ch:summary}.

The author attempted to write this dissertation to be readable by junior
high energy physicists interested in lattice gauge theory. Therefore
there is a collection of Appendices containing extra background.
A brief introduction to statistical analysis in in Appendix~\ref{ap:prb}.
To keep the discussion of this dissertation focused, some calculational 
details are postponed to Appendix~\ref{ap:calc}. 

%% file: chapter2.tex
\chapter{Preliminaries}\label{ch:preliminaries}

LGT is introduced in Section~\ref{sec:LGT}
by reviewing local gauge symmetries in QFT, regularizing a pure
gauge theory on the lattice, discussing the true continuum limit,
and introducing finite temperature. In Section~\ref{sec:refscales}, 
reference scales are defined, and systematic error within the context
of scale setting is explored. Topological observables are introduced
in Section~\ref{sec:topinvar}, and effects of topology barriers
are considered.

\section{Lattice gauge theory}\label{sec:LGT}

Physically, QFT is defined on a 4D Minkowskian space-time. In LGT 
the 4D space-time is instead equipped with a Euclidean metric, which is 
related to the original metric via a {\it Wick rotation}
\begin{equation}
  t\to i\tau.
\end{equation} 
Therefore we will work with a Euclidean metric
and use downstairs summation indices. 
We will also use {\it natural units}
$\hbar=c=k_B=1$. In natural units, every physical
quantity has units of some power of length. For example
time has units of length, while energy, mass, and momentum
have units of inverse length. 
We first work in the continuum, then 
discretize the theory by defining the lattice.

\subsection{Local gauge symmetries}

Local gauge symmetries play a central role in the SM. 
Starting from a Lagrangian that depends on the derivatives of
some field, the requirement of local gauge invariance suggests that
we introduce a {\it gauge field}. This gauge field allows one to 
define a {\it covariant derivative} whose transformation law will 
respect the local gauge symmetry.
Excitations of the gauge field are gauge bosons,
which are the force-carrying particles of the SM.

As an example consider $N_c$ complex scalar fields $\phi_{i}(x)$ equipped with 
a global $\SU(N_c)$ symmetry. The Lagrangian is
\begin{equation}\label{eq:LagrM}
  \Lagr_{M}=-\partial_{\mu}\phi^{\dagger}(x)\partial_{\mu}\phi(x)
                +m^{2}\phi^{\dagger}(x)\phi(x),
\end{equation}
where $\phi(x)$ is the $N_c$-dimensional vector formed by these fields.
$\Lagr_{M}$ becomes invariant under local $\SU(N_c)$ transformations, i.e. 
transformations of the form
\begin{equation}\label{eq:fxform}
  \phi(x)\to U(x)\phi(x),
\end{equation}
where $U(x)\in \SU(N_c)$, when one replaces the partial derivative by the 
covariant derivative $D_\mu$, which transforms as
\begin{equation}\label{eq:dxform}
  D_\mu(x)\to U(x) D_{\mu}(x)U^{\dagger}(x).
\end{equation}
We define
\begin{equation}\label{eq:covariantderivative}
  D_{\mu}(x)\equiv\partial_{\mu}+A_{\mu}(x), \qquad 
       A_{\mu}(x)\equiv-igA^{a}_{\mu}(x)T^{a},
\end{equation}
where $g$ is the bare coupling constant, $A_{\mu}(x)$ is the gauge field, and 
$T^{a}$, $a=1,\dots,N^2-1$, are the generators of the $\SU(N_c)$ Lie algebra 
$\mathfrak{su}(N_c)$. For notational convenience we now suppress dependence
on $x$. Using this
definition of $D_\mu$, the gauge fields must change according to
\begin{equation}
  A_{\mu}\to U A_{\mu}U^{\dagger}
          -\big(\partial_{\mu}U\big)U^{\dagger}.
\end{equation}
The gauge field becomes dynamic by adding the kinetic part
\begin{equation}\label{eq:lagrkin}
  \Lagr_G=\frac{1}{4} F_{\mu\nu}^{a}F_{\mu\nu}^{a}
         =-\frac{1}{2g^2}\tr F_{\mu\nu}F_{\mu\nu},
\end{equation}
where 
\begin{equation}\label{eq:fieldstrength}
          F_{\mu\nu}^a\equiv \partial_\mu A_\nu^a-\partial_\nu A_\mu^a
                        +gf^{abc}A_\mu^b A_\nu^c,\qquad
  F_{\mu\nu}\equiv-igF_{\mu\nu}^aT^a=\left[D_\mu,D_\nu\right],
\end{equation} 
and $f^{abc}$ are the structure constants of $\SU(N_c)$.
$\Lagr_{G}$ is also invariant 
under the transformation of eqs.~\eqref{eq:fxform} and
\eqref{eq:dxform}. Taken altogether, the gauge-invariant, dynamical, 
scalar theory is described by the Lagrangian
\begin{equation}
  \Lagr=-\big(D_\mu\phi\big)^\dagger D_\mu\phi
                +m^2\phi^\dagger\phi
                -\frac{1}{2g^2}\tr F_{\mu\nu}F_{\mu\nu}.
\end{equation}
We would like
to point out that the definitions~\eqref{eq:covariantderivative} and
\eqref{eq:fieldstrength} are somewhat different than the convention
of many QFT books such as Srednicki~\cite{srednicki_quantum_2007} or
Peskin and Schroeder~\cite{peskin_introduction_1995}. An advantage of
the convention we have taken, which is also used in, for instance, Montvay and
M\"unster~\cite{montvay_quantum_1994}, is that one can explicitly see the
dependence of the Lagrangian~\eqref{eq:lagrkin} on the coupling. 

In this dissertation we will be primarily interested in a theory with 
$\Lagr_G$ only and gauge group $\SU(2)$; such a theory is referred
to as {\it pure} $\SU(2)$. $\SU(2)$ is the simplest,
phenomenologically interesting, non-Abelian gauge group. 
Often the gauge particles of pure $\SU(2)$ theories are referred
to as ``gluons," even though $N_c\neq 3$. Because it
is non-Abelian, it has nonzero structure constants, which means it
contains self-interactions of the form $AAA$ and $AAAA$. 
%These self-interactions are responsible for confinement, which we discuss
%further in the next section. 
For the purpose of a lattice study,
it is useful to look at a non-Abelian theory, which
has a well-defined continuum limit.

\subsection{Lattice regularization}\label{sec:latreg}
We now define QFT on a lattice. Let $N_{1},N_{2},N_{3},N_{4}
\in \mathbb{N}$. The {\it lattice} ${\L}$ is defined by
\begin{equation}
  {\L}\equiv\{x\,|\,x_\mu=an_\mu,\,n_\mu\le N_\mu,\,
      \mu=1,2,3,4\}.
\end{equation}
Here $a$ is called the {\it lattice spacing}. 
After our Wick rotation, we identify $N_1$, $N_2$, and $N_3$
as the extensions of the lattice in the spatial directions,
and $N_4$ is taken to be the extension in the Euclidean time 
direction. Matter fields and gauge 
transformations are defined on the {\it sites} $x\in\L$. We shall 
take the lattice to have periodic boundary conditions (BCs), i.e.
\begin{equation}\label{eq:PBC}
  x+aN_{\mu}\hat{\mu}=x,
\end{equation}
where $\hat{\mu}$ is the unit vector in the direction indicated by $\mu$. 
Since the lattice is discrete, one must replace partial derivatives 
by finite differences, 
\begin{equation}
  \partial_\mu f(x)\to\Delta_{\mu}f(x)\equiv\frac{f(x+a\hat{\mu})-f(x)}{a},
\end{equation}
and similarly replace integrals with sums,
\begin{equation}\label{eq:inttosum}
  \int d^4x\to a^4\sum_x.
\end{equation}
Moreover the BCs~\eqref{eq:PBC} imply for every direction
that the momentum is discretized as
\begin{equation}
  p_\mu=\frac{2\pi}{a}\frac{n_\mu}{N_\mu},
\end{equation}
which means that momentum space integrals must also be replaced by
sums
\begin{equation}
  \int\frac{d^4p}{(2\pi)^4}\to
  \frac{1}{a^4N_1N_2N_3N_4}\sum_p. 
\end{equation}
Putting QFT on a lattice regularizes the theory. 
To see this, consider a field $\phi$ defined on the lattice. 
Its Fourier transform
\begin{equation}
  \widetilde{\phi}(p)=a^4\sum_xe^{-ipx}\phi(x)
\end{equation}
is periodic in momentum space, which gives us the correspondence
$p_\mu\leftrightarrow p_\mu+2\pi/a$. Hence we can restrict
momenta to the {\it first Brillouin zone},
\begin{equation}
 -\frac{\pi}{a}<p_\mu\le\frac{\pi}{a}
\end{equation}
and one obtains a UV cutoff $|p_\mu|\le\pi/a$.

\begin{figure}[t]
  \centering
  \includegraphics[width=0.9\linewidth]{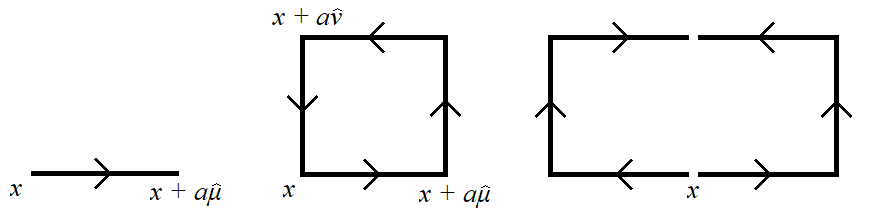}
  \caption{Left: A link variable. Middle: A plaquette. 
           Right: A staple matrix in 2D.}
  \label{fig:links}
\end{figure}

Now we define the building blocks necessary to construct paths on the 
lattice. The directed {\it link} connects $x$ with the neighboring point 
$x+a\hat{\mu}$, and its corresponding {\it link variable} 
$U_\mu(x)\in \SU(N_c)$ is defined by
\begin{equation}
  U_\mu(x)=e^{-aA_{\mu}(x)},
\end{equation}
where $A_{\mu}(x)\in\su(N_c)$. A link variable is
depicted in Fig.~\ref{fig:links} (left).
We associate to any path $\mathcal{C}$ 
the ordered product of its link variables $U(\mathcal{C})$. 
If we follow a path and then reverse our steps, we should end
up back where we started; hence
\begin{equation}
  U_{-\mu}(x+a\hat{\mu})U_\mu(x)=\id.
\end{equation}
Furthermore $U^\dagger(x)U(x)=\id$, so we can see the effect
of the dagger on link variables:
\begin{equation}
  U_\mu^\dagger(x)=U_{-\mu}(x+a\hat{\mu}).
\end{equation}
Let $\mathcal{C}_{x}$ be a path on the lattice that originates and 
terminates at the point $x$. The corresponding {\it Wilson loop} is 
defined by $\tr U(\mathcal{C}_{x})$. 
Under local gauge transformations, link variables 
transform as
\begin{equation}
  U_\mu(x)\to\Lambda^\dagger(x)U_\mu(x)\Lambda(x+a\hat{\mu}),\qquad
  \Lambda(x)\in\SU(2),
\end{equation}
which ensures the gauge invariance of Wilson loops.
A {\it plaquette}, shown in Figure~\ref{fig:links} (middle), is the 
smallest Wilson loop, an oriented square of side length $a$ with 
corresponding link variable 
\begin{equation}
  U^\square_{\mu\nu}(x)=U_\mu(x)U_\nu(x+a\hat{\mu})
                        U^\dagger_\mu(x+a\hat{\nu})U^\dagger_\nu(x).
\end{equation}
Every link variable in 4D LGT is part of six plaquettes. The remaining
three edges of any particular plaquette are shaped like
a staple; therefore we call the combination
\begin{equation}\label{eq:staple}
  U^\sqcup_\mu(x)=
  \sum_{\nu\ne\mu}\left[U_\nu(x)U_\mu(x+a\hat{\nu})U^\dagger_\nu(x+a\hat{\mu})
  +U^\dagger_\nu(x-a\hat{\nu})U_\mu(x-a\hat{\nu})
                   U_\nu(x-a\hat{\nu}+a\hat{\mu})\right]
\end{equation}
the {\it staple matrix}. A 2D staple matrix is shown in
Fig.~\ref{fig:links} (right); alternatively one can view it
as one of the three terms in the sum~\eqref{eq:staple}.

Plaquettes are used to construct the gauge invariant $\SU(N_c)$ {\it Wilson 
action} \cite{wilson_confinement_1974}, given by
\begin{equation}\label{eq:wilsonaction}
    S_W\equiv\beta\sum\limits_{x,\mu<\nu}\left(1
         -\frac{1}{N_c}\Re\tr U^\square_{\mu\nu}(x)\right).
\end{equation}
The factor $\beta$ is given this name in analogy to the inverse temperature 
in statistical mechanics. Using the Campbell-Baker-Hausdorff formula, 
one can show
\begin{equation}
  U^\square_{\mu\nu}(x)=
    \exp\left[-a^2F_{\mu\nu}(x)+\mathcal{O}\big(a^3\big)\right].
\end{equation}
After some algebra, the connection between the Wilson action and the action
corresponding to eq.~\eqref{eq:lagrkin} becomes clear. We find
\begin{equation}
  S_{W}=-\frac{\beta}{4N_c}\sum\limits_{x}a^4\tr F_{\mu\nu}(x)
         F_{\mu\nu}(x)+\mathcal{O}\big(a^5 \big).
\end{equation}
In the limit $a\to0$, the Wilson action coincides with the action
$S_G=\int d^4x\,\Lagr_G$ when one identifies 
\begin{equation}
  \beta=\frac{2N_c}{g^2}.
\end{equation}
Because of this identification, $\beta$ is also (besides $g$) sometimes 
referred to as the {\it coupling constant}.

We close this subsection with a remark about confinement.
Let $\mathcal{C}_{RT}$ be a rectangular loop on the
lattice of side lengths $R$ and $T$ and let $W(\mathcal{C}_{RT})$ be the
corresponding Wilson loop. 
Then the {\it static quark potential} $V(R)$ is defined by
\begin{equation}
  V(R)\equiv-\lim_{T\to\infty}\frac{1}{T}\log W(\mathcal{C}_{RT})  
\end{equation}
and gives the energy of the gauge field due to two color sources separated by
a distance $R$. The {\it string tension} $\sigma$ is defined by
\begin{equation}
  \sigma\equiv\lim_{R\to\infty}\frac{1}{R}V(R).
\end{equation}
If the string tension is non-vanishing, then the potential scales linearly 
with $R$ in the large $R$ limit; 
this phenomenon has been observed in LGT 
simulations~\cite{montvay_quantum_1994}.
Thus we see one of the major successes
of LGT: it proffers an explanation of confinement.  

\subsection{The renormalization group and the continuum limit}

In the limit $a\to0$, physical quantities $P$ should agree with experimental
results, which means they should become independent of $a$, ``forgetting"
about the lattice structure. Since $P$ depends in general also on $g$,
this means that changes in $a$ have to be compensated by changes in $g$
to keep the physics constant. More precisely, it must be that 
\begin{equation}
  \lim_{a\to0}P\Big(g(a),a\Big)=P_0
\end{equation}
where $P_0$ is the physical quantity's experimental value.
Callan~\cite{callan_broken_1970} and
Symanzik~\cite{symanzik_small_1970,symanzik_small-distance-behaviour_1971}
independently formulated the requirement of constant physics as
a differential equation
\begin{equation}\label{eq:RGflow}
  \left(\frac{\partial}{\partial\log a}
        +\frac{\partial g}{\partial\log a}
         \,\frac{\partial}{\partial g}\right)P=0. 
\end{equation}
(The RHS of this equation is more precisely
$\mathcal{O}\big((a/\xi)^2\log(a/\xi)\big)$ for a lattice system
with correlation length $\xi$~\cite{montvay_quantum_1994}.)
Equation~\eqref{eq:RGflow} relates to a semi-group of scale 
changing transformations called the {\it renormalization group} (RG). 
The coefficient of the second term is called the {\it beta function},
\begin{equation}\label{eq:betafunction}
  \beta\equiv-\frac{\partial g}{\partial\log a},
\end{equation}
and it measures how the bare coupling $g$ must change when $a$ changes.
The use of the symbol $\beta$ here is unfortunately a convention; it
is not to be confused with the coupling constant. It is usually clear
from context what is meant. In practice $\beta$ can be determined
from perturbation theory. An explicit dependence of $g$ on $a$ is then
determined by solving the differential equation~\eqref{eq:betafunction}. 

For example the pure $\SU(N_c)$ lattice beta function has been calculated 
up to 3-loop order in perturbation theory. It is given by
\begin{equation}\label{eq:blatgs}
  \beta_{L}(g)=-b_{0}g^{3}-b_{1}g^{5}-b_{2}^{L}g^{7}
                   +\mathcal{O}\big(g^{9}\big)
\end{equation}
where
\begin{equation}\label{eq:blat3loop}
  b_{0}= \frac{11}{3}\frac{N_c}{16\pi^2}, \qquad
  b_{1}= \frac{34}{3}\Bigg(\frac{N_c}{16\pi^2}\Bigg)^{2},\qquad
  b_{2}^{L}=\Bigg(-366.2+\frac{1433.8}{N_c^2}-\frac{2143.0}{N_c^4}\Bigg)
            \Bigg(\frac{N_c}{16\pi^2}\Bigg)^{3}
\end{equation}
have been calculated at one-loop 
\cite{gross_d.j._ultraviolet_1973,politzer_reliable_1973}, two-loop 
\cite{belavin_calculation_1974,caswell_asymptotic_1974,jones_two-loop_1974},
and three-loop order~\cite{alle_three-loop_1997}, respectively.
The constants $b_0$ and $b_1$ are universal in the sense that they do not 
depend on the regularization scheme; however $b_2$ does depend on the
regularization scheme, with $b_2^L$ being the value using lattice
regularization. The RG equation on the lattice is
\begin{equation}
  \beta_{L}(g)=-a\frac{dg}{da},
\end{equation}
and its solution is given by
\begin{equation}\label{eq:latRGEsoln}
  a\Lambda_{L}
   =\exp\Bigg(\int^{g}\frac{dg'}{\beta_{L}(g')}\Bigg) \\
   =f_{as}\big(g^2\big)
   \equiv f_{as}^0\big(g^2\big)\sum\limits_{i=0}^{\infty}
     q_i\,g^{2i},
\end{equation}
where $q_0=1$, the other $q_i$ are coefficients that can be, in 
principle, calculated perturbatively, and
\begin{equation}\label{eq:f0}
   f_{as}^0\big(g^2\big)\equiv\exp\Bigg(-\frac{1}{2b_{0}g^2}\Bigg)
    (b_{0}g^{2})^{-b_{1}/2b_{0}^{2}}.
\end{equation}
In fact from eq.~\eqref{eq:blatgs} and \eqref{eq:blat3loop}, one obtains
\begin{equation}\label{eq:q1value}
  q_1=\frac{b_{1}^{2}-b_{2}^{L}b_{0}}{2b_{0}^{3}}=
  \begin{cases}
     0.08324 & \text{for}\ \SU(2) \\
     0.18960 & \text{for}\ \SU(3).
  \end{cases}
\end{equation}
The integration constant $\Lambda_{L}$ has units of mass and is called the
{\it lattice $\Lambda$-parameter}.
From eq.~\eqref{eq:latRGEsoln} one sees that
\begin{equation}\label{eq:llat}
  \Lambda_L=\lim_{g\to0}\frac{1}{a}f_{as}^0\big(g^2\big).
\end{equation}

\begin{figure}
  \centering
  \includegraphics[width=0.9\linewidth]{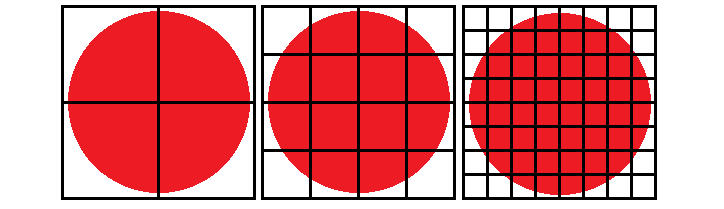}
  \caption{A schematic representation of the continuum limit. The
           red object represents some physical quantity. As the
           images progress to the right, the lattice spacing decreases
           relative to the physical length, and the bare coupling
           becomes weaker.}
  \label{fig:climit}
\end{figure}

The fact that pure $\SU(N_c)$ theory has a negative beta function
\eqref{eq:blatgs} has a profound physical implication. In particular 
when we invert eq.~\eqref{eq:latRGEsoln} keeping only universal
terms, we find
\begin{equation}\label{eq:rgesoln}
  g(a)^{-2}=b_0\log\left(a^{-2}\Lambda_L^{-2}\right) 
            +\frac{b_1}{b_0}\log\log\left(a^{-2}\Lambda_L^{-2}\right)
            +\mathcal{O}\left(1/\log\left(a^2\Lambda_L^2\right)\right).
\end{equation}
Two consequences are that the coupling $g(a)$ is driven to zero
as $a$ approaches zero (UV cutoff), which is known as 
{\it asymptotic freedom}, while at low energies, $g(a)$ becomes too 
large for reliable perturbative analysis.

From eq.~\eqref{eq:latRGEsoln} we see that taking $g\to0$ drives $a\to0$.
However the limit $g\to0$ is not enough to ensure a well-defined continuum
limit. The physical size of the lattice is proportional to $a^4$, and hence
collapses to zero unless we also increase the number of sites. Therefore
we extrapolate to the continuum limit by calculating our observable
of interest at different values of the coupling constant, with the 
extensions $N_1$, $N_2$, $N_3$, and $N_4$ chosen so that the physical 
size of the lattice is large enough for a reliable calculation of 
the observable of interest. A schematic representation is shown in
Figure~\ref{fig:climit}. We note that two kinds of systematic uncertainty 
arise in this context. Namely, to what extent do finite lattice
spacing (which limits the smallest wavelength) and finite lattice
size (which limits the largest wavelength) affect our results?
These questions are discussed in detail in 
Section~\ref{sec:refscales}.

\subsection{Finite temperature}\label{sec:finitetemp}
We now restrict
our attention to lattices that have extension $N_1=N_2=N_3\equiv N_s$
and $N_4\equiv N_\tau$.
Expectation values of physical observables $X$ are given in 4D,
Euclidean, pure $\SU(2)$ LGT at zero temperature by
\begin{equation}\label{eq:0Tintegral}
  \ev{X}=\frac{1}{Z}\int\mathcal{D}U\,e^{-S(U)}X(U),
\end{equation}
where the action is related to the Lagrangian by
\begin{equation}
  S=\int d^4x\,\Lagr,
\end{equation}
$Z$ is the {\it partition function}
\begin{equation}
  Z\equiv\int\mathcal{D}U\,e^{-S(U)},
\end{equation}
and the integration measure, called the {\it Haar} or {\it Hurwitz} measure, is
\begin{equation}
  \int\mathcal{D}U\equiv\int\prod_{x,\mu}dU_\mu(x).
\end{equation}
The quantities $X$ and $S$ appearing in the integral~\eqref{eq:0Tintegral}
are functionals of the configuration $U$, and this integral is
called a {\it functional integral}. The Haar measure is a product of
measures, one measure per link, each running over all possible values of
the link; in other words, the Haar measure runs over all possible
configurations. The functional integral is therefore 
a weighted average of the observable $X$ over all possible configurations,
each configuration receiving a weighting factor $\int\mathcal{D}U\,e^{-S}/Z$. 

The functional integral for a 3D, 
pure $\SU(2)$ LGT system in contact with a thermal reservoir at 
temperature $T$ has the same structure, except that the corresponding 
action is
\begin{equation}\label{eq:Tintegral}
  S(T)=\int_0^{1/T}dx_4\int d^3x\,\Lagr,
\end{equation}
and the Haar measure runs over fields that are periodic in the
$x_4$ direction. Because the functional integral for both systems is
formally the same, we interpret a 4D system with $N_s\gg N_\tau$ as
a 3D system at finite temperature, with $x_4$ running along a 
temperature direction rather than a time direction. 
The continuum limit of the finite temperature system corresponds
to $a\to0$ with $aN_s$ and $aN_\tau$ fixed. The {\it physical
temperature} is seen to be
\begin{equation}
  T=\frac{1}{aN_\tau}.
\end{equation}

\section{Reference scales}\label{sec:refscales}

Lattice computations deliver dimensionless quantities $L=\ell/a$, where 
$\ell$ is some physical length. The requirement that the theory has a 
well-defined continuum limit means that for two length scales 
$\ell_i$ and $\ell_j$
\begin{equation}
  r_{ij}\equiv\frac{\ell_i}{\ell_j}=\lim_{a\to 0}\frac{L_i}{L_j}
        \equiv\lim_{a\to 0}R_{ij},
\end{equation}
i.e. in the continuum limit, length ratios attain their
physical values.
Continuum limit extrapolations of a particular length $\ell_i$ therefore
depend on how one determines $R_{ij}$ and on the choice of the
{\it reference scale} or {\it reference length} $\ell_j$. 
Choosing a reference scale to use for continuum limit extrapolation
is called {\it scale setting}, and commonly one says
``we set the scale with $\ell_j$."

Calculation of the constants $R_{ij}$ is prone to nontrivial statistical and
systematic errors because they come from MCMC simulations performed on 
finite lattices with nonzero spacing. Therefore it is desirable to set
the scale with a quantity that is computable with low numerical effort,
has small systematic uncertainties, and good statistical precision.
Controlling systematic error is
discussed in Section~\ref{sec:sysa}, while the
discussion of statistical error is postponed to Chapter~\ref{ch:MCMC}. We
begin by introducing some reference scales. 

\subsection{Defining reference scales}\label{sec:definescales}

One choice of scale in this project is the {\it deconfining phase 
transition temperature}
\begin{equation}
  T_c=\frac{1}{a(\beta_c)\,N_\tau}.
\end{equation}
For $T<T_c$ gluons are bound into  glueballs, while 
at higher temperatures $T>T_c$ they exist in a gluon plasma. 
The deconfining phase transition is a second-order phase transition for 
$\SU(2)$ (see Engels et al.~\cite{engels_critical_1996} and
references therein) and a first-order transition for $\SU(N_c)$ when $N_c>2$. 
The order parameter for this transition is the {\it Polyakov loop}, 
\begin{equation}
  P(\vec{x})=\tr\prod_\tau U_4(\vec{x},\tau),
\end{equation}
which is a straight 
Wilson loop of length $N_\tau$ that is parallel to the Euclidean time 
axis and closes due to the periodic BCs. In practice, 
we determine $\beta_c$ by looking at plots of the Polyakov loop 
susceptibility,
\begin{equation}
  \chi=\ev{|P|^2}-\ev{|P|}^2, \qquad\qquad 
    P\equiv\sum\limits_{\vec{x}}P(\vec{x}),
\end{equation}
as a function of $\beta$ and estimating (in the infinite volume limit)
where it diverges. Numerical estimates
of $T_c$ are prone to systematic error because the simulations are
performed at finite lattice size while $T$ is only sharp in the infinite
volume limit. It is therefore necessary to extrapolate, for fixed $N_{\tau}$,
the dependence of $\beta_c(N_\tau)$ on the spatial size $N_s$ to the
infinite volume limit $N_s\to\infty$. Inverting $\beta_c(N_\tau)$ gives
our first length scale $N_\tau(\beta)$, which we call the
{\it deconfinement scale}.

A reference scale due to L\"uscher 
\cite{luscher_properties_2010} involves using 
the {\it gradient flow}. We begin by introducing a
fictitious {\it flow time} $t$ and evolve the system
according to the evolution equation
\begin{equation}\label{eq:Wflow}
  \dot{V}_\mu(x,t)=-g^2V_\mu(x,t)\,\partial_{x,\,\mu}S[V(t)]
\end{equation} 
with initial condition
\begin{equation}
  V_\mu(x,0)=U_\mu(x).
\end{equation}
In the above, the $\SU(N_c)$ link derivatives are defined by
\begin{equation}
  \partial_{x,\,\mu}f(V)\equiv
     i\sum_aT^a\frac{d}{ds}f\left(e^{itX^a}V\right)\Big|_{t=0},
    \qquad
    X^a(x',\mu')\equiv
    \begin{cases}
       T^a & \text{if} (x',\mu')=(x,\mu)\\
       0   & \text{otherwise.}
    \end{cases}
\end{equation}
L\"uscher showed that the gradient flow averages the gauge field
$A_\mu$ over a sphere with mean-square radius $\sqrt{8t}$ in 4D. Hence
$t$ has dimension length squared, and $\sqrt{8t}$ is interpreted
as the {\it smoothing range} of the flow.
From eq.~\eqref{eq:Wflow} we see that the gradient flow lowers the action. 
For pure $\SU(2)$ the link derivative of the action takes the simple form
\begin{equation}\label{eq:SU2gflow}
  g^2\partial_{x,\mu}S(V)=\frac{1}{2}
          \left(V_\mu^\Box(x)-V_\mu^\Box(x)^\dagger\right).
\end{equation}

After choosing an energy density discretization $E$ (for example
one might use the Wilson action) a scale is defined by choosing an
appropriate, fixed, dimensionless {\it target value} $y$ and integrating
the gradient flow equation until 
\begin{equation}\label{eq:tardef}
  y=t^2 E(t).
\end{equation}
As a function of $\beta$, a {\it gradient scale}
\begin{equation}\label{eq:s}
  s(\beta)=\sqrt{t(\beta)}
\end{equation}
scales like a length, provided that
\begin{enumerate}
  \item lattice sizes are chosen so that $N_{\min}\gg \sqrt{8t}$,
        where $N_{\min}=\min\,N_i$
        for simulations on an $N_1N_2N_3N_4$ lattice;
  \item the target values are large enough so that $\sqrt{8t}\gg 1$ 
        for the smallest used flow time; and
  \item the values of $\beta$ are large enough to be in the
        $\SU(2)$ scaling region.
\end{enumerate}
In contrast to the deconfinement scale, the computation of 
a gradient scale does not require fits or extrapolations.
The only remaining ambiguity is how to choose a target value.

An alternative to the gradient flow that is similar and 
algorithmically simpler is known as {\it cooling}. 
Cooling was introduced as part of an investigation of
topological charge in the 2D O(3) sigma model \cite{berg_dislocations_1981}.
Bonati and D'Elia showed that using cooling as a smoothing technique produces
similar results for topological observables as the gradient flow
for pure $\SU(3)$ LGT \cite{bonati_comparison_2014}.
In pure SU(2) a standard cooling step is
\begin{equation}\label{eq:cool}
  V_\mu(x,n_c)=\frac{V^\sqcup_\mu(x,n_c-1)}
                    {\sqrt{\det V_\mu^\sqcup(x,n_c-1)}},
\end{equation}
where $n_c$ is the number of cooling steps. 
The update \eqref{eq:cool} minimizes the local contribution to the
action, so that the ``cooling flow" decreases the action. Like with
the gradient flow, one picks a target value and iterates 
eq.~\eqref{eq:cool} until
\begin{equation}
  y=t_c^2 E(t_c),
\end{equation}
and a {\it cooling scale} is given by
\begin{equation}\label{eq:u}
  u(\beta)=\sqrt{t_c(\beta)}.
\end{equation}

\subsection{Continuum limit extrapolation and finite size scaling}
\label{sec:sysa}

One desires to know the ratio $r_{ij}$ of two scales in
the continuum limit. In principle this could be estimated by simulating
very near to the continuum limit, where $a\ll1$. 
The continuum limit of LGT is defined in the vicinity of a second order
phase transition in the bare coupling. Because the correlation length
diverges near critical points, subsequent configurations become more
correlated, and it requires more configurations to obtain effectively
independent data. This is called {\it critical slowing down}. 
In practice, one therefore calculates
$R_{ij}$ at multiple $\beta$ (hence multiple $a$) and extrapolates
the continuum limit result based on these data. We now discuss
two possible fitting forms for continuum limit extrapolation. 

Using the Wilson action, ratios of observables that have
units of length are known to scale as 
\begin{equation}\label{eq:stdscaling}
  R_{ij}\equiv\frac{L_i}{L_j}=\frac{\ell_i}{\ell_j}\Big(1+
                     \mathcal{O}\big(a^2\Lambda_L^2\big)\Big).
\end{equation}
In the continuum limit, ratios
of lengths approach their continuum limit values.
Sometimes corrections depending on $a$, such as in the equation above, are
referred to as {\it lattice artifacts}. In general the approach to the
continuum limit is thought to have lattice artifacts of power $p$
(RG considerations show that these $a^p$ artifacts are modified by
powers of logarithms~\cite{montvay_quantum_1994}) where $p$ 
depends on the lattice discretization.
The Wilson action in particular has $p=2$.
Equation~\eqref{eq:stdscaling} suggests a two-parameter fit of the form
\begin{equation}\label{eq:stdscalingfit}
  R_{ij}
               =r_{ij}+c_{ij}\,\left(\frac{1}{L_j}\right)^2,
\end{equation}
where $r_{ij}$ and $c_{ij}$ are the 
fit parameters. We will refer to this behavior as {\it standard scaling}. 

Another possibility for continuum limit extrapolation uses
the asymptotic scaling relation~\eqref{eq:latRGEsoln}
\begin{equation}\label{eq:asr}
  a\Lambda_L=f_{as}(\beta).
\end{equation}
We start by noting that the scale
$L_i$ calculated on the lattice is some function of the spacing,
so it can be expanded as a power series in $a$:
\begin{equation}
  L_i=\frac{c_i}{a\Lambda_L}\left(1+\sum\limits_{k=1}^{\infty}
                           \alpha_{i\,k}(a\Lambda_L)^k\right),
\end{equation}
where the $\alpha_{i\,k}$ are expansion coefficients.
Allton suggested using this equation to fit the approach to the continuum
limit~\cite{allton_lattice_1997}. Inserting eq.~\eqref{eq:asr}
into the above power series yields
\begin{equation}
  L_i=\frac{c_i}{f_{as}(\beta)}
              \left(1+\sum\limits_{k=1}^{\infty}
                    \alpha_{i\,k}\,f_{as}(\beta)^{\,k}\right).
\end{equation}
In practice $f_{as}$ is only known up to three loops, so we must
truncate it at some order $m$. Furthermore to have a finite number
of fit parameters, we must truncate the power series at some order
$n$. Hence, the approach of a length to the continuum limit can
be fit according to 
\begin{equation}\label{eq:lenasf}
  L_i=\frac{c^{mn}_i}{f^{m}_{as}(\beta)}
                    \left(1+\sum\limits_{k=1}^{n}
                    \alpha^{mn}_{i\,k}\,f^{m}_{as}(\beta)^{\,k}\right),
\end{equation}
where upper indices $m$ and $n$ are attached to quantities that
will change if $m$ or $n$ change. The fit parameters are
$c^{mn}_i$ and the $\alpha_{ik}^{mn}$. 

In general, asymptotic scaling would allow $\mathcal{O}(a)$ corrections. 
In order to ensure non-perturbative corrections are 
$\mathcal{O}\big(a^2\big)$, we improve on Allton by demanding that 
all scales have the
same $k=1$ term $\alpha^{mn}_{i,1}$; then terms of order $a$ cancel 
in the ratio. Using eq.~\eqref{eq:lenasf} along with this restriction, 
one obtains
\begin{equation}\label{eq:ascalingfit}
  R_{ij}=r_{ij}+\sum_{k=2}^n\kappa^{mn}_{i\,k}
                  \,f^m_{as}\left(L_j\right)^{\,k},
\end{equation}
where the fit parameters are now $r_{ij}$ and the $\kappa_{i\,k}^{mn}$.
One can switch the domain of $f_{as}^m$ from $\beta$ to the
reference $L_j$ using, for instance, eq.~\eqref{eq:lenasf}.
The continuum limit estimate $r_{ij}$ also depends on $m$ and $n$, but
we have suppressed these indices for clearer comparison with
the standard scaling fit~\eqref{eq:stdscalingfit}. We will refer
to the behavior of eq.~\eqref{eq:lenasf} or \eqref{eq:ascalingfit}
as {\it asymptotic scaling}.

If we carry out a naive continuum limit without changing the
extension of the lattice, its physical volume collapses to zero.
Ideally, calculations would be performed in the {\it thermodynamic limit},
where $N_s\to\infty$ and $N_\tau\to\infty$, and then take the limit
$a\to0$. In practice, the infinite volume observable is determined
by simulating at fixed $\beta$ on lattices of several sizes, then
extrapolating to the thermodynamic limit.
For some observables, the dependence on finite lattice size is known
from theory. For example the critical coupling constant $\beta_c(N_\tau)$
is known~\cite{engels_critical_1996} to depend on $N_s$ as
\begin{equation}
  \beta_c(N_\tau,N_s)=\beta_c(N_\tau)+a_1(N_\tau)N_s^{a_2(N_\tau)}.
\end{equation}
The $N_s=\infty$ result $\beta_c(N_\tau)$ can then
be extracted from a fit of the three parameters $\beta_c(N_\tau)$,
$a_1(N_\tau)$, and $a_2(N_\tau)$.

\section{Topological invariants}\label{sec:topinvar}

\subsection{Topological charge and instantons}\label{sec:topchargeandinstant}

This section follows Chapter 93 of Srednicki~\cite{srednicki_quantum_2007}; 
more details can be found there.
We start by considering classical, pure $\SU(2)$ gauge theory
\begin{equation}\label{eq:topaction}
  \Lagr=-\frac{1}{2g^2}F_{\mu\nu}F_{\mu\nu}
\end{equation}
at fixed $x_4$, focusing for the moment on $U$ that are time-independent.
Let $U\equiv U(\vec{x})\in\SU(2)$, and set the BC $U(\infty)=U_0$
for some constant matrix $U_0$. 
The {\it topological winding number} 
or {\it Pontryagin index} of the map $U$ is
\begin{equation}\label{eq:wn}
  n\equiv\frac{1}{24\pi^2}\int d^3x\,\epsilon_{ijk}
    \tr U\,\partial_iU^\dagger U\,\partial_jU^\dagger
        U\,\partial_kU^\dagger.
\end{equation}
The winding number is invariant under coordinate changes since the Jacobian of
the measure cancels the Jacobian of the partial derivatives.
Given the BC, it is also invariant under smooth deformations of $U$, 
which follows from integration by parts.

The quantity~\eqref{eq:wn} is called a winding number because it counts
the number of times the mapping $U$ ``winds around" or ``covers" 
the integration region. Let us see how this works in the present case.
The integration region is the 3D surface of space-time, which is
homeomorphic to the 3-sphere $S^3$. A point $\hat{x}\in S^3$
is specified by two polar angles $\chi$ and $\psi$ and an azimuthal
angle $\phi$ as
\begin{equation}
  \hat{x}=\colvec{4}{\s_\chi\s_\psi\co_\phi}{\s_\chi\s_\psi\s_\phi}
                    {\s_\chi\co_\psi}{\co_\chi}.
\end{equation}
Then the mapping $U:S^3\to\SU(2)$ given by
\begin{equation}\label{eq:uexplicit}
U(\hat{x})=\left(\begin{array}{cc}
             \co_\chi+i\s_\chi\co_\psi     & i\s_\chi\s_\psi e^{-im\phi}\\
             i\s_\chi\s_\psi e^{im\phi}& c_\chi-i\s_\chi\co_\psi 
            \end{array}\right),
\end{equation}
has winding number $m$. Intuitively, one can see this in the following manner:
Any $\SU(2)$ matrix can be written in terms of four real components as 
\begin{equation}
  U=a_4\id+i\vec{a}\cdot\vec{\sigma},
\end{equation}
where $a_\mu a_\mu=1$. The vector corresponding to the 
map~\eqref{eq:uexplicit} is
\begin{equation}
  \hat{a}=\colvec{4}{\s_\chi\s_\psi\co_{m\phi}}{\s_\chi\s_\psi\s_{m\phi}}
                    {\s_\chi\co_\psi}{\co_\chi}.
\end{equation}
We see that if we sweep through $\phi$, $\hat{x}$ sweeps over $S^3$ once
while $\hat{a}$ sweeps over $S^3$ $m$ times. Plugging the 
mapping~\eqref{eq:uexplicit} into eq.~\eqref{eq:wn} we find $n=m$,
confirming that the integral extracts the winding number.

In QFT, Noether's theorem tells us that to each continuous symmetry
of the Lagrangian there exists a corresponding conserved charge. 
Similarly we can identify a charge for each topological invariant 
of a system. Since $n$ is invariant under smooth deformations, it 
is a topological invariant, so it is sometimes referred to as a
{\it topological charge}, and represented by $Q$ instead of $n$.

Consider two maps $U$ and $U'$ that are gauge transformations of
zero and with different winding numbers. Since the winding number
is a topological invariant, the only way to deform $U$ to $U'$ is
to pass through configurations with $F_{\mu\nu}\neq0$; in other
words, there is an energy barrier between $U$ and $U'$. 
The corresponding quantum theory therefore has degenerate
vacuum states characterized by their winding numbers.

We will now discuss the topology of gauge field
configurations defined on all space-time. Let $r=(x_\mu x_\mu)^{1/2}$.
We require that
\begin{equation}
  A_\mu(x)\to U(x)\partial_\mu U^\dagger(x)
\end{equation}
as $r\to\infty$ to keep the action finite. (Infinite actions are
exponentially suppressed in the path integral.) The 3D integration
region will be the surface of space-time at infinity.
In addition to the BC $U(\infty)=U_0$, we specify $U$ at $x_4=-\infty$
to have winding number $n_-$ and $U$ at $x_4=+\infty$ to have
winding number $n_+$. The entire boundary is homeomorphic to
$S^3$, and the winding number of $U$ is
\begin{equation}
  Q\equiv n_+-n_-, 
\end{equation}
where the relative minus sign is due to the surfaces at $x_4=\pm\infty$
having opposite orientation. By viewing the integrand of eq.~\eqref{eq:wn} 
as the surface integral over a 4D region, defining the 
{\it Chern-Simons current}
\begin{equation}
  J_\mu^{CS}\equiv 2\epsilon_{\mu\nu\rho\sigma}\tr
    \left(a_\nu F_{\rho\sigma} +\frac{2}{3}A_\nu A_\rho A_\sigma\right),
\end{equation}
and applying Gauss's theorem, one can identify the winding number as 
an integral over the four-divergence of $J_\mu^{CS}$. We find
\begin{equation}\label{eq:Q}
  Q=\frac{1}{16\pi^2}\int d^4x\tr\dual{F_{\mu\nu}}F_{\mu\nu}
   \equiv \int d^4x\,q,
\end{equation}
where
\begin{equation} 
\dual{F}_{\mu\nu}=\frac{1}{2}\epsilon_{\mu\nu\rho\sigma}F_{\rho\sigma}
\end{equation}
is the {\it dual} field strength tensor. The quantity
$q$ is called the {\it topological charge density}. 

With eq.~\eqref{eq:Q} we can find vacuum solutions to the Euclidean
field equations
\begin{equation}
  D_\mu F_{\mu\nu}=0.
\end{equation}
The trick is to construct a lower bound on the action. Then if we can
find a solution saturating the bound, it must solve the field equations,
since it minimizes the action. This is called a {\it Bogomolny bound}.
Using eq.~\eqref{eq:topaction}, we find
\begin{equation}\label{eq:bogo}
  S\geq 8\pi^2|Q|/g^2,
\end{equation}
which becomes saturated when
\begin{equation}\label{eq:efeq}
  \dual{F_{\mu\nu}}=(\text{sign}\;n)F_{\mu\nu}.
\end{equation}
We arrive at an explicit solution to the above equation using
the map~\eqref{eq:uexplicit} with $Q=1$ $(m=1)$. We make the ansatz
\begin{equation}
  A_\mu(x)=f(r)U(\hat{x})\partial_\mu U^\dagger(\hat{x})
\end{equation}
where $f(\infty)=1$ to match the BC, and $f(0)=0$ so that $A_\mu$
is well-defined at the origin. Then this is a solution of
eq.~\eqref{eq:efeq} when 
\begin{equation}
  f(r)=\frac{r^2}{r^2+R^2}.
\end{equation}
This solution is called the {\it instanton}~\cite{belavin_pseudoparticle_1975}
and the integration constant $R$ is called the {\it instanton size}. 

The instanton mediates between vacuum configurations at Euclidean times
$\infty$ and $-\infty$ with winding numbers $n_+$ and $n_-$.
When $Q=-1$ we have an {\it anti-instanton}.
When $|Q|>1$, the mediating solution is constructed of multiple 
instantons or anti-instantons. When separations are large compared
to their sizes, we call this a {\it dilute gas} of instantons
or anti-instantons. From eq.~\eqref{eq:bogo}
we see that each instanton or anti-instanton contributes
$8\pi^2/g^2$ to the Bogomolny bound.

The {\it topological susceptibility} is defined as
\begin{equation}
  \chi_Q\equiv\int d^4x\,\ev{q(x)q(0)},
\end{equation}
where $q$ is the topological charge density of eq.~\eqref{eq:Q}.
The topological susceptibility gives evidence that the topological
structure of the underlying gauge fields has phenomenological significance.
In particular, by performing a calculation in the large $N_c$ limit, 
Witten and Veneziano~\cite{witten_current_1979,veneziano_u1_1979} showed
that at $N_c=\infty$ the $\eta'$ mass is related to 
the topological susceptibility through
\begin{equation}
  m_{\eta'}^2+m_\eta^2-2m_K=\frac{4N_f\chi_Q}{f_\pi^2},
\end{equation}
where $m_\eta$ is the $\eta$ mass, $m_K$ is the mass of the kaon, 
$N_f$ is the number of fermion flavors, and $f_\pi$ is the
pion decay constant. This mechanism can be used to explain the 
$\eta-\eta'$ mass difference. 
Plugging experimental values into the above
formula for $N_f=3$, one finds
\begin{equation}\label{eq:chivalue}
  \chi_Q\approx(180~\text{MeV})^4.
\end{equation}
While a conventional derivation of the Witten-Veneziano formula depends
on large $N_c$, lattice calculations for pure $\SU(2)$ and
pure $\SU(3)$ land relatively close to eq.~\eqref{eq:chivalue}.

\subsection{Topological charge on the lattice}

%It may seem counterintuitive that a well-defined notion of topology
%exists at all on the lattice. By taking an open cover of the
%lattice as the manifold (e.g., one can associate a hypercube to
%each site, keeping periodic BCs) one can define a topological charge on the 
%lattice; the non-trivial topology is contained in transition functions
%connecting cells of the open cover. For example 
%L\"uscher~\cite{luscher_topology_1982} showed
%the existence of a well-defined topological charge, which approaches 
%$Q$ in the continuum limit, for configurations with a small action
%density, i.e., with
%\begin{equation}
%  \max\,\tr\left(\id-U^\Box\right)<\epsilon
%\end{equation}
%for some small $\epsilon>0$. The maximum is taken over all plaquettes.
%Configurations not satisfying this inequality are called {\it exceptional}. 
%Since
%\begin{equation}
%  \ev{\tr\left(\id-U^\Box\right)}=\frac{3}{8}g^2+\mathcal{O}\big(g^4\big),
%\end{equation} 
%exceptional configurations become suppressed as the lattice spacing
%decreases. In order for a configuration of one topological charge to
%tunnel to another, it must pass through such an exceptional configuration;
%hence as $g$ decreases (as $\beta$ increases), it becomes increasingly 
%more difficult for a configuration to tunnel out of its topological sector. 
%This phenomenon is sometimes called {\it topological freezing}.
%Besides L\"uscher's paper, we recommend Kronfeld's
%review~\cite{kronfeld_topological_1988} for more detail.

Definitions of topological charge on the lattice can be found in reviews
such as the review by Kronfeld~\cite{kronfeld_topological_1988}.
For our definition of topological charge, we follow the example of 
eq.~\eqref{eq:Q} using the rule~\eqref{eq:inttosum}. It is reasonable to 
measure a topological charge on the lattice by
\begin{equation}\label{eq:QL}
  Q_L=a^4\sum_x q_L(x),
\end{equation}
where the sum is over all lattice sites and
\begin{equation}\label{eq:QLdensity}
  q_L(x) = -\frac{1}{2^9\pi^2}\sum\limits_{\mu\nu\rho\sigma=\pm 1}^{\pm 4}
         \tilde{\epsilon}_{\mu\nu\rho\sigma}
         \tr U^\Box_{\mu\nu}(x)U^\Box_{\rho\sigma}(x).
\end{equation}
Here $\tilde{\epsilon}=\epsilon$ for positive indices while
$\tilde{\epsilon}_{\mu\nu\rho\sigma}=
  -\tilde{\epsilon}_{(-\mu)\nu\rho\sigma}$ for negative indices.
The summation over backwards indices along with the definition of
$\tilde{\epsilon}$ ensures $q_L$ has negative parity.
The restriction of generated configurations to a subset with some fixed
topological charge is what we mean by {\it topological sector}.
The lattice expression for the topological susceptibility is
\begin{equation}
  \chi_L=a^4\sum_x\ev{q_L(x)q_L(0)}=\frac{1}{N^4}\ev{Q_L^2},
\end{equation}
where we have assumed a geometry $N\equiv N_1=N_2=N_3=N_4$
and utilized the translational invariance due to periodic BCs.

Lattice gauge theories typically experience
local fluctuations of the gauge fields, which are produced stochastically. 
These fluctuations blur the topological structure of the lattice, and 
must therefore be stripped away from the configuration before measuring 
$Q_L$. The signal is considerably improved by {\it smoothing}, where
one replaces each link by a local average of links; $Q_L$ is then
constructed on the smoothed field. 

Standard cooling minimizes the local contribution
to the action, which forces a gauge field to take a more typical
(smoother) value given its neighbors.
As mentioned earlier, the gradient flow averages the gauge field over 
a neighborhood, and therefore also has a smoothing effect. 
Ideally, these methods work because they make local modifications,
which therefore leave the global topological charge relatively intact.
A delicate issue with these smoothing algorithms is that they can destroy 
physical instantons; in fact after protracted cooling, a lattice will
eventually be brought to $Q_L=0$. This happens because certain
{\it exceptional configurations} or {\it dislocations} do not allow
for a well-defined topological charge. A lattice can then change
its topological charge by passing through these exceptional configurations.
In practice, one cools just enough
that topological observables become {\it quasi-stable}, i.e. just
enough that they do not change after many additional cooling sweeps.

%% file: chapter3.tex
\chapter{MCMC Simulations}\label{ch:MCMC}

As discussed in Section~\ref{sec:finitetemp}, expectation values of 
physical observables $X$ in pure $\SU(2)$ LGT
are given by functional integrals
\begin{equation}\label{eq:exobs}
  \ev{X}=\frac{1}{Z}\int\mathcal{D}U\,e^{-S(U)}X(U).
\end{equation}
Even though the integral~\eqref{eq:exobs} is well-defined on a lattice
because there are 
finitely many sites, it is not feasible to evaluate it numerically; even
relatively small lattices have $4\times10^4$ links. The goal of an MCMC
simulation is to estimate $\ev{X}$ by randomly generating configurations,
distributed with probability $e^{-S}$,
and on each configuration, making a measurement $X_i$. The average 
\begin{equation}\label{eq:arithmeticaverage}
  \bar{X}=\frac{1}{\nconf}\sum_{i=1}^{\nconf} X_i
\end{equation}
serves as the estimator.

In Section~\ref{sec:MCMCintro} we introduce MCMC simulations as they 
are applied to the project. Section~\ref{sec:statana} summarizes
some of the tools needed to statistically analyze the generated data; 
a more detailed presentation of probability and statistics is given in 
Appendix~\ref{ap:prb}. 
The final Section~\ref{sec:implement} provides 
details of how our simulation is implemented on the computer.
Further details can be found in, for instance, Berg~\cite{berg_markov_2004} 
and Gattringer and Lang~\cite{gattringer_quantum_2010}. 

\section{Markov chain Monte Carlo}\label{sec:MCMCintro}

To generate our configurations, we start from some arbitrary configuration
$C_0$ and construct a stochastic sequence of configurations. 
Configuration $C_i$ is generated based on
configuration $C_{i-1}$, which we call an {\it update} or {\it Monte Carlo
step}. The result is a {\it Markov chain}
\begin{equation}
  C_0\to C_1\to C_2\to...
\end{equation}
of configurations. 

{\it Markov chain Monte Carlo} (MCMC) is characterized by the probability 
$W^{CC'}\equiv\pr{C'|C}$, the probability to jump to configuration 
$C'$ given that the system started in configuration $C$.
The MCMC {\it transition matrix}
\begin{equation}
  W\equiv\Big(W^{CC'}\Big)
\end{equation}
is constructed to bring the system to {\it equilibrium}.
In equilibrium, the chain should have no sinks or sources of probability,
which means that the probability of jumping into a configuration $C'$
should be the same as jumping out of $C'$. This property
is called {\it balance}
\begin{equation}\label{eq:balance}
    \sum\limits_CW^{CC'}\pr{C}=
    \sum\limits_CW^{C'C}\pr{C'},
\end{equation}
with the LHS representing the total probability to end up in $C'$ and
the RHS representing the probability to transition out of $C'$.
If $W$ satisfies
\begin{enumerate}
  \item {\it ergodicity}, i.e.
        \begin{equation}\label{eq:ergodicity}
          \pr{C}>0\;\;\text{and}\;\;\pr{C'}>0\;\;\Rightarrow\;\;
          \exists\;n\in\mathbb{N}\;\;\text{s.t.}\;\;\big(W^n\big)^{CC'}>0;
        \end{equation}
  \item {\it normalization}, i.e.
        \begin{equation}
          \sum\limits_{C'}W^{CC'}=1;
        \end{equation}
  \item and balance, 
\end{enumerate}
then the Markov process is guaranteed to bring the ensemble toward 
equilibrium. Using normalization, one finds from eq.~\eqref{eq:balance}
\begin{equation}
    \sum\limits_CW^{CC'}\pr{C}=\pr{C'},
\end{equation}
which shows that the equilibrium distribution is a fixed point of
the Markov chain. The first property, ergodicity, guarantees that it
is possible to transition from $C$ to $C'$ in a finite number of steps.
In realistic simulations, it is important that the $n$ appearing in
eq.~\eqref{eq:ergodicity} is not too large. For example 
the Markov chain may have difficulty connecting different topological sectors
in configuration space.

\subsection{Update: Metropolis and heat bath}
In this and the following subsection,
we omit the Lorentz index and space-time point from link variables to
avoid clutter.  We use $U$ to indicate the link to be updated, 
$U^\sqcup$ to indicate the staple matrix attached to $U$, and
$U'$ to indicate a trial link. We will use the Boltzmann
distribution $\pr{C}\propto e^{-S_C}$.

One trivial way to satisfy the balance condition~\eqref{eq:balance} is
to find an update that satisfies it term-by-term. For such an update, 
\begin{equation}\label{eq:detailedbalance}
    W^{CC'}\pr{C}=W^{C'C}\pr{C'}.
\end{equation}
This property is known as {\it detailed balance}.
One of the most well-known Monte Carlo updates satisfying detailed
balance is the {\it Metropolis algorithm}~\cite{metropolis_equation_1953}. 
In the Metropolis algorithm, a trial configuration $C'$ is selected 
with some probability distribution $\prt{C'|C}$. Then $C'$
is accepted with likelihood
\begin{equation}\label{eq:metupdate}
  \pr{C\to C'}=\min\left[1,\frac{\prt{C|C'}e^{-S_{C'}}}
    {\prt{C'|C}e^{-S_C}}\right],
\end{equation}
where $S_C$ is the action corresponding to $C$. 
If $C'$ is rejected, the unchanged configuration is counted
in the Markov chain. Using the fact that the total probability
to transition from $C$ to $C'$ is $W^{CC'}=\prt{C'|C}\pr{C\to C'}$, one
can show that this update satisfies detailed balance.

Another update is the {\it heat bath} (HB). In our
simulations, a new configuration is generated from an old one by updating one
link. For the $\SU(2)$ HB algorithm, the trial link distribution is
\begin{equation}\label{eq:PTdist}
  d\prt{U'}\propto dU'\exp\left(\frac{\beta}{2}\,\tr\,
  U'U^\sqcup\right)
\end{equation}
and the transition probability is
\begin{equation}
  \pr{C\to C'}=\min\left[1,e^{-(S_{C'}-S_C)}\right].
\end{equation}
This construction also satisfies detailed balance. The new configuration 
is automatically accepted whenever it lowers the action, and
increases in the action are exponentially suppressed.
HB updates ensure local equilibrium, but they often 
take more CPU time. For $\SU(2)$ the guarantee of local equilibrium
turns out to be more impactful, so heat bath updates are more efficient
than general Metropolis updates.

Single link Metropolis or HB updates of links carried out in a systematic 
(as opposed to random) order fulfill balance, but do not
fulfill detailed balance.

\subsection{Update: Over-relaxation}

An additional useful update for $\SU(2)$ is the 
{\it over-relaxation} (OR) update. 
Adler introduced OR algorithms \cite{adler_over-relaxation_1981} and 
they were further developed by Creutz \cite{creutz_overrelaxation_1987} 
and others. The idea of the OR algorithm is to speed up relaxation 
by generating a group element ``far away" from $U$ without
destroying equilibrium, which is here achieved by keeping the action constant.

More precisely let $U\in\SU(N_c)$ and suppose we have some method of
choosing another link variable $U_0$ that maximizes
the action for this staple.  We assume that this method of selection has no
dependence on $U$.  Pick some element $V\in\SU(N_c)$ such that $U=VU_0$; viewed
in this way, $U$ is ``on one side of $U_0$," and the element 
``on the other side" is $U'=V^{-1}U_0$.  Note that
\begin{equation}
  V=U U_0^{-1},
\end{equation}
which implies
\begin{equation}
  U' = U_0 U^{-1} U_0.
\end{equation}
This manner of constructing a new link variable $U'$, which generates
a group element ``far away" from $U$ without changing the action, 
is what we mean by over-relaxation.

In principle an OR update should be more efficient than a Monte
Carlo update. This is because we chose the new link variable to be two
group elements away from the old one, thrusting us further
along configuration space. However unlike Metropolis updates, OR updates 
only sample the subspace of constant action, and are therefore not ergodic. 
Hence to ensure an approach to equilibrium, they must be supplemented with, 
for instance, HB updates.

We implement the $\SU(2)$ OR update by
\begin{equation}
  U\to U'=\frac{1}{\det U^\sqcup}\left(U^\sqcup UU^\sqcup\right)^\dagger.
\end{equation}
It is easily seen that this update does not change the $\SU(2)$ Wilson 
action, which means the proposal is always accepted. This simple 
behavior is special to $\U(1)$ and $\SU(2)$ LGT. Its usefulness is 
extended to $\SU(N_c)$ when $N_c>2$ via the method of Cabibbo and Marinari
\cite{cabibbo_new_1982}.

\section{Statistical analysis}\label{sec:statana}

Since $C_i$ is generated based on $C_{i-1}$, measurements on subsequent
configurations are correlated. In our simulations, these
correlations are reduced in two ways:
\begin{enumerate}
  \item Subsequent configurations are separated by multiple updating sweeps;
        and then
  \item configurations are grouped into $\nconf$ {\it blocks} or {\it bins}.
\end{enumerate}
The final measurements $X_i$ used in data analysis are obtained by averaging
within each block.
To check whether the final data are effectively independent, one can use
the {\it integrated autocorrelation time}. 
For statistically independent measurements, we expect the variance 
$\sigma^2_{\bar{X}}$ of $\bar{X}$ to be
\begin{equation}
  \sigma^2_{\bar{X}}=\frac{\sigma^2}{\nconf}
\end{equation}
due to the Central Limit Theorem. In practice, however, one finds 
\begin{equation}
  \sigma^2_{\bar{X}}=\frac{\sigma^2}{\nconf}\tauint.
\end{equation}
The factor $\tauint$ is the integrated autocorrelation time. It is the 
ratio between the estimated variance of the sample
mean and what this variance would have been if the data were independent.
For effectively independent data, $\tauint=1$.

So, the final measurements are drawn from some
distribution with mean $\ev{X}$ and variance $\sigma^2$
and are effectively independent. The
estimator $\bar{X}$ of the mean is the average~\eqref{eq:arithmeticaverage}, 
while the unbiased estimator $\bar{\sigma}^2$ of the variance is
\begin{equation}\label{eq:estimatorvariance}
  \bar{\sigma}^2=\frac{1}{\nconf-1}\sum_{i=1}^{\nconf}
      \left(X_i-\bar{X}\right)^2.
\end{equation}
An estimator is {\it biased} if its mean for finite $\nconf$ 
does not agree with the exact result;
the {\it bias} is the difference. Generally, problems with bias emerge whenever
one wishes to estimate some non-linear function $f$ of the mean $\ev{X}$.
Naively one might guess
\begin{equation}
  \bar{f}_\text{bad}=\frac{1}{\nconf}\sum_{i=1}^{\nconf}f(X_i)
\end{equation}
as an estimator; however it can be shown that the bias of 
$\bar{f}_\text{bad}$ is $\mathcal{O}(1)$,
i.e. it never converges to the exact result.
An estimator for $f(\ev{X})$ that converges to its true value is
\begin{equation}
  \bar{f}=f(\bar{X});
\end{equation}
in particular, the bias of this estimator is $\mathcal{O}(1/\nconf)$.
Therefore in the large $\nconf$ limit, the bias vanishes faster than the 
statistical error bar.

We have introduced a way to estimate the mean and variance of some
operator, as well as a way to estimate the mean of some function of
that operator. Now we need a way to estimate the error bar of that
function. We cannot use
\begin{equation}
  \bar{\sigma}^2_{\bar{f}}=\frac{\bar{\sigma}^2_{\bar{f}}}{\nconf}
    =\frac{1}{\nconf\,(\nconf-1)}\sum_{i=1}^{\nconf}
     \left(f(X_i)-\bar{f}\right)^2
\end{equation}
because $f(X_i)$ is not a valid sample point.
%If the $X_i$ are independent and fluctuations of the $X_i$
%are close enough to the mean that $f$ is approximately linear, 
One could analytically produce an error bar for $\bar{f}$ 
using error propagation. However when the function is complicated, 
error propagation becomes extremely unwieldy.

{\it Jackknifing} allows one to extract a mean and error bar,
and it is straightforward to implement; 
therefore it makes sense to use the jackknife method generally.
The idea of jackknifing is to throw away the first measurement, 
leaving $\nconf-1$ resampled values. Then we resample again, this
time throwing out the second point, and so on. The resulting
{\it jackknife bins} are
\begin{equation}
  X_{J,i}=\frac{1}{\nconf-1}\sum_{j\neq i}X_j.
\end{equation}
The jackknife estimator for $f(\ev{x})$ is then
\begin{equation}
  \bar{f}_J=\frac{1}{\nconf}\sum_{i=1}^{\nconf}f(X_{J,i}),
\end{equation}
while the estimator for the variance of $\bar{f}_J$ is
\begin{equation}
  \bar{\sigma}^2_{f_J}=\frac{\nconf-1}{\nconf}\sum_{i=1}^{\nconf}
    \left(f(X_{J,i})-\bar{f}_J\right)^2.
\end{equation}

In many instances, we will need to compare two estimates 
of the same quantity against each other and decide whether the difference 
between them is significant. This can happen, for example, if we want to
compare another group's results with our own. Let their result
be $\bar{X}$ with uncertainty $\sigma_{\bar{X}}$ and ours be
$\bar{Y}$ with uncertainty $\sigma_{\bar{Y}}$. Then the probability
that these two estimates differ by at least $D$ is
\begin{equation}
  q=\pr{|\bar{X}-\bar{Y}|>D}
    =1-\erf\left(\frac{D}{\sqrt{2
     \left(\sigma_{\bar{X}}^2+\sigma_{\bar{Y}}^2\right)}}\right)
\end{equation}
assuming $\bar{X}$ and $\bar{Y}$ are normally distributed with
the same mean. This is called a {\it Gaussian difference test}. 
The quantity $q$ is called the {\it q-value}.
In practice we take $q\leq0.05$ to be an indication of a possible 
discrepancy between $\bar{X}$ and $\bar{Y}$, keeping in mind that 
$q\leq0.05$ by chance one out of twenty times.

In practice, the true variances $\sigma_{\bar{X}}$ and $\sigma_{\bar{Y}}$
are not known. If one wishes to use the estimators $\bar{\sigma}_{\bar{X}}$
and $\bar{\sigma}_{\bar{Y}}$ instead, one can perform a {\it Student
difference test} or {\it t-test} to investigate whether the discrepancy 
$D$ is due to chance. Suppose the estimate $\bar{X}$ comes from 
$M_{\text{conf}}$ data, while $\bar{Y}$ comes from $\nconf$ data.
Assume $\sigma_{\bar{X}}=\sigma_{\bar{Y}}$, which
happens when the sampling methods used are identical.
We introduce the random variable
\begin{equation}
  t=\frac{D}{\bar{\sigma}_D},
\end{equation}
where $D=\bar{X}-\bar{Y}$, and
\begin{equation}
  \bar{\sigma}^2_D=\left(\frac{1}{M_{\text{conf}}}+\frac{1}{\nconf}\right)
                   \frac{(M_{\text{conf}}-1)\,\bar{\sigma}_{\bar{X}}^2
                    +(\nconf-1)\,\bar{\sigma}_{\bar{Y}}^2}
                   {M_{\text{conf}}+\nconf-2}.
\end{equation}
Then the probability that these two estimates differ by at least $D$ is
\begin{equation}
 q=2
 \begin{cases}
 \,I\left(z,\frac{\nu}{2},\frac{1}{2}\right) & \text{for }t\leq 0, \\
 \,1-\frac{1}{2}\,I\left(z,\frac{\nu}{2},\frac{1}{2}\right) & \text{otherwise},
 \end{cases}
\end{equation}
where $I$ is the incomplete beta function, $\nu=M_{\text{conf}}+\nconf-2$,
and
\begin{equation}
  z=\frac{\nu}{\nu+t^2}.
\end{equation}

To estimate finite size corrections and carry out continuum limit
extrapolations, we need a way to fit data to curves.
Consider a sample of $\nsimil$ Gaussian, independent data points $(X_i,Y_i)$,
where the $Y_i$ have standard deviations $\sigma_i$ and the $X_i$ have
no errors. For instance, if one is interested in a continuum limit
extrapolation, the $X_i$ are $\beta$ values while the $Y_i$ are
ratios of scales evaluated at that $\beta$.
We model these data with a fit that depends on some set
of $M$ parameters
\begin{equation}
  y=y(x;a),
\end{equation}
where $a=(a_1,...,a_M)$ is the vector of these parameters. Our goal
is to estimate the $a_j$.
Assuming that $y(x;a)$ is the exact law for the data, the probability
distribution for the measurements $Y_i$ is 
\begin{equation}
  f(y_1,...,y_{\nsimil})=\prod_{i=1}^{\nsimil}\frac{1}{\sqrt{2\pi}\sigma_i}
      \exp\left[\frac{-(y_i-y(x_i;a))^2}{2\sigma_i^2}\right].
\end{equation}
The probability that the data fall within a region near what was observed is
\begin{equation}
  \text{P}=\prod_{i=1}^{\nsimil}\frac{1}{\sqrt{2\pi}\sigma_i}
      \exp\left[\frac{-(y_i-y(x_i;a))^2}{2\sigma_i^2}\right]dy_i.
\end{equation}
Our strategy for determining the correct fit will be to find the vector $a$
that maximizes the above probability. This happens when
\begin{equation}
  \chi^2(a)\equiv\sum_{i=1}^{\nsimil}\frac{(y_i-y(x_i;a))^2}{2\sigma_i^2}
\end{equation}
is minimized. This strategy is an example of a {\it maximum likelihood method}.

We now describe an iterative method to search for the minimum of $\chi^2$.
Let $a_n$ be the vector of parameters for the $n^{\text{th}}$ iteration.
As long as $a$ is in a small enough neighborhood of $a_n$, we can safely
approximate
\begin{equation}\label{eq:NRapprox}
  \chi^2(a)\approx\chi^2(a_n)+(a-a_n)\cdot b
           +\frac{1}{2}(a-a_n)\,A\,(a-a_n),
\end{equation}
where the coefficients of the vector $b$ and the $M\times M$ matrix $A$ 
are given by the first and second derivatives of $\chi^2$ evaluated at $a_n$.
In the {\it Newton-Raphson method}, the next iteration $a_{n+1}$ is
determined from the condition $\nabla\chi^2(a)|_{a=a_{n+1}}=0$,
which yields
\begin{equation}\label{eq:NR}
  a_{n+1}=a_n-A^{-1}b.
\end{equation}
If the approximation~\eqref{eq:NRapprox} is not good, one can instead move
a small step in the direction of the gradient by
\begin{equation}\label{eq:SD}
  a_{n+1}=a_n-c\,b,
\end{equation}
where $c$ is a constant that is small enough not to overshoot direction
of steepest descent. This is an example of a {\it steepest descent method}.
The Levenberg-Marquardt 
method~\cite{levenberg_method_1944,marquardt_algorithm_1963}, which is 
our method of choice, varies smoothly between~\eqref{eq:NR} and 
\eqref{eq:SD}. Steepest descent is used far from the minimum, and 
then it switches to the Newton-Raphson method when the minimum is approached.

\section{Computer implementation}\label{sec:implement}

Now that we have introduced the general idea of MCMC, along with some specific
updating schemes, and complications for statistical analysis, we are ready 
to discuss the computer implementation. 

As mentioned earlier, we design the simulation using {\it local updates}, which
means we update the links one at a time. This is done in a systematic order,
because there is some computational advantage compared to updating in a
random order~\cite{berg_markov_2004}. 
An updating {\it sweep} updates every link on the lattice once. 
To maximize efficiency while maintaining ergodicity, our updating sweeps 
have a combination of HB and OR updating. We call this a 
{\it Monte Carlo Over-relaxation} (MCOR) sweep.

An MCMC simulation of LGT broadly consists of three essential steps:
\begin{enumerate}
  \item {\it Initialization}: The first thing to do is get everything ready for
        the simulation. This includes initializing the random number generator,
        and setting up an initial configuration.
  \item {\it Equilibration}: To avoid over-sampling rare configurations, 
        one must perform many sweeps to bring the system to its equilibrium 
        distribution. The structure of this section looks like 
        \begin{verbatim}
        do from n=1 to n=nequi 
          call MCOR update
        end do
        \end{verbatim}
  \item {\it Measurements}: All observables of interest are measured on the
        equilibrated configurations. To help reduce correlations between
        measurements, multiple updating sweeps are performed in between.
        This section is structured as
        \begin{verbatim}
        do from n=1 to n=nmeasurements
          do from n=1 to n=ndiscarded
            call MCOR update
          end do
          take measurement
        end do
        \end{verbatim}
\end{enumerate}

For simulations like ours, it may take months (or years!) 
for a single-processor MCMC simulation to
generate enough data to get reasonable error bars. Therefore it is
advantageous to divide the lattice into smaller sublattices, updating
simultaneously on each sublattice, passing relevant information between the
sublattices whenever necessary. {\it Parallelizing} in this way offers a speed
up factor somewhat less than the number of sublattices used. A standard way to
parallelize code is to use the Message Passing Interface (MPI). MPI allows
for efficient exchange of information between processors and is easily included
in Fortran or C programs.

\begin{figure}
  \centering
  \includegraphics[width=0.45\textwidth,height=0.45\textheight,keepaspectratio]
                   {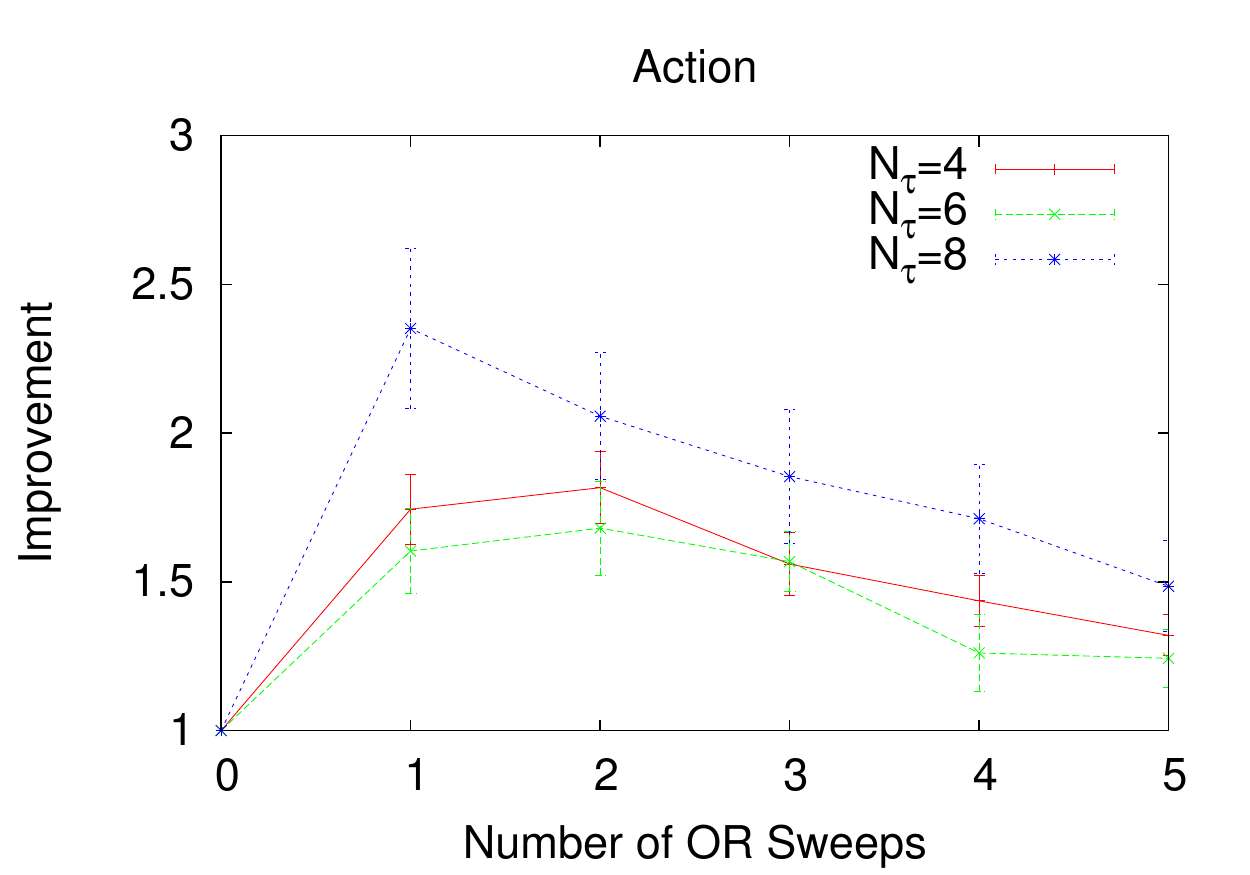}
  \includegraphics[width=0.45\textwidth,height=0.45\textheight,keepaspectratio]
                   {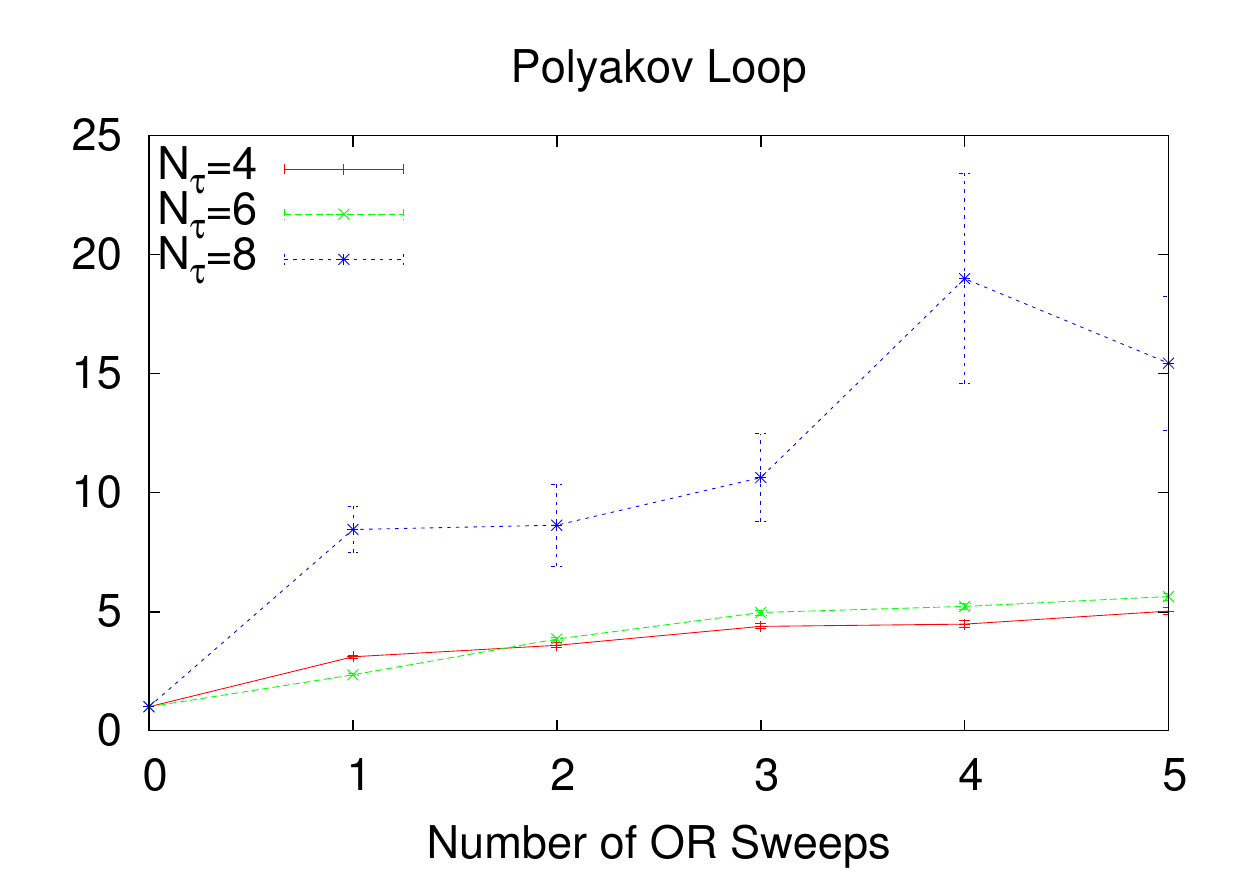}
  \caption{Left: Improvement factor for action as a function of the number
           of OR sweeps. Right: Improvement factor for Polyakov loops. 
           The error bars of the Polyakov loop are magnified by a factor of 
           10 to increase visibility.}
  \label{fig:ORdiag}
\end{figure}

One may wish to optimize the number of OR sweeps. 
To do this we looked at the action and Polyakov loops for
$8^3\times4$, $12^3\times6$, and $16^3\times8$ lattices and calculated
the improvement ratio
\begin{equation}\label{eq:impv}
  I=\frac{\tauint(0)}{\tauint(n)}\frac{t(0)}{t(n)},
\end{equation}
where $\tauint(n)$ and $t(n)$ are, respectively, the integrated 
autocorrelation time and CPU time for a simulation using one HB update 
and $n$ OR updates per sweep. Figure~\ref{fig:ORdiag} 
shows the improvements for the action (left) and Polyakov loops (right). 
The action improvement seems to peter out after the first OR sweep, 
while the Polyakov loop improvement increases up to at least four OR sweeps. 
Therefore using two OR sweeps is a good
compromise for these observables.

\begin{figure}
  \centering
  \includegraphics[width=0.6\textwidth]{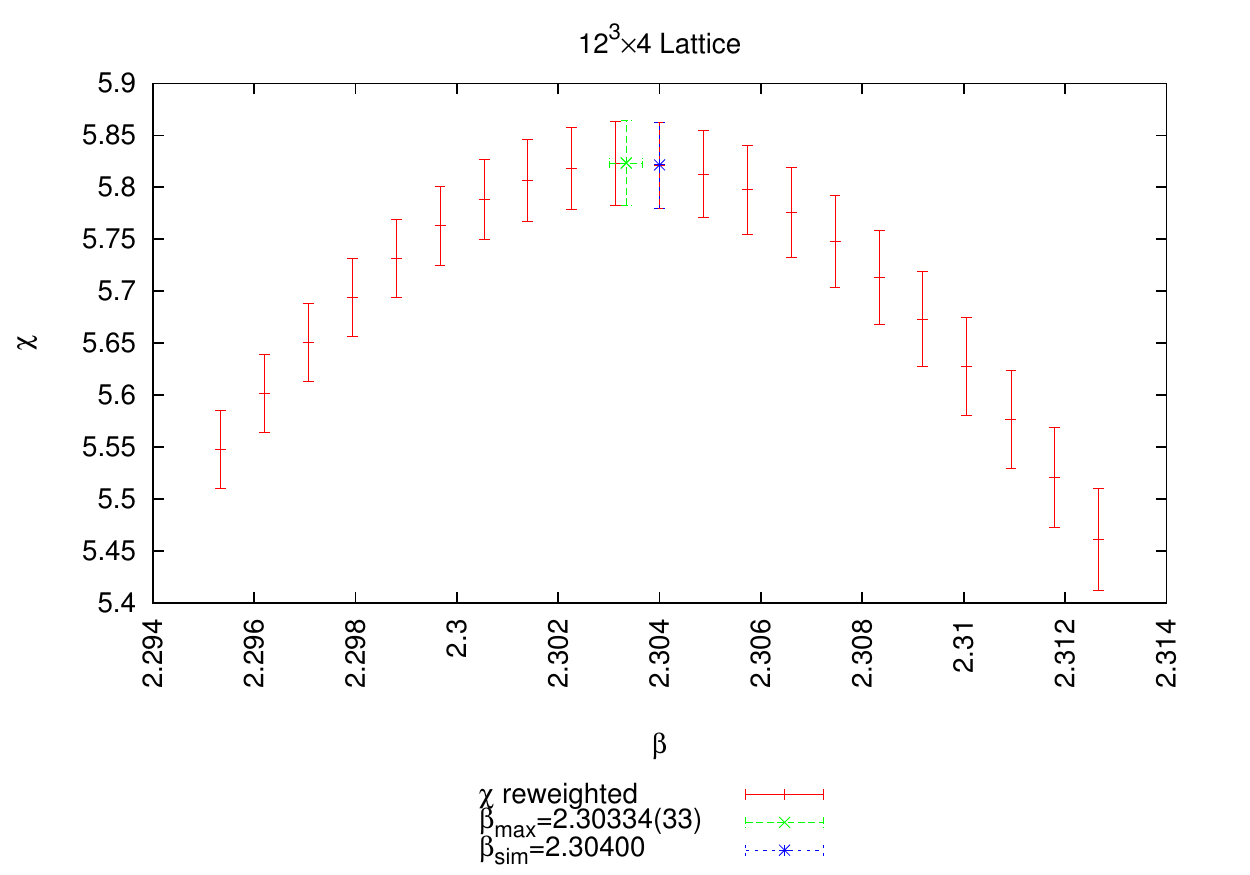}
  \caption{Example reweighting curve for the Polyakov loop susceptibility
           of a pure $\SU(2)$ $12^3\times4$ lattice. The blue line 
           indicates the simulation point $\beta$, while the red lines 
           indicate reweighted estimates calculated using eq.~\eqref{eq:RW}
           at various $\beta'$. The green point shows the
           estimate of the $\beta$ maximizing $\chi$, along
           with its error bar.}
  \label{fig:rwexample}
\end{figure}

The goal of some simulations is to determine phase transition points. 
Close to these points, on a finite lattice, the susceptibility of the 
relevant order parameter attains its maximum. The most straightforward 
strategy of estimating this maximum is to run multiple simulations 
in the vicinity of the transition point. Because this strategy 
requires multiple runs, it is inefficient. 
{\it Reweighting} (see~\cite{ferrenberg_new_1989} and
references therein) is an efficient alternative. Consider
the expectation value of an observable $X$ calculated at $\beta'$. We have
\begin{equation}\label{eq:RW}
\begin{aligned}
  \ev{X}_{\beta'}&=Z_{\beta'}^{-1}\int d\phi\,e^{-\beta'E(\phi)}X(\phi)
                  e^{(\beta-\beta)E(\phi)}\\
                 &=Z_{\beta'}^{-1}\int d\phi\,e^{(\beta-\beta')E(\phi)}
                  X(\phi)e^{-\beta E(\phi)}\\
                 &=Z_{\beta'}^{-1}Z_{\beta}\ev{e^{(\beta-\beta')E}X}_\beta\\
                 &=\ev{\frac{Z_\beta}{Z_{\beta'}}e^{(\beta-\beta')E}X}_\beta.
\end{aligned}
\end{equation}
We can calculate the expectation value in the last line
using data from a time series generated at $\beta$, and this gives us an
estimate for $\ev{X}_{\beta'}$. Reweighting is only useful when
$E\Delta\beta=\mathcal{O}(1)$. Provided that the critical parameter $\beta_c$
is sufficiently close to the simulation point $\beta$, it suffices
to have only one simulation, then estimate the maximum by reweighting to
multiple nearby $\beta'$. An example reweighting curve is shown in 
Figure~\ref{fig:rwexample}.

Our simulations were performed on the FSU HEP theory cluster, as well as
at the National Energy Research Scientific Computing Center (NERSC) 
using HEP and nuclear physics computing grants.
The FSU HEP cluster consists of 16 nodes, each with 4 Intel Core i7 CPU
processors, and each processor supports 2 threads. The HEP cluster is
well-suited for simulations of our smaller lattices, and we used it
extensively. However there is no MPI communication between nodes, 
so simulations can efficiently use at most 8 processes. It is desirable 
for larger lattices to use many more processes, and when this is necessary, 
we turn to NERSC. 
NERSC's supercomputer Cori lets us use up to 1,932 Intel Xeon Haswell
nodes with 32 cores each, allowing for up to 61,824 processes. Using 8,000
processors on Cori, we were able to simulate an $80^3\times8$ lattice with high
statistics in less than two days of real time. Summing over all simulations
we have run on NERSC, we have carried out 14.9 million raw machine
hours (about 1,700 years) of single-processor calculation.

%% file: chapter4.tex
\chapter{Comparison of Scaling Violations}\label{ch:comparison}

We investigate three types of reference scale: the deconfinement scale,
the gradient scale, and the cooling scale. The goals of this investigation
are to compare the computational efficiency of these scales, determine
whether they experience seriously distinct scaling behavior, and estimate
the systematic error accrued from the choice of fitting form for continuum
limit extrapolation. Altogether we examine thirteen scales: the deconfinement
scale, which we label $L_0$; six gradient scales $L_1-L_6$; and
six cooling scales $L_7-L_{12}$.

Our results are obtained by analyzing configurations generated
by MCMC simulation at NERSC and on the FSU HEP computer cluster. 
The statistics are reported in units of MCOR sweeps. 
One MCOR sweep updates each link in a 
systematic order using the Fabricius-Haan-Kennedy-Pendleton heat bath 
algorithm~\cite{fabricius_heat_1984,kennedy_improved_1985} then, in the
same order, twice by over-relaxation~\cite{adler_over-relaxation_1981}.
The lattice is checkerboard updated~\cite{barkai_can_1982} and,
using MPI Fortran, divided into sublattices that are updated in parallel. 
Lattice sizes are reported as $N_s^3\times N_\tau$. Statistical error
bars are reported in the last two digits of each measurement, in
parentheses.

This chapter covers our investigation of the continuum limit of
the aforementioned scales~\cite{berg_deconfinement_2017,berg_estimates_2018}.
In Section~\ref{sec:ntsc} we report our numerical results for the deconfinement
scale, which we used to guide our choice of target values for the gradient
and cooling scales. Sections~\ref{sec:gradsc} and \ref{sec:coolsc} give 
our results for six gradient scales and six cooling scales, respectively. 
Scaling and asymptotic scaling behavior of these altogether thirteen 
reference lengths are analyzed in Section~\ref{sec:scalingsc}. 
Our findings are summarized in Section~\ref{sec:summarysc}.

\section{Deconfinement length numerical results}\label{sec:ntsc}

To obtain results for the deconfinement length, we use lattices with 
$N_s\geq2N_\tau$ because temperature definitions are only sharp in
the $N_s\to\infty$ limit. The deconfinement length $L_0$ is extracted
using the following procedure:
\begin{enumerate}
\item Simulations are carried out at a coupling constant $\beta_{\text{sim}}$ 
      expected to be near the critical point, given the lattice size. 
\item The location of pseudo-critical coupling constants $\beta_c(N_\tau,N_s)$ 
      and their error bars are then estimated by reweighting the Polyakov
      loop susceptibility curve.
\item After repeating this process for multiple space-like sizes $N_s^3$,
      the critical coupling 
      \begin{equation}
        \beta_c(N_\tau)\equiv\beta_c(N_\tau,\infty)
      \end{equation}
      is extrapolated from the three-parameter fit
      \begin{equation}\label{eq:3paramtc}
        \beta_c(N_\tau,N_s)=\beta_c(N_\tau)+a_1(N_\tau)N_s^{a_2(N_\tau)}.
      \end{equation}
\item The deconfinement length for the coupling constant $\beta_c$ is 
      then $L_0(\beta_c)=N_\tau(\beta_c)$.
\end{enumerate}

Table~\ref{tab:bcvalues} collects our data for pseudo-critical coupling
constants for lattices with $N_\tau$ up to 12 and $N_s$ up to 80. The
statistics assembled ranges between $2^{18}-2^{23}$ MCOR sweeps,
with an exceptional $2^{25}$ MCOR sweeps for the $40^3\times12$ lattice.
The range in MCOR sweeps depended somewhat on what passed through the
NERSC scavenger queue.
To produce error bars, the time series is grouped into 32 or more
bins, we reweight in each bin, and the bins are then jackknifed.

\begin{table}
\caption{Pseudo-critical coupling constants $\beta_c(N_s,N_\tau)$.}
\begin{tabularx}{\linewidth}{LCCCCR} \hline\hline
$N_s$&$N_\tau=4$  &$N_\tau=6$  &$N_\tau=8$  &$N_\tau=10$&$N_\tau=12$\\
\hline
8    &2.30859(53) &            &            &           &\\
12   &2.30334(33) &2.43900(33) &            &           &\\
16   &2.30161(30) &            &2.52960(90) &           &\\
18   &            &2.43096(43) &            &           &\\
20   &2.30085(17) &2.42973(11) &            &2.59961(52)&\\
24   &2.30060(16) &2.42873(35) &2.51678(43) &2.58909(49)&2.66317(91)\\
28   &2.30025(19) &2.427939(74)&            &2.58497(26)&\\
30   &            &2.427690(87)&            &           &\\
32   &2.299754(99)&            &2.51296(20) &2.58270(27)&2.64450(39)\\
36   &            &2.427274(67)&            &2.58117(13)&2.64223(33)\\
40   &2.299593(74)&            &2.51192(12) &2.58046(26)&2.64039(26)\\
44   &            &2.426827(67)&2.51150(11) &2.58002(17)&2.63925(24)\\
48   &2.299452(83)&2.426756(64)&2.51119(11) &2.57941(15)&2.63839(27)\\
52   &            &            &2.51130(11) &2.57949(23)&2.63744(19)\\
56   &2.299435(29)&2.426605(62)&2.511096(85)&2.57876(18)&\\
60   &            &2.426596(55)&            &           &\\
64   &            &            &2.510635(83)&2.57851(15)&\\
72   &            &            &2.510716(72)&           &\\ 
80   &            &            &2.510517(79)&           &\\
\hline\hline
\end{tabularx}
\label{tab:bcvalues}
\end{table}

\begin{figure}
  \centering
  \includegraphics[width=0.65\linewidth]{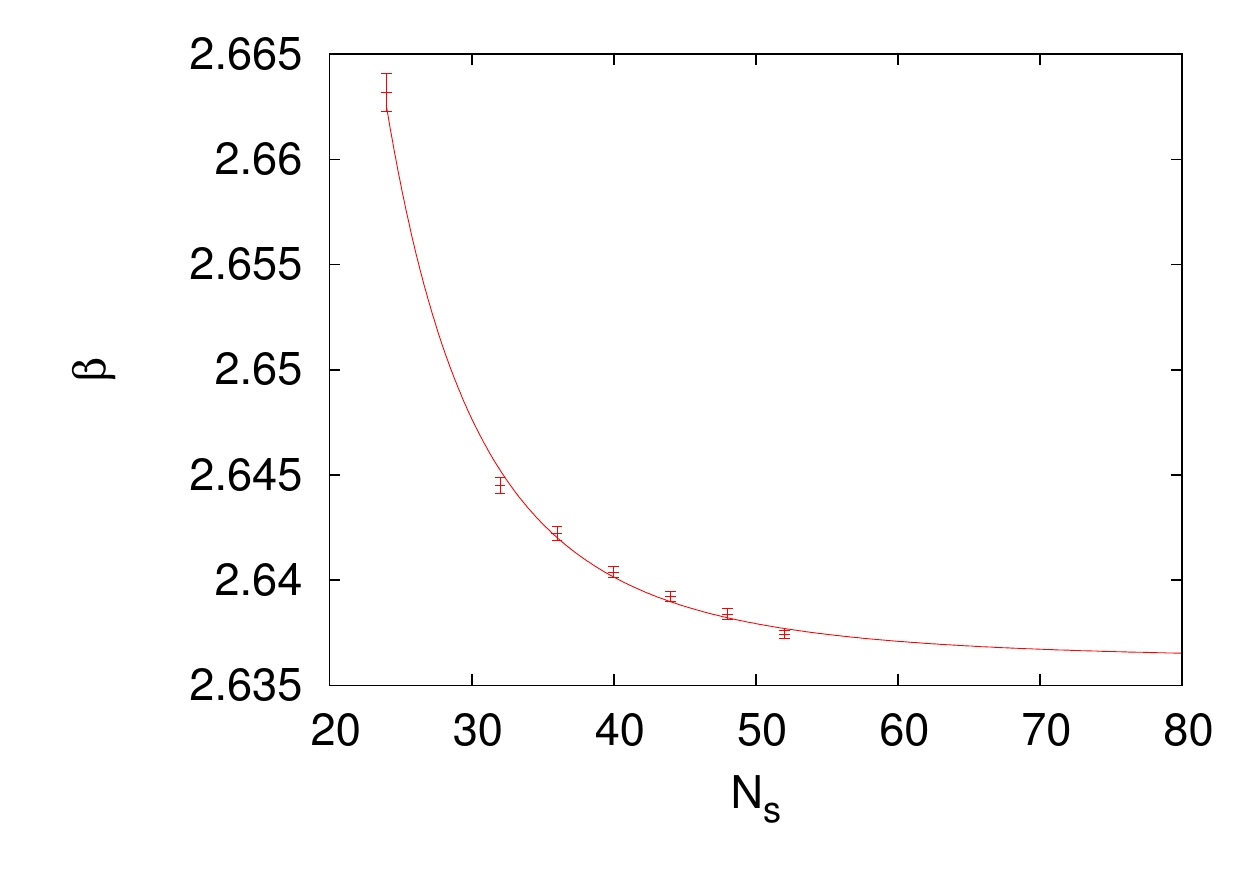}
  \caption{Three-parameter fit~\eqref{eq:3paramtc} for
           $N_\tau=12$.}
  \label{fig:tcfss}
\end{figure}

\begin{table}
\caption{Critical coupling constants $\beta_c(N_\tau)$ and corresponding
         deconfinement lengths $L_0(\beta)$.}
\begin{tabularx}{\linewidth}{LCCR} \hline\hline
$N_\tau$ &$\beta_c(N_\tau)$ &$q$  &$L_0(\beta)$\\\hline
4        &2.299188(61)      &0.56 &4.00000(63)\\
6        &2.426366(52)      &0.73 &6.0000(11)\\
8        &2.510363(71)      &0.14 &8.0000(19)\\
10       &2.57826(14)       &0.29 &10.0000(45)\\
12       &2.63625(35)       &0.06 &12.000(13)\\
\hline\hline
\end{tabularx}
\label{tab:L0}
\end{table}

Critical coupling constants and their corresponding deconfinement lengths
are reported in Table~\ref{tab:L0}. The three-parameter 
fit~\eqref{eq:3paramtc} is carried out using the
Levenberg-Marquardt approach; the corresponding goodness-of-fit $q$
is reported in the third column. Figure~\ref{fig:tcfss} gives
an example finite size fit for $N_\tau=12$; the remaining fits
are included in Appendix~\ref{ap:suppfigs}.
Error bars are attached to the 
deconfinement length using the equation
\begin{equation}\label{eq:L10eb}
  \triangle L_0\ =\ \frac{L_0}{L_{10}^{1,3}(\beta_c)}\,\left[
  L_{10}^{1,3}(\beta_c)-L_{10}^{1,3}(\beta_c-\triangle\beta_c)\right]\,,
\end{equation}
where the cooling length $L_{10}^{1,3}(\beta)$ is introduced in
Section~\ref{sec:scalingsc}. Equation~\eqref{eq:L10eb} is justified 
because $L_0$ error bars depend only mildly on the choice of 
the interpolation of its scaling behavior.

Let us contextualize the results Table~\ref{tab:L0} by comparing these
critical coupling estimates with other pure $\SU(2)$ results. Previously Engels 
et al.~\cite{engels_critical_1996} studied $N_\tau=4$ with volumes up to
$N_s^3=26^3$ and showed that it falls into
the 3D Ising universality class. Their estimate $\beta_c(4)=2.29895(10)$ is
somewhat lower than ours, with the Gaussian difference test giving $q=0.042$. 
Lucini et al.~\cite{lucini_high_2004} present estimates $\beta_c(4)=2.2986(6)$,
$\beta_c(6)=2.4271(17)$, and $\beta_c(8)=2.5090(6)$, for which Gaussian
difference tests against our estimates give $q=0.33$, $q=0.67$, and $q=0.022$,
respectively; we see good agreement for the $N_\tau=4$ and $N_\tau=6$
estimates and some tension with their slightly lower $N_\tau=8$ estimate.
We seem to have the only results for $N_\tau=10$ and $N_\tau=12$, which
appear to be the largest $N_\tau$ for which pure $\SU(2)$ deconfinement
temperatures have been calculated. 

\begin{figure}
  \centering
  \includegraphics[width=0.489\linewidth]{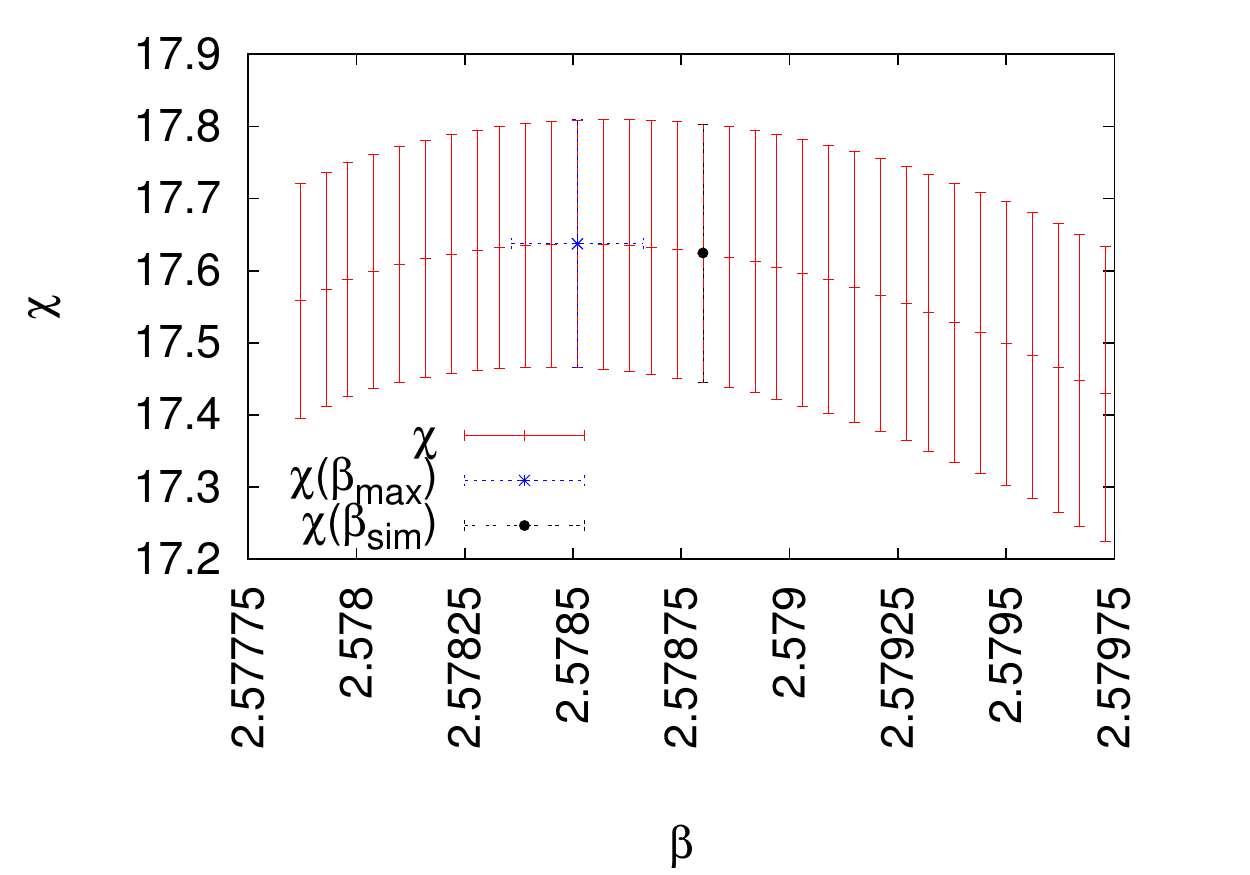}
  \includegraphics[width=0.489\linewidth]{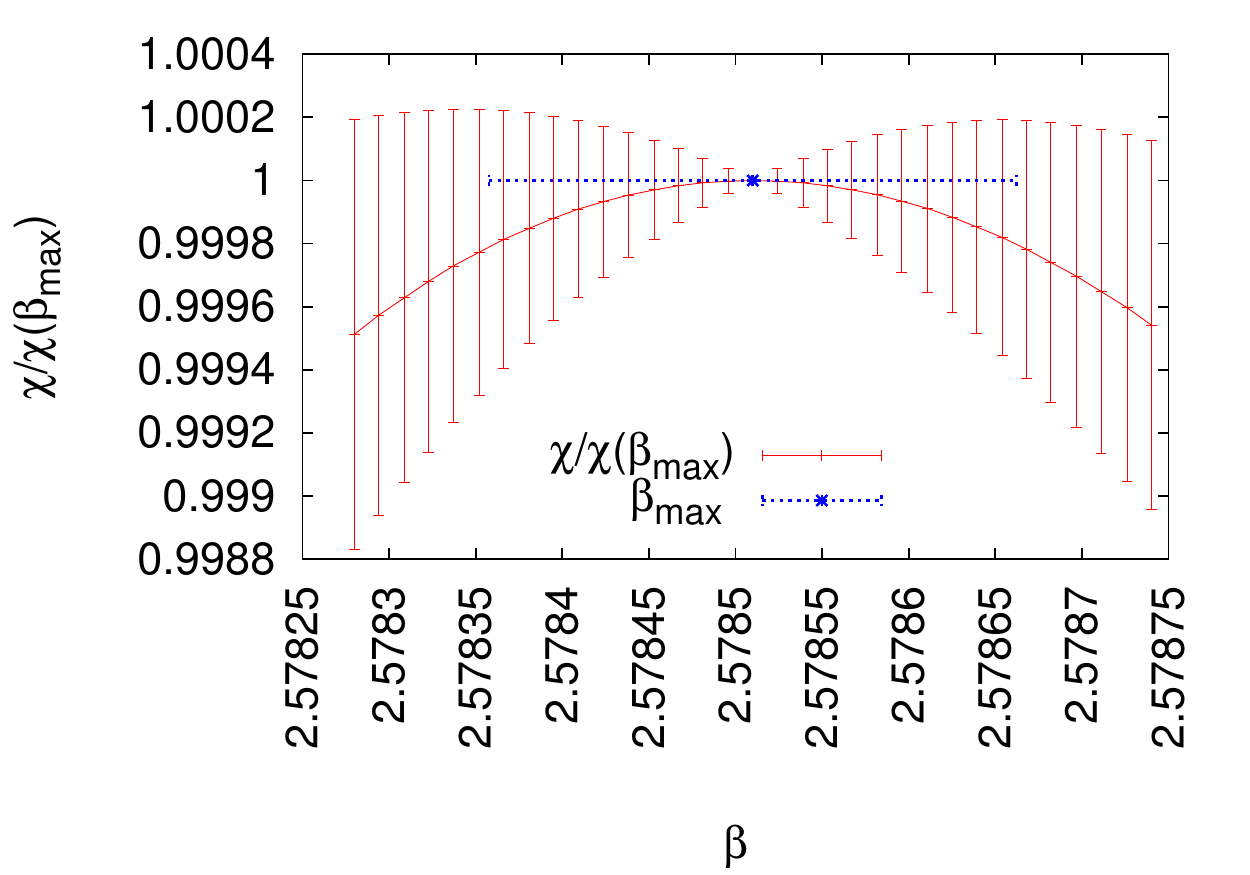}
  \caption{Left: Reweighted Polyakov loop susceptibility curve on 
           a $64^3\times 10$ lattice simulated at $\beta=2.5788$. Right:
           Susceptibility curve with maximum value divided out in each
           jackknife bin.}
  \label{fig:chirat}
\end{figure}

As a technical note, our reweighting curves for $N_\tau=10$ and $N_\tau=12$
are rather flat near the maximum susceptibility $\chi_{\text{max}}$ within
large error bars. This can be seen for our $64^3\times10$ lattice in
Figure~\ref{fig:chirat} (left). The
astonishingly accurate estimates $\beta_c(N_s,N_\tau)$ given in 
Table~\ref{tab:bcvalues} are due to correlations between the
error bars of the reweighted Polyakov loop susceptibilities. 
Dividing out the maximum
value $\chi_{\text{max}}$ in each jackknife bin leads us to
Figure~\ref{fig:chirat} (right), which makes the small error bar
of the estimate pseudo-critical coupling estimate plausible.

\section{Gradient length numerical results}\label{sec:gradsc}

Our numerical results rely on MCMC simulations for the $\beta$ values
and lattice sizes given in Table~\ref{tab:gradscales}. 
In each run $128=2^7$ configurations were generated,
and on each of them, the gradient flow was performed. To implement the
$\SU(2)$ gradient flow on the computer, we use the $\SU(2)$ relationship
\eqref{eq:SU2gflow} and integrate the flow equation~\eqref{eq:Wflow} 
numerically. Following Ref.~\cite{luscher_properties_2010} we applied 
a Runge-Kutta scheme with
$\epsilon=0.01$ and
\begin{equation}
  Z_i=\epsilon Z(W_i),~~~
  Z(W_i)=\frac{1}{2}\left(W_i-W_i^\dagger\right)^\dagger,~~~
  W_0=U_\mu(x).
\end{equation}

\begin{table}
\caption{Gradient length scales.}
\begin{tabularx}{\linewidth}{LCCCCCCR} 
\hline\hline
$\beta$&Lattice &$L_1$   &$L_2$      &$L_3$
                &$L_4$   &$L_5$      &$L_6$\\ \hline
2.3&   $8^4$  &1.361(13)  &1.361(13)  &1.359(15)
              &1.897(24)  &1.897(24)  &1.900(25)\\
   &   $12^4$ &1.3538(52) &1.3538(50) &1.2955(88)
              &1.8905(84) &1.8897(83) &1.824(12)\\
   &   $16^4$ &1.3593(28) &1.3589(27) &1.2756(75)
              &1.8963(48) &1.8956(48) &1.807(11)\\
2.43&  $12^4$ &2.126(20)  &2.115(20)  &2.038(20)
              &2.849(34)  &2.842(33)  &2.771(34)\\
    &  $16^4$ &2.0961(91) &2.0848(90) &1.964(14)
              &2.791(15)  &2.784(15)  &2.653(20)\\
    &  $24^4$ &2.1066(41) &2.0952(40) &1.974(11)
              &2.8044(66) &2.7968(65) &2.644(15)\\
    &  $28^4$ &2.1023(30) &2.0911(30) &1.9666(98)
              &2.7994(48) &2.7920(47) &2.645(13)\\
2.51&  $16^4$ &2.730(21)  &2.715(21)  &2.603(23)
              &3.586(34)  &3.575(34)  &3.436(34)\\
    &  $20^4$ &2.766(15)  &2.750(15)  &2.585(20)
              &3.653(25)  &3.642(25)  &3.453(29)\\
    &  $28^4$ &2.7590(73) &2.7428(73) &2.570(14)
              &3.624(12)  &3.613(12)  &3.406(19)\\
2.574& $20^4$ &3.389(26)  &3.369(26)  &3.166(28)
              &4.437(39)  &4.423(39)  &4.178(44)\\
     & $24^4$ &3.395(17)  &3.374(17)  &3.175(22)
              &4.429(26)  &4.415(26)  &4.171(29)\\
     & $32^4$ &3.406(11)  &3.385(11)  &3.193(17)
              &4.454(15)  &4.440(15)  &4.219(22)\\
     & $40^4$ &3.4103(72) &3.3896(71) &3.149(16)
              &4.458(12)  &4.444(11)  &4.175(21)\\
2.62&  $24^4$ &3.993(28)  &3.968(28)  &3.711(35)
              &5.252(46)  &5.233(45)  &4.916(49)\\
    &$24^348$ &3.947(22)  &3.923(21)  &3.699(26)
              &5.135(33)  &5.119(33)  &4.868(38)\\
    &  $28^4$ &3.950(20)  &3.926(20)  &3.704(24)
              &5.145(30)  &5.129(30)  &4.849(32)\\
    &  $40^4$ &3.954(10)  &3.9293(99) &3.672(19)
              &5.156(16)  &5.140(16)  &4.827(26)\\
2.67&  $28^4$ &4.680(33)  &4.651(33)  &4.350(39)
              &6.131(53)  &6.110(53)  &5.740(60)\\
    &  $32^4$ &4.651(27)  &4.622(27)  &4.350(33)
              &6.057(40)  &6.038(40)  &5.719(46)\\
    &  $40^4$ &4.622(17)  &4.593(17)  &4.297(24)
              &6.020(27)  &6.000(27)  &5.645(32)\\
2.71&  $32^4$ &5.217(37)  &5.185(37)  &4.867(42)
              &6.776(55)  &6.754(55)  &6.357(56)\\
    &  $36^4$ &5.252(33)  &5.220(33)  &4.852(42)
              &6.831(50)  &6.809(50)  &6.401(57)\\
    &  $40^4$ &5.199(22)  &5.167(22)  &4.817(27)
              &6.773(32)  &6.751(32)  &6.334(39)\\
2.751&$32^364$&5.879(35)  &5.843(34)  &5.466(39)
              &7.642(51)  &7.617(51)  &7.179(57)\\
     & $36^4$ &5.893(38)  &5.856(38)  &5.465(48)
              &7.659(60)  &7.633(59)  &7.161(68)\\
     & $40^4$ &5.909(34)  &5.872(34)  &5.457(41)
              &7.694(50)  &7.668(50)  &7.211(59)\\
2.816& $44^4$ &7.092(48)  &7.049(47)  &6.530(54)
              &           &           &\\
2.875& $52^4$ &8.510(64)  &8.456(65)  &7.883(68)
              &           &           &\\
\hline\hline
\end{tabularx}
\label{tab:gradscales}
\end{table}

To optimize our
use of computational resources, we allocated our CPU time in approximately 
equal parts to generation of configurations and to the gradient flow. 
Subsequent configurations are separated by $2^{11}$ to $3\times2^{12}$ 
MCOR sweeps,
where the increase from $2^{11}$ to larger numbers of MCOR sweeps is
due to the number of gradient sweeps needed to reach
the target values. The dividing line from $2^{11}$ to $2^{12}$
sweeps is between $\beta=2.574$ and $\beta=2.62$, and from $2^{12}$ to
$2^{13}$ between $\beta=2.67$ and $\beta=2.71$. We estimated integrated
autocorrelation times $\tau_{\rm int}$ using software of
Ref.~\cite{berg_markov_2004} for the time series of 128 measured scale 
values and found all $\tauint$ compatible with the lower bound~1, 
where the unit is set by the number of sweeps between the configurations.
This gives evidence that our data are statistically independent.
Error bars are calculated using the jackknife method with respect
to these 128 configurations. The lattices are hypercubic
($N_\tau=N_s$) with the exception of $24^3\times 48$ and
$32^3\times 64$, which were generated to compare with
Ref.~\cite{luscher_properties_2010}.

\begin{figure}
  \centering
  \includegraphics[width=0.30\textwidth,height=0.30\textheight,keepaspectratio]
                   {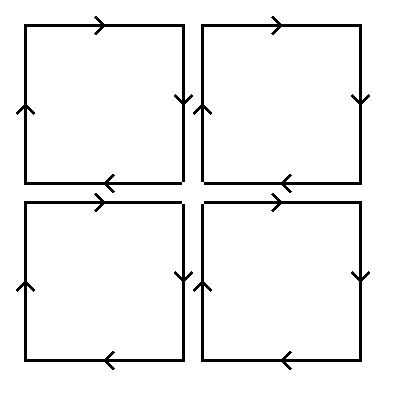}
  \caption{Configuration of plaquettes used for symmetric definition of
           energy density. The $\mu-\nu$ plane lies in the paper, with 
           $\hat\mu$ to the right and $\hat\nu$ upward. All of the 
           plaquettes begin and terminate at $x$, which is in the center.}
  \label{fig:symme}
\end{figure}

Let us now discuss the definitions of our gradient scales. Each gradient
scale is characterized by an energy density and a target value.
We parameterize lattice expectation values 
of plaquette matrices by
\begin{equation}
  \ev{U^\Box(t)}_L=a_0(t)\id+i\sum_{i=1}^3a_i(t)\sigma_i.
\end{equation}
To follow our gradient and cooling flows, we use three discretizations of the
energy density
\begin{equation}
  E_0\equiv 2(1-a_0),~~~
  E_1\equiv \sum_{i=1}^3a_i^2,~~~\text{and}~~~
  E_4\equiv\frac{1}{16}\sum_{i=1}^3\left(a_i^{ul}+a_i^{ur}+a_i^{dl}+a_i^{dr}
                                    \right)^2,
\end{equation}
where $E_4$ is L\"uscher's energy density~\cite{luscher_properties_2010}
that averages over four plaquettes in a fixed $\mu\neq\nu$ plane. The
superscripts of $a_i$ stand for up ($u$), down ($d$), left ($l$), and
right ($r$); the configuration of plaquettes for this definition is
shown in Figure~\ref{fig:symme}. The definition $E_0$ is the Wilson
action density. The definitions $E_0$ and $E_1$ will be highly correlated since
$1=a_\mu a_\mu$. All definitions become $\sim F_{\mu\nu}F_{\mu\nu}$
in the continuum limit. We introduce the notation $s_i$ to indicate
a gradient scale that uses the energy density $E_i$.

\begin{figure}
  \centering
  \includegraphics[width=0.489\linewidth]{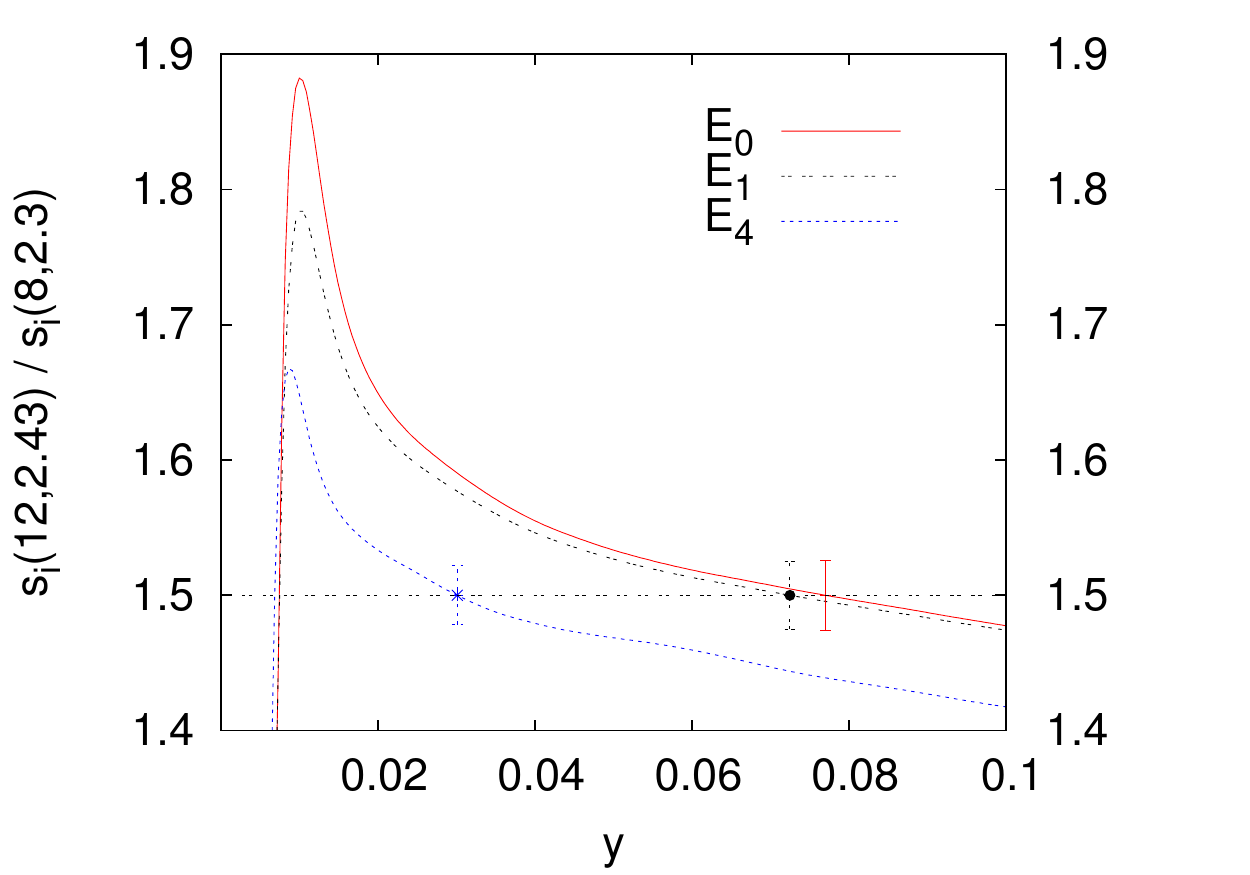}
  \includegraphics[width=0.489\linewidth]{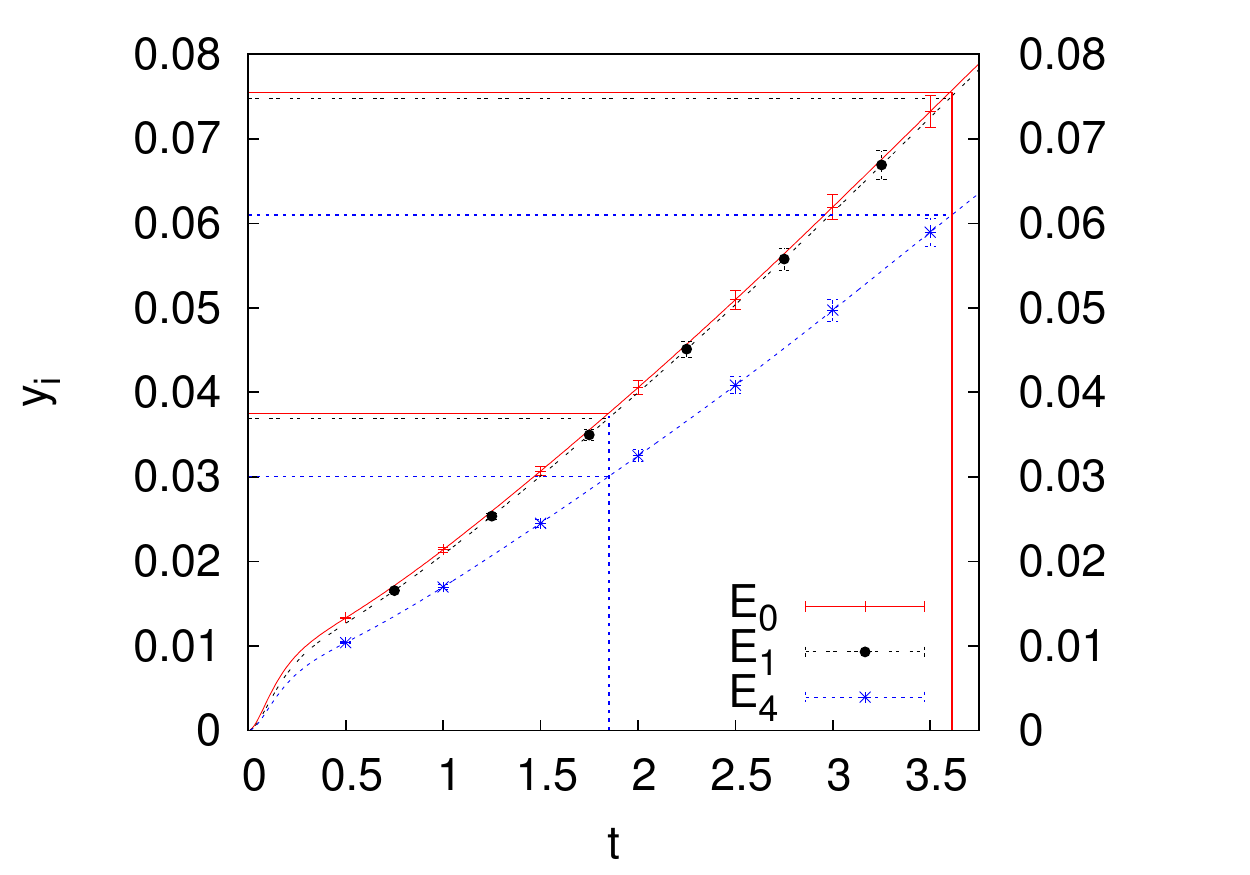}
  \caption{Left: Gradient flow ratios as function of $y$. The horizontal
           line indicates the deconfinement ratio 1.5.
           Right: Gradient flow of an $8^4$ lattice at $\beta=2.3$.}
  \label{fig:gfrat}
\end{figure}

Next we define target values. 
Our strategy was to choose target values so that initial estimates of the
scales $s_i$ agree with the deconfinement scale for small $\beta$.
More precisely we use target values satisfying
\begin{equation}
  \frac{s_i(N=12,\beta=2.43)}{s_i(N=8,\beta=2.3)}
  \approx\frac{N_\tau(\beta=2.43)}{N_\tau(\beta=2.3)}
  =\frac{6}{4}
  =1.5,
\end{equation}
where the left approximate equality holds due to scaling.
For instance from eq.~\eqref{eq:stdscaling} we expect
\begin{equation}
  N_\tau(a)=s(a)\left(\frac{N_\tau}{s}+\mathcal{O}\left(a^2\right)\right),
\end{equation}
so that by considering two lattice spacings $a_1$ and $a_2$ one finds
\begin{equation}
  \frac{N_\tau(a_1)}{N_\tau(a_2)}
  =\frac{s(a_1)}{s(a_2)}\Big(1+\mathcal{O}\left(a_1^2\right)
                               +\mathcal{O}\left(a_2^2\right)\Big).
\end{equation}

Figure~\ref{fig:gfrat} (left) plots the gradient scale ratio against the 
target value. We see essentially two intersections with 1.5, the first 
coming from the $E_4$ curve and another coming from the $E_0$ and
$E_1$ curves, which practically agree. 
Figure~\ref{fig:gfrat} (right) plots the function $t^2E_i$ against the flow
time. Picking initially $y_4^1$, the target value corresponding to the
aforementioned $E_4$ intersection, defines a flow time, indicated by the
vertical dotted blue line at $t=1.85$. This flow time is then used to define
two more target values $y_0^1$ and $y_1^1$, determined by following the vertical
dotted blue line up until it intersects with the $E_0$ and $E_1$ curves.
Similarly, picking initially $y^2_0$ (or equivalently $y^2_1$) delivers a
target value from the $E_0$ intersection in the left figure,
then two more target values $y^2_1$ and $y^2_4$ from the vertical solid 
red line at $t=3.61$ in the right figure. Altogether we consider the
six gradient flow target values
\begin{gather} \label{eq:gfy1}
   y^1_0=0.0376,~~y^1_1=0.0370,~~y^1_4=0.030,\\
   \label{eq:gfy2}
   y^2_0=0.0755,~~y^2_1=0.0748,~~y^2_4=0.061.
\end{gather}
A gradient length scale $s_i^j$ is obtained according to 
eq.~\eqref{eq:tardef} and \eqref{eq:s} when the gradient flow hits the 
target value $y^i_j$. 
For later convenience we define
\begin{equation}
  L_1\equiv s^1_0,~~~~~
  L_2\equiv s^1_1,~~~~~
  L_3\equiv s^1_4,~~~~~
  L_4\equiv s^2_0,~~~~~
  L_5\equiv s^2_1,~~~~~
  L_6\equiv s^2_4.
\end{equation}

Our MCMC estimates for these scales are reported in Table~\ref{tab:gradscales}.
We see the strong correlation between scales defined using $E_0$ and $E_1$,
often being identical within error. To control for finite size
effects, these scales are simulated for multiple lattice sizes. 
For the largest lattices, finite size effects are negligible, with differences
between scales calculated on the largest lattice and on the second largest
lattice being comparable to or smaller than the statistical error.
Gradient scales at $\beta=2.816$ and $\beta=2.875$ were not simulated 
for smaller lattices because results for the 
cooling scale give evidence that these lattices are already large enough 
for finite size effects to be negligible (see Table~\ref{tab:coolscales1}.)
For these two lattices, the allocated gradient flow was too short to reach its
$y^2_i$ targets.

As mentioned in the previous section, each simulation for the deconfinement 
length took at least $2^{18}$ MCOR sweeps, requiring as many as
$2^{23}$ MCOR sweeps for large $\beta$. By contrast our longest gradient
flow simulation required $128\times2^{13}=2^{20}$ MCOR sweeps.
Furthermore finite size scaling extrapolations are
necessary in order to obtain a reliable estimate for the deconfinement
scale, usually requiring 10 or so simulations to achieve the
desired error bars. The gradient length, meanwhile, is already well-defined 
without requiring $N_s\gg N_\tau$, so that arguably only one 
simulation at each $\beta$ is necessary.
Taking achieved error bars, lattice sizes, and number of simulations
needed into account, using the gradient scale over the deconfinement
scale amounts to a two to three order of magnitude improvement. 
For instance at $\beta=2.62$ a gradient scale can be estimated with
at worst a relative error of $5\times10^{-4}$ on a $40^4$ lattice
using $2^{19}$ MCOR sweeps. Meanwhile the nearby deconfinement length
$N_\tau=12$ required $2^{24}$ MCOR sweeps on four lattices that are
very roughly half as large as $40^4$ to achieve a relative error
of $10^{-3}$. Putting this together, the gradient scale at this
spacing is at least 256 times as efficient.

\section{Cooling length numerical results}\label{sec:coolsc}

\begin{figure}
  \centering
  \includegraphics[width=0.489\linewidth]{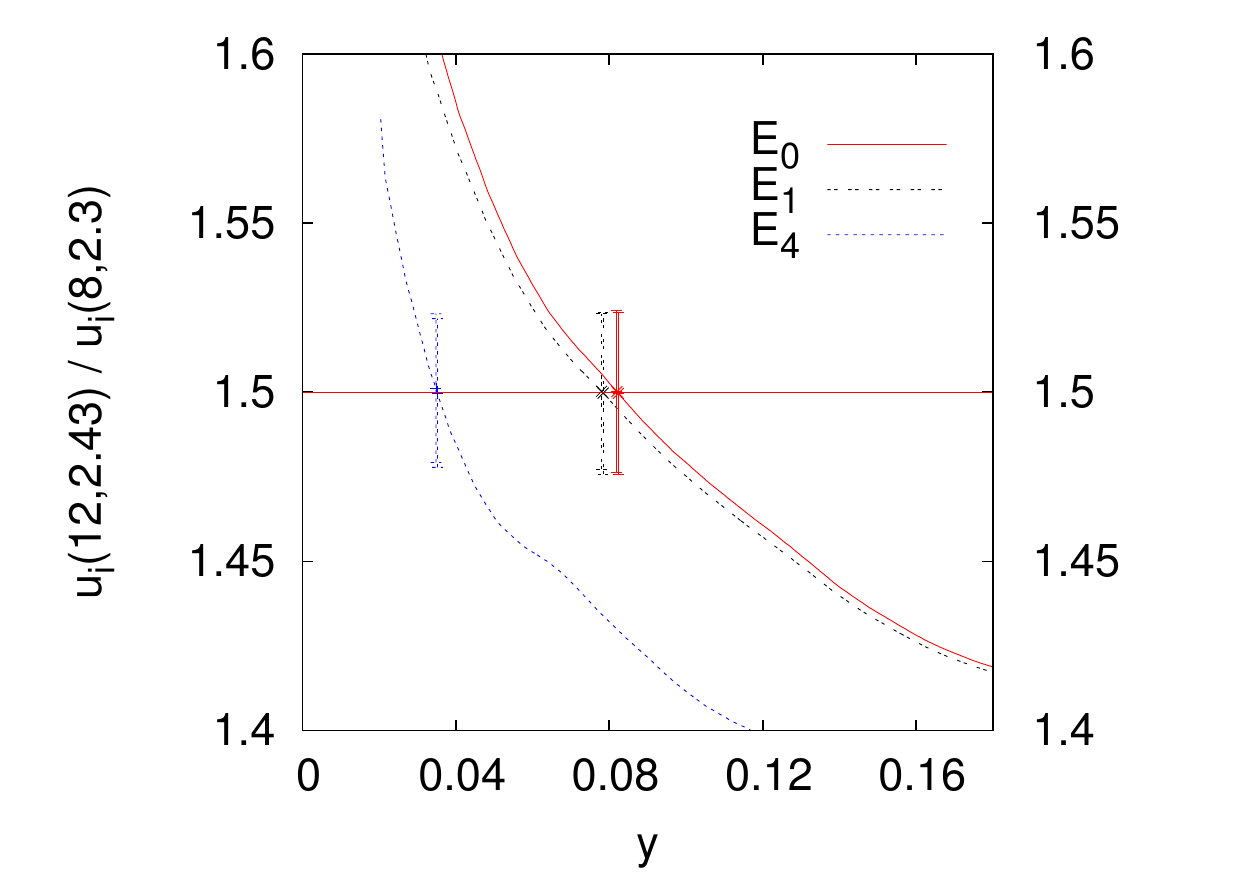}
  \includegraphics[width=0.489\linewidth]{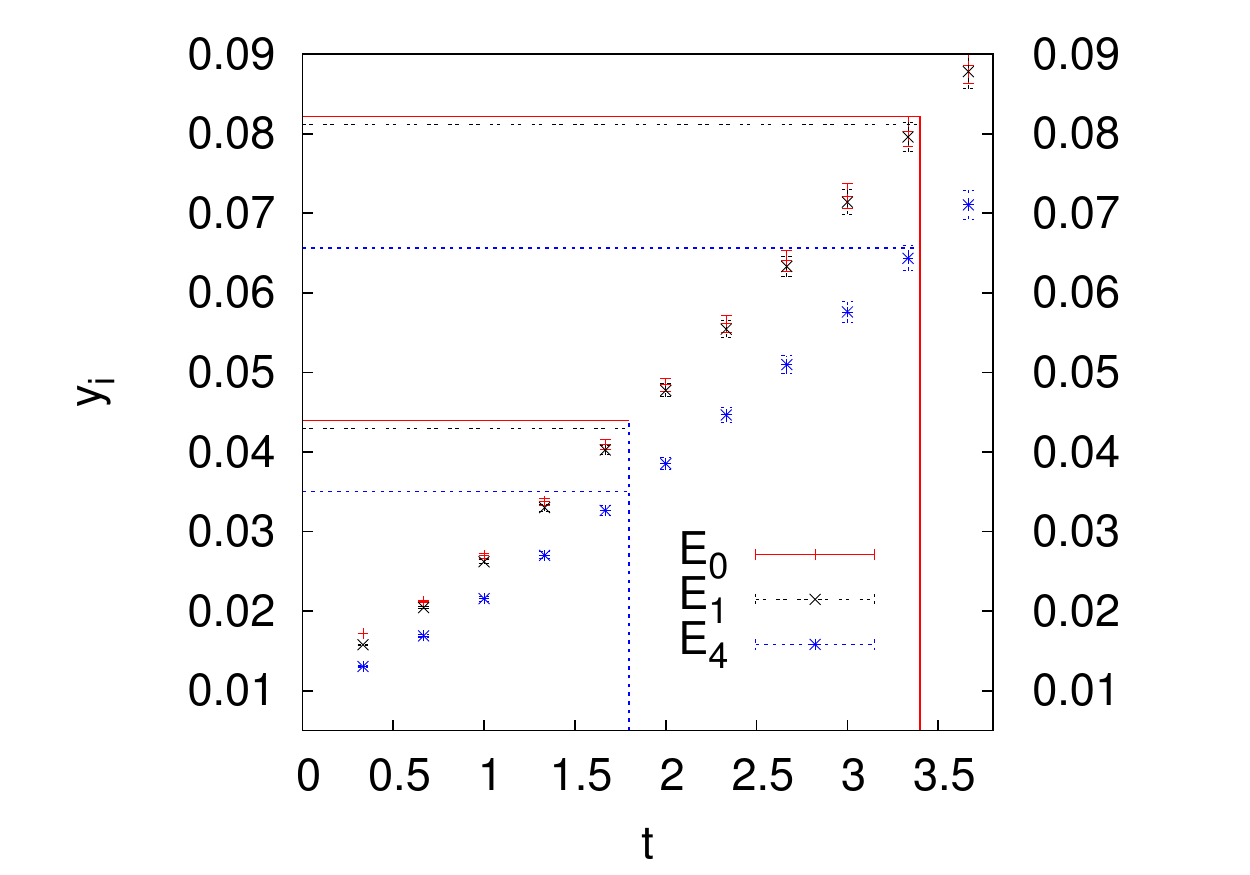}
  \caption{Left: Cooling flow ratios as function of $y$. The horizontal line
           indicates the deconfinement ratio 1.5.
           Right: Cooling flow of an $8^4$ lattice at $\beta=2.3$.}
  \label{fig:cfrat}
\end{figure}

Bonati and D'Elia~\cite{bonati_comparison_2014} showed that $n_c$ cooling
sweeps corresponds to a flow time
\begin{equation}
  t=n_c/3.
\end{equation}
If $n_g$ denotes the number of sweeps of the gradient flow algorithm, then
$t=\epsilon\,n_g=0.01\,n_g$, so the above relation implies
\begin{equation}
  n_g=33.\bar{3}\,n_c,
\end{equation}
i.e. one cooling sweep traverses the same flow time as $33.\bar{3}$
gradient sweeps. Combined with the fact that a gradient
sweep is computationally more intensive than a cooling sweep due to
the Runge-Kutta, one expects the cooling flow to reach its target value 
at least 34 times faster than the gradient flow.

The cooling flow~\eqref{eq:cool} is performed on the same configurations 
as the gradient flow. Cooling sweeps are performed in the same systematic 
order as our MCMC sweeps. As another check of statistical independence,
we calculated on our largest lattices the topological charge~\eqref{eq:QL}
of each configuration, using the cooling flow to smooth them, and looked
at $\tauint$ for the time series of 128 topological charges. These
$\tauint$ were found to be statistically compatible with 1, confirming
again the statistical independence of these configurations.
The topological charge was defined at 100 cooling sweeps, which
may be too low to be metastable for our smallest lattices, but is
sufficient for the purpose of checking statistical independence.
More details are given in Chapter~\ref{ch:top}.

To determine target values, we follow the same approach as with the
gradient flow. The analogue to Figure~\ref{fig:gfrat} is given
in Figure~\ref{fig:cfrat}. Due to the large cooling steps, gaps between the
points are clearly visible. The intersection of target value lines and flow
time lines in Figure~\ref{fig:cfrat} (right) are determined using linear
interpolation. We find target values
\begin{eqnarray}\label{cyi01}
  y^{1}_0&=&0.0440,~~y^{1}_1=0.0430\,,~~y^{1}_4=0.0350,
  ~~~~\\ \label{cyi02} 
  y^{2}_0&=&0.0822,~~y^{2}_1=0.0812\,,~~y^{2}_4=0.0656,
  ~~~~
\end{eqnarray} 
where a superscript 1 again indicates target values obtained from the $E_4$
ratio curve crossing 1.5 in Figure~\ref{fig:cfrat} (left), and the superscript
2 indicates target values obtained from the $E_0$ ratio curve. These target
values deliver cooling length scales $u_i^j$ according to eq.~\eqref{eq:u}. 
For later convenience we define
\begin{equation}
  L_7   \equiv u^1_0,~~~~~
  L_8   \equiv u^1_1,~~~~~
  L_9   \equiv u^1_4,~~~~~
  L_{10}\equiv u^2_0,~~~~~
  L_{11}\equiv u^2_1,~~~~~
  L_{12}\equiv u^2_4.
\end{equation}

Our MCMC estimates for these scales are reported in Table~\ref{tab:coolscales1}.
Again we see evidence that finite size effects are not detectable within our
statistics for the largest lattices, and that scales defined with densities
$E_0$ and $E_1$ give almost identical results. 

\begin{table}
\caption{Cooling length scales.}
\begin{tabularx}{\linewidth}{LCCCCCCR} 
\hline\hline
$\beta$&Lattice&$L_7$     &$L_8$      &$L_9$
               &$L_{10}$  &$L_{11}$   &$L_{12}$\\\hline
2.3  & $ 8^4$ & 1.342(12) & 1.337(12) & 1.342(14)
              & 1.846(22) & 1.844(22) & 1.843(22)\\
     & $12^4$ & 1.3391(47)& 1.3343(45)& 1.2730(85)
              & 1.8241(74)& 1.8217(72)& 1.743(12)\\
     & $16^4$ & 1.3433(24)& 1.3385(23)& 1.2575(74)
              & 1.8307(39)& 1.8282(39)& 1.728(10)\\
2.43 & $12^4$ & 2.111(19) & 2.092(18) & 2.013(20)
              & 2.769(29) & 2.759(29) & 2.669(32)\\
     & $16^4$ & 2.0837(90)& 2.0653(90)& 1.951(13)
              & 2.725(14) & 2.715(14) & 2.572(18)\\
     & $24^4$ & 2.0929(38)& 2.0744(38)& 1.947(11)
              & 2.7395(57)& 2.7287(57)& 2.561(14)\\
     & $28^4$ & 2.0892(28)& 2.0707(28)& 1.9446(95)
              & 2.7317(43)& 2.7212(42)& 2.565(12)\\
2.51 & $16^4$ & 2.728(19) & 2.703(19) & 2.587(23)
              & 3.531(30) & 3.516(30) & 3.370(31)\\
     & $20^4$ & 2.753(14) & 2.727(14) & 2.567(20)
              & 3.571(23) & 3.555(23) & 3.359(27)\\
     & $28^4$ & 2.7522(68)& 2.7267(66)& 2.548(15)
              & 3.552(10) & 3.5371(99)& 3.315(18)\\
2.574& $20^4$ & 3.396(25) & 3.365(24) & 3.157(26)
              & 4.356(37) & 4.337(37) & 4.084(38)\\
     & $24^4$ & 3.389(16) & 3.357(16) & 3.155(22)
              & 4.352(24) & 4.333(24) & 4.080(29)\\
     & $28^4$ & 3.422(13) & 3.390(13) & 3.168(18)
              & 4.405(20) & 4.386(29) & 4.123(25)\\
     & $32^4$ & 3.4001(97)& 3.3686(95)& 3.153(17)
              & 4.374(14) & 4.355(14) & 4.100(21)\\
     & $40^4$ & 3.4048(69)& 3.3730(67)& 3.137(17)
              & 4.377(11) & 4.358(10) & 4.074(20)\\
2.62 & $24^4$ & 3.988(26) & 3.949(26) & 3.717(32)
              & 5.157(40) & 5.133(39) & 4.836(44)\\
     &$24^348$& 3.949(20) & 3.912(19) & 3.688(25)
              & 5.070(30) & 5.047(29) & 4.788(34)\\
     & $28^4$ & 3.952(19) & 3.915(19) & 3.680(23)
              & 5.059(28) & 5.037(28) & 4.751(30)\\
     & $40^4$ &3.9509(95) & 3.9137(93)& 3.645(22)
              & 5.068(15) & 5.045(15) & 4.725(26)\\
2.67 & $28^4$ & 4.676(32) & 4.631(31) & 4.314(39)
              & 6.021(46) & 5.993(46) & 5.603(58)\\
     & $32^4$ & 4.644(27) & 4.600(26) & 4.282(31)
              & 5.950(38) & 5.923(38) & 5.532(42)\\
     & $40^4$ & 4.618(17) & 4.574(16) & 4.298(26)
              & 5.910(25) & 5.884(25) & 5.536(33)\\
2.71 & $28^4$ & 5.232(41) & 5.184(40) & 4.829(47)
              & 6.675(58) & 6.645(57) & 6.228(67)\\
     & $32^4$ & 5.216(36) & 5.167(35) & 4.833(41)
              & 6.656(51) & 6.626(51) & 6.208(55)\\
     & $36^4$ & 5.256(31) & 5.207(31) & 4.803(42)
              & 6.724(48) & 6.692(48) & 6.223(58)\\
     & $40^4$ & 5.203(21) & 5.154(21) & 4.794(28)
              & 6.656(31) & 6.626(30) & 6.188(38)\\
2.751& $28^4$ & 5.880(82) & 5.824(78)& 5.487(74)
              & 7.55(13)  & 7.52(13) & 7.07(11)\\
     &$32^364$& 5.874(32) & 5.819(32) & 5.437(37)
              & 7.515(49) & 7.481(48) & 7.010(52)\\
     & $36^4$ & 5.892(36) & 5.836(35) & 5.478(49)
              & 7.531(53) & 7.497(53) & 7.033(66)\\
     & $40^4$ & 5.913(32) & 5.857(32) & 5.434(40)
              & 7.576(46) & 7.541(46) & 7.038(54)\\
2.816& $28^4$ & 8.247(27) & 8.167(26)& 7.561(25)
              & 10.48(35) & 10.44(35)& 9.72(34)\\
     & $40^4$ & 7.089(58) & 7.021(58)& 6.517(68)
              & 9.076(84) & 9.034(84)& 8.426(92)\\
     & $44^4$ & 7.105(45) & 7.039(45) & 6.511(55)
              & 9.056(65) & 9.015(64) & 8.349(73)\\
2.875& $40^4$ & 8.55(11)  & 8.464(10) & 7.885(97)
              & 10.98(16) & 10.93(16) &10.21(16)\\
     & $44^4$ & 8.637(93) & 8.554(92) & 7.912(89)
              & 11.11(15) & 11.06(15) &10.29(15)\\
     & $52^4$ & 8.514(60) & 8.433(59) & 7.825(68)
              &10.879(87) &10.830(86) & 10.122(92)\\
2.928& $40^4$ &10.90(30)  & 10.79(29) & 9.89(27)
              &13.99(42)  &13.92(42)  &12.87(40)\\
     & $44^4$ &10.01(16)  & 9.92(16)  & 9.18(14)
              &12.78(23)  &12.72(23)  &11.82(21) \\
     & $52^4$ &9.940(88)  & 9.846(87) & 9.112(93)
              &12.72(13)  &12.67(13)  &11.76(13)\\
     & $60^4$ &9.835(67)  & 9.742(66) & 9.053(70)
              &12.561(97) &12.503(96) &11.653(95)\\
\hline\hline
\end{tabularx}
\label{tab:coolscales1}
\end{table}

\section{Scaling and asymptotic scaling behavior}\label{sec:scalingsc}

We analyze the approach of ratios of the length scales $L_0-L_{12}$
to the continuum limit. We first fit using standard scaling, then
asymptotic scaling. Additionally we provide estimates of systematic
uncertainty from the choice of continuum limit fitting form. All gradient 
and cooling scale results rely on the largest lattice at each $\beta$, since 
finite size effects are not detectable within statistical
error for these sizes. Results from $\beta=2.928$ are not included
in the following analysis, which was carried out before simulations
at this coupling constant finished.

\subsection{Standard scaling}

We begin with standard
scaling using eq.~\eqref{eq:stdscalingfit}
\begin{equation}
  R_{ij}=\frac{L_i}{L_j}=r_{ij}+c_{ij}\left(\frac{1}{L_j}\right)^2.
\end{equation}
This is a linear fit in the squared lattice spacing with fit parameters
$r_{ij}$ and $c_{ij}$, $r_{ij}$ being the continuum limit estimate for the
ratio $R_{ij}$. Table~\ref{tab:scaleratios} reports these continuum limit
estimates for various scale combinations. The first column labels the numerator
$L_i$ and the top row labels the denominator $L_j$. The scales $L_2$, $L_5$,
$L_8$, and $L_{11}$ are omitted from the table, since they use the
discretization $E_1$, which essentially agrees with $E_0$. For example
$r_{10,11}=0.995397(24)$. Data points from $\beta=2.3$ were omitted from
fits with $q<0.05$, as they may not be deep enough in the scaling region.
After applying this cut, these fits satisfy $0.11\leq q\leq0.98$.
The deconfinement fit relies on all five points from Table~\ref{tab:L0}
with goodness-of-fit $q=0.25$.

\begin{table}
\caption{Continuum limit estimates of ratios $r_{ij}$ from scaling.}
\begin{tabularx}{\linewidth}{LCCCR}
\hline\hline
$i\,\backslash\, j$& $L_1$ & $L_4$  & $L_7$ & $L_{10}$ \\ \hline
$L_0$   & 2.8896(71) & 2.2290(46) & 2.8855(68) & 2.2618(42)\\
$L_1$   &            & 0.77382(61)& 0.99845(38)& 0.78433(43)\\
$L_3$   & 0.9250(19) & 0.7163(17) & 0.9241(19) & 0.7264(16)\\
$L_4$   & 1.2943(11) &            & 1.29135(99)& 1.01520(49)\\
$L_6$   & 1.2090(26) & 0.9346(20) & 1.2081(27) & 0.9490(21)\\
$L_7$   & 1.00156(38)& 0.77398(79)&            & 0.78570(50)\\
$L_9$   & 0.9222(21) & 0.7141(19) & 0.9213(20) & 0.7243(17)\\
$L_{10}$& 1.27509(70)& 0.98508(47)& 1.27300(80)&            \\
$L_{12}$& 1.1835(24) & 0.9164(21) & 1.1825(24) & 0.9292(19)\\
\hline\hline
\end{tabularx}
\label{tab:scaleratios}
\end{table}

\begin{figure}
\centering
\includegraphics[width=0.9\linewidth]{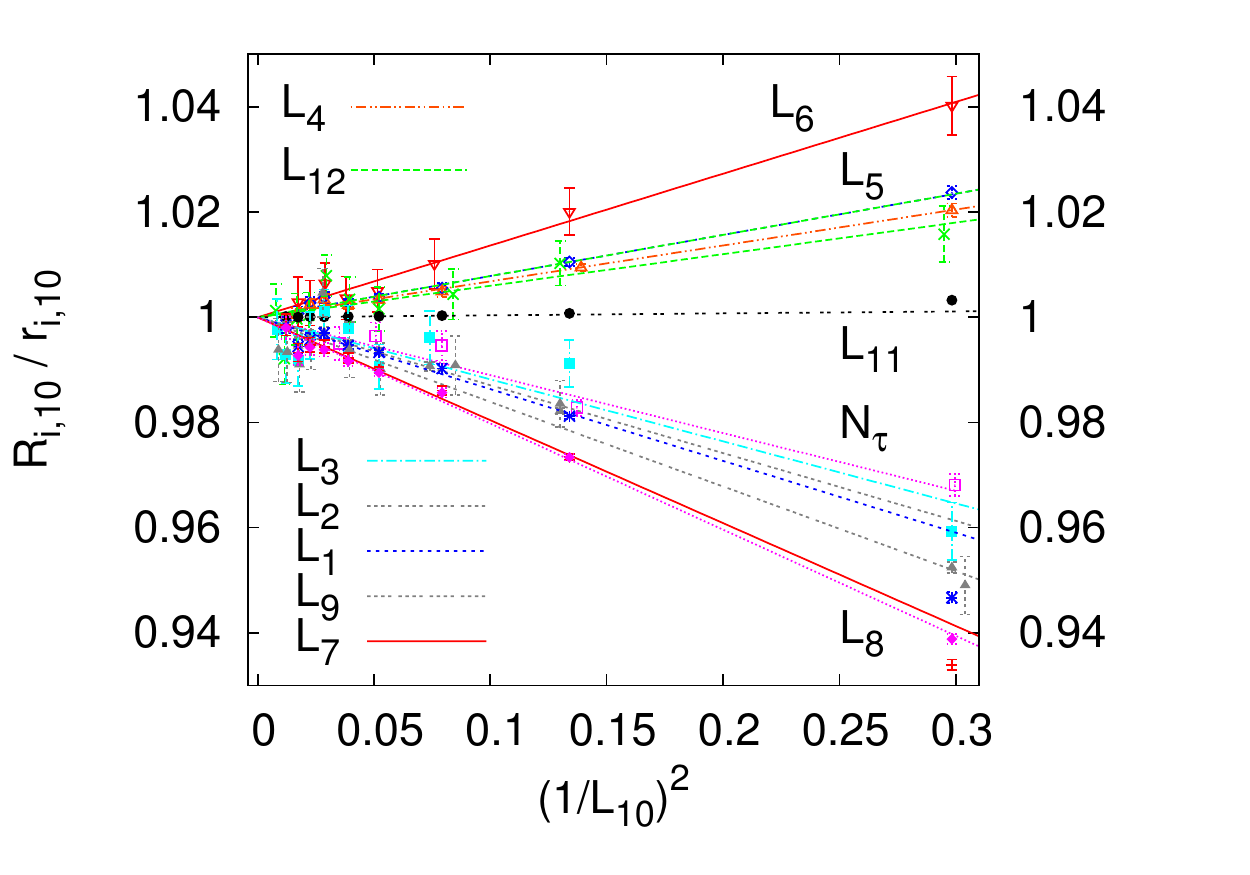}
\caption{Scaling corrections of order $a^2$ for ratios $L_i/L_{10}$.
  Some data are slightly shifted for better visibility.
  Some labels are attached to the lines and others put into the legend. The
  top-bottom order in the legend matches the top-bottom order in the plot.} 
\label{fig:LiLj}
\end{figure}

\begin{figure}
\centering
\includegraphics[width=0.489\linewidth]{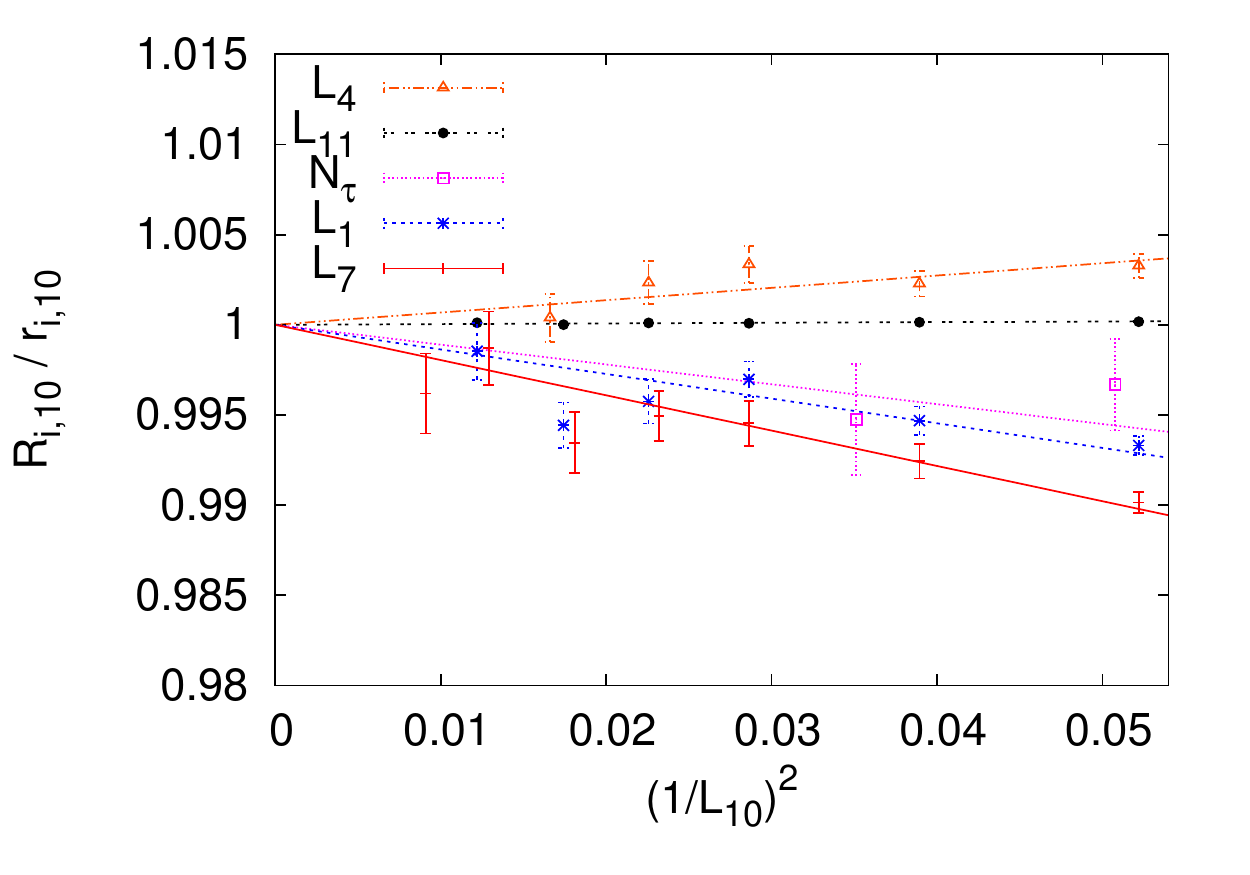}
\includegraphics[width=0.489\linewidth]{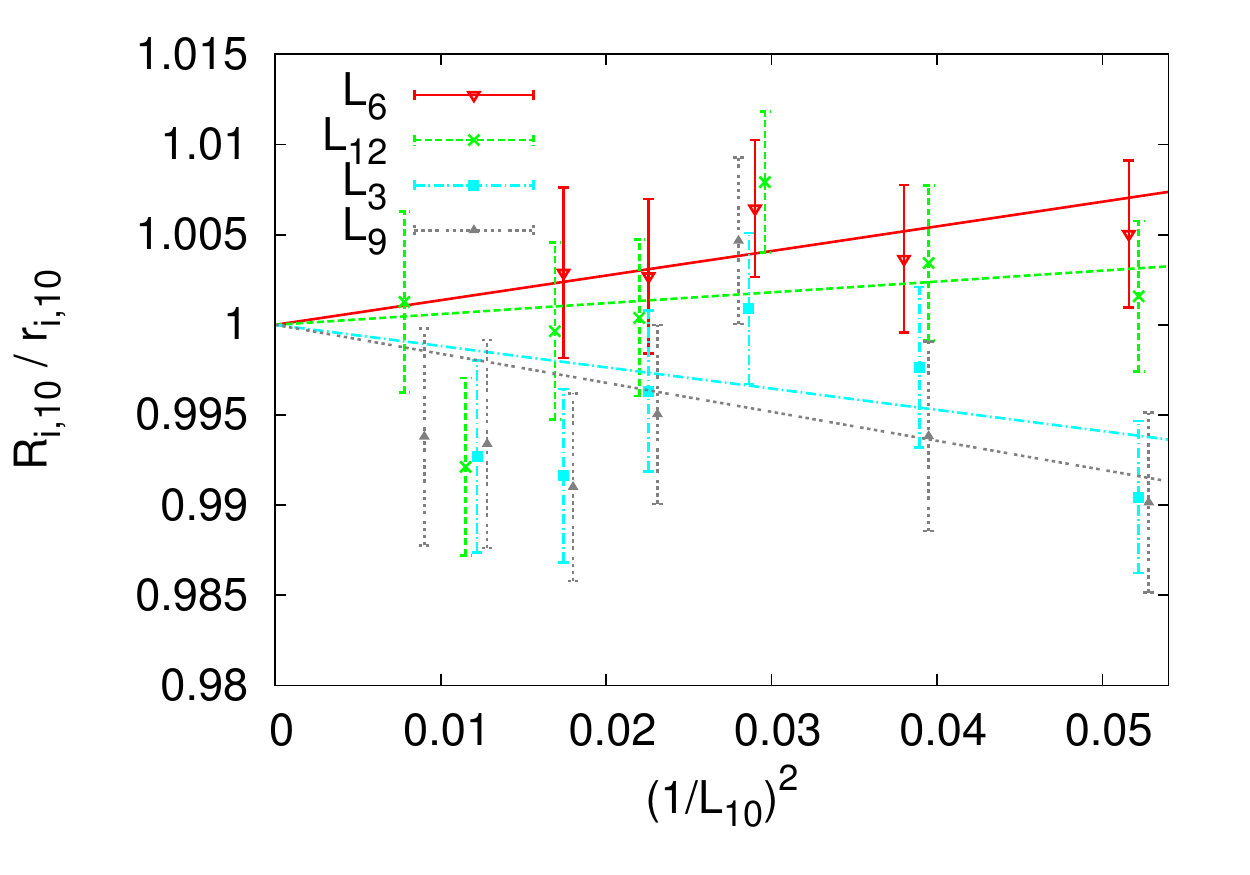}
\caption{Left: Enhancement of the scaling fits of Figure~\ref{fig:LiLj} for 
  scales using $E_0$ as well as $L_{11}$ and the deconfinement scale.
  Right: Enhancement of the scaling fits of Figure~\ref{fig:LiLj} for
  scales using $E_4$.}
\label{fig:LiLj1+2}
\end{figure}

To compare scaling corrections between the ratios, we rescale $R_{ij}$ with
the extrapolation $r_{ij}$ and choose $\ell_{10}$ as a reference scale.
We chose $\ell_{10}$ for aesthetic reasons: the fits distribute rather
evenly about $R_{i,10}/r_{i,10}=1$ with this choice.
A collection of these fits is shown in Figure~\ref{fig:LiLj}. The abscissa
ranges up to $(1/L_{10})^2\approx0.3$, which corresponds to $\beta=2.3$.
Goodness-of-fit cuts were made for the scales $L_{11}$, $L_2$, $L_1$, and
$L_7$; correspondingly in the figure, one can see the deviations of their
$\beta=2.3$ points from the fit lines. The $L_{11}$ scale is close to 1
throughout, again because $L_{10}$ and $L_{11}$ rely on the $E_0$ and $E_1$
densities and have the same target value. There is clear overlap between
cooling and gradient scales; for example cooling scales $L_{10}-L_{12}$ fall
within the spread of gradient scales $L_1-L_6$, which shows that cooling scales
do not suffer significant scaling violations compared to gradient scales. At
$(1/L_{10})^2\approx0.3$ we read off scaling violations of about 10\%.

Figure~\ref{fig:LiLj1+2} shows enhancements of Figure~\ref{fig:LiLj} for two
scale sets, deep in the scaling region. The abscissa ranges up to
$(1/L_{10})^2\approx0.05$, which corresponds to $\beta=2.574$. Both figures
include two gradient and two cooling scales. Comparing the relative sizes of
their error bars shows that there is no discernible loss of precision using
cooling scales over gradient scales. Figure~\ref{fig:LiLj1+2} (left) 
features gradient and cooling scales relying on $E_0$, with the exception 
of $L_{11}$, which relies on $E_1$ and is included because $L_{10}$ was 
taken as reference.
Figure~\ref{fig:LiLj1+2} (right) features scales relying on $E_4$. These scales
clearly exhibit larger error bars than scales using the $E_0$ density. Since
both sets of scales show similar scaling violations, and since $E_0$ has the
simplest definition, we recommend using $E_0$ over the other two densities
for the purpose of defining gradient and cooling scales in pure $\SU(2)$. 

\subsection{Asymptotic scaling}

Next we consider asymptotic scaling fits~\eqref{eq:lenasf} of the length scales
\begin{equation}\label{eq:lenasf2}
  L_i=\frac{c^{mn}_i}{f^{m}_{as}(\beta)}
                    \left(1+\sum\limits_{k=1}^{n}
                    \alpha^{mn}_{i\,k}\,f^{m}_{as}(\beta)^{\,k}\right).
\end{equation}
Since the pure $\SU(2)$ beta function is only known to three-loop order on the
lattice, we consider only $m=0,\,1$. We arrive at definitions
\begin{equation}
  f^{\,0}_{as}(\beta)=C^0\left(\frac{4b_0}{\beta}\right)^{-b_1/2b_0^2}
              \exp\left(-\frac{\beta}{8b_0}\right)~~~~\text{and}~~~~
  f^1_{as}(\beta)=\frac{C^1}{C^0}f_{as}^0(\beta)
               \left(1+\frac{4q_1}{\beta}\right),
\end{equation}
where $b_0$, $b_1$, and $q_1$ are the constants from 
eqs.~\eqref{eq:blat3loop} and~\eqref{eq:q1value}. We have 
also introduced normalization constants $C^0$ and $C^1$ to enforce
for convenience
\begin{equation}
  f_{as}^m(2.3)=1.
\end{equation}

Estimates of normalization constants for asymptotic scaling fits of
gradient and cooling scales are collected in Table~\ref{tab:ascalingnorm}. 
As explained in Section~\ref{sec:sysa}, we demand the
same $\alpha^{mn}_{i,1}$ for all scales. Using the $E_0$ and $E_4$
scales, these coefficients were determined by a maximum likelihood
approach, varying $\alpha^{mn}_{i,1}$ and minimizing $q$ by bisection. 
$E_1$ scales are left out because they would just amplify the weight
of the $E_0$ scales. We find 
\begin{equation}
  \alpha^{1,2}_{i,1}=-0.6209,~~~~~~
  \alpha^{0,3}_{i,1}=-0.38157,~~~~~~\text{and}~~~~~~
  \alpha^{1,3}_{i,1}=-0.32536.
\end{equation}
On a technical note, we eliminate
the normalization constants $c_i^{m,n}$ from the search for the $\chi^2$
minimum by treating them as functions of the $\alpha^{mn}_{ik}$ parameters
\cite{berg_asymptotic_2015}. This stabilizes the minimization considerably, 
for which we used the Levenberg-Marquardt approach.

\begin{table}
\caption{Normalization constants $c_i^{mn}$ for asymptotic scaling fits
         of gradient and cooling scales, along with the corresponding 
         goodness-of-fit.}
\begin{tabularx}{\linewidth}{LCCCCCR} \hline\hline\noalign{\vskip 1mm}
$L_i$ & $c_i^{1,2}$&  $q$&$c_i^{0,3}$ & $q$& $c_i^{1,3}$& $q$\\ \hline
$L_1$ & 2.2481(32)& 0.04& 2.1937(64)& 0.91& 2.1083(61)&0.91\\
$L_2$ & 2.2311(32)& 0.03& 2.1812(64)& 0.92& 2.0961(60)&0.92\\
$L_3$ & 2.0743(56)& 0.17& 2.022(11)& 0.66& 1.9432(98)&0.67\\
$L_4$ & 2.8945(54)& 0.08& 2.846(11)& 0.98& 2.735(11)&0.98\\
$L_5$ & 2.8835(53)& 0.04& 2.837(11)& 0.98& 2.727(11)&0.98\\
$L_6$ & 2.7068(85)& 0.95& 2.658(18)& 0.95& 2.555(17)&0.95\\
$L_7$ & 2.2498(30)& 0.02& 2.1996(61)& 0.93& 2.1138(57)&0.94\\
$L_8$ & 2.2254(30)& 0.01& 2.1807(60)& 0.92& 2.0956(57)&0.93\\
$L_9$ & 2.0664(58)& 0.16& 2.018(11)& 0.69& 1.9397(99)&0.69\\
$L_{10}$ & 2.8501(46)& 0.02& 2.8037(91)& 0.89& 2.6942(86)&0.89\\
$L_{11}$ & 2.8357(45)& 0.01& 2.7914(89)& 0.88& 2.6824(85)&0.89\\
$L_{12}$ & 2.6485(74)& 0.26& 2.599(14)& 0.52& 2.498(13)&0.52\\ \hline\hline
\end{tabularx}
\label{tab:ascalingnorm}
\end{table}

\begin{table}
\caption{Normalization constants $c_i^{mn}$ for asymptotic scaling fits
         of the deconfinement length, along with the corresponding 
         goodness-of-fit.}
\begin{tabularx}{\linewidth}{LCCCCCR} \hline\hline\noalign{\vskip 1mm}
$L_i$ & $c_i^{1,3}$&  $q$  & $c_i^{0,4}$& $q$ & $c_i^{1,4}$& $q$ \\ \hline
$L_0$ & 6.6682(56)& 0.00& 6.114(29) & 0.71& 5.892(27) & 0.68\\ \hline\hline
\end{tabularx}
\label{tab:ascalingnormntau}
\end{table}

Fitting the gradient and cooling scales with only one
additional parameter, $\alpha_{i,2}^{1,2}$, the normalization constants
$c_i^{1,2}$ of column two are obtained. Most $q$-values of these
fits are too low, so we allowed one more fit parameter,
$\alpha_{i,3}^{m,3}$. The results are shown in columns four and six with
$m=0,\,1$. The $q$-values for these fits would be suspiciously high 
if they were statistically independent. But as they all
rely on the same data set, correlations can explain that a whole series
of fits exhibits $q>0.5$, mostly close to 0.9. Notably, consistent fits
due to adding the parameter $\alpha_{i,3}^{m,3}$ come at the price of
roughly doubled error bars compared to those of column two. 
In Table~\ref{tab:ascalingnormntau} we collect normalization constants for the
deconfinement scale. The deconfinement scale requires an additional fit
parameter $\alpha_{0,4}^{m,4}$ to obtain acceptable $q$-values. 
This is accompanied by some instability discussed later. 
We conclude that $n=3$ is essentially the smallest number of terms in 
the power series expansion needed to obtain acceptable $q$-values.

\begin{figure}
  \centering
  \includegraphics[width=0.95\linewidth]{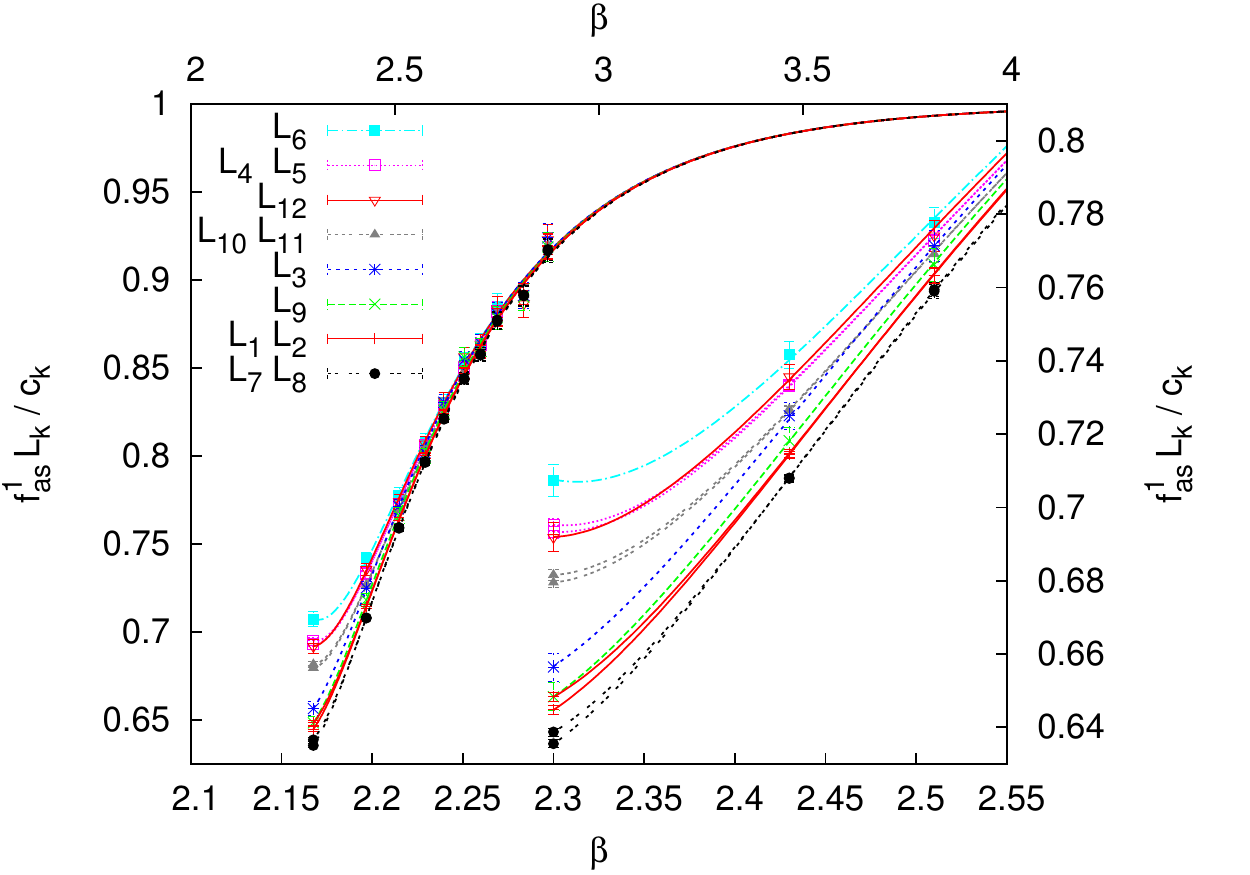}
  \caption{Asymptotic scaling corrections for lengths $L_i$. The top
     abscissa and left ordinate correspond to
     the top set of curves, while the bottom abscissa and right ordinate
     correspond to the bottom set of curves, which is an enlargement of
     the top curves for low $\beta$ values.}
  \label{fig:ascaling}
\end{figure}

Using $m=1$ instead of $m=0$ for the asymptotic scaling
function decreases the $c^{mn}_{ik}$ values of Tables~\ref{tab:ascalingnorm}
and \ref{tab:ascalingnormntau} by slightly less than 4\%. More
prominent is the decrease between 6.7\% to 9\% from column two to
column six, which comes from allowing one more free parameter. 
We take these decreases as an indication that the remaining truncation 
error may be as large as~10\%.

We now consider asymptotic scaling fits of 
gradient and cooling scales with $m=1$ and $n=1$.
Figure~\ref{fig:ascaling} plots eq.~\eqref{eq:lenasf2} with $m=1$ and $n=3$
against $\beta$, with the asymptotic scaling behavior divided out. With this
normalization the curves approach 1 in the continuum limit.
The curves on the left use the top abscissa and left ordinate. The curves on
the right are an enhancement of the left curves for the lowest three $\beta$.
These curves use the bottom abscissa and right ordinate.
At $\beta=4$ all fits have almost reached the
asymptotic value~1. At $\beta=2.3$ asymptotic scaling
violations are seen to range from 28\% to 37\%. The relative differences
reach only $0.72/0.63\approx 1.14$, consistent with the ratio $1.04/0.93
\approx 1.12$ observed at $(1/L_{10})^2\approx0.3$ in Fig.~\ref{fig:LiLj}.

For a more direct comparison with scaling, we compute ratios of length scales
using asymptotic scaling~\eqref{eq:ascalingfit}
\begin{equation}\label{eq:ascalingfit2}
  R_{ij}=r_{ij}+\sum_{k=2}^n\kappa^{mn}_{i\,k}
                  \,f^m_{as}\left(L_j\right)^{\,k},
\end{equation}
where $\kappa^{mn}_{i,1}=0$ because all scales are assumed to have the same
first order term in eq.~\eqref{eq:lenasf2}. Except for the
deconfinement length scale $L_0$, which is statistically independent
from the other scales, we can not use error propagation. Therefore 
for the gradient and cooling scales we calculate $R_{ij}$ in jackknife 
bins built from the individual runs. 

\begin{table}
\caption{Continuum limit estimates of ratios $r_{ij}$ from
         asymptotic scaling. The asterisks indicate scales that required an
         additional fit parameter for an acceptable $q$ value.}
\begin{tabularx}{\linewidth}{LCCCR}
\hline\hline
$i\,\backslash\,j$& $L_1$& $L_4$    & $L_7$       & $L_{10}$    \\ \hline
$L_0$ (as)& 2.795(16)& 2.154(14) & 2.787(15) & 2.187(13)\\
$L_0$   &*2.914(15) & 2.2393(52) &*2.903(14) & 2.2692(48)\\
$L_1$   &             &*0.7703(12) & 0.99808(34)&*0.78185(77)\\
$L_3$   & 0.9240(20) & 0.7187(19) & 0.9221(20) & 0.7275(17)\\
$L_4$   &*1.2996(21) &             &*1.2957(27) & 1.01373(57)\\
$L_6$   & 1.2000(31) & 0.9334(23) & 1.1972(32) & 0.9465(24)\\
$L_7$   & 1.00188(34)&*0.7728(16) &             &*0.78419(88)\\
$L_9$   & 0.9214(22) & 0.7171(21) & 0.9197(22) & 0.7255(18)\\
$L_{10}$&*1.2795(13) & 0.98638(55)&*1.2760(15) &            \\
$L_{12}$& 1.1786(26) & 0.9167(24) & 1.1760(26) & 0.9283(20)\\
\hline\hline
\end{tabularx}
\label{tab:ascaleratios}
\end{table}

Results for the continuum limit extrapolations $r_{ij}$ using $m=1$ are 
given in Table~\ref{tab:ascaleratios}. One free parameter 
$\kappa^{1,2}_{i,2}$, in addition to the continuum estimate $r_{ij}$, suffices
to deliver in more than half of the cases $0.13\leq q\leq0.99$.
For the other cases, indicated by an asterisk in the table, one more free
parameter $\kappa^{1,3}_{i,3}$ is also needed. For these ratios
the goodness-of-fit falls within the range $0.45\leq q\leq 0.75$. 
Error bars of asymptotic scaling estimates are similar to the standard 
scaling estimates of Table~\ref{tab:scaleratios}, except for the starred 
estimates, whose error bars are approximately twice as large.
It is reassuring that the estimates of $r_{ij}$ from 
Tables~\ref{tab:scaleratios} and \ref{tab:ascaleratios} never differ
by more than roughly 1\%, which is nevertheless up to an order of magnitude 
larger than the statistical errors. Statistical uncertainties of ratios can
be extremely small due to correlations between the estimators. 
We conclude that the two fitting approaches supplement each other
and give insight to systematic errors one might expect due to 
choice of continuum limit fitting form.

\begin{figure}
\centering
\includegraphics[width=0.8\linewidth]{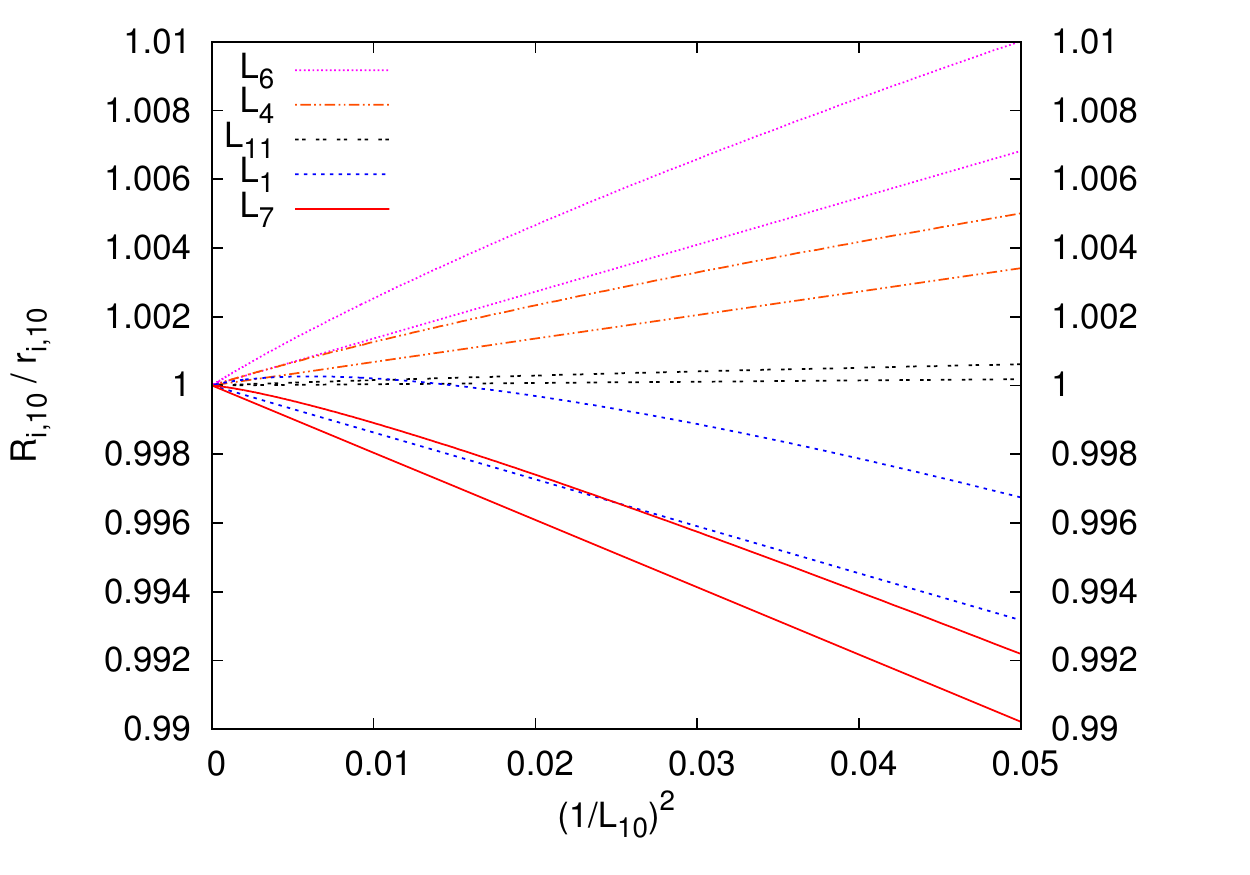}
\caption{Direct comparison between representative scaling fits and asymptotic
   scaling fits deep in the scaling region. Slightly curved fits of the pairs
   belong to the asymptotic scaling form. Data and error bars are omitted.} 
\label{fig:compare}
\end{figure}

In Figure~\ref{fig:compare} we plot the normalized ratio $R_{i,10}/r_{i,10}$
for both standard scaling and asymptotic scaling fits against the
squared lattice spacing. Straight line fits are standard scaling fits, while
slightly curved fits are asymptotic scaling fits. The abscissa ranges up
to $(1/L_{10})^2\approx0.05$, which corresponds to $\beta=2.574$. 
At this spacing, systematic error due to choice of fitting form alone 
seems not to exceed about 0.6\%. The combined systematic error due to choice
of scale and continuum limit fitting form is read off to be
around 2\%.

\begin{figure}
\centering
\includegraphics[width=0.8\linewidth]{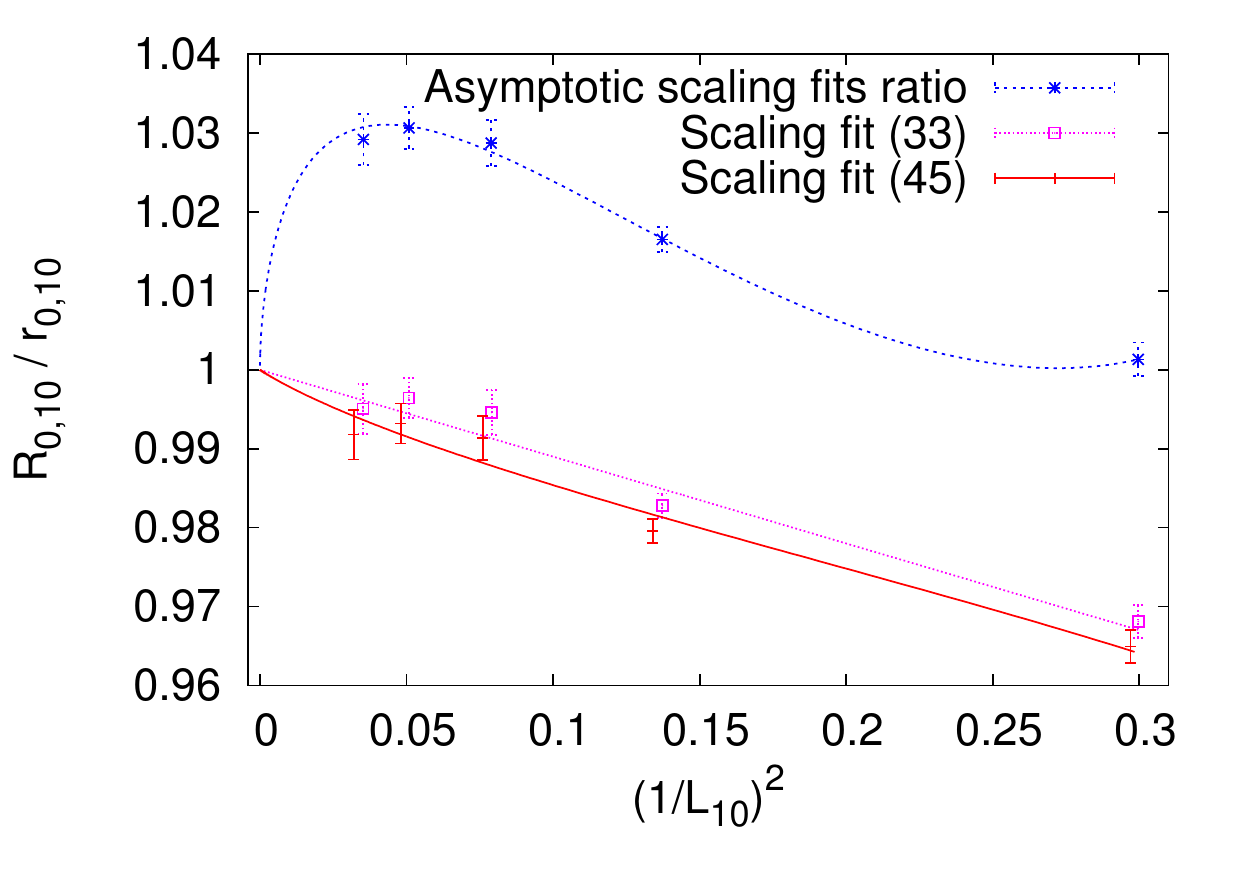}
\caption{Three fits of the deconfinement length to the continuum limit.} 
\label{fig:fanciful}
\end{figure}
Let us now discuss the instabilities of the $L_0$ fit mentioned earlier.
In the $L_0$ (as) row of Table~\ref{tab:ascaleratios} we report estimates
obtained from using the constants of the sixth column of
Table~\ref{tab:ascalingnorm} and error propagation. Compared with the
standard scaling estimates of Table~\ref{tab:scaleratios}, we find a systematic
decrease between 3.2\% and 3.6\%. This is larger than the statistical error,
which never exceeds 0.6\%. Since the asymptotic scaling fit for $L_0$ needs
four parameters to fit just five data points, one may suspect over-fitting.
As a tie-breaker, we perform the fit~\eqref{eq:ascalingfit2} for jackknifed
ratios $R_{0,j}$, $j=1,\,4,\,7,\,10$, and obtain the estimates of the
$L_0$ row of Table~\ref{tab:ascaleratios}. Systematic differences between
Table~\ref{tab:scaleratios} are now down to less than 1\%.

The normalized ratio $R_{0,10}/r_{0,10}$ is plotted for three different
fits in Figure~\ref{fig:fanciful}. The bottom curve corresponds to
the eq.~\eqref{eq:ascalingfit2} using jackknifed ratios. The next lowest
fit is the straight line scaling fit from Figure~\ref{fig:LiLj}.
The top curve is obtained by dividing the $L_0$ fit from column six
of Table~\ref{tab:ascalingnormntau} by the $L_{10}$ fit of
column six of Table~\ref{tab:ascalingnorm}. As suspected, this fit
looks rather strange. One should keep in mind that absolute differences between
these three fits are small. Systematic errors at $\beta=2.3$ are read
off to be less than 4\%.

\section{Summary}\label{sec:summarysc}

We calculated the pure $\SU(2)$ deconfinement temperature out to larger $\beta$
than has been done in previous literature using reweighting curves of Polyakov 
loop susceptibilities. Dividing out the
maximum susceptibility in each jackknife bin verifies that small error
bars in the critical coupling constant are reasonable.

We calculated six gradient scales and six cooling scales, distinguished
by choice of energy density operator and target value. Reasonable
target values were determined by requiring that initial estimates of gradient
or cooling scales agree with the deconfinement scale for low $\beta$.
Measured in CPU time, gradient scales are at least two orders of 
magnitude faster to calculate than deconfinement scales. 
Cooling scales take at least a factor 34 less CPU 
time than gradient scales; however in this case, the generation of
configurations takes the same CPU time for both. 
Looking at scaling fits, cooling scales fall within
the spread of gradient scales, showing that cooling and gradient scales do
not exhibit seriously distinct scaling behavior. We find no loss of precision
using cooling over gradient scales. Therefore cooling scales
are viable alternatives to gradient scales for the purpose of scale setting.

The approach to the continuum limit was fitted using scaling fits and 
asymptotic scaling fits. For scaling fits, we find scaling violations 
of about 10\% at $\beta=2.3$; these violations are reduced to less 
than 5\% at $\beta=2.46$, deeper in the scaling region. For asymptotic 
fits, enforcing a common fit parameter yields the expected 
$\mathcal{O}(a^2)$ corrections to ratios of scales. Systematic error 
of normalization constants
due to distinct truncations of asymptotic fits are estimated to be 
up to roughly 10\%. This drops out in ratios, and at $\beta=2.574$, 
combined systematic error of length ratios due to 
reference scale and fitting form is around 2\%. Continuum limit 
estimates of ratios differ systematically by at most 1.3\%, 
but this is still larger than statistical errors.

Our suggestion is that cooling scales may offer a computationally more
efficient alternative to gradient scales in physically realistic
theories as well. One may test this at some coupling constant values
and, if confirmed, continue with the cooling scale.

%% file: chapter5.tex
\chapter{Topology in Pure SU(2) LGT}\label{ch:top}

Here we present detailed analysis of the topological susceptibility 
and reinforce that standard cooling can be used to obtain stable 
topological sectors. We estimate finite size corrections of the 
topological susceptibility and come up with a continuum limit 
extrapolation. 

Topological freezing has often been a point of concern, even though fixed
topological sectors imply a bias of only $1/V$ for local operators
\cite{brower_qcd_2003,aoki_finite_2007}. Generally, topological
freezing can not be ignored, since some observables are known to
have dependence on $Q$~\cite{delia_phase_2013}
and hence require the topological sectors to be well-sampled.
This is an active area of research; for instance L\"uscher
and Shaefer~\cite{luscher_lattice_2011} proposed that topological freezing
can be alleviated using open boundary conditions, and even more
recently, L\"uscher~\cite{luscher_stochastic_2018} has suggested the 
use of master-field simulations. 
Therefore, we investigate whether there are
statistically significant differences between cooling scales that are
restricted to different topological sectors. In our investigation
we find our lattices are large enough that the $1/V$ bias is swallowed
by statistical uncertainty.

This chapter focuses on our study of cooling scales and topological
observables~\cite{berg_topological_2018}.
In Section~\ref{sec:smoothingcool} we discuss our data for the topological
charge using cooling as a smoothing algorithm. The following 
Section~\ref{sec:chi2} presents new data for cooling scales 
along with an estimate of the pure $\SU(2)$ topological susceptibility. 
In Section~\ref{sec:coolcor} we search for correlations between cooling scales
and topological sectors. A summary is given in the final
Section~\ref{sec:qsumm}.

\section{Smoothing using standard cooling}\label{sec:smoothingcool}

Our discretization of the topological charge density
\begin{equation}
  q_L(x) = -\frac{1}{2^9\pi^2}\sum\limits_{\mu\nu\rho\sigma=\pm 1}^{\pm 4}
         \tilde{\epsilon}_{\mu\nu\rho\sigma}
         \tr U^\Box_{\mu\nu}(x)U^\Box_{\rho\sigma}(x),
\end{equation}
which is given in eq.~\eqref{eq:QLdensity}, follows the
field-theoretical definition
\begin{equation}
  q(x)=\frac{1}{16\pi^2}\tr\dual{F_{\mu\nu}}(x)F_{\mu\nu}(x).
\end{equation}
The topological charge in lattice units is then given by
\begin{equation}
  Q_L=\sum_x q_L(x).
\end{equation}
Measurements of the topological charge on lattice configurations
generated by Monte Carlo suffer from lattice artifacts, which we suppressed by
standard cooling. Provided the lattice spacing is fine enough, and the physical
volume is large enough, one reaches metastable configurations after many
cooling sweeps; a topological charge relatively free of lattice artifacts can
be assigned to such configurations. The obtained topological charge values
still suffer from discretization effects, which can be absorbed by a
normalization factor through the following procedure~\cite{debbio__2002}: 
Picking a suitable number $m_c$ of cooling sweeps, one makes the replacement
\begin{equation}\label{eq:qnorm}
  Q_L\to Q^{m_c}\equiv A^{m_c}Q_L,
\end{equation}
where $A^{m_c}$ is determined by minimizing
\begin{equation}\label{eq:minimize}
  \ev{\left(A^{m_c}Q_L-\text{nint}\,A^{m_c}Q_L\right)^2}.
\end{equation}
Here, the expected value is taken over all configurations with a fixed
$\beta$ and lattice size. An exact mapping onto integers is obtained by
\begin{equation}
  Q^{m_c}_I\equiv\text{nint}\,Q^{m_c},
\end{equation}
which is the definition we use to identify topological sectors.

\begin{table}
\caption{Overview of our largest lattices at each $\beta$.
         Integrated autocorrelation times of $Q^{100}$ and normalization 
         constants $A^{m_c}$ are given.
         The last column reports the stability of the charge sectors
         under the next 1048 cooling sweeps after $n_c=1000$.}
\begin{tabularx}{\linewidth}{LCCCCCR}
\hline\hline\noalign{\vskip 1mm}
Lattice & $\beta$& $\tauint$ & $A^{100}$& $A^{1000}$& $A^{2048}$
        & \% stable\\
\hline
$16^4$& 2.300 & 1.26(24) & 1.202 & 1.178 & 1.155 & 61.7\\ 
$28^4$& 2.430 &          & 1.258 & 1.128 & 1.129 & 60.9\\
$28^4$& 2.510 & 1.01(21) & 1.148 & 1.127 & 1.124 & 66.4\\
$40^4$& 2.574 & 1.49(48) & 1.159 & 1.117 & 1.113 & 58.6\\
$40^4$& 2.620 & 0.91(22) & 1.135 & 1.111 & 1.110 & 78.1\\
$40^4$& 2.670 & 0.92(26) & 1.131 & 1.110 & 1.108 & 83.6\\
$40^4$& 2.710 & 0.85(22) & 1.131 & 1.107 & 1.105 & 87.5\\
$40^4$& 2.751 & 1.68(51) & 1.113 & 1.108 & 1.108 & 94.5\\
$44^4$& 2.816 & 1.59(35) & 1.111 & 1.105 & 1.101 & 89.1\\
$52^4$& 2.875 & 1.17(27) & 1.112 & 1.100 & 1.098 & 96.9\\
$60^4$& 2.928 &          & 1.106 & 1.107 & 1.097 & 96.1\\
\hline\hline
\end{tabularx}
\label{tab:normalization}
\end{table}

For this study we use hypercubic lattices with $N_s=N_\tau\equiv N$. 
For each $\beta$
and lattice size up to $52^4$, we generated 128 configurations separated 
by $2^{11}-3\times2^{12}$ MCOR sweeps, as outlined in Section~\ref{sec:gradsc}.
This large separation between configurations guarantees that subsequent
measurements of the topological charge are effectively independent.
For example the third column of Table~\ref{tab:normalization} shows
the integrated autocorrelation times of $Q^{100}$ for the time series
of 128 configurations. The error bars are relatively large due to the 
small number of 128 data points. Within this limitation, $\tauint$ is
seen to be statistically compatible with 1. Statistical fluctuations allow
for $\tauint<1$. Therefore the $Q_I$ are effectively
independent, so that topological freezing is not
an issue for our data.

On each configuration we performed 2048 cooling sweeps and applied the
minimization~\eqref{eq:minimize} with multiplicative constants $A^{m_c}$
defined at $m_c=100$, $m_c=1000$, and $m_c=2048$. The data are given
in columns 4, 5, and 6 of Table~\ref{tab:normalization}. When $\beta$
is small, the constants $A^{m_c}$ amount to corrections of at most 26\%.
For our largest $\beta$ values and lattices, these corrections are down
to about 10\%, with little dependence on $m_c$.

\begin{figure}
\centering
\includegraphics[width=0.48\linewidth]{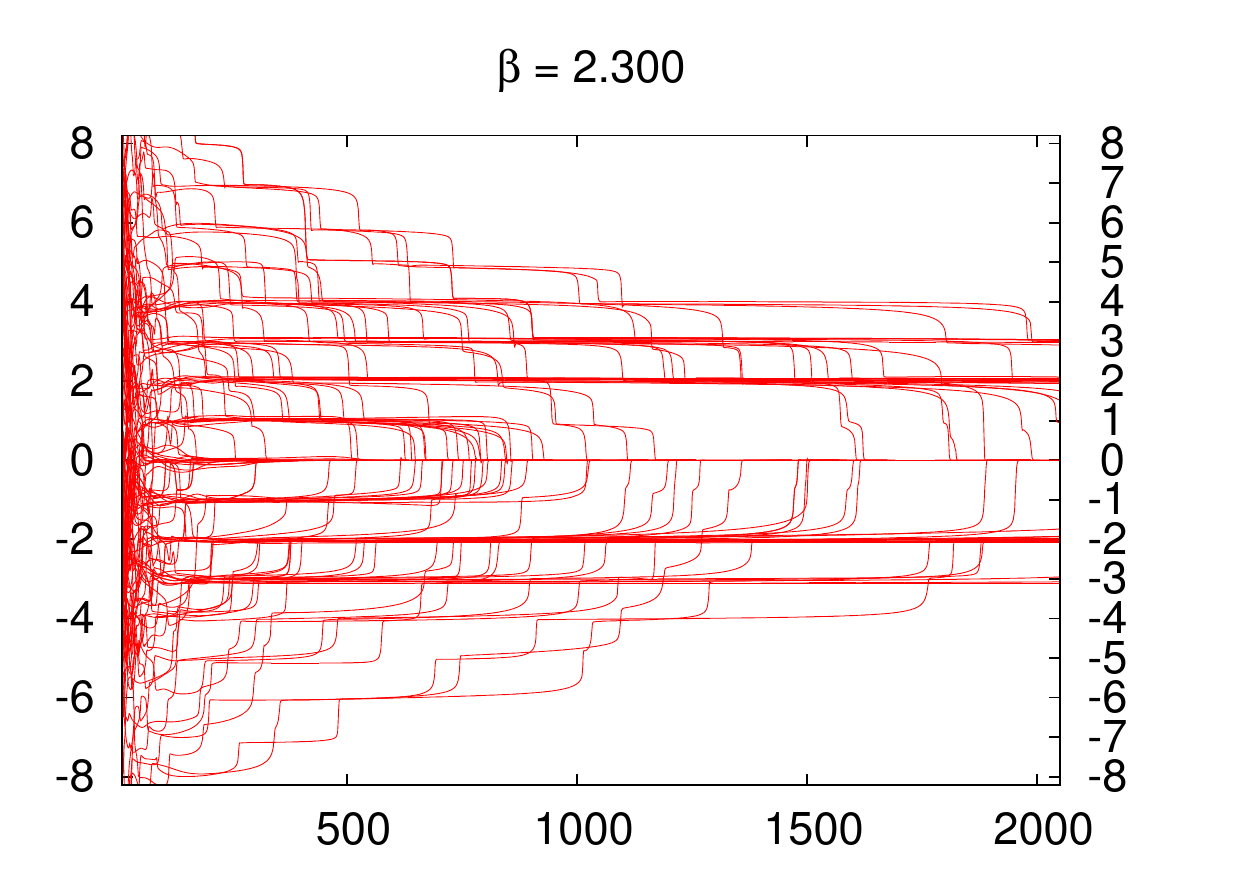}
\includegraphics[width=0.48\linewidth]{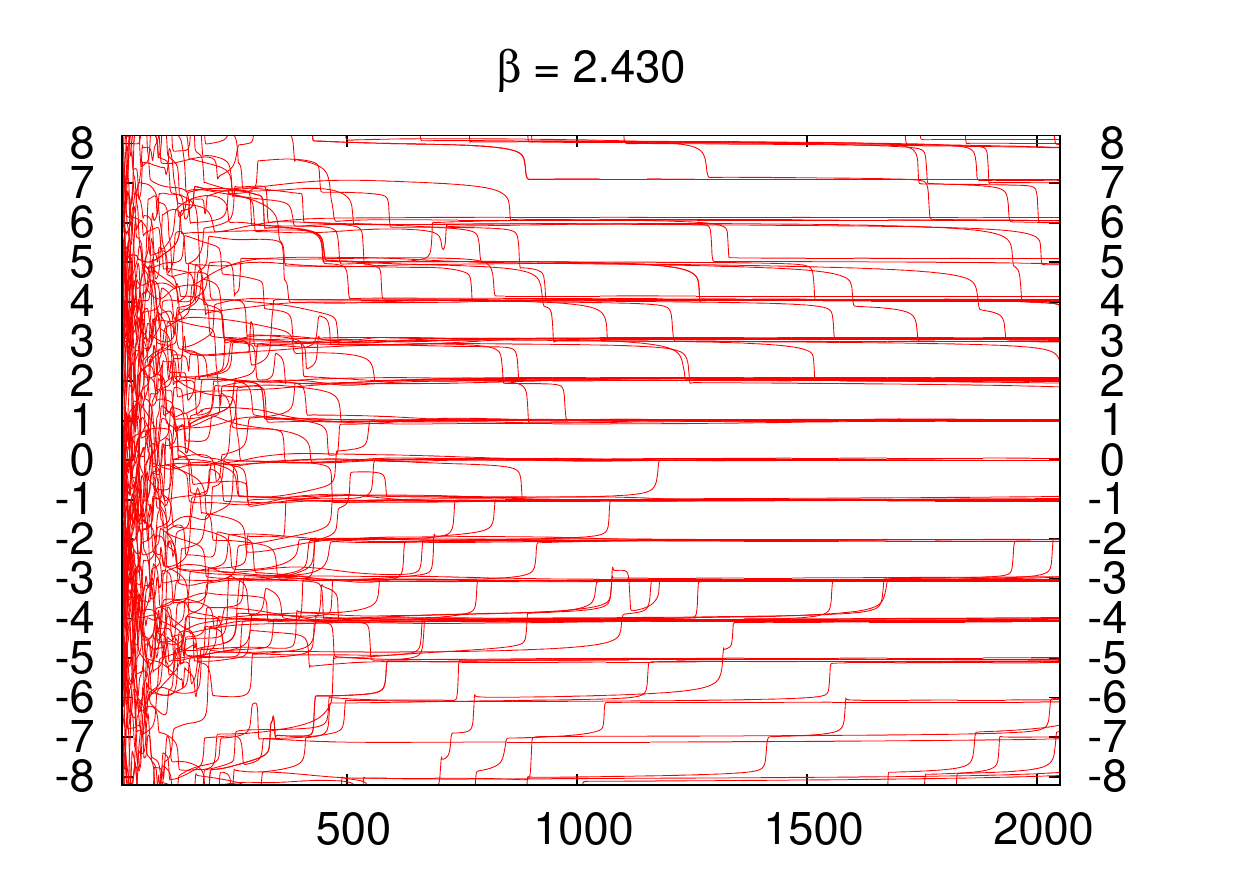}
\includegraphics[width=0.48\linewidth]{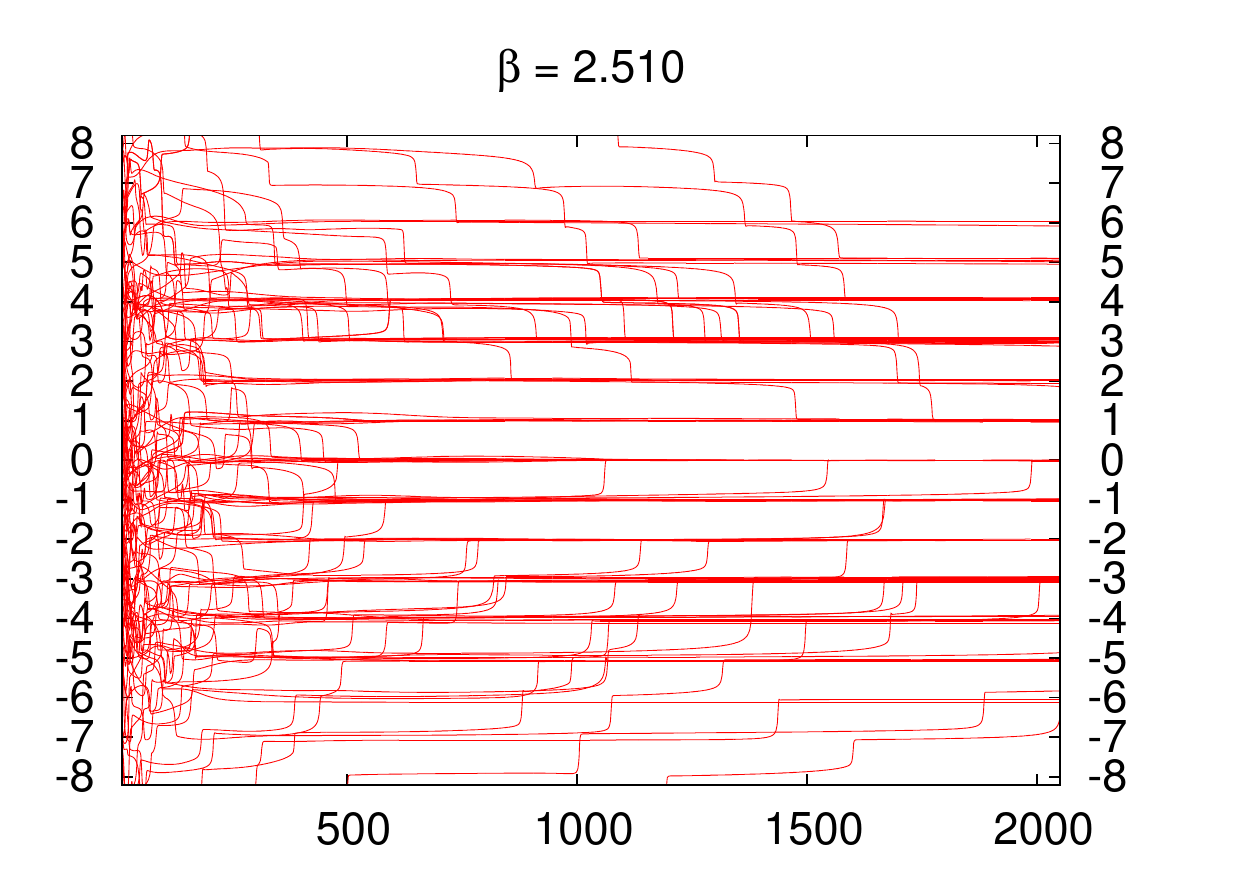}
\includegraphics[width=0.48\linewidth]{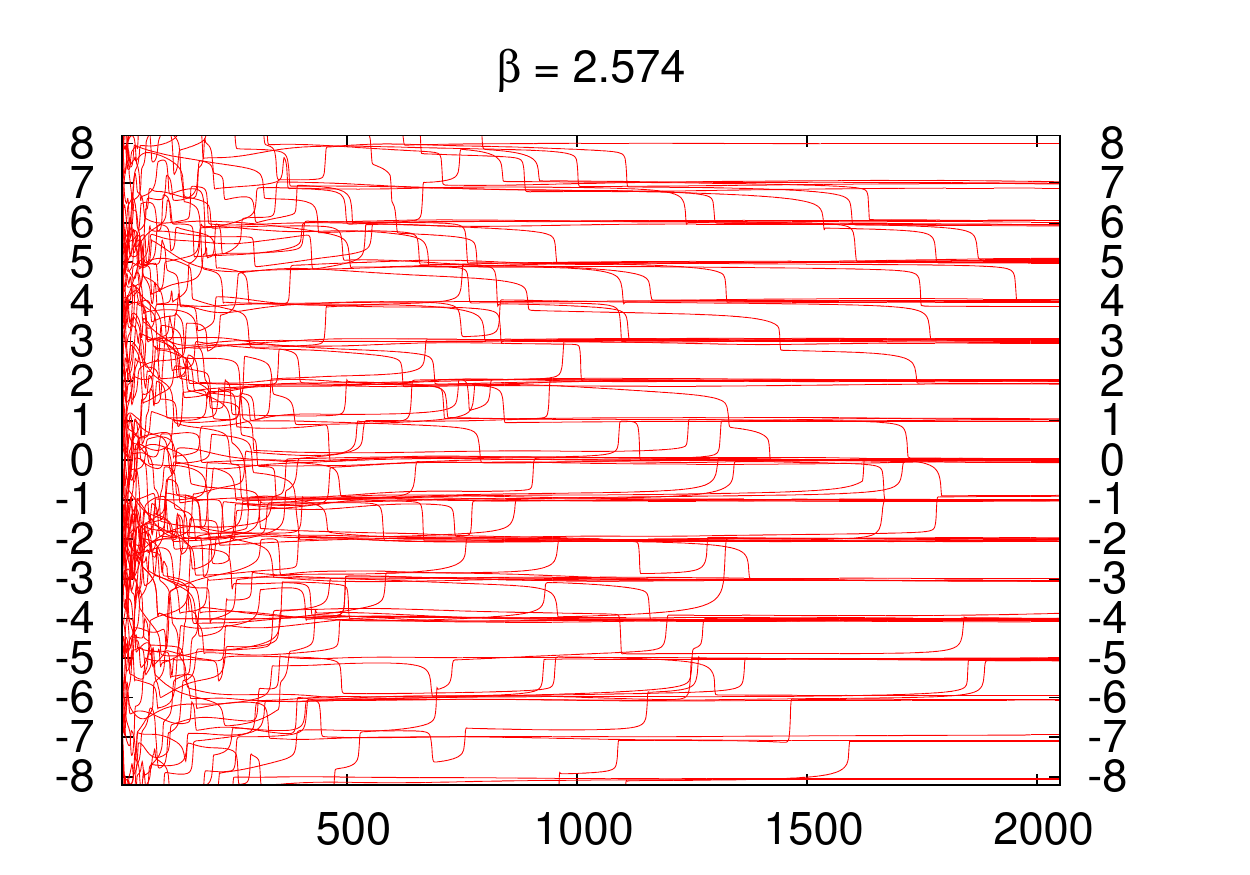}
\includegraphics[width=0.48\linewidth]{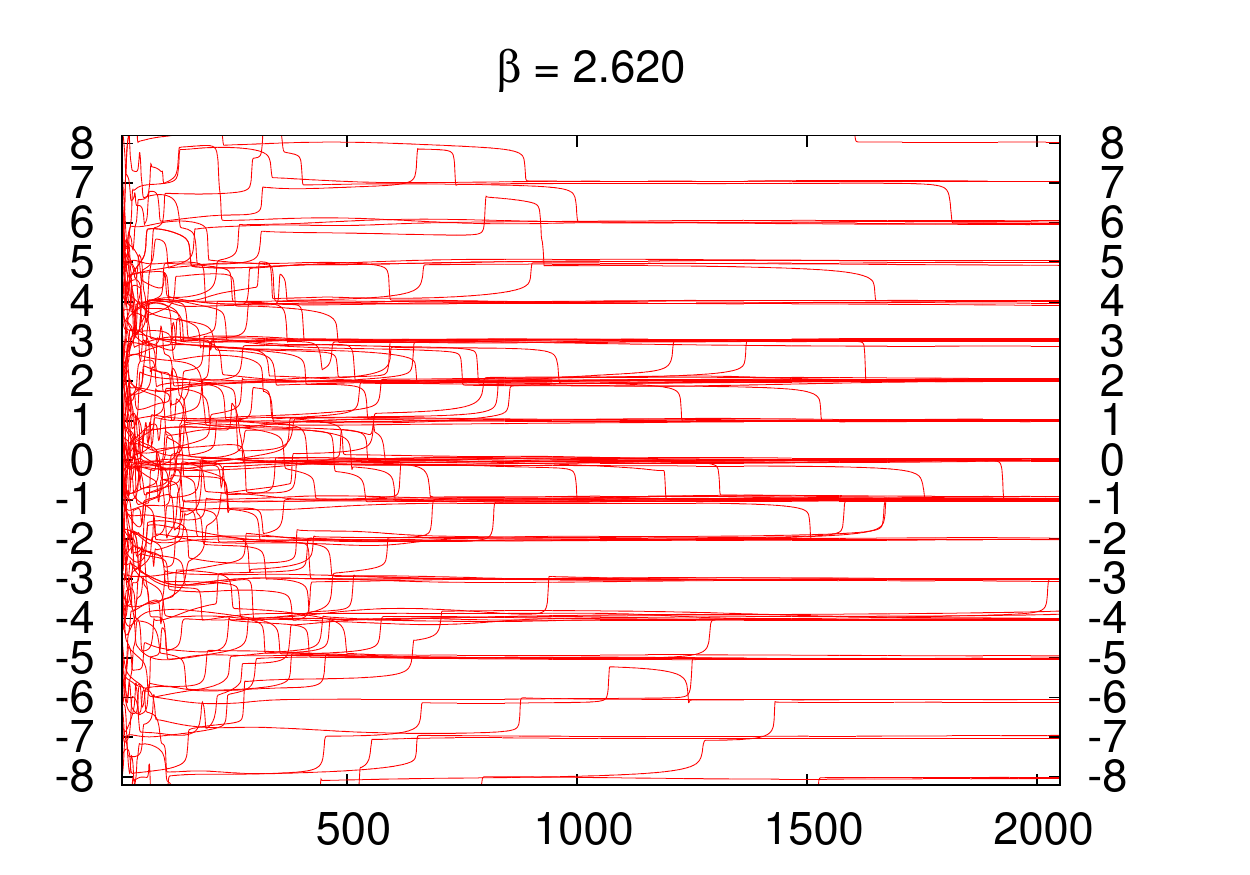}
\includegraphics[width=0.48\linewidth]{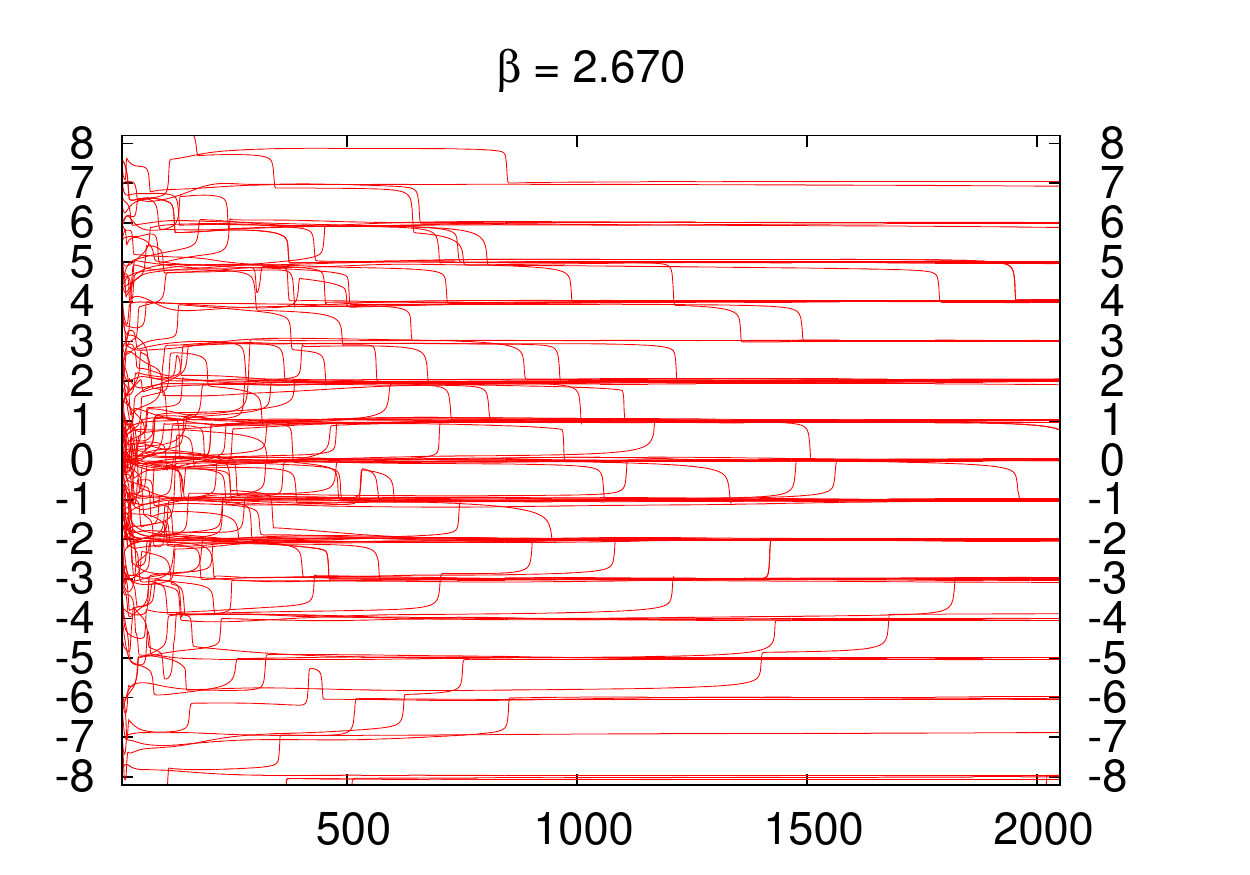}
\caption{Topological charge trajectories. Each line follows the 
         charge history of one configuration. The number of cooling 
         sweeps $n_c$ is on the abscissa, and $Q^{2048}(n_c)$ is on 
         the ordinate. The data come from our largest lattice
         at each $\beta$ value.}
\label{fig:cooltraj}
\end{figure}

\begin{figure}\ContinuedFloat
\centering
\includegraphics[width=0.48\linewidth]{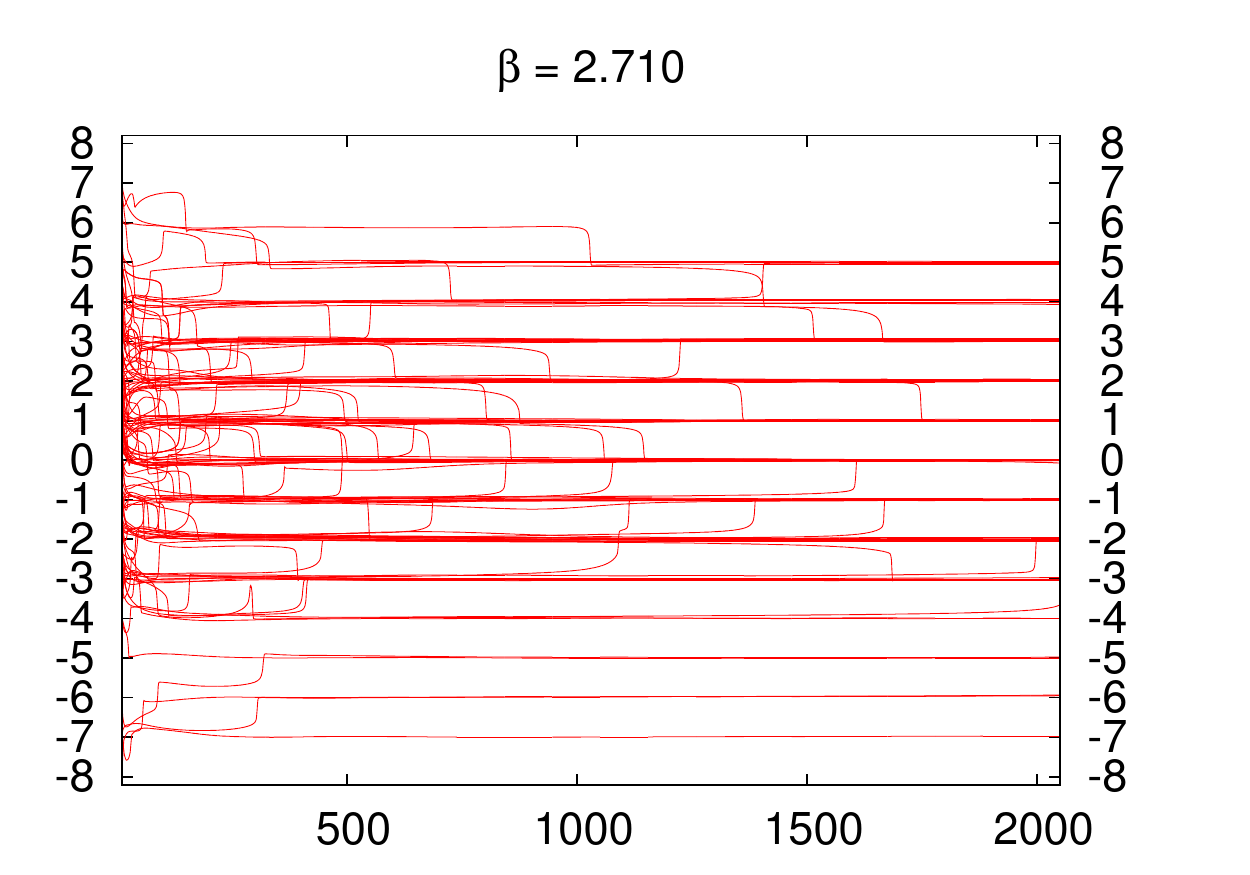}
\includegraphics[width=0.48\linewidth]{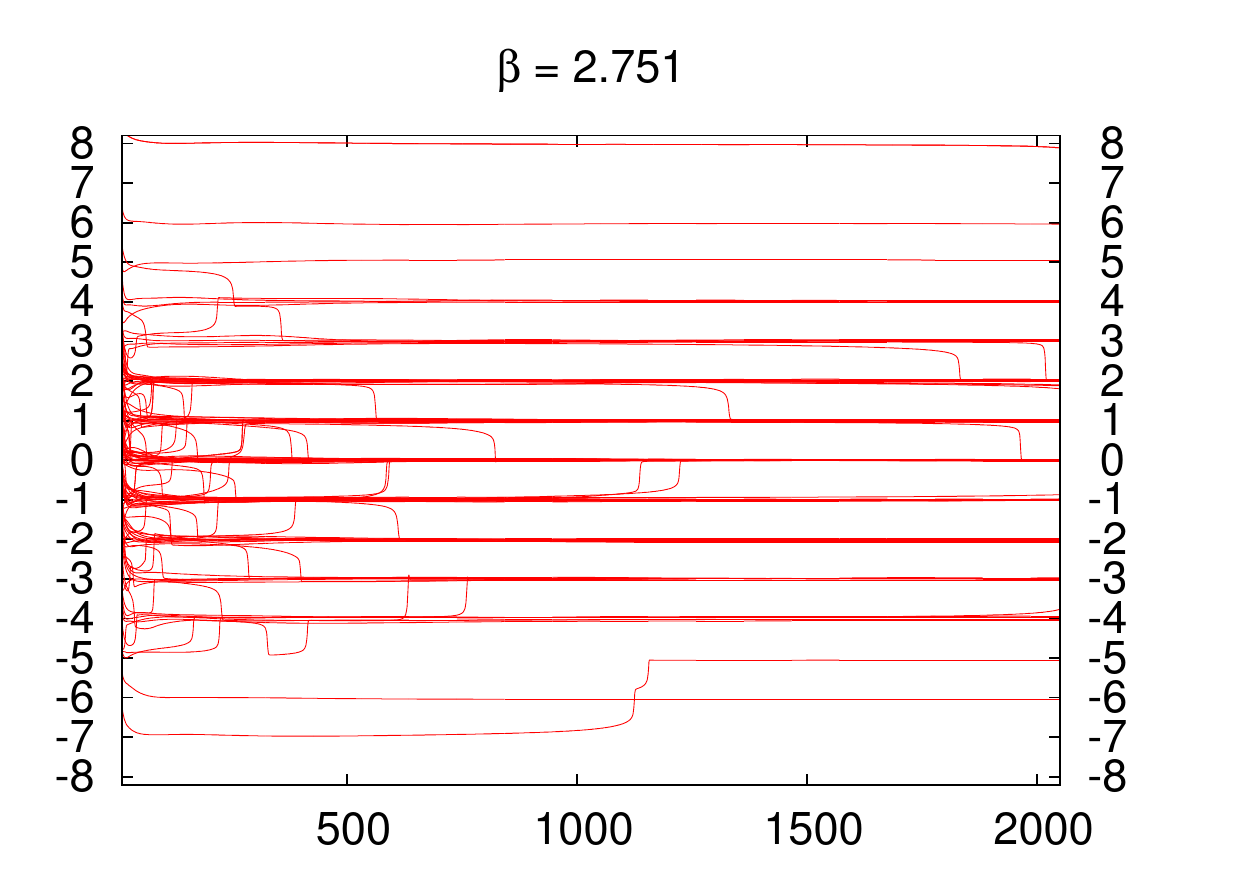}
\includegraphics[width=0.48\linewidth]{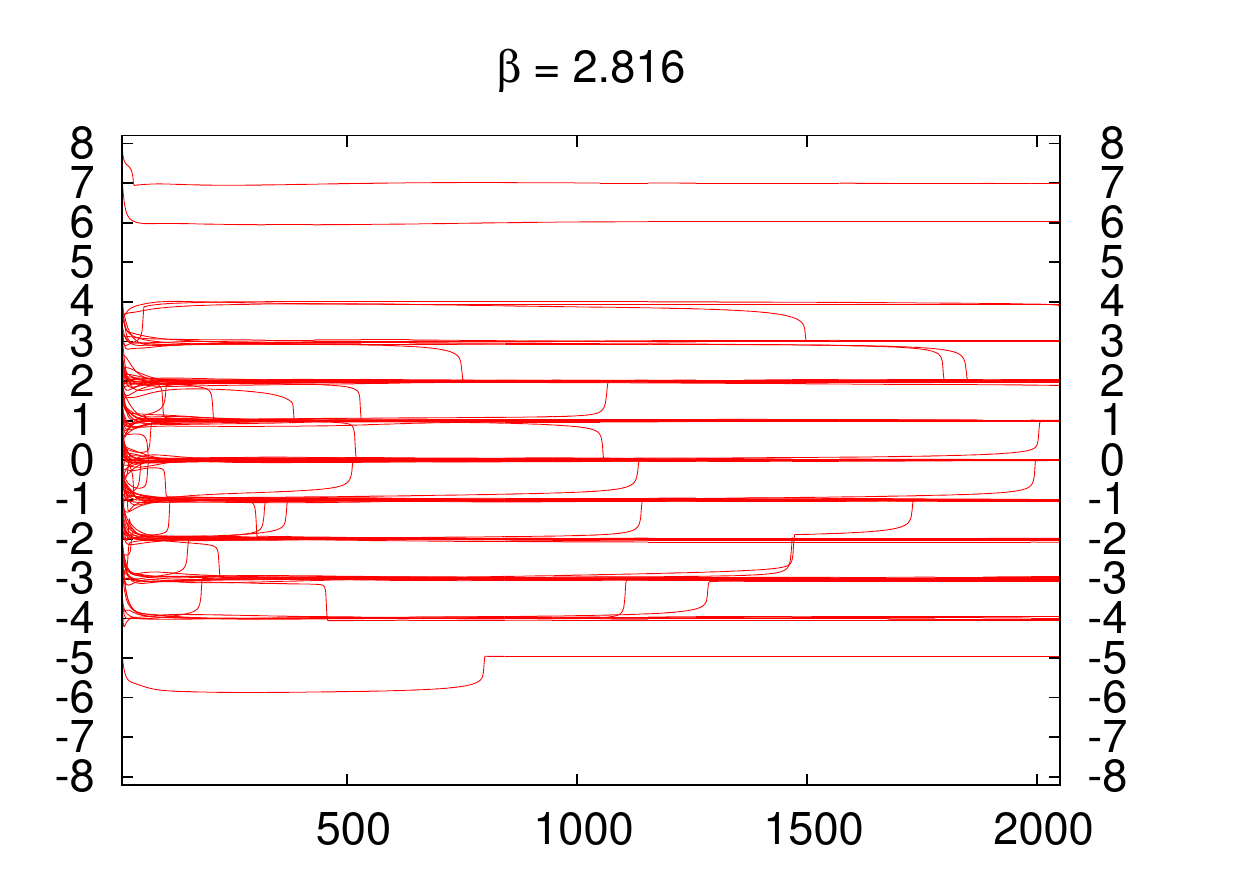}
\includegraphics[width=0.48\linewidth]{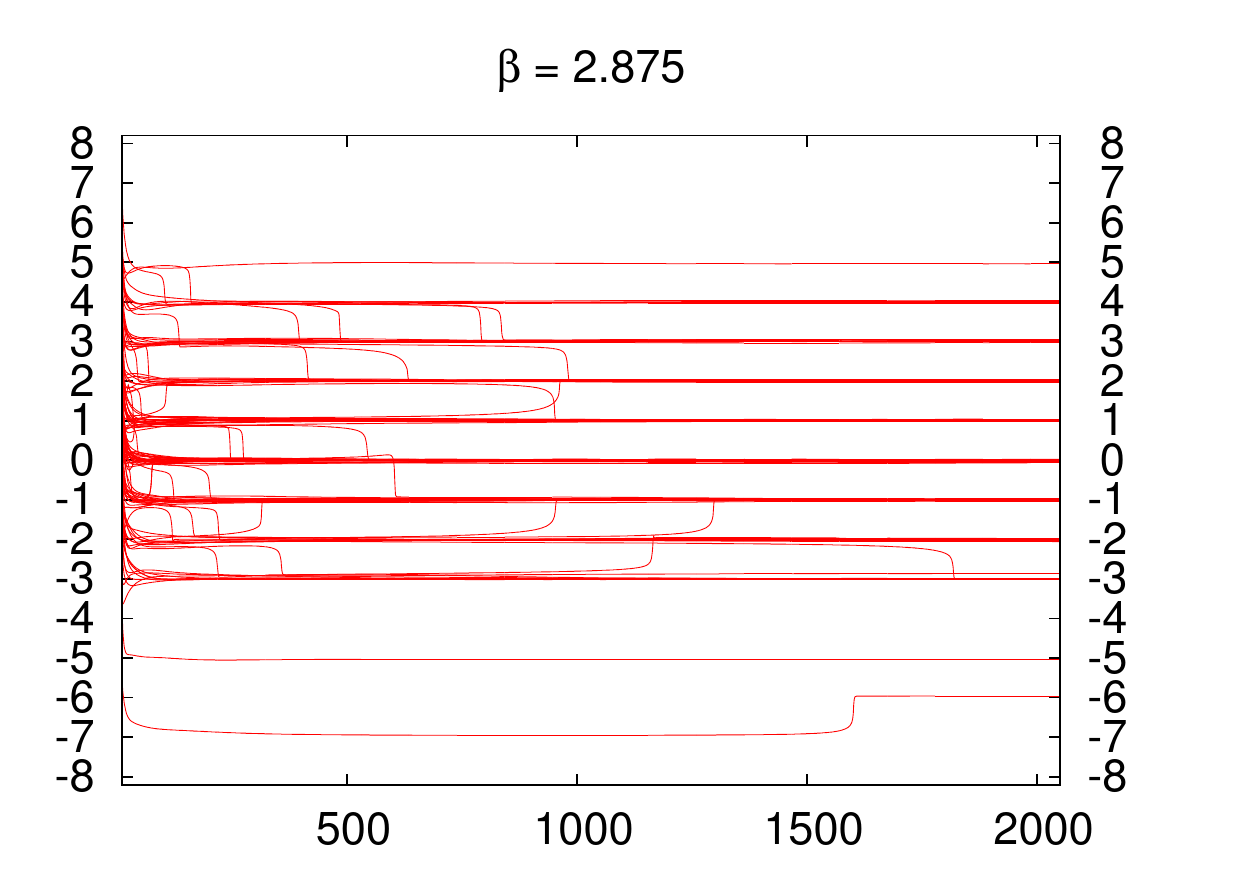}
\includegraphics[width=0.48\linewidth]{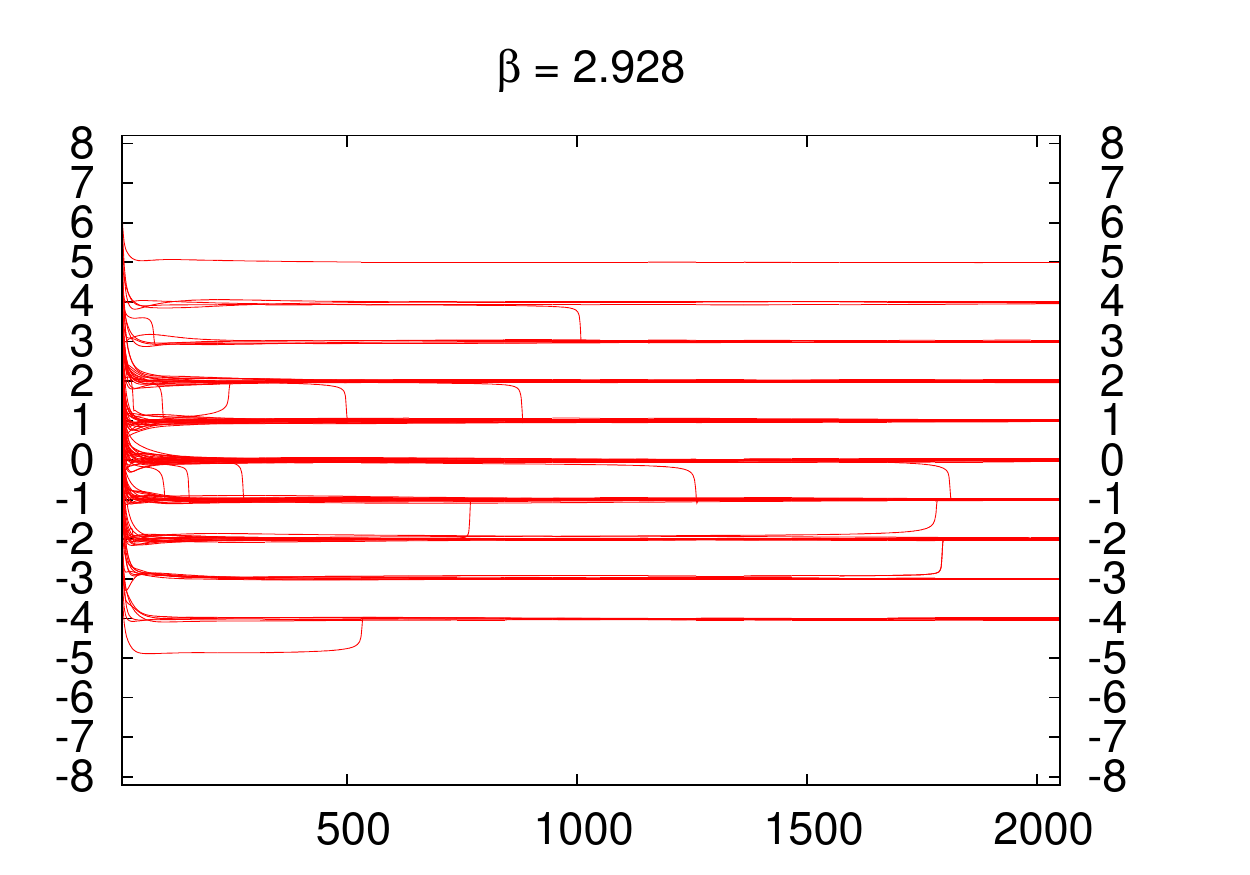}
\caption[]{Continued.}
\label{fig:cooltrajII}
\end{figure}

When approaching the continuum limit, the topological charge
must be defined at a fixed, large enough number $n_c$ of cooling 
sweeps~\cite{vicari__2009}. This number can agree with the number 
of sweeps $m_c$ used for the normalization, but it need not 
necessarily be identical. Therefore our charges 
$Q^{m_c}(n_c)$ have two labels.
In Figures~\ref{fig:cooltraj} and~\ref{fig:cooltrajII} we plot
$Q^{2048}(n_c)$ against $n_c$. Each plot corresponds to a different
$\beta$ using our largest lattice, and each line follows the topological
charge history of a configuration under cooling. These topological charge 
trajectories help to identify a fixed $n_c$ for which 
the charge is metastable. We chose to plot $Q^{2048}$ instead of $Q_I^{2048}$
to emphasize how good the mapping of eqs.~\eqref{eq:qnorm} and
\eqref{eq:minimize} is. Furthermore examining trajectories
of $Q^{2048}_I$ instead of $Q^{2048}$ does not affect our conclusion
for choosing $n_c$.

For the $\beta=2.3$ plot in Figure~\ref{fig:cooltraj}, it is
clear that a good choice for $n_c$ does not exist;
there is always a considerable number of transitions between topological
sectors, with the charges visibly cascading to zero.
Nevertheless using these $n_c=100$ and $n_c=1000$ data gives acceptable results
for the continuum limit extrapolation of the susceptibility.
The situation improves as the lattice becomes finer, with the
density of transitions in the figures decreasing. 
In addition we see that with increasing $\beta$,
it becomes easier to remove dislocations using some initial
cooling sweeps. For the largest
four or five lattices, shown in Figure~\ref{fig:cooltrajII}, we find
very few transitions over a large range of $n_c$, in particular
for $n_c\geq1000$. 

\begin{table}
\caption{Histograms of $|Q_I^{2048}(1000)|$ for the $\beta$ values and
lattices of Table~\ref{tab:normalization}.}
\begin{tabularx}{\linewidth}{lCCCCCCCCCCCCCCCCR}
\hline\hline
$\beta$& 0& 1& 2& 3& 4& 5& 6& 7& 8& 9&10&11&12&13&14&15&17\\ \hline
 2.300 &57& 4&36&20& 6& 4& 1& 0& 0& 0& 0& 0& 0& 0& 0& 0& 0\\
 2.430 & 6&22&15&15&22&10&10& 5& 7& 7& 2& 2& 1& 2& 0& 2& 0\\
 2.510 &11&21&17&23&19&17& 7& 3& 1& 5& 2& 1& 0& 0& 1& 0& 0\\
 2.574 &11&12&19&14&14&12&10&12& 5& 2& 5& 6& 3& 0& 2& 0& 1\\
 2.620 &13&18&23&19&13&13& 7& 5& 2& 3& 5& 3& 3& 1& 0& 0& 0\\
 2.670 &12&28&31&11&15&12& 9& 3& 3& 3& 1& 0& 0& 0& 0& 0& 0\\
 2.710 &20&30&33&23&11& 7& 3& 1& 0& 0& 0& 0& 0& 0& 0& 0& 0\\
 2.751 &28&37&31&16&11& 1& 2& 1& 1& 0& 0& 0& 0& 0& 0& 0& 0\\
 2.816 &24&42&32&18& 9& 1& 1& 1& 0& 0& 0& 0& 0& 0& 0& 0& 0\\
 2.875 &29&40&27&24& 5& 2& 0& 1& 0& 0& 0& 0& 0& 0& 0& 0& 0\\
 2.928 &26&49&30&12&10& 1& 0& 0& 0& 0& 0& 0& 0& 0& 0& 0& 0\\
\hline\hline
\end{tabularx} 
\label{tab:chargehist} 
\end{table}

The number $n_c=1000$ is significantly larger than what one
might have expected from previous literature; for example in Figure~3
of Ref.~\cite{bonati_comparison_2014}, the topological charge on
a $20^4$ pure $\SU(3)$ lattice at $\beta=6.2$ is defined after only
21 standard cooling sweeps. One might therefore be concerned about
the destruction of physical instantons. To give a worst-case scenario
estimate for this systematic effect, one can look at the charge histograms
of Table~\ref{tab:chargehist} along with the cooling trajectories. The total
topological charge content at $n_c=1000$ for each lattice can be determined
from the histograms. After removing initial dislocations, if we assume
that every transition toward $Q=0$ signals the destruction of a
physical instanton, which is the worst case, we find a reduction of the 
total topological charge content between $\sim$4\% for $\beta=2.928$ 
up to at most $\sim$10\% for $\beta=2.71$ and $\beta=2.751$. This
systematic effect is further suppressed for the topological
susceptibility, which depends on the square of $Q$.

To determine a lattice spacing fine enough to deliver reliable
topological sectors, we also examine the stability of the cooling 
trajectories using data given in the last column of 
Table~\ref{tab:normalization}. This column reports the fraction of 
configurations that changed charge between $n_c=1000$ and $n_c=2048$. 
Starting from about $\beta=2.574$ we see a gradually improving trend, up to
statistical fluctuations. If one desires that roughly 90\% of configurations
are metastable, we must require $\beta\gtrsim2.75$. It is also important
that the physical size of the lattice is large enough to accommodate
physical instantons. 

\section{Calculation of the topological susceptibility}\label{sec:chi2}

To investigate the scaling behavior of the topological susceptibility
\begin{equation}
  \chi=\frac{1}{\nconf}\frac{1}{N^4}\sum_{i=1}^{\nconf}\ev{Q_i^2}
\end{equation}
and correct for finite size effects, we use lattices at multiple
$\beta$, and for each $\beta$, lattices of multiple sizes. We
determined $\chi$ at $m_c=n_c=100$ and $m_c=n_c=1000$. 
For each lattice we generated $\nconf=128$
configurations.

\begin{table}
\caption{Topological susceptibility defined at 1000 and 100
         cooling sweeps. The asterisk denotes lattices that are too
         small to deliver reliable estimates.}
\begin{tabularx}{\linewidth}{LCCCCR} \hline\hline
 & & 1000 & & 100 & \\
$\beta$ &Lattice& $\chi^{1/4}$ & $L_{10}\,\chi^{1/4}$
          & $\chi^{1/4}$ & $L_{10}\,\chi^{1/4}$ \\ \hline
2.300& $16^4$ &0.0903(28) &0.1654(52)&0.1231(35) &0.2253(64)\\
2.430& $28^4$ &0.0834(27) &0.2276(72)&0.1023(33) &0.2790(89)\\
2.510& $28^4$ &0.0744(25) &0.2642(86)&0.0821(26) &0.2917(90)\\
2.574&$16^4$* &0.0510(37) &0.232(16) &0.0667(21) &0.3033(82)\\
     & $28^4$ &0.0601(18) &0.2647(77)&0.0653(26) &0.288(11)\\
     & $40^4$ &0.0609(19) &0.2666(80)&0.0677(21) &0.2963(92)\\
2.620&$16^4$* &0.0291(32) &0.169(17) &0.0562(20) &0.3272(63)\\
     & $28^4$ &0.0537(16) &0.2740(76)&0.0570(16) &0.2912(79)\\
     & $40^4$ &0.0557(19) &0.2821(93)&0.0582(19) &0.2950(94)\\
2.670&$16^4$* &0.026(26)  &0.21(21)  &0.0419(25) &0.332(13)\\
     & $28^4$ &0.0467(15) &0.2811(81)&0.0477(15) &0.2873(83)\\
     & $40^4$ &0.0484(16) &0.2860(90)&0.0511(17) &0.3020(96)\\
2.710& $16^4$*&  0        & 0        &0.0345(25) &0.341(75)\\
     & $28^4$ &0.0444(16) &0.2966(97)&0.0460(17) &0.307(11)\\
     & $40^4$ &0.0404(12) &0.2692(77)&0.0416(13) &0.2772(82)\\
2.751& $28^4$ &0.0387(15) &0.2925(96)&0.0399(16) &0.3010(98)\\
     & $40^4$ &0.0381(15) &0.286(11) &0.0385(15) &0.290(11)\\
2.816& $28^4$ &0.0305(15) &0.3195(97)&0.0327(18) &0.343(14)\\
     & $40^4$ &0.0324(12) &0.294(10) &0.0328(12) &0.298(10)\\
     & $44^4$ &0.0332(12) &0.3010(96)&0.0336(12) &0.3045(96)\\
2.875&$28^4$* &0.0227(16) &0.333(12) &0.0390(17) &0.3512(94)\\
     & $40^4$ &0.02748(89)&0.3017(87)&0.02800(96)&0.3074(93)\\
     & $44^4$ &0.02681(92)&0.2980(92)&0.0270(11) &0.300(11)\\
     & $52^4$ &0.02760(92)&0.3002(97)&0.02822(93)&0.3070(97)\\
2.928&$28^4$* &0.0173(17) &0.345(12) &0.0173(17) &0.345(14)\\
     & $40^4$ &0.0235(11) &0.3287(97)&0.0235(11) &0.3286(98)\\
     & $44^4$ &0.02492(77)&0.3185(75)&0.02534(85)&0.3239(84)\\
     & $52^4$ &0.02359(81)&0.3002(94)&0.02360(80)&0.3003(93)\\
     & $60^4$ &0.02297(70)&0.2885(84)&0.02313(72)&0.2906(87)\\
\hline\hline
\end{tabularx}
\label{tab:suscept}
\end{table}

Results for $\chi^{1/4}$ are given in
Table~\ref{tab:suscept}. We set the scale with the cooling length
$L_{10}$, defined in Chapter~\ref{ch:comparison}. Due to scaling,
one expects the dimensionless product $L_{10}\chi^{1/4}$ to 
approach a constant in the continuum limit. Estimates of 
$L_{10}\chi^{1/4}$ are also reported in Table~\ref{tab:suscept}.
To obtain error bars for each quantity, $\chi^{1/4}$ and $L_{10}\chi^{1/4}$
are calculated in 128 jackknife bins on every lattice.
Smaller lattices for $\beta=2.3$, 2.43, and 2.51 were not examined,
since the lattices listed in Table~\ref{tab:suscept} were previously 
found to be large enough to neglect finite size corrections.

\begin{figure}
\centering
\includegraphics[width=0.7\linewidth]{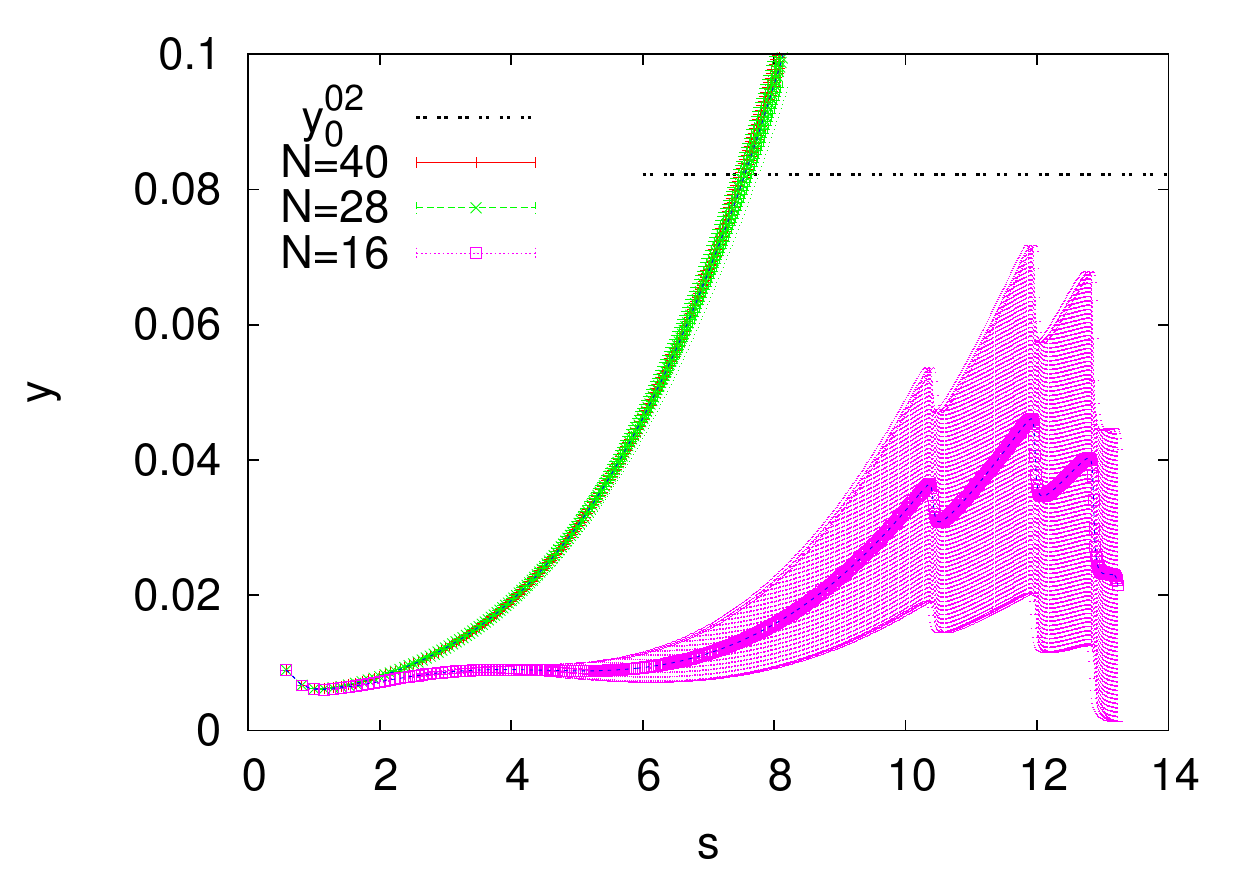}
\caption{Cooling trajectories with error bars at $\beta=2.751$ for 
         different lattice sizes. The square root of the cooling
         flow time $\sqrt{n_c}$ is on the abscissa, while
         the target function $n_c^2\,E_0$ is on the ordinate.
         The dashed line indicates the $L_{10}$
         target value $y_0^2=0.0822$. For the $16^4$ lattice,
         the $L_{10}$ trajectory fails to attain its target,
         while the $N=40$ and $N=28$ trajectories fall on top
         of one another.}
\label{fig:breakcool}
\end{figure}

\begin{figure}
\centering
\includegraphics[width=0.7\linewidth]{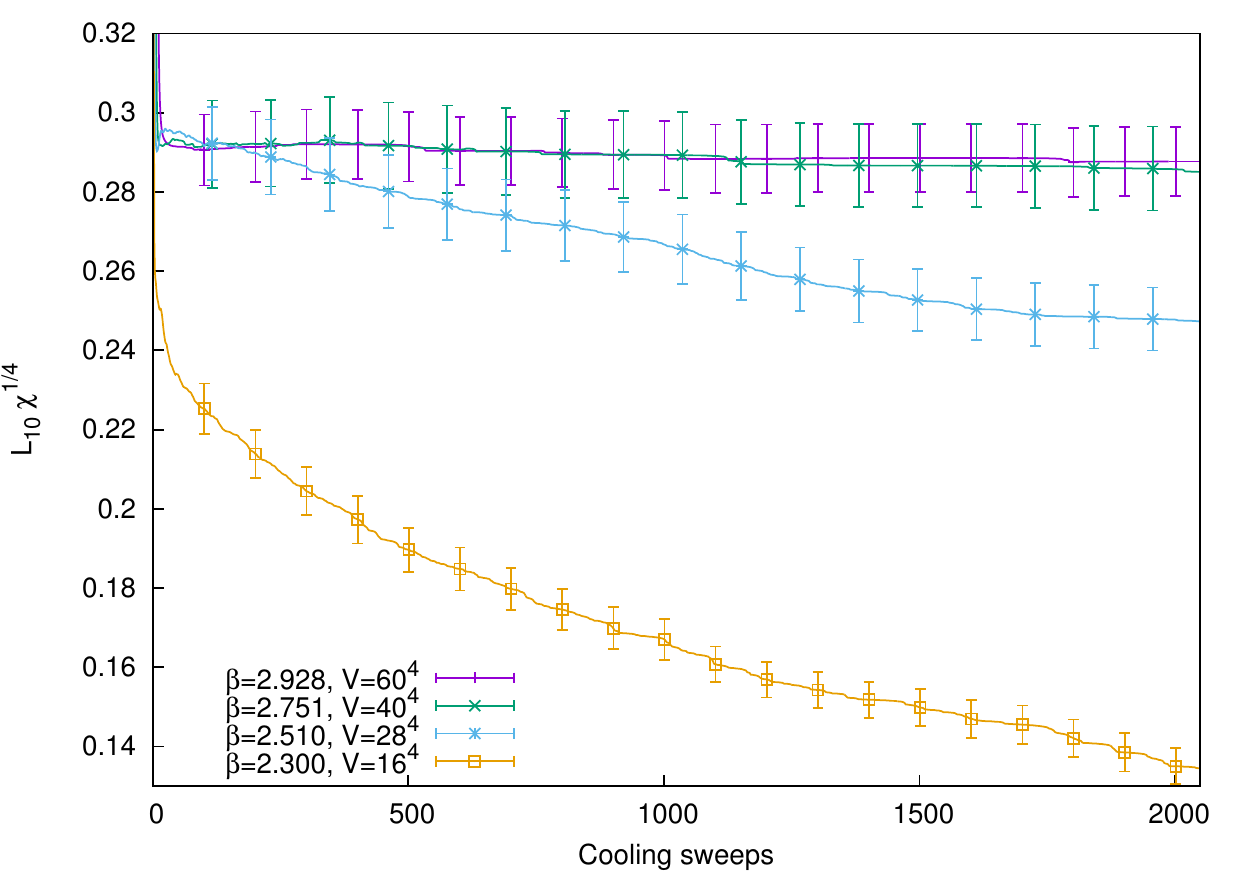}
\caption{Cooling trajectories for $L_{10}\chi^{1/4}$ for different
         $\beta$ on their largest lattices. Error bars are sparsely
         plotted for better visibility.}
\label{fig:s1o4}
\end{figure}

Lattices marked with an asterisk in the second column were too small
to deliver reliable data. 
For instance, the topological susceptibility
was found to be 0 for the $16^4$ lattice at $\beta=2.71$, because the
charge was 0 on every configuration. 
This indicates that the physical
size of the lattice is too small to accommodate instantons. 
Additionally $L_{10}$ breaks down when the physical size of the
lattice is too small. For $16^4$ lattices, this happens for
$\beta\gtrsim2.751$, and the effect is illustrated in 
Figure~\ref{fig:breakcool}. In this figure, the trajectories
of the target function $n_c^2\,E_0$ are given
as a function of the cooling time. While the trajectories
for the $40^4$ and $28^4$ fall on top of one another, the
$16^4$ trajectory fails to reach the $L_{10}$ target
value $y_0^2=0.822$. 

As another check for metastability and retention of physical instantons,
we examine the behavior of $L_{10}\chi^{1/4}$ under cooling.
Figure~\ref{fig:s1o4} plots $L_{10}\chi^{1/4}$ against the
number of cooling sweeps for two of our lowest $\beta$ and two
of our largest $\beta$. Error bars are plotted only every 100 sweeps
to increase visibility. The $\beta=2.928$ and $\beta=2.751$
trajectories fall on top of one another. For both lattices, any
decrease is relatively minute, and dwarfed entirely by the
statistical error. 
This gives another indication that we have achieved metastable
topological sectors on our finest lattices, with almost no
destruction of physical instantons, provided the physical size is
large enough. 
By contrast, $L_{10}\chi^{1/4}$ is seen to decrease almost monotonically 
throughout the entire cooling process on our coarsest lattices, 
with the situation greatly improving as $\beta$ increases.
This is because on coarser lattices, there exists a higher fraction
of exceptional configurations in configuration space, making it easier
for the cooling process to lower the action by changing the topological 
charge.

\begin{table}
\caption{Results of finite size fits for $L_{10}\chi^{1/4}$.}
\begin{tabularx}{\linewidth}{LCCCR}
\hline\hline
         & 1000               &     & 100                &     \\
 $\beta$ & $L_{10}\chi^{1/4}$ & $q$ & $L_{10}\chi^{1/4}$ & $q$ \\\hline
2.928 &0.273(12) &0.92 &0.275(12) &0.72 \\
2.875 &0.298(19) &0.78 &0.305(20) &0.57 \\
2.816 &0.289(11) &0.48 &0.285(12) &0.40 \\
2.751 &0.287(11) &     &0.290(11) & \\
2.71  &0.2692(77)&     &0.2772(82)& \\
2.67  &0.2860(90)&     &0.3020(96)& \\
2.62  &0.2821(93)&     &0.2950(94)& \\
2.574 &0.2666(80)&     &0.2963(92)& \\
\hline\hline
\end{tabularx} 
\label{tab:fsschi} 
\end{table}

\begin{figure}
\centering
\includegraphics[width=0.5\linewidth,angle=270]{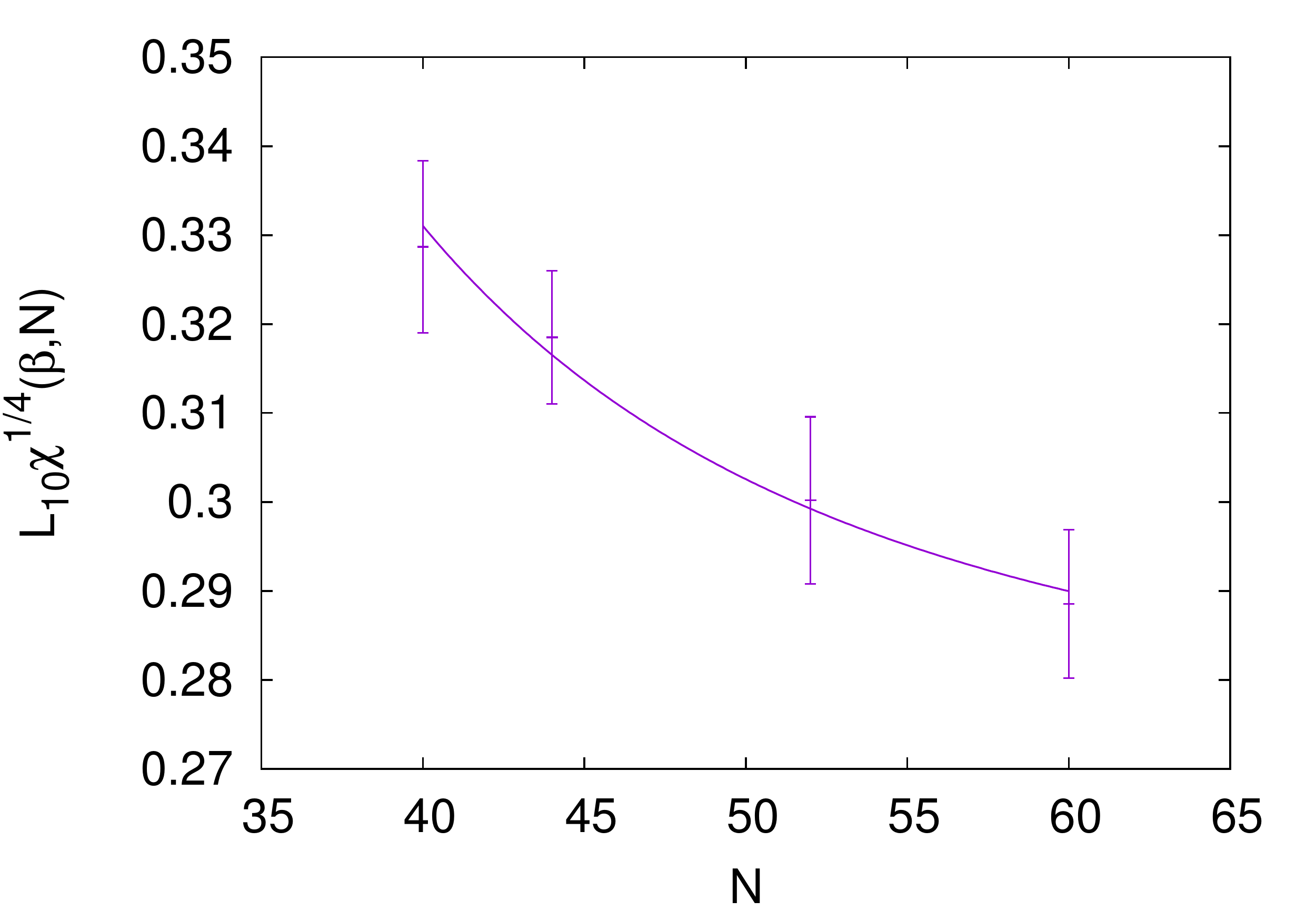}
\caption{Example finite size scaling fit of $L_{10}\chi^{1/4}$ for
         $\beta=2.928$.}
\label{fig:fsschi}
\end{figure}

\begin{figure}
\centering
\includegraphics[width=0.7\linewidth]{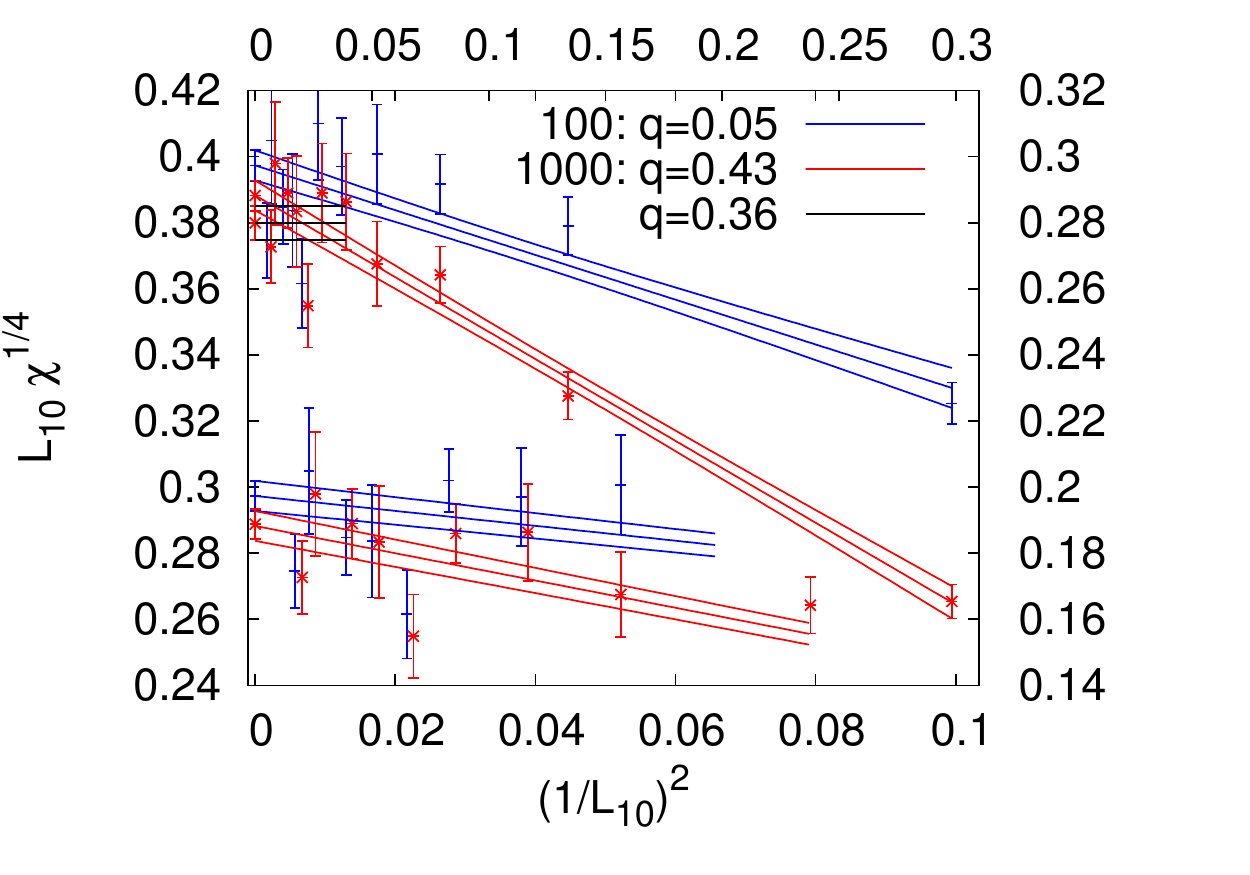}
\caption{Scaling of $L_{10}\chi^{1/4}$. The upper part of the figure uses
         the top abscissa and right ordinate. The lower part of the figure
         is an enhancement of the scaling fits deeper in the scaling
         region. It uses the bottom abscissa and left ordinate. The black
         lines give a fit to a constant. The goodness-of-fit is 
         reported in the key.}
\label{fig:BAV}
\end{figure}

We now turn to finite size scaling analysis for $L_{10}\chi^{1/4}$.
For this purpose we employ a two-parameter fit
\begin{equation}
  L_{10}\chi^{1/4}(\beta,V)=L_{10}\chi^{1/4}(\beta)+\frac{\alpha}{V},
\end{equation}
where the fit parameters are $L_{10}\chi^{1/4}(\beta)$ and $\alpha$,
and $V=N^4$.
This can be viewed as an effective fit, chosen partly because the
bias of the susceptibility is expected to be $1/V$, and partly because
for some $\beta$ we have only two reliable lattice sizes.
We fit the data in Table~\ref{tab:suscept} modulo the unreliable lattices.
Results of the finite size fit are given in Table~\ref{tab:fsschi}, 
and are seen to be consistent with the fit form. For $\beta=2.574$, 2.62, 2.67,
2.71, and 2.751, we have two-parameter fits with only two data points,
so there is no goodness-of-fit to report.
For $\beta=2.3$, 2.43, and 2.51, the
result from the single lattice listed in Table~\ref{tab:suscept} will
be used for the scaling analysis.
An example finite size fit for $\beta=2.928$ is shown in
Figure~\ref{fig:fsschi}.

In Figure~\ref{fig:BAV} we show different continuum limit
fits of the thus obtained data.
The upper part of the figure uses the upper abscissa and right ordinate,
while the lower inlay uses the bottom abscissa and left ordinate.
Using the $L_{10}\,\chi^{1/4}$ estimates down to $\beta=2.3$, linear
fits to  $a^2$ scaling corrections given by $1/(L_{10})^2$ are shown
in the upper part of the figure along with their error bar ranges, while
the lower part shows an enhancement. 
The continuum limit extrapolations are
\begin{eqnarray} \label{eq:L10s1o4L1000}
  L_{10}\,\chi^{1/4}&=&0.2882(46),~q=0.43~{\rm for}~n_c=1000,
  ~~\\ \label{eq:L10s1o4L0100}
  L_{10}\,\chi^{1/4}&=&0.2961(49),~q=0.05~{\rm for}~n_c=100.
\end{eqnarray}

Although the fits to $a^2$ scaling corrections work well, one may question
whether the $L_{10}\,\chi^{1/4}$ results at $\beta=2.3$ and 2.43 and to
some extent also at $\beta=2.51$ and 2.574 are really reliable. In short,
one could argue in favor or against taking out all $\beta$ values for
which the susceptibility after $n_c=100$ cooling sweeps is significantly
larger than after $n_c=1000$ cooling sweeps. Taking them out and fitting
the remaining points to $L_{10}\,\chi^{1/4}=constant$, one obtains the
estimates
\begin{eqnarray} \label{eq:L10s1o4C1000}
  L_{10}\,\chi^{1/4}&=&0.2799(51),~q=0.36~{\rm for}~n_c=1000,
  \\ \label{eq:L10s1o4C0100}
  L_{10}\,\chi^{1/4}&=&0.2844(54),~q=0.25~{\rm for}~n_c=100.
\end{eqnarray}
To avoid overloading Figure~\ref{fig:BAV}, the fit to a constant is only
indicated for $n_c=1000$ in the upper part of the figure in black.
Averaging eq.~\eqref{eq:L10s1o4L1000} with \eqref{eq:L10s1o4C1000}, and
eq.~\eqref{eq:L10s1o4L0100} with \eqref{eq:L10s1o4C0100}, we obtain
\begin{eqnarray} \label{eq:L10s1000}
  L_{10}\,\chi^{1/4}&=&0.2841(49)~~{\rm for}~~n_c=1000\,,
  \\ \label{eq:L10s0100}
  L_{10}\,\chi^{1/4}&=&0.2903(52)~~{\rm for}~~n_c=100\,.
\end{eqnarray}
To relate $\chi^{1/4}$ to physical scales, we use
$(T_c\,L_{10})^{-1}=2.2618(42)$, which is taken from 
Table~\ref{tab:scaleratios}.
Propagating the statistical errors, we obtain from
eqs.~\eqref{eq:L10s1000} and~\eqref{eq:L10s0100}
\begin{eqnarray} \label{eq:sTc1000}
  \chi^{1/4}/T_c&=&0.643(12)~~{\rm for}~~n_c=1000\,,\\ 
  \label{eq:sTc0100}
  \chi^{1/4}/T_c&=&0.657(12)~~{\rm for}~~n_c=100\,.
\end{eqnarray}
In the literature $\chi^{1/4}$ for SU(2) LGT has been reported in units
of the square root of the string tension $\sqrt{\sigma}$. The 
most accurate estimate of
$T_c/\sqrt{\sigma}$ appears to be $T_c/\sqrt{\sigma}=0.7091\,(36)$
from Lucini {\it et. al.}~\cite{lucini_high_2004}, which is 
consistent with the earlier value $T_c/\sqrt{\sigma}=0.69\,(2)$ of 
Fingberg {\it et. al.}~\cite{fingberg_scaling_1993}. 
Using the former estimate along with
propagation of uncertainty, our estimates \eqref{eq:sTc1000} 
and \eqref{eq:sTc0100} convert to
\begin{eqnarray} \label{eq:soss1000}
  \chi^{1/4}/\sqrt{\sigma}&=&0.4557(83)~~{\rm for}~~n_c=1000\,,
  \\ \label{eq:soss0100}
  \chi^{1/4}/\sqrt{\sigma}&=&0.4655(88)~~{\rm for}~~n_c=100\,.
\end{eqnarray}

\begin{table}
\caption{Estimates of the topological susceptibility in units of
         the square root of the string tension. The third and fourth
         columns give Gaussian difference tests with our
         $n_c=1000$ and $n_c=100$ estimates, respectively.}
\begin{tabularx}{\linewidth}{LCCR} \hline\hline\noalign{\vskip 1mm}
(year) [Reference]&$\chi^{1/4}/\sqrt{\sigma}$&$q_{1000}$&$q_{100}$\\ \hline
(1997) \cite{de_forcrand_topology_1997} & 0.501(45)  & 0.32& 0.44  \\
(1997) \cite{degrand_topological_1997}  & 0.528(21)  & 0.00& 0.01  \\
(1997) \cite{alles_topology_1997}       & 0.480(23)  & 0.32& 0.56  \\
(2001) \cite{lucini_su_2001}            & 0.4831(56) & 0.01& 0.09  \\
(2001) \cite{lucini_su_2001}            & 0.4745(63) & 0.07& 0.40  \\
(2001) \cite{lucini_su_2001}            & 0.4742(56) & 0.06& 0.40  \\ 
\hline\hline
\end{tabularx} 
\label{tab:pastchi}
\end{table}

Past estimates for $\chi^{1/4}/\sqrt{\sigma}$ are compiled in
Table~\ref{tab:pastchi}. The last two columns report Gaussian difference
tests between our results~\eqref{eq:soss1000} and \eqref{eq:soss0100}
and the corresponding literature result. Both of our estimates are
lower than the literature estimates, which is not surprising since
$\chi$ decreases with increasing $n_c$. Past results for
$\chi^{1/4}/\sqrt{\sigma}$ relied on smaller lattices and $\beta$
for which only small $n_c$ can be used. So it appears that even 
$n_c=100$ is too small. We favor our $n_c=1000$
results~\eqref{eq:sTc1000} and \eqref{eq:soss1000}.

\begin{table}
\caption{Cooling scales in topological sectors. The data come from
         our largest available lattice at each $\beta$ value. The
         second column labels the topological charge group, and
         the third column gives the number of configurations
         in each group.}
\begin{tabularx}{\linewidth}{lCcCCCCCR} \hline\hline\noalign{\vskip 1mm}
$\beta$ & $|Q|$&$\nconf$& $L_7$   & $L_8$     & $L_9$          
                  	& $L_{10}$  & $L_{11}$  & $L_{12}$  \\ \hline
2.928	& 0 	& 26    & 9.85(15)  & 9.76(15)  & 9.07(15) 
			& 12.61(23) & 12.55(23) & 11.66(21) \\
	& 1 	& 49	& 9.93(13)  & 9.83(13)  & 9.06(13) 
			& 12.74(18) & 12.68(17) & 11.66(18) \\
	&$\ge2$ & 53 	& 9.750(92) & 9.650(90) & 9.040(97) 
			& 12.39(14) & 12.34(14) & 11.64(14) \\
2.875	& 0	& 29	& 8.64(16)  & 8.55(16)  & 7.89(19) 
			& 11.16(25) & 11.11(25) & 10.31(24) \\
	& 1	& 40	& 8.58(12)  & 8.50(12)  & 7.86(12) 
			& 11.02(17) & 10.97(17) & 10.15(18) \\
	&$\ge2$ & 59 	& 8.416(73) & 8.338(72) & 7.771(89) 
			& 10.68(10) & 10.633(99)& 10.02(12) \\
2.816	& 0	& 24	& 7.281(99) & 7.212(98) & 6.68(12)  
			& 9.32(15)  & 9.27(15)  & 8.63(16)  \\
	& 1	& 42	& 7.103(75) & 7.036(74) & 6.540(93) 
			& 9.06(12)  & 9.02(12)  & 8.41(12)  \\
	&$\ge2$ & 62	& 7.044(66) & 6.979(65) & 6.435(80) 
			& 8.964(91) & 8.924(91) & 8.22(11)  \\
2.751	& 0	& 28	& 5.878(70) & 5.822(69) & 5.381(66) 
			& 7.55(11)  & 7.52(11)  & 7.006(95) \\
	& 1	& 37	& 5.895(63) & 5.840(62) & 5.416(75) 
			& 7.542(96) & 7.507(95) & 7.10(11)  \\
	&$\ge2$ & 63	& 5.882(43) & 5.828(43) & 5.382(51) 
			& 7.491(61) & 7.456(62) & 6.920(65) \\
2.710	& 0	& 20	& 5.277(66) & 5.227(65) & 4.803(59) 
			& 6.750(90) & 6.720(90) & 6.185(97) \\
	& 1	& 30	& 5.229(48) & 5.179(47) & 4.825(73) 
			& 6.707(77) & 6.676(73) & 6.267(92) \\
	&$\ge2$ & 78	& 5.175(24) & 5.127(24) & 4.781(34) 
			& 6.615(34) & 6.585(34) & 6.161(45) \\
\hline\hline
\end{tabularx} 
\label{tab:sectorscales}
\end{table}

\section{Dependence of cooling scales on topological sector}\label{sec:coolcor}

For $\beta\ge 2.71$ we calculated cooling scales with lattice sizes
given according to Table~\ref{tab:normalization}, then grouped
them by according to the topological sectors 
with charges $Q$ calculated at $n_c=m_c=1000$.
We performed Student difference tests between $L_i(Q_1)$ and $L_i(Q_2)$
for $7\leq i\leq12$ and $Q_1\neq Q_2$, then looked at resulting
$q$-values to determine whether a scale gives different results
when calculated in different sectors.

In this way, we determined that within the available statistics, all 
scales with $Q\ge 2$ are consistent with each other, and similarly all scales 
with $Q\le-2$ agree. Because of this agreement, and because sectors with
$Q>2$ commonly have fewer than 10 configurations belonging to them, we
regrouped cooling scales according to sectors
$Q\le -2$, $Q=-1$, $Q=0$, $Q=1$ and $Q\ge 2$.
After this regrouping, we still find no statistically significant
differences when comparing a cooling scale calculated in sectors
$Q$ and $-Q$. To again increase the statistics for the $Q\ne0$ sectors, 
we therefore combined them into $|Q|=1$ and $|Q|\ge 2$.
This regrouping achieves reasonable statistics; the results
for the cooling scales are given in Table~\ref{tab:sectorscales}. 
The scales $L_7$ and $L_8$, as well as
$L_{10}$ and $L_{11}$, almost agree because the fluctuations of
the operators $E_0$ and $E_1$ are strongly correlated and almost
identical. Therefore they are averaged in the following.

\begin{figure}
\centering
\includegraphics[width=0.6\linewidth]{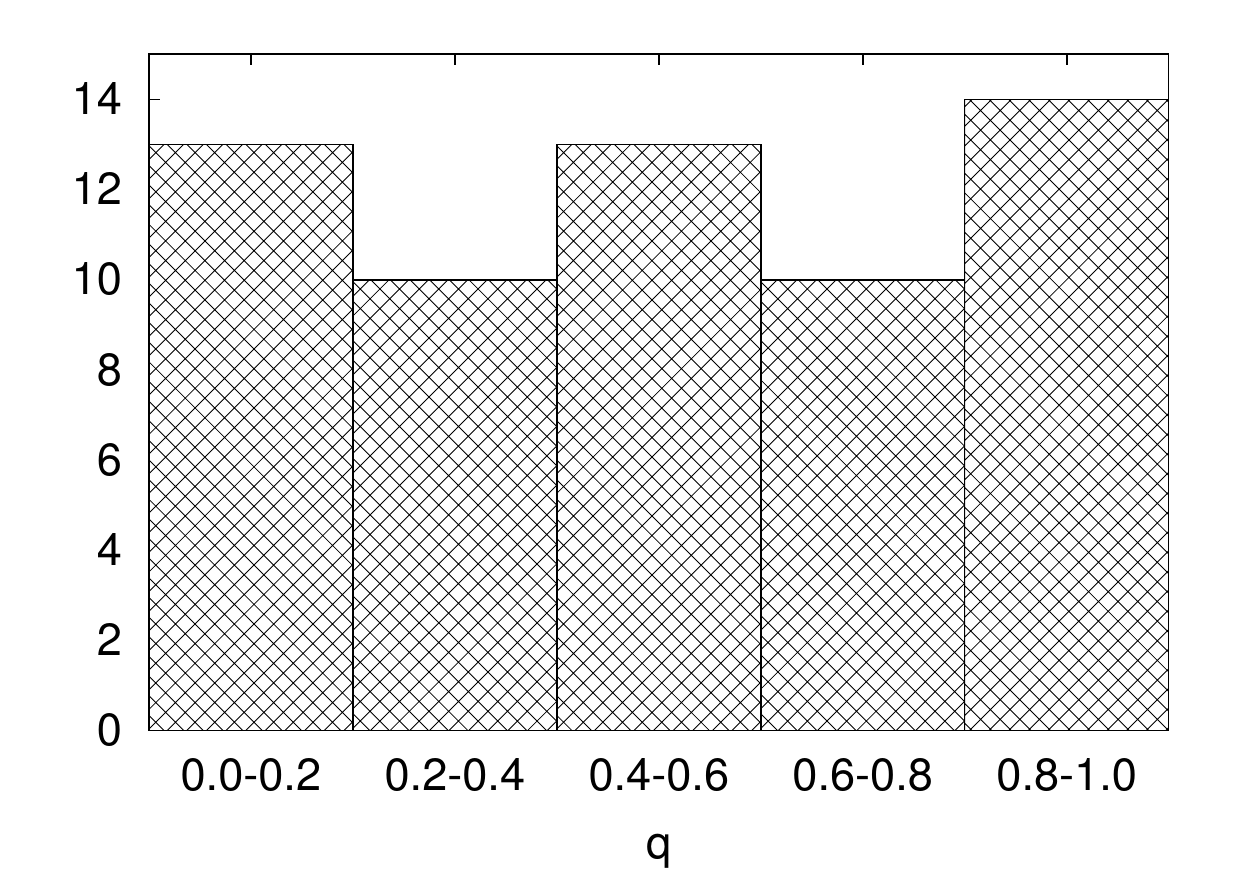}
\caption{Histogram of Student difference tests comparing cooling scales 
         between different topological sectors.}
\label{fig:qhistogram}
\end{figure}

A histogram of the $q$-values of the remaining $4\times 15=60$ Student
difference tests for the scales of Table~\ref{tab:sectorscales} is shown
in Figure~\ref{fig:qhistogram}. When the compared data are statistically
independent, rely on the same estimator, and are drawn from a Gaussian
distribution, the Student difference tests return uniformly distributed
random numbers $q$ in the range $0<q<1$, which is consistent with
Figure~\ref{fig:qhistogram}. Furthermore, their mean value comes out to
be $\overline{q}=0.508(40)$ in agreement with the expected $0.5$.
If there are still some residual correlations between our $q$-values,
this would have decreased the error bar, because the number of
independent $q$ would have been counted too high, while each of them
still fluctuates like a uniformly distributed random number in the
interval (0,1). A Kolmogorov test between the distribution of
Figure~\ref{fig:qhistogram} and a uniform distribution
yields $q_{\text{Kolm}}=0.11$, which further supports the $q$-values
being normally distributed.
Taken altogether, we find convincing evidence that the $1/V$ bias
expected for our scales due to topological freezing disappears within
our statistical noise.

\section{Summary}\label{sec:qsumm}
We calculated the topological charge for pure $\SU(2)$ LGT using standard
cooling for larger $\beta$ values and lattices than has been done previously.
We find stable topological sectors for $\beta\gtrsim2.75$ and lattices
large enough to support physical instantons, with metastability
for $n_c\approx 1000$, which is larger than what one may have expected
from past studies. For these lattices, destruction of instantons appears
not to be an issue. From these data, we obtain the estimates
\eqref{eq:sTc1000} and \eqref{eq:soss1000} for the topological
susceptibility, which are surprisingly close to previous results.
This may be a lucky accident due to extrapolations performed on
systems that are too small, as illustrated by the $n_c=1000$ versus
$n_c=100$ fits of Figure~\ref{fig:BAV}. Using
$\sqrt{\sigma}=400~\text{MeV}$ as reference, this yields 
$\chi\approx180~\text{MeV}$, which is
close to the large $N_c$ prediction~\eqref{eq:chivalue}.

Within our statistics, we find no observable correlations between
cooling scales and topological charge sectors. Our number of statistically
independent configurations is of a typical size as used for scale setting.
Due to the relatively low computational cost of generating pure $\SU(2)$ 
configurations, it is perhaps not surprising that topological freezing is
not a problem; indeed other pure $\SU(2)$ studies seem to also sample
the topological charge quite well~\cite{hirakida_thermodynamics_2018}.
We can safely conclude that topological freezing is no concern for pure 
$\SU(2)$ cooling scales at this level of precision.

%% file: chapter6.tex
\chapter{Summary and Conclusions}\label{ch:summary}

We carried out a detailed investigation of pure $\SU(2)$ LGT on large
lattices and at large $\beta$. We picked pure $\SU(2)$ because it is
computationally simple compared to a more physically realistic
model like QCD with $N_f\geq 2$. This means we can achieve high
precision with moderate computing power. Being a non-Abelian gauge
group, pure $\SU(2)$ exhibits asymptotic freedom, meaning it has
a well-defined continuum limit, making it ideal for lattice study.
Pure $\SU(2)$ is thus a useful testing ground for new methods. 

We calculated the pure $\SU(2)$ deconfinement temperature, i.e.,
followed the scaling behavior of the associated length, out to larger $\beta$
and with larger lattices than has been done previously. These were
extensive simulations, with our largest lattice being $80^3\times8$. 
In our study, we used the initial (small lattice) $N_\tau$ scaling behavior 
to fix the initial scaling behavior of gradient and cooling
reference lengths. Of course, results for $T_c$ on fine lattices 
are also interesting for current pure $\SU(2)$ 
thermodynamics studies~\cite{hirakida_thermodynamics_2018}. 

We investigated six cooling
scales by comparing their scaling behavior to six gradient scales
and the deconfinement length. We found no distinct scaling behavior
for the cooling scales and no loss of precision, in agreement with
a suggestion by Bonati and D'Elia. Calculating gradient scales is
two to three orders of magnitude faster than calculating
the deconfinement scale. The cooling flow
progresses 34 times faster than the gradient flow. 
This is of possible interest to QCD calculations, especially whenever
scale setting becomes a significant source of systematic error.

Next, we studied the approach of these length scales to the continuum limit
using asymptotic scaling fits and standard
scaling. For the asymptotic scaling fits, we modified an approach
introduced by Allton to enforce the expected
$\mathcal{O}\big(a^2\big)$ behavior of length ratios. Relative differences
between results from different fit forms serve as an estimate for the
systematic error. Similarly comparing results
from different scales gives an estimate of systematic error due to choice of
reference scale. Deep in the scaling region, at $\beta=2.574$, total systematic
error due to both is around 2\%. This can be viewed as a warning to
QCD investigations that one may need very fine lattices to bring 
systematic error of this type close to 1\%. Continuum limit estimates 
of length ratios differ systematically by up to 1.3\%, which is again 
clearly relevant when one aims at 1\% precision.
Along this vein, it may be worthwhile to investigate asymptotic scaling
fits in physically realistic theories.

We investigated pure $\SU(2)$ topological charge sectors, and the performance
of standard cooling as a smoothing algorithm. 
Provided that the lattice is fine enough and has a
large enough physical size, we find little to no evidence of
destruction of instantons. 
It takes roughly $n_c\approx1000$ cooling sweeps
to attain stable sectors, which is surprisingly large given past studies.
Topological freezing is not a problem for
pure $\SU(2)$ cooling scales, where it is possible to have enough 
sweeps between measurements to escape topological sectors. 

We performed a continuum limit extrapolation for the
topological susceptibility. This extrapolation relies on estimates at
each $\beta$ that take finite size corrections into account, and
the susceptibility was determined at larger $\beta$ than done previously.
Our favored estimate 
\begin{equation}
 \chi^{1/4}/T_c=0.643(12),
\end{equation} 
calculated at $n_c=1000$, is somewhat smaller than past estimates. 
Its value in physical units, approximately $180~\text{MeV}$, is close to the
large $N_c$ estimation.

Finally we calculated cooling scales in different topological sectors.
We found no evidence of correlations within our statistics. 
So even if there were significant topological freezing 
in this theory, it would not matter for cooling scales, when the
statistics are comparable to those used in typical investigations.
Presumably this is also true for pure $\SU(3)$, but this needs to be 
investigated. Ultimately our study indicates that cooling scales are
more efficient than gradient scales. High precision can be reached, and there
is no need to worry too much about topological freezing.

%% file: appendix6.tex
\chapter{Supplementary Figures}\label{ap:suppfigs}
% N_tau=4
\begin{figure}[H]
\vskip -10pt% Needed because latex is stupid
  \centering
  \includegraphics[width=0.489\linewidth]{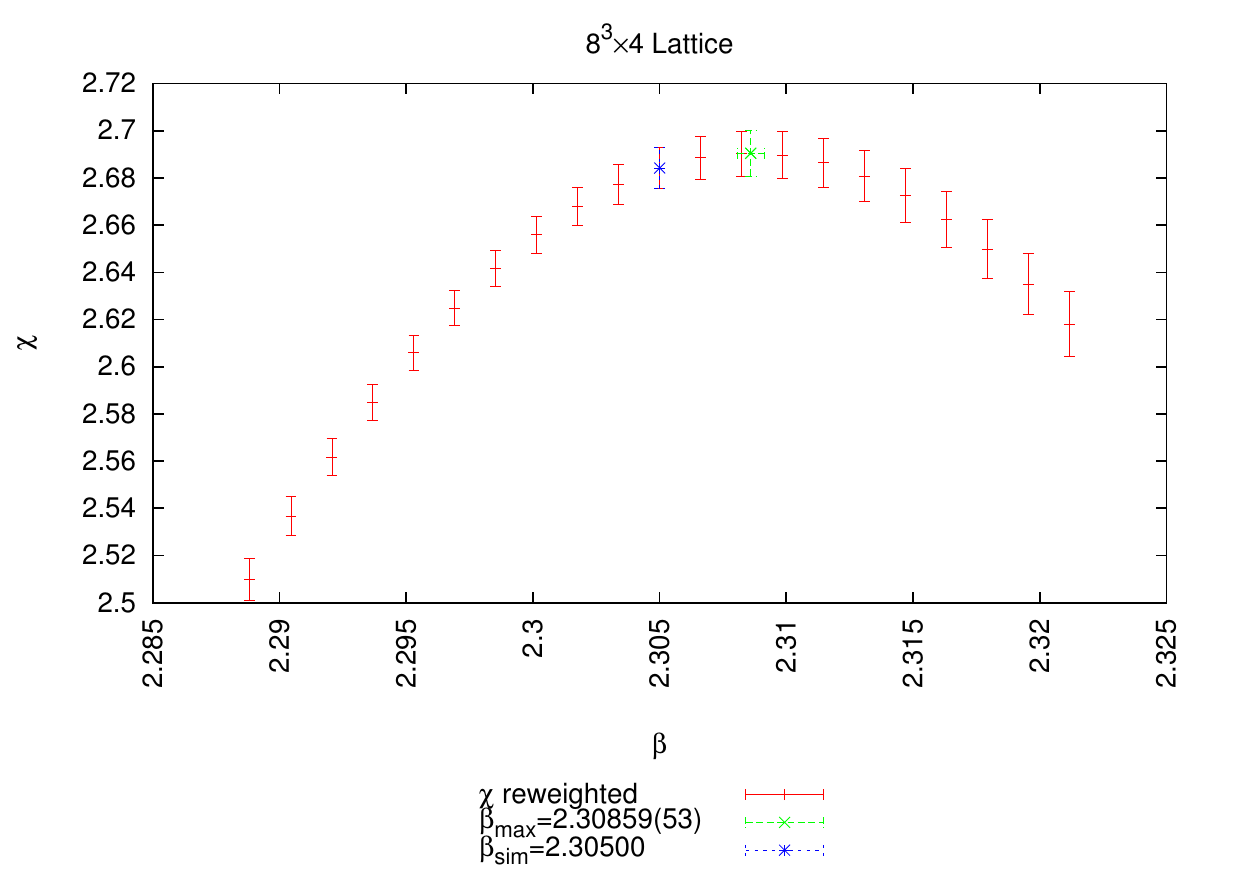}
  \includegraphics[width=0.489\linewidth]{figures/tcplots/nt04/CP12-eps-converted-to.pdf}
  \includegraphics[width=0.489\linewidth]{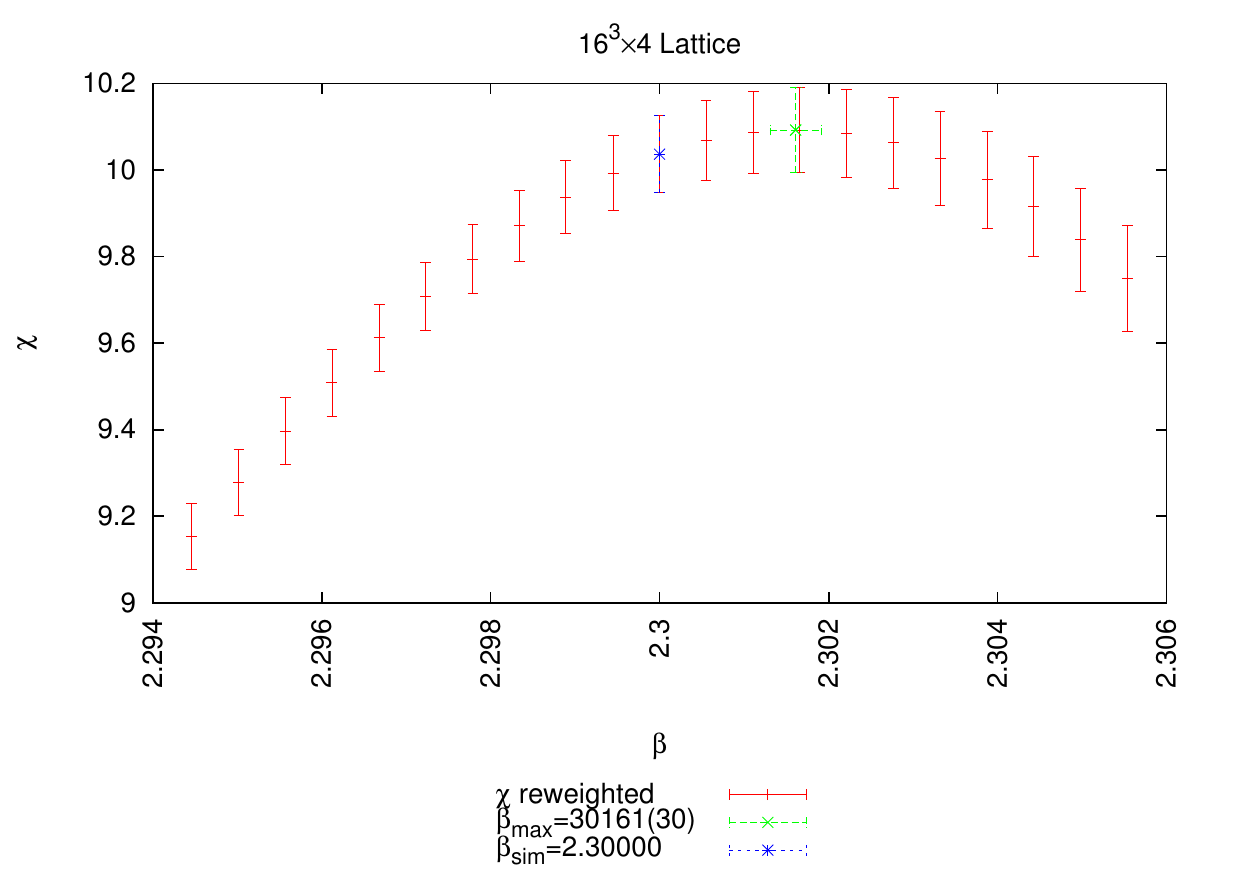}
  \includegraphics[width=0.489\linewidth]{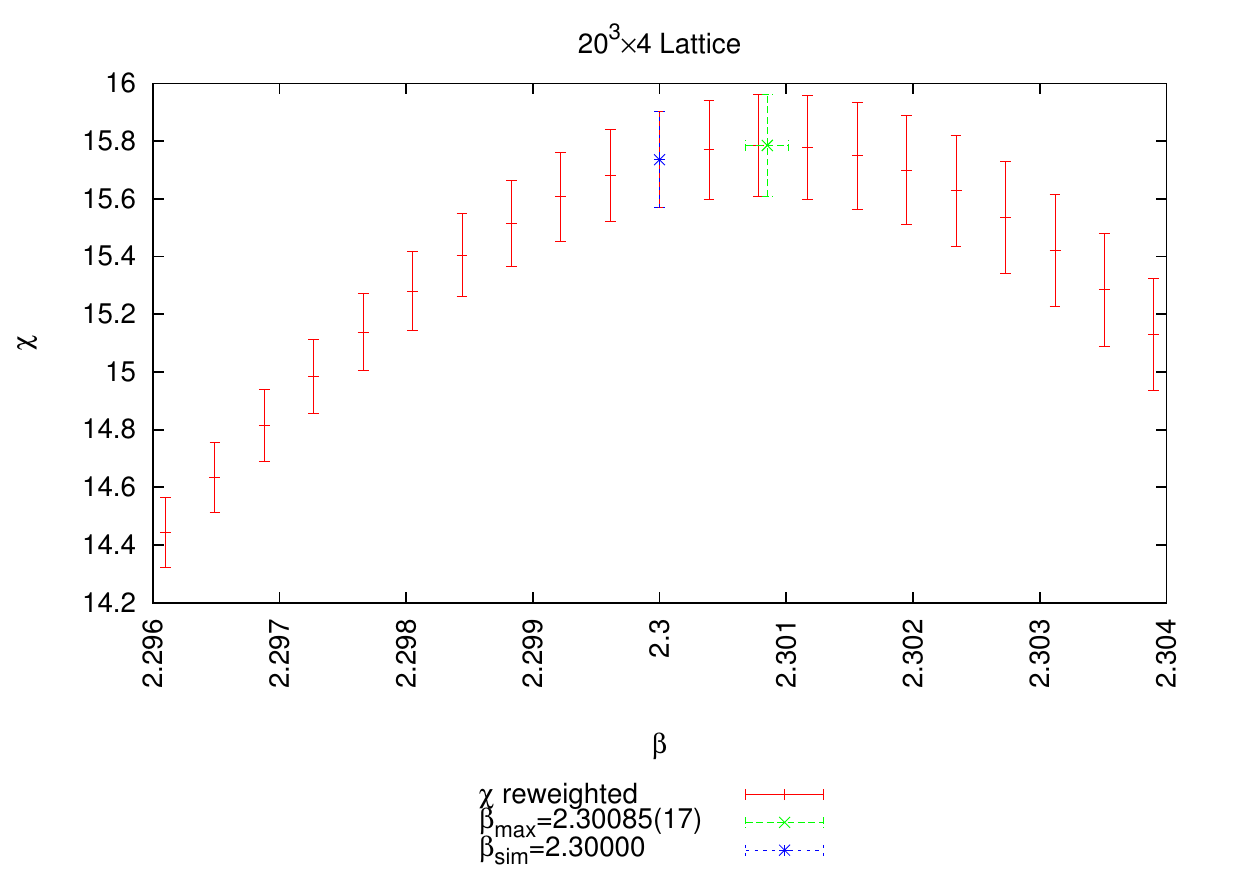}
  \includegraphics[width=0.489\linewidth]{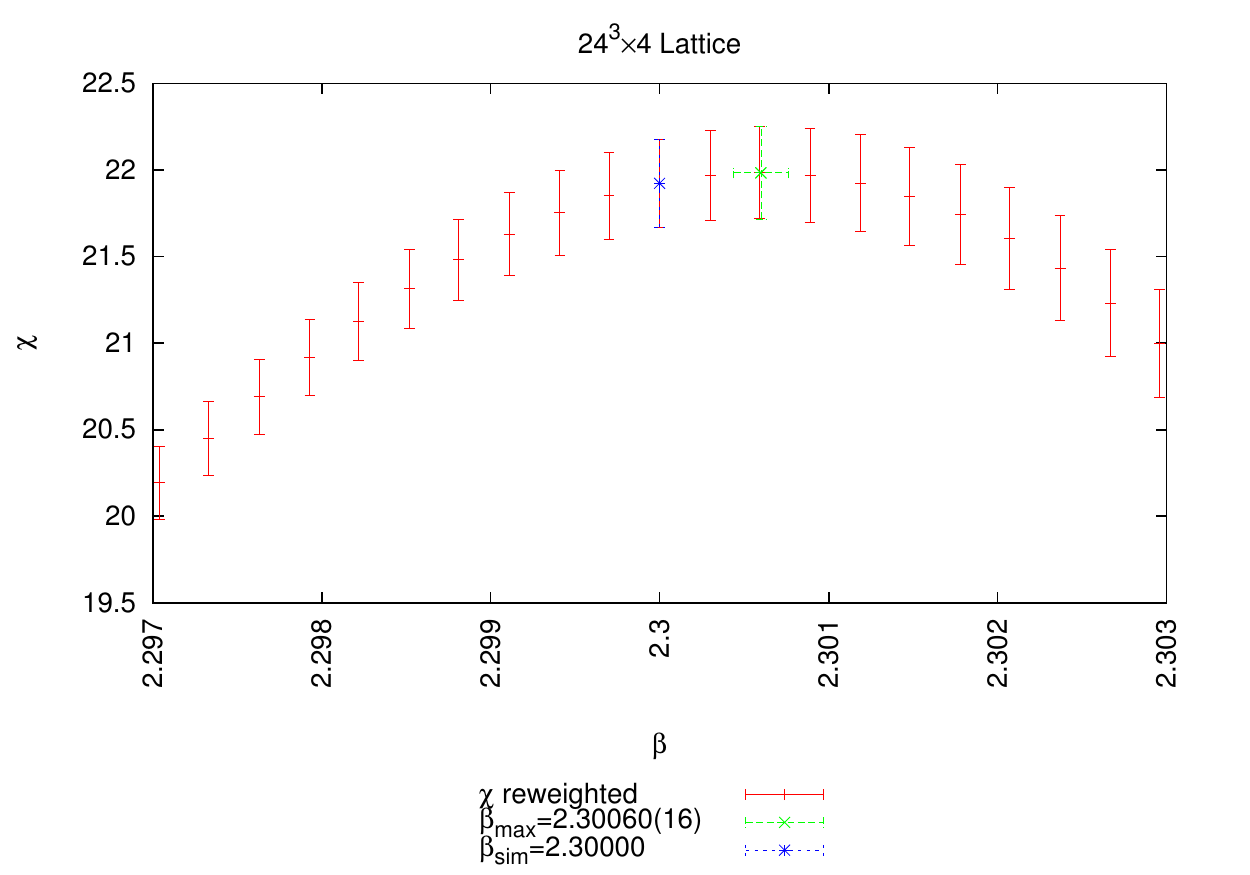}
  \includegraphics[width=0.489\linewidth]{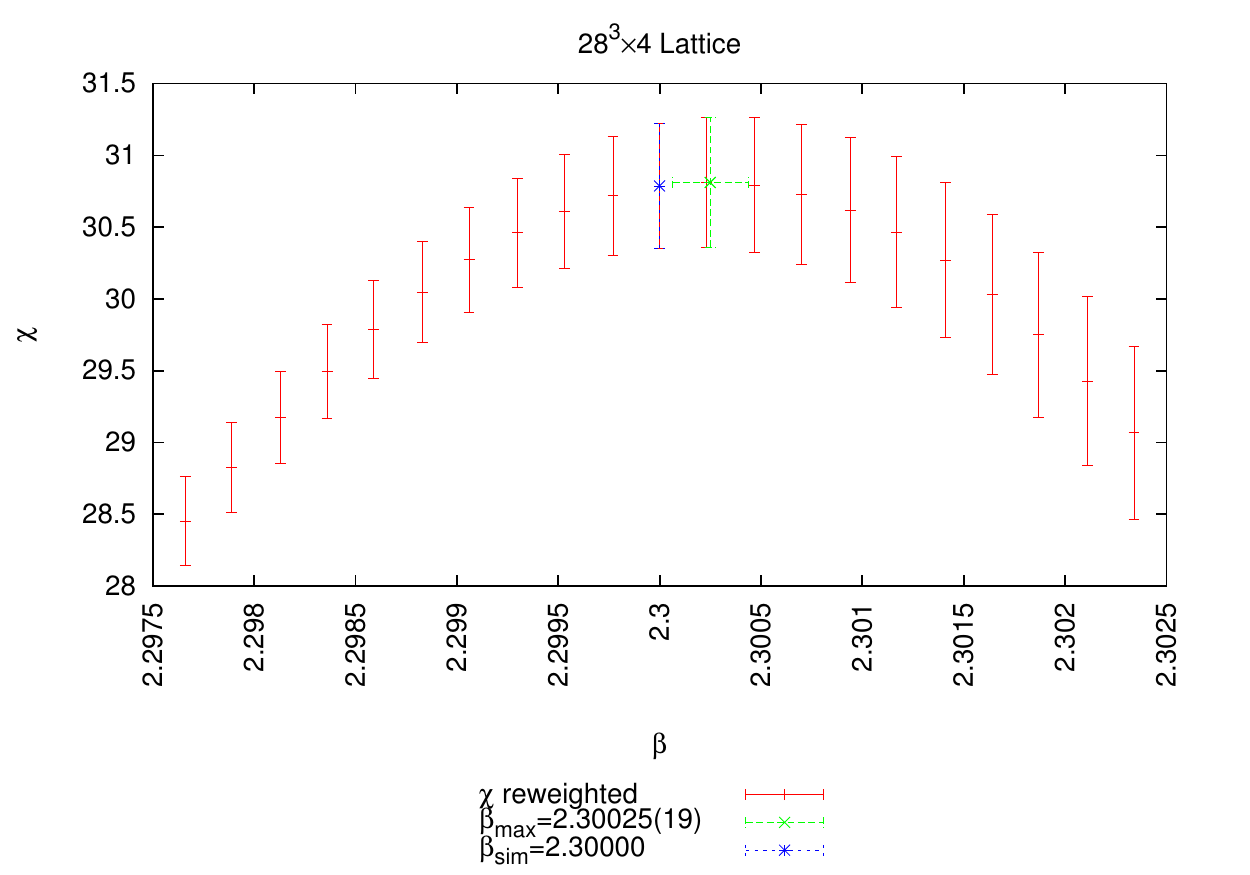}
  \caption{$N_\tau=4$ Polyakov loop reweighting.}
\end{figure}
\begin{figure}\ContinuedFloat
  \centering
  \includegraphics[width=0.489\linewidth]{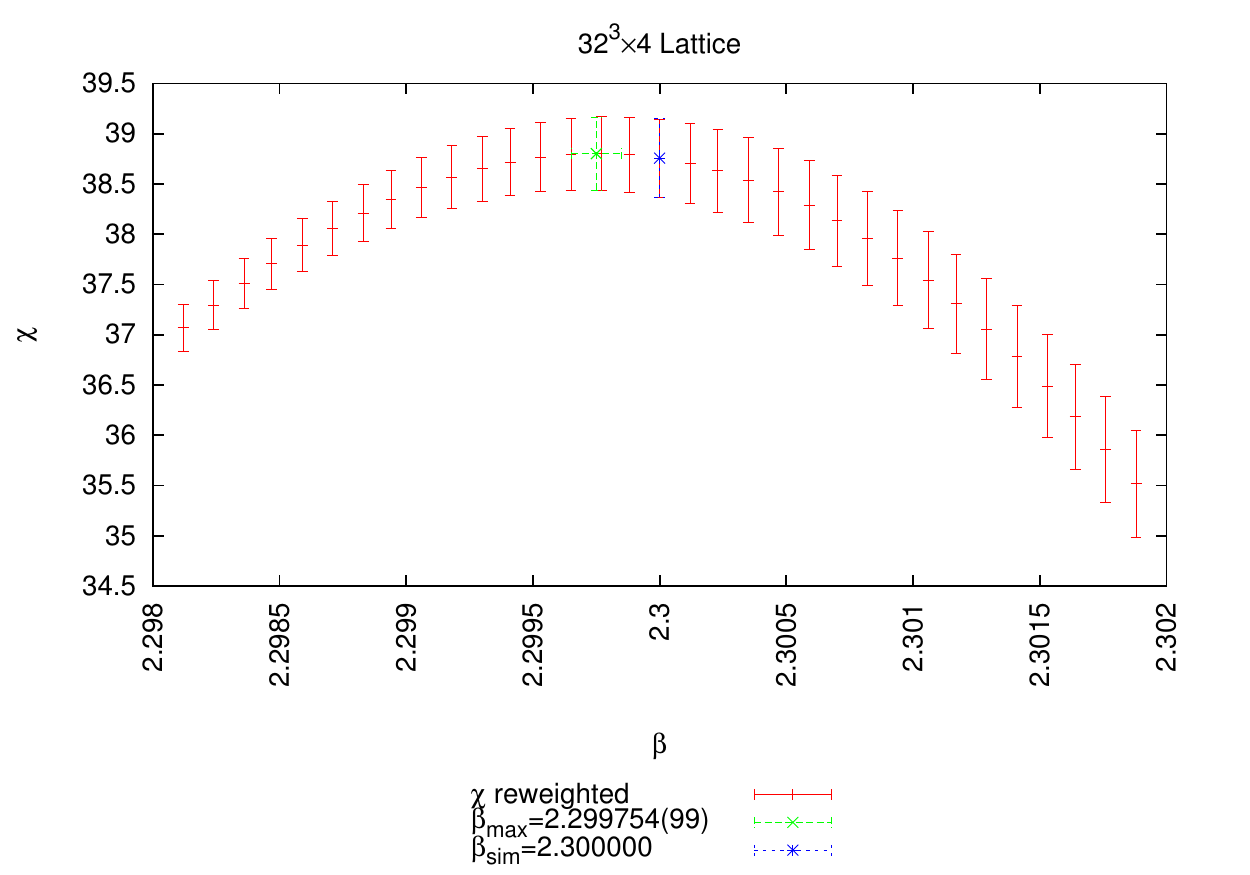}
  \includegraphics[width=0.489\linewidth]{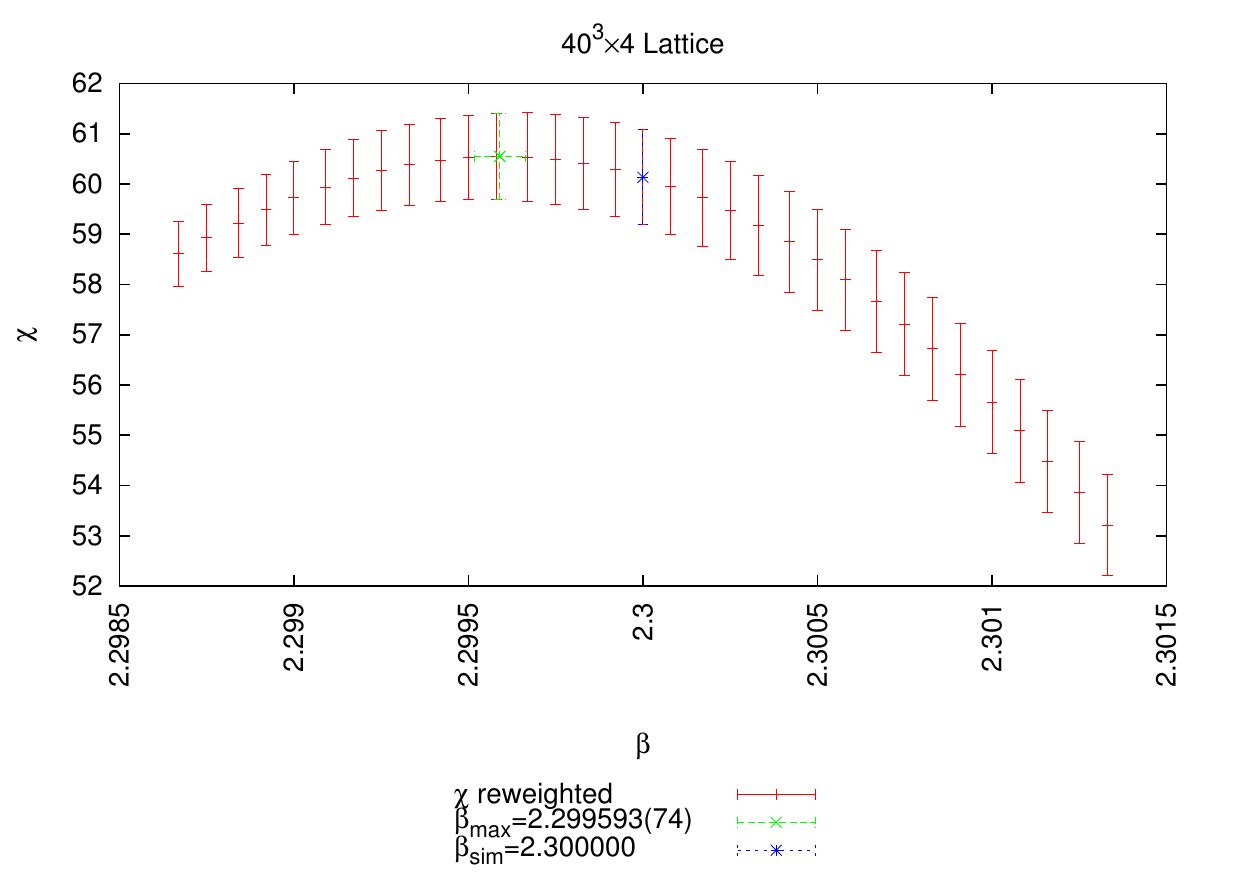}
  \includegraphics[width=0.489\linewidth]{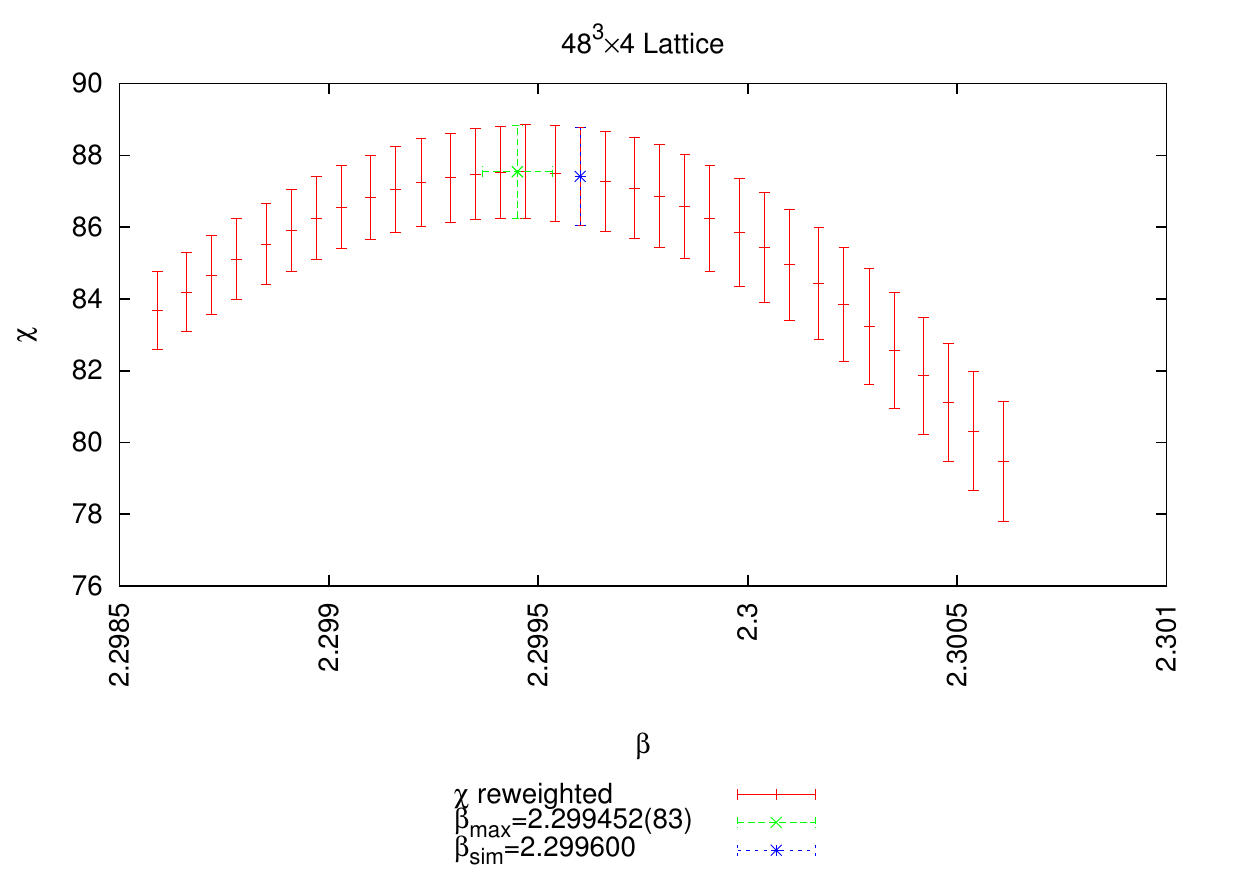}
  \includegraphics[width=0.489\linewidth]{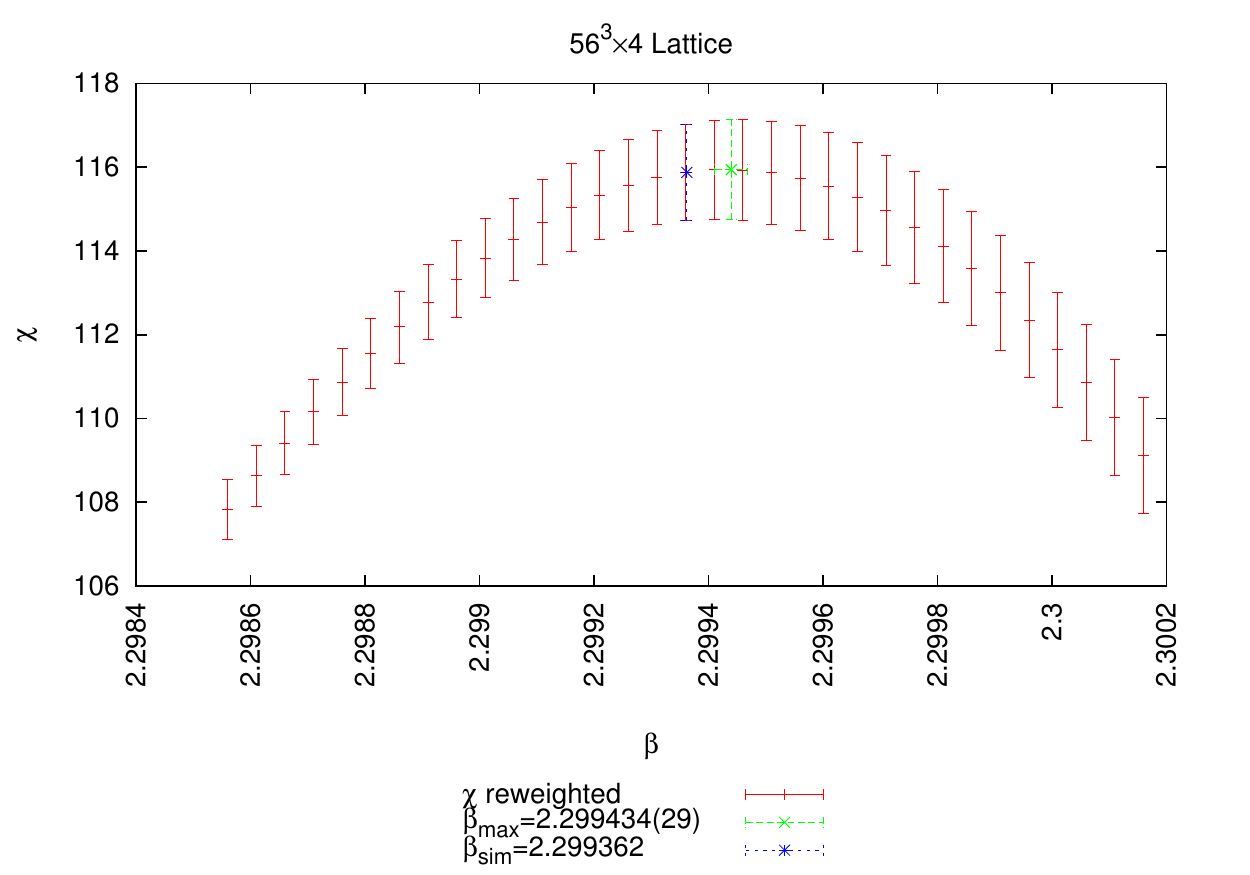}
  \caption[]{Continued.}
\end{figure}

% N_tau=6
\begin{figure}
  \centering
  \includegraphics[width=0.489\linewidth]{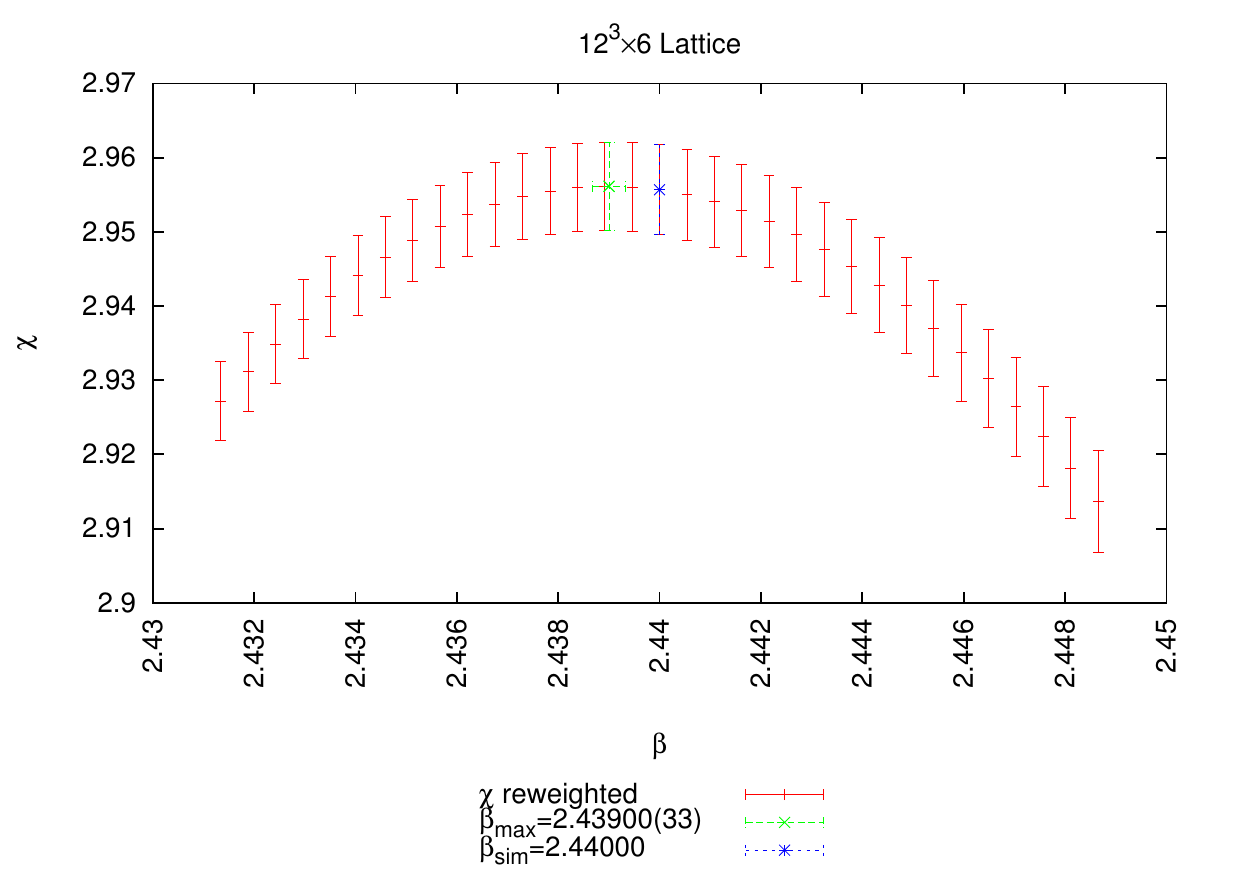}
  \includegraphics[width=0.489\linewidth]{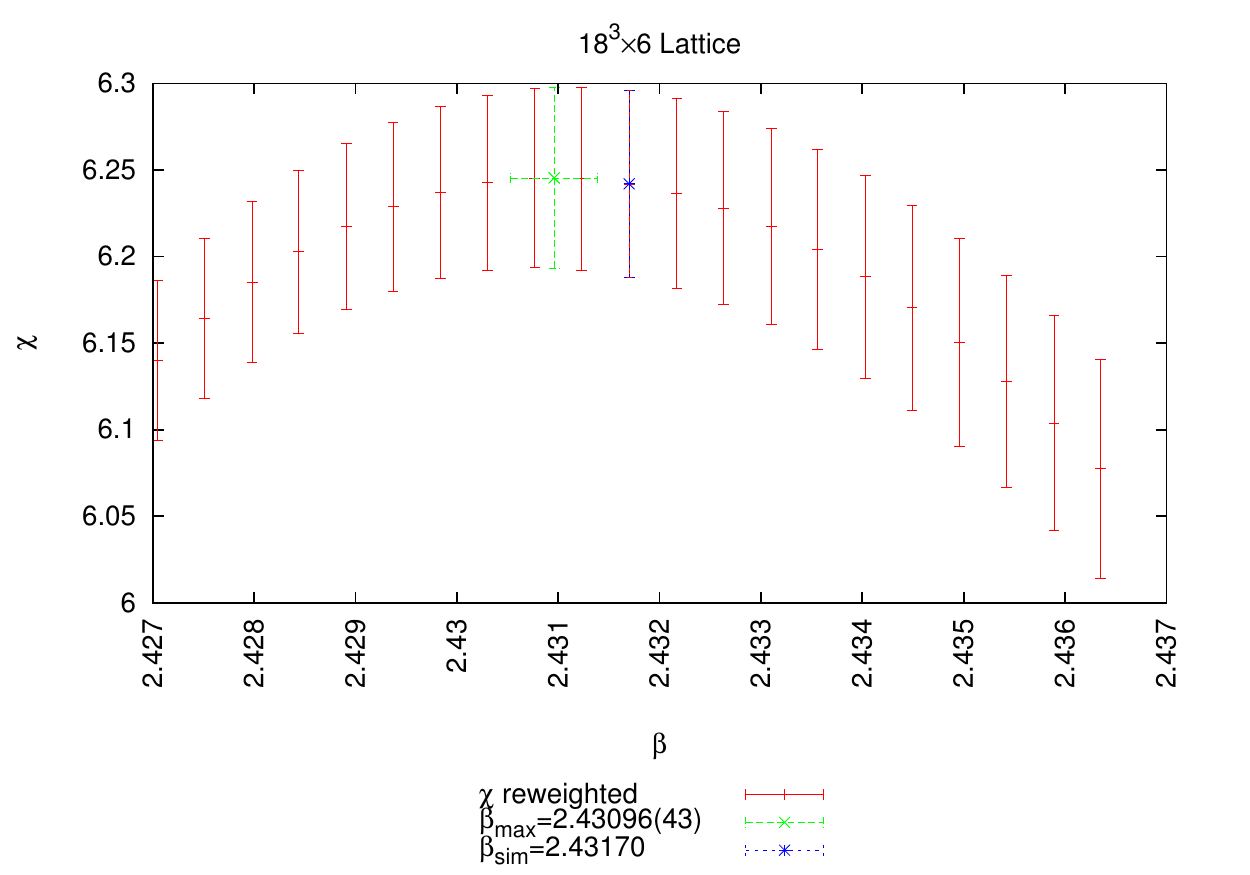}
  \caption{$N_\tau=6$ Polyakov loop reweighting.}
\end{figure}
\begin{figure}\ContinuedFloat
  \centering
  \includegraphics[width=0.489\linewidth]{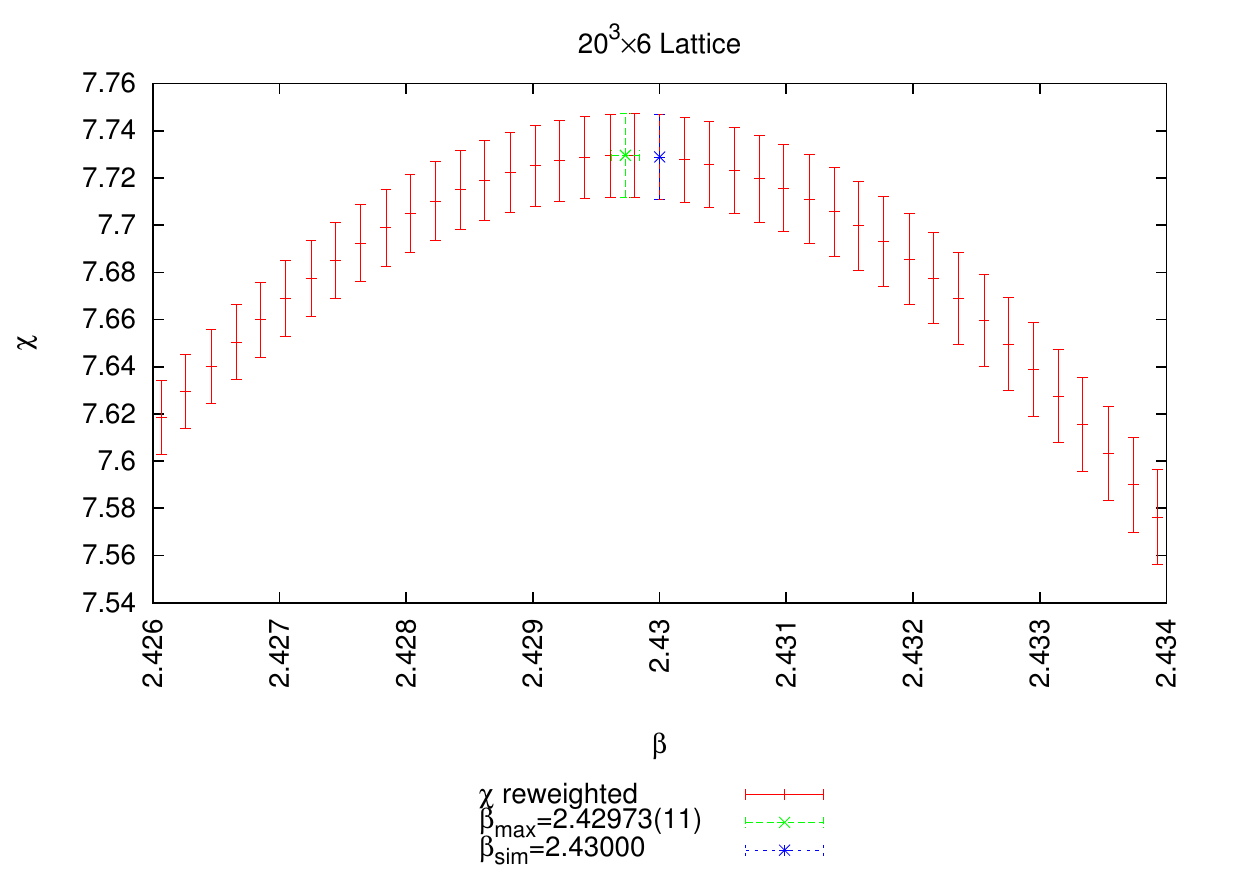}
  \includegraphics[width=0.489\linewidth]{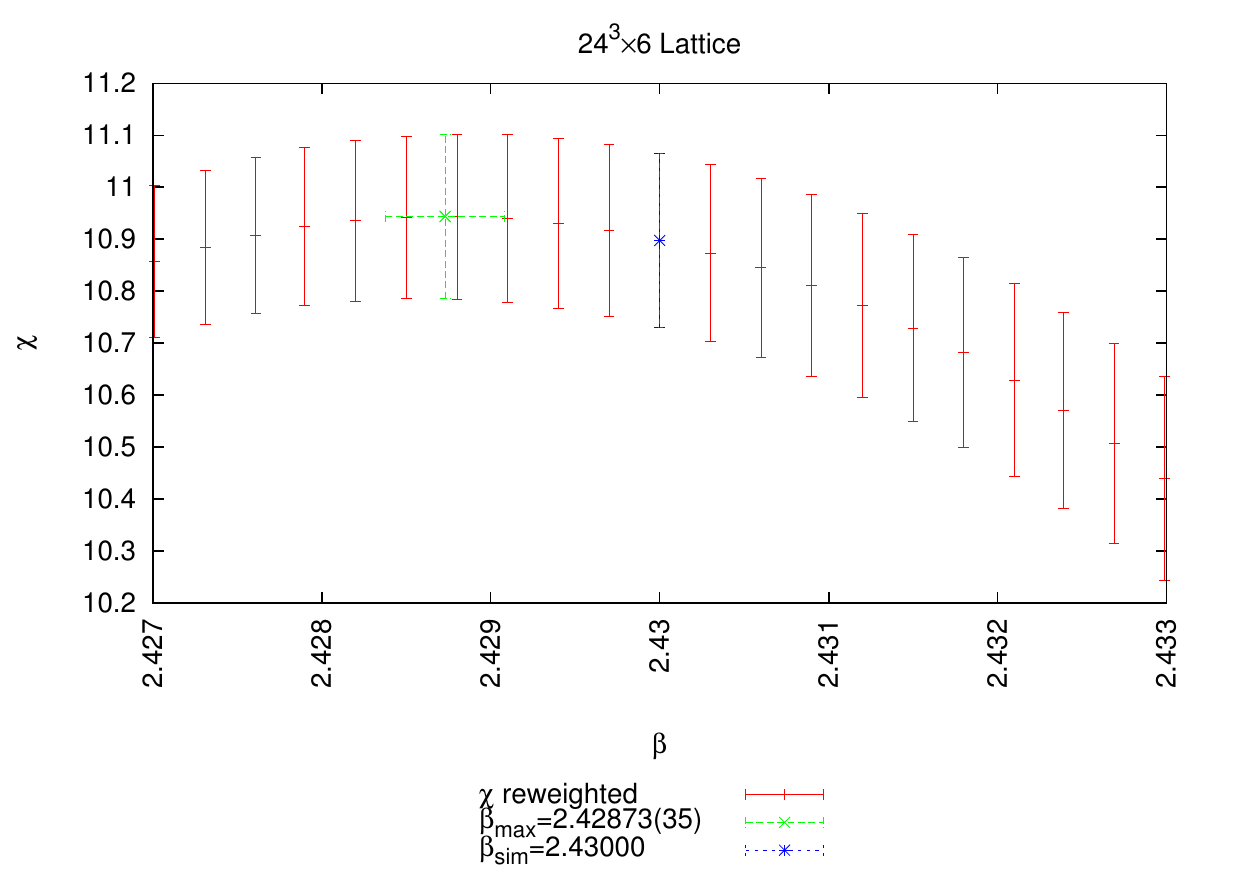}
  \includegraphics[width=0.489\linewidth]{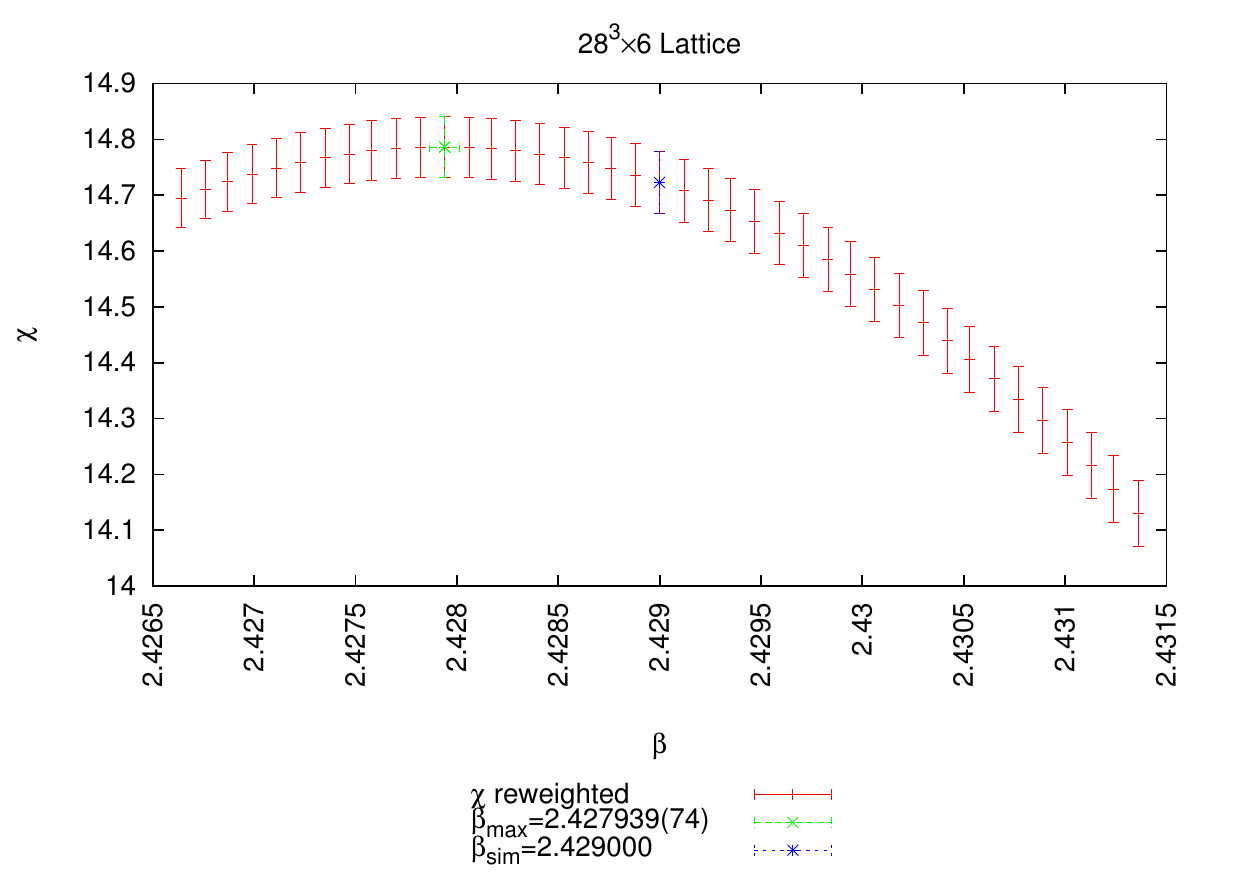}
  \includegraphics[width=0.489\linewidth]{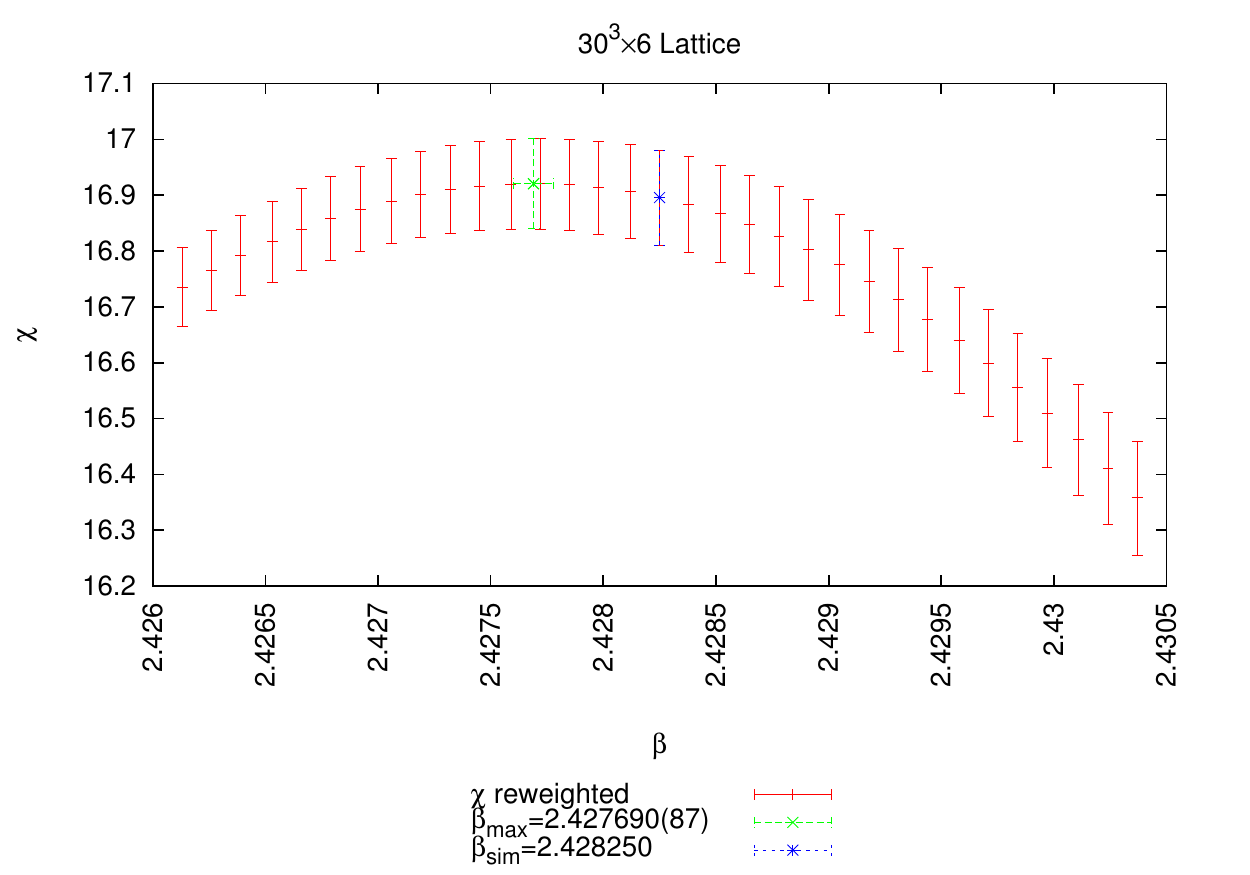}
  \includegraphics[width=0.489\linewidth]{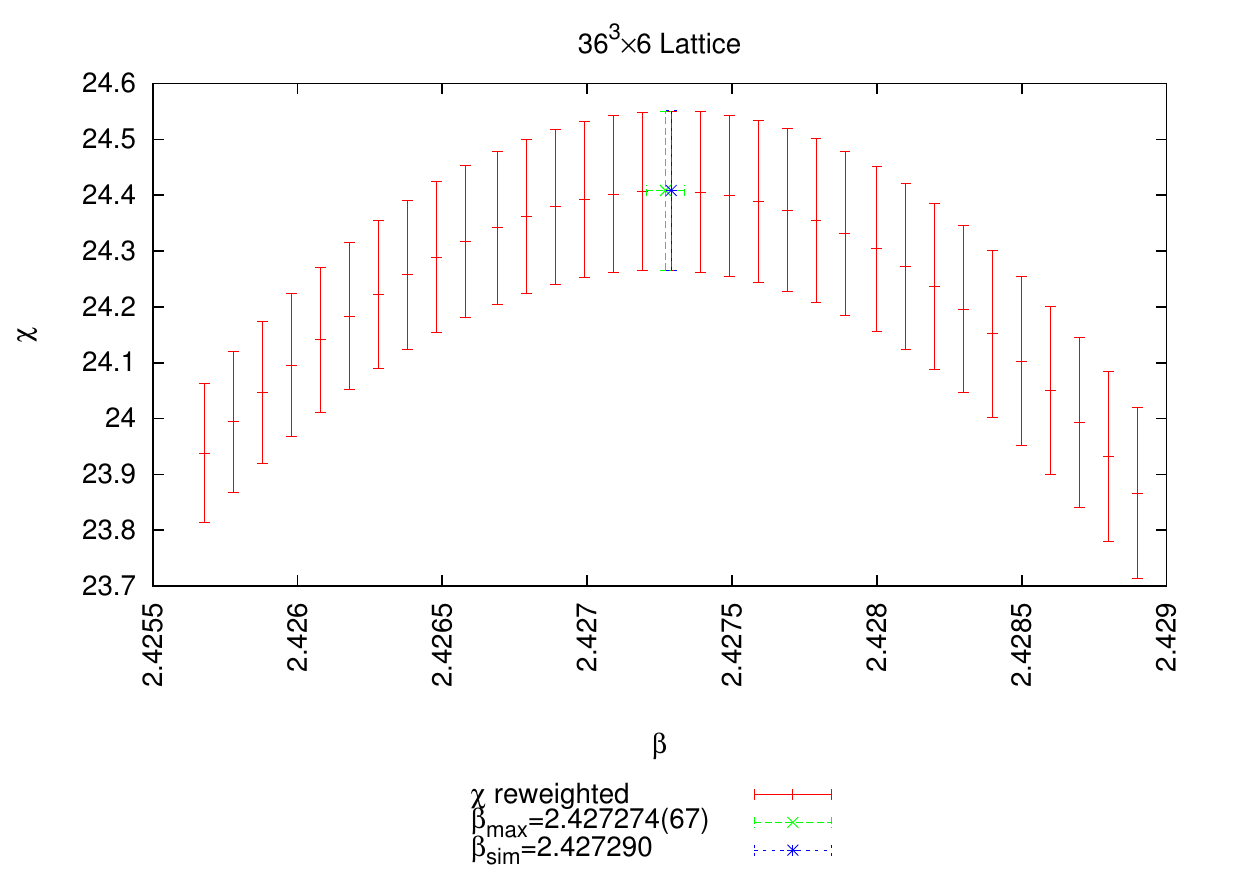}
  \includegraphics[width=0.489\linewidth]{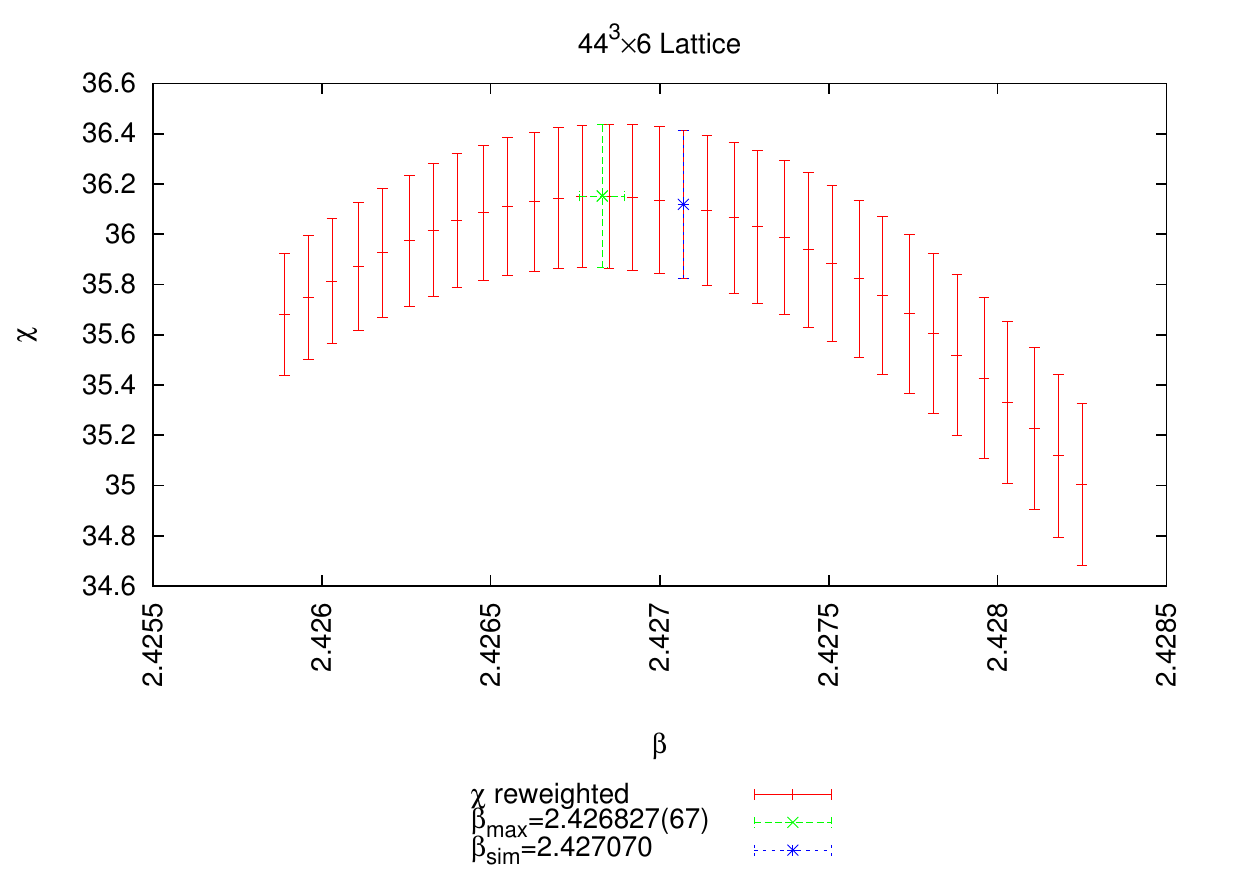}
  \caption[]{Continued.}
\end{figure}
\begin{figure}\ContinuedFloat
  \centering
  \includegraphics[width=0.489\linewidth]{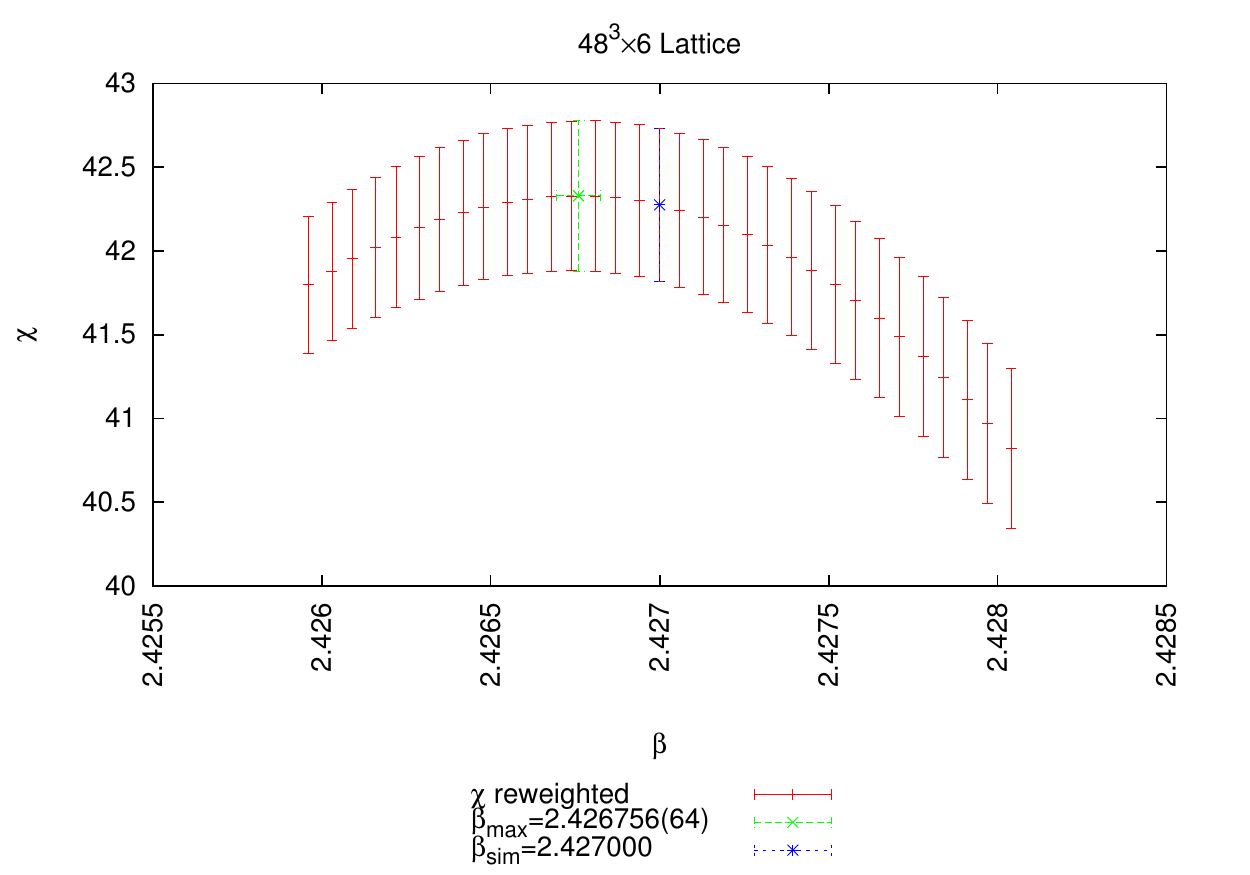}
  \includegraphics[width=0.489\linewidth]{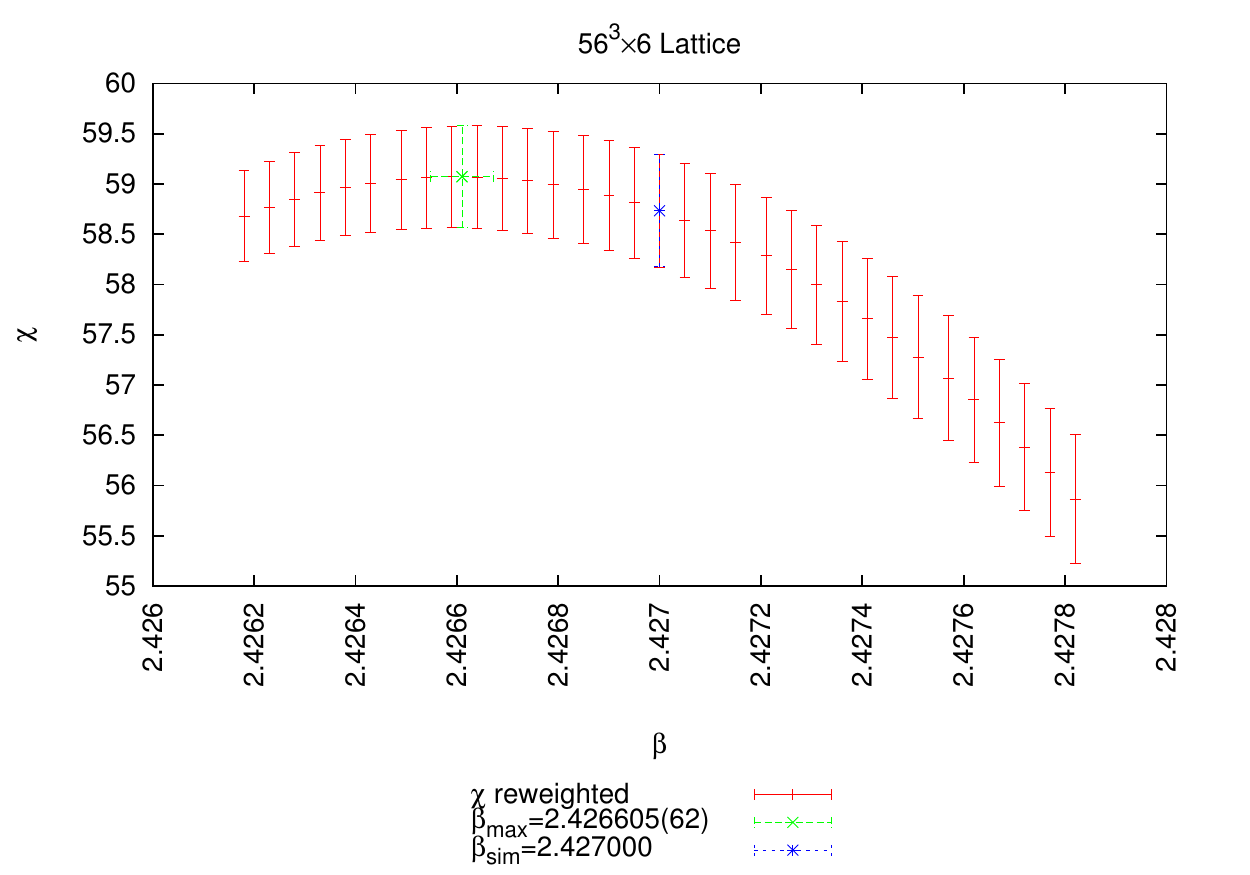}
  \caption[]{Continued.}
\end{figure}

% N_tau=8
\begin{figure}
  \centering
  \includegraphics[width=0.489\linewidth]{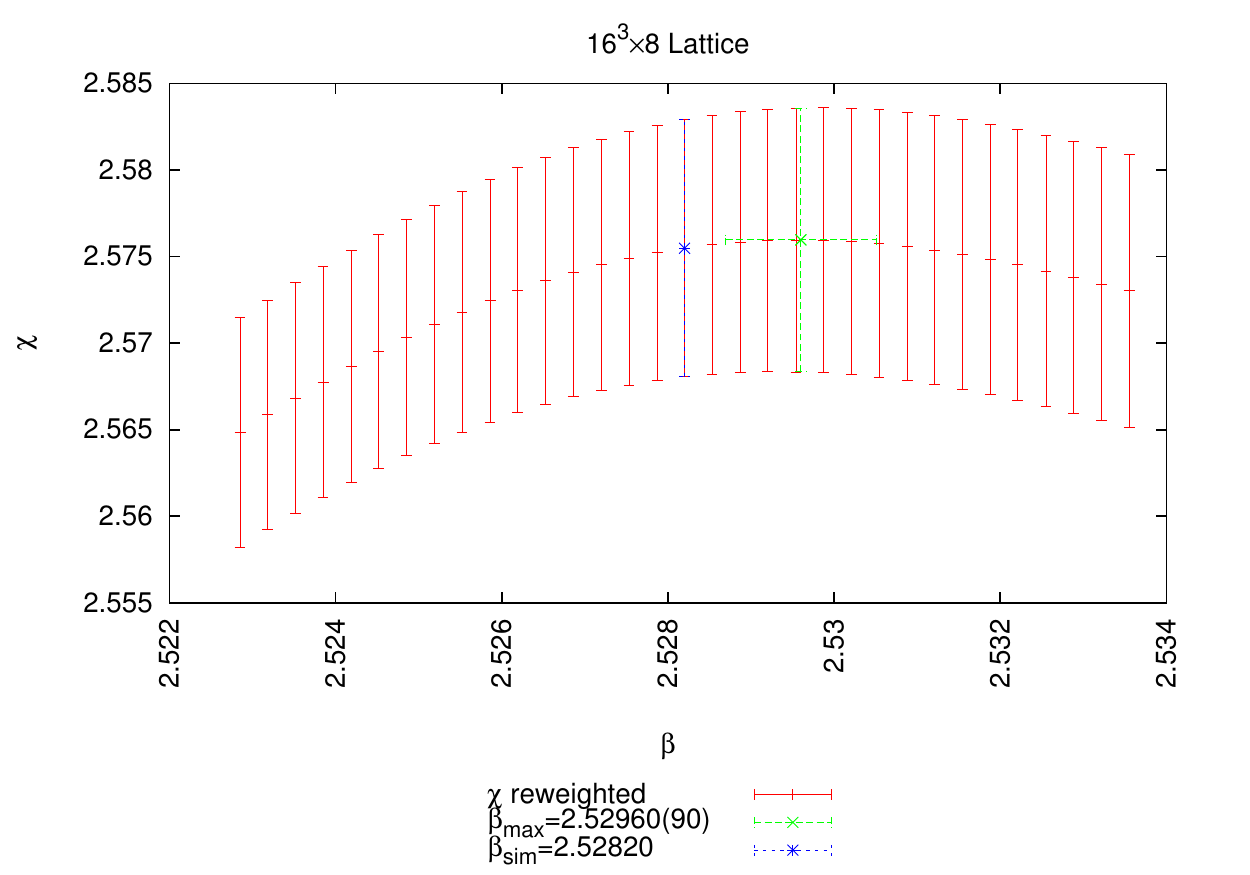}
  \includegraphics[width=0.489\linewidth]{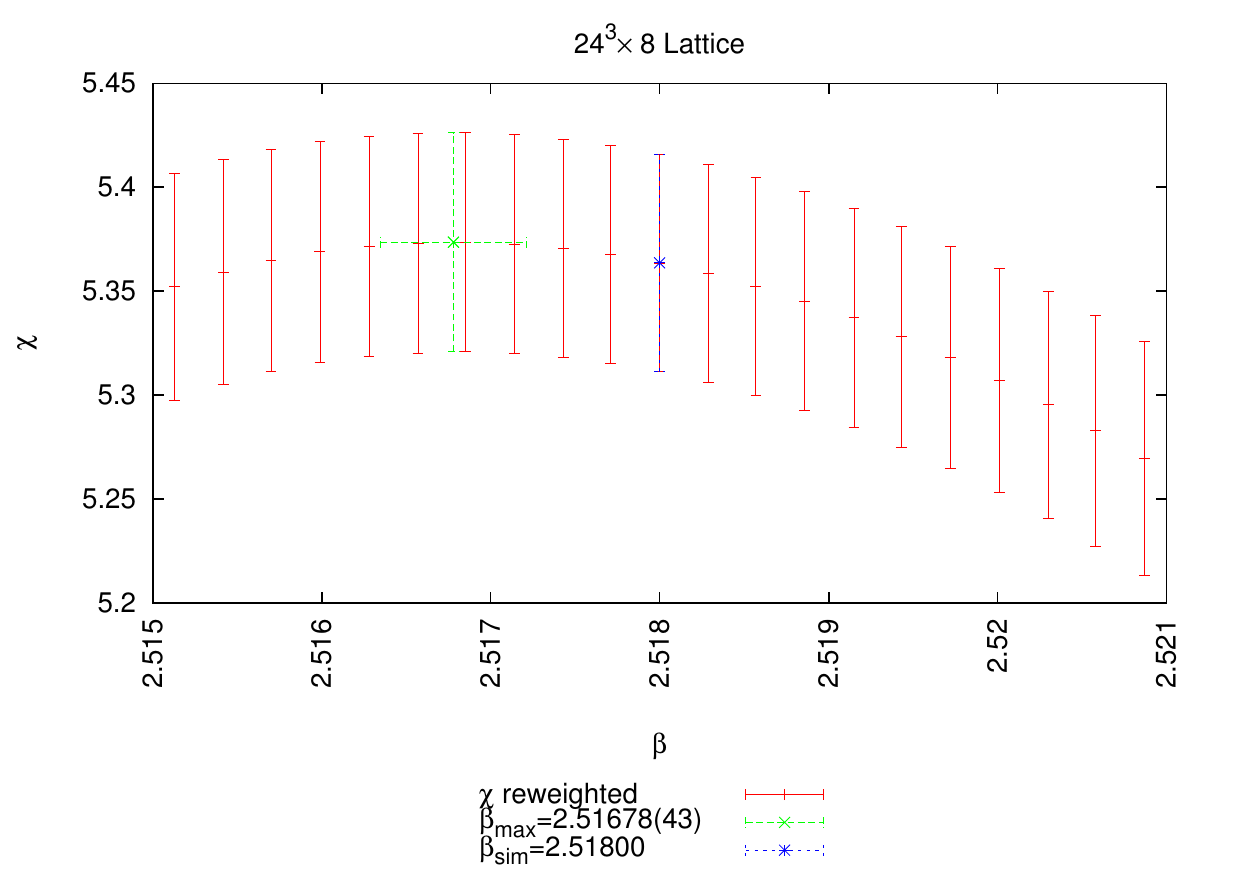}
  \includegraphics[width=0.489\linewidth]{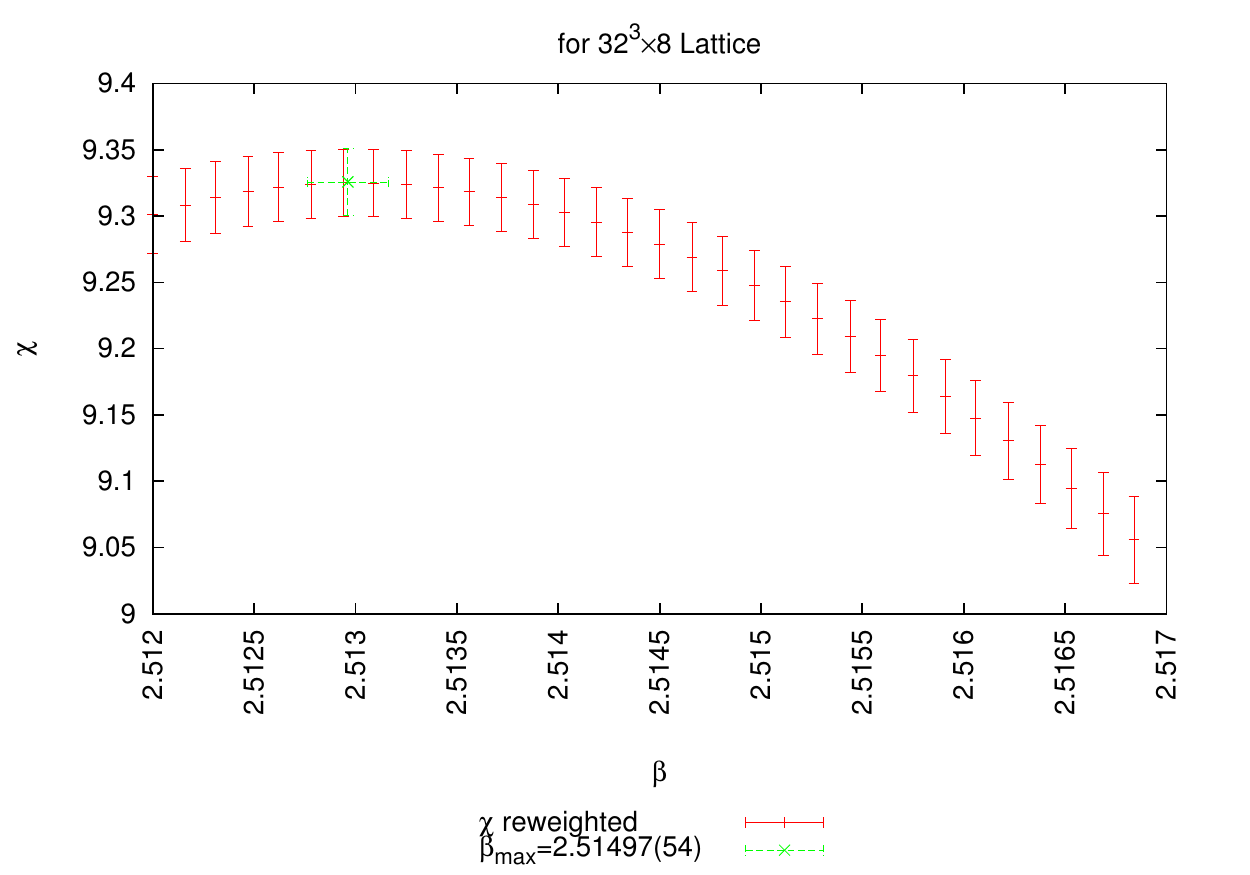}
  \includegraphics[width=0.489\linewidth]{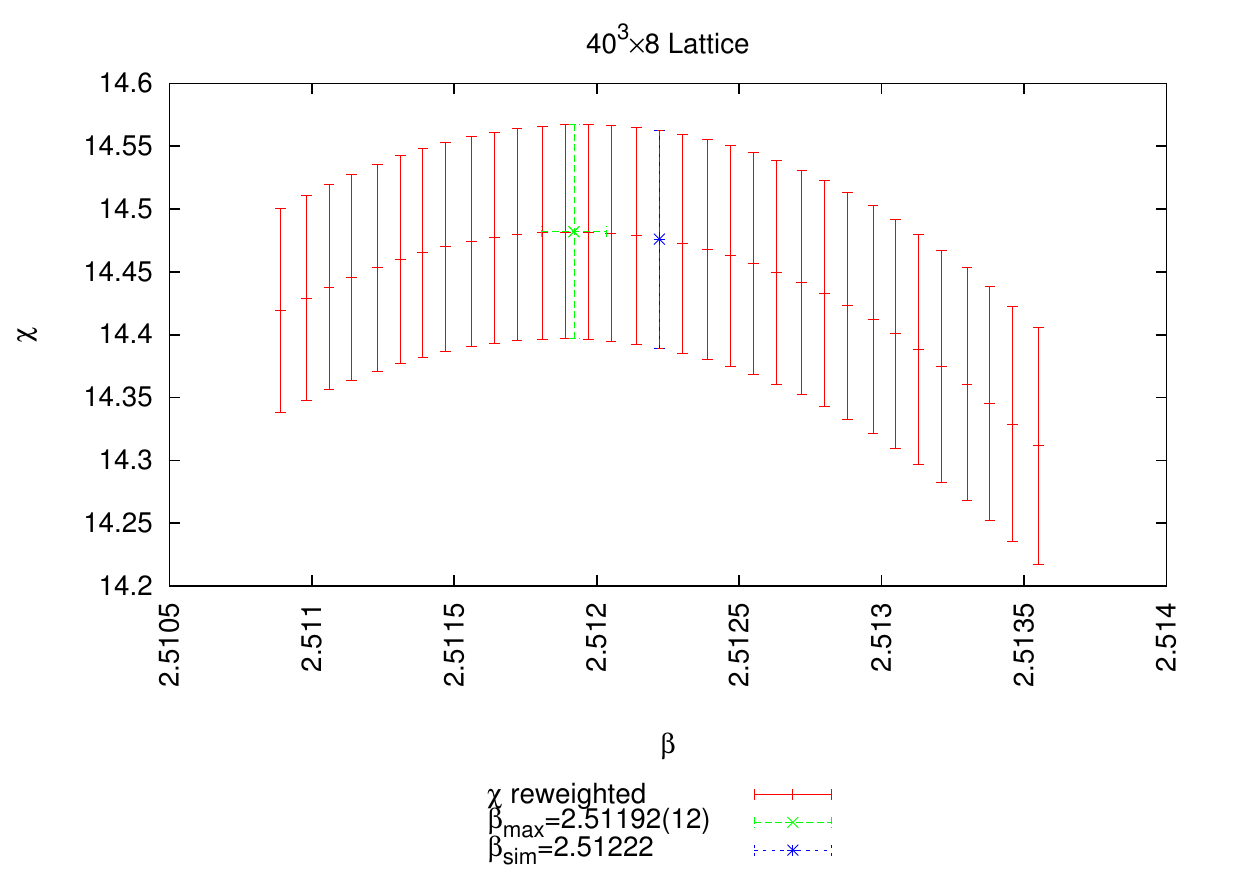}
  \caption{$N_\tau=8$ Polyakov loop reweighting. 
           Figures without a simulation point have results combined from 
           data generated at multiple nearby simulation points.}
\end{figure}
\begin{figure}\ContinuedFloat
  \centering
  \includegraphics[width=0.489\linewidth]{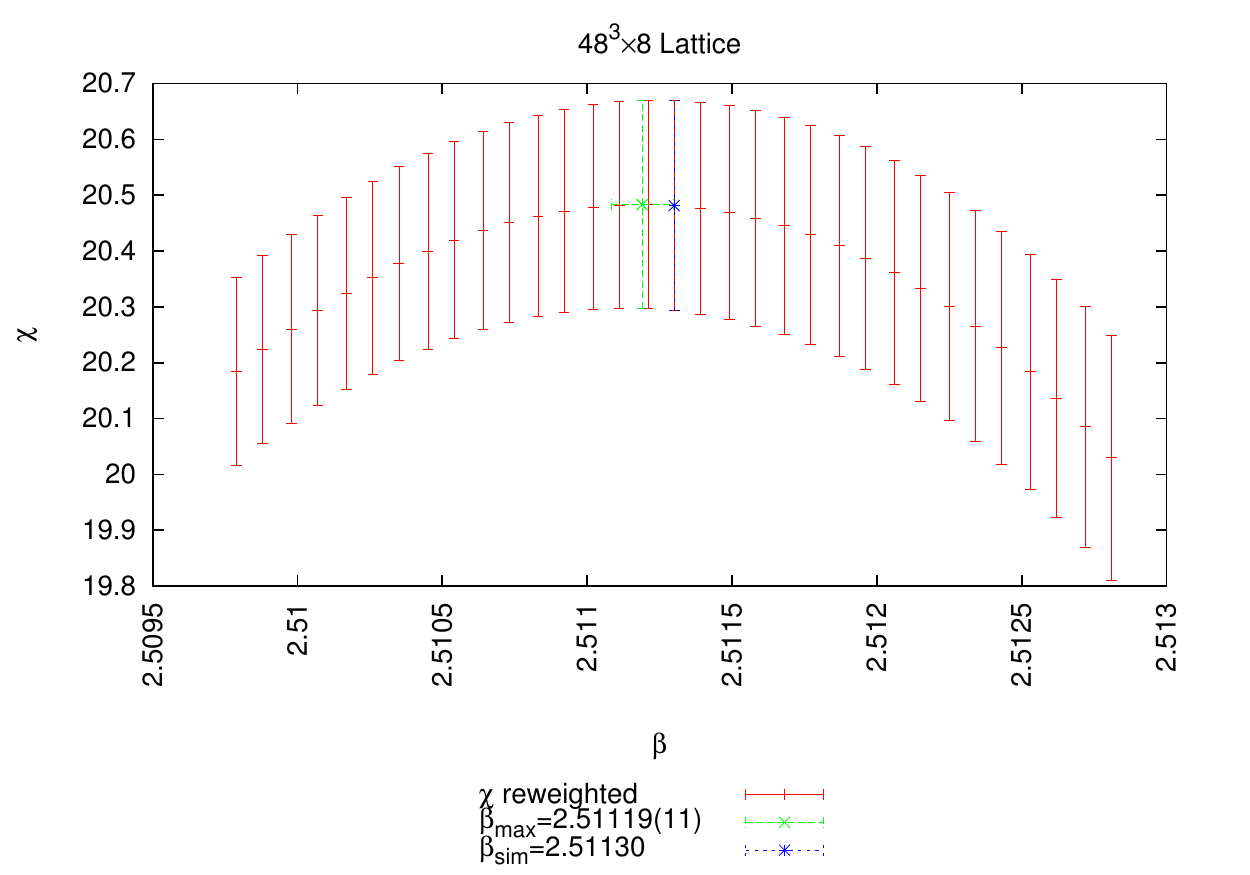}
  \includegraphics[width=0.489\linewidth]{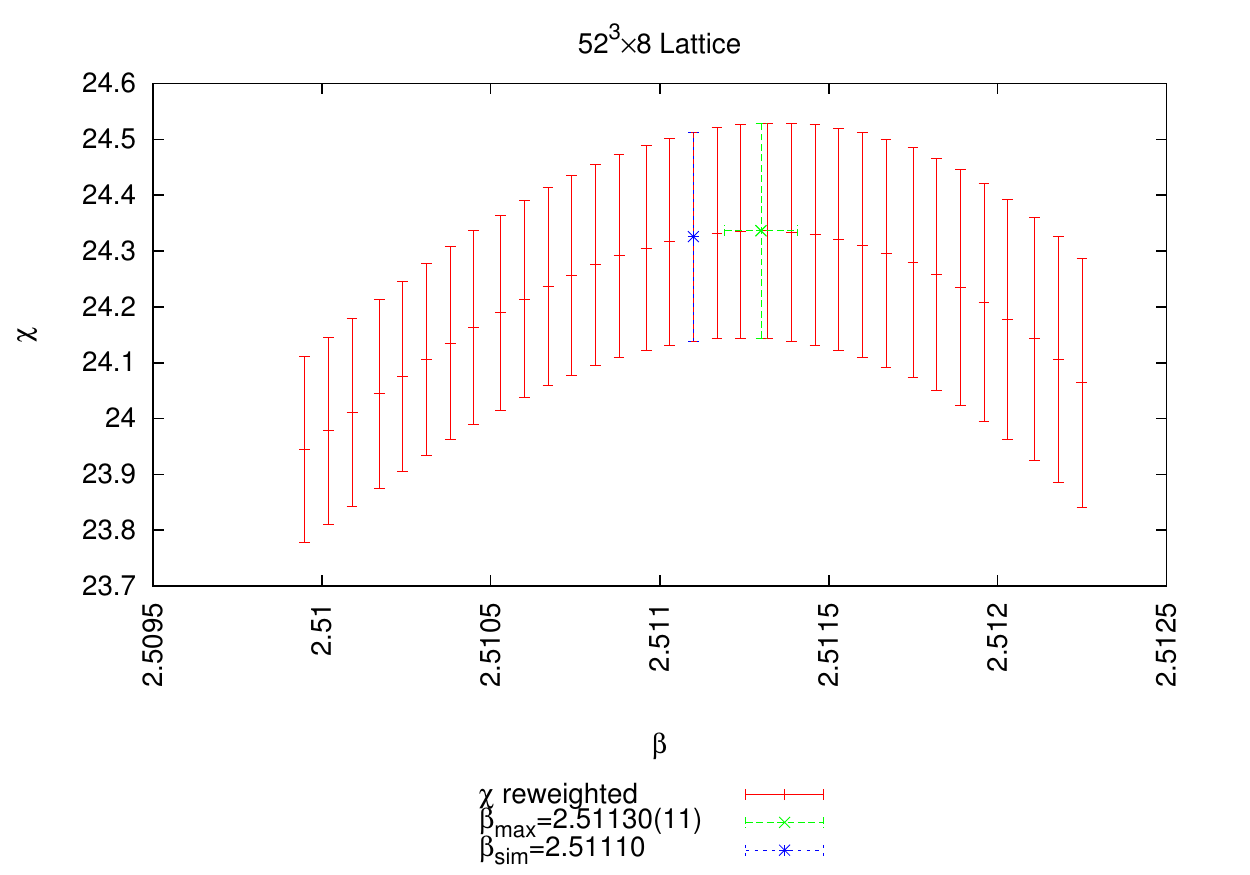}
  \includegraphics[width=0.489\linewidth]{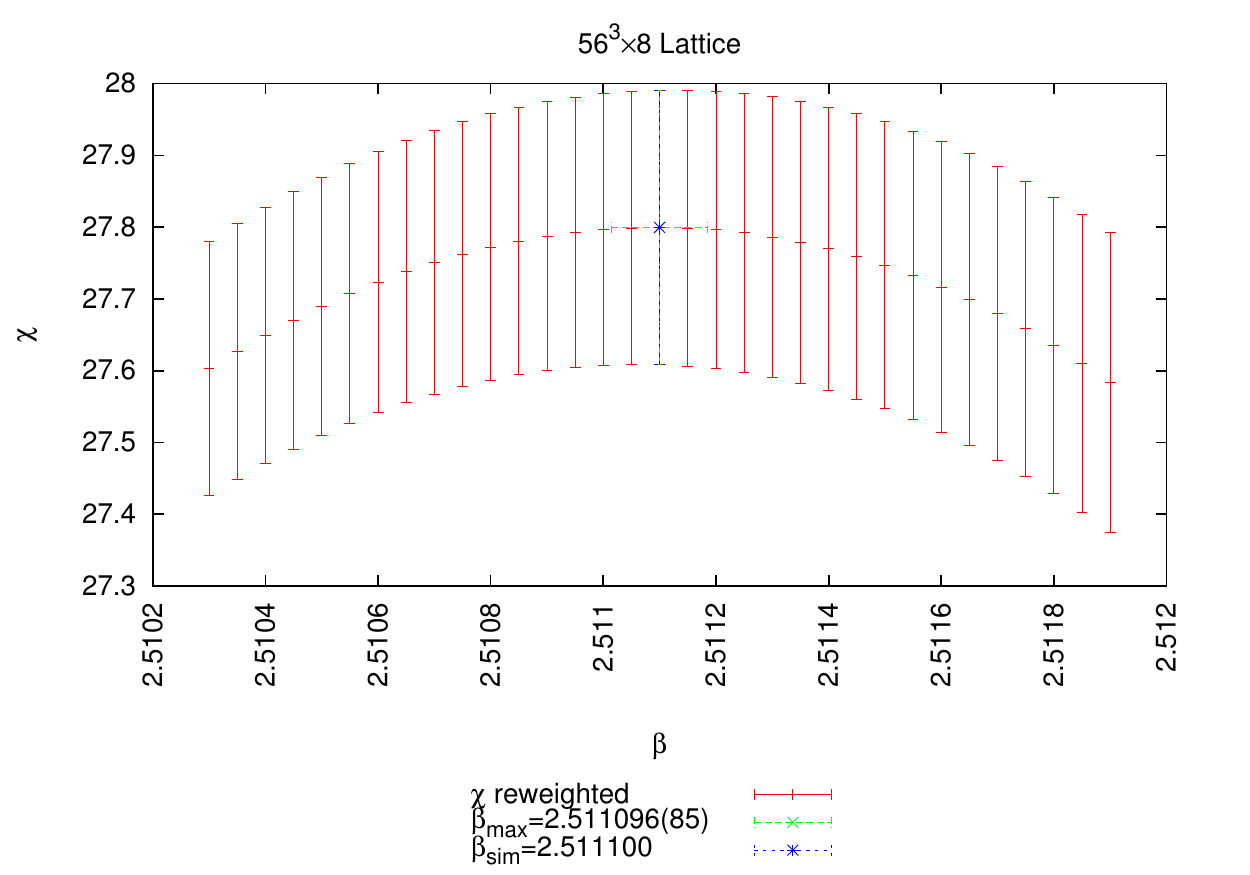}
  \includegraphics[width=0.489\linewidth]{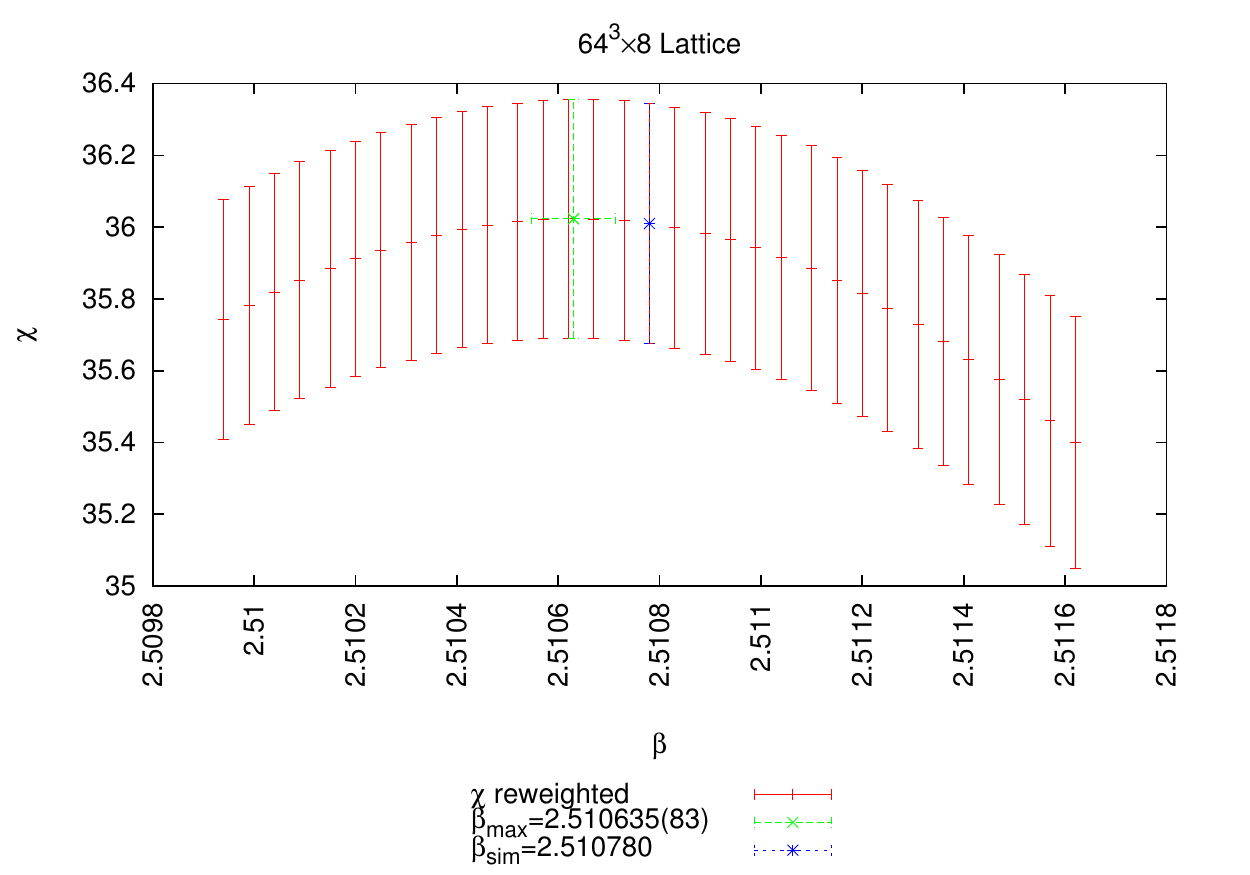}
  \includegraphics[width=0.489\linewidth]{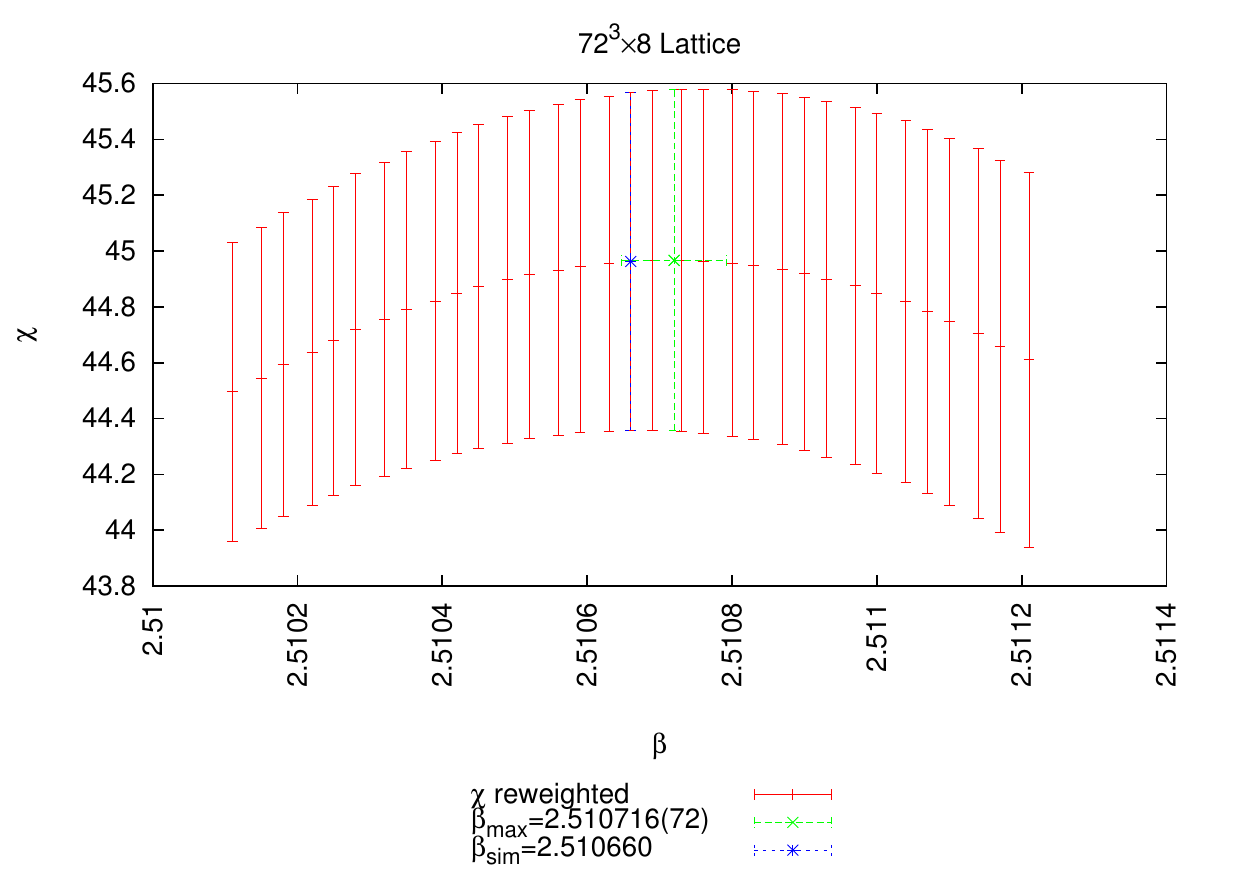}
  \includegraphics[width=0.489\linewidth]{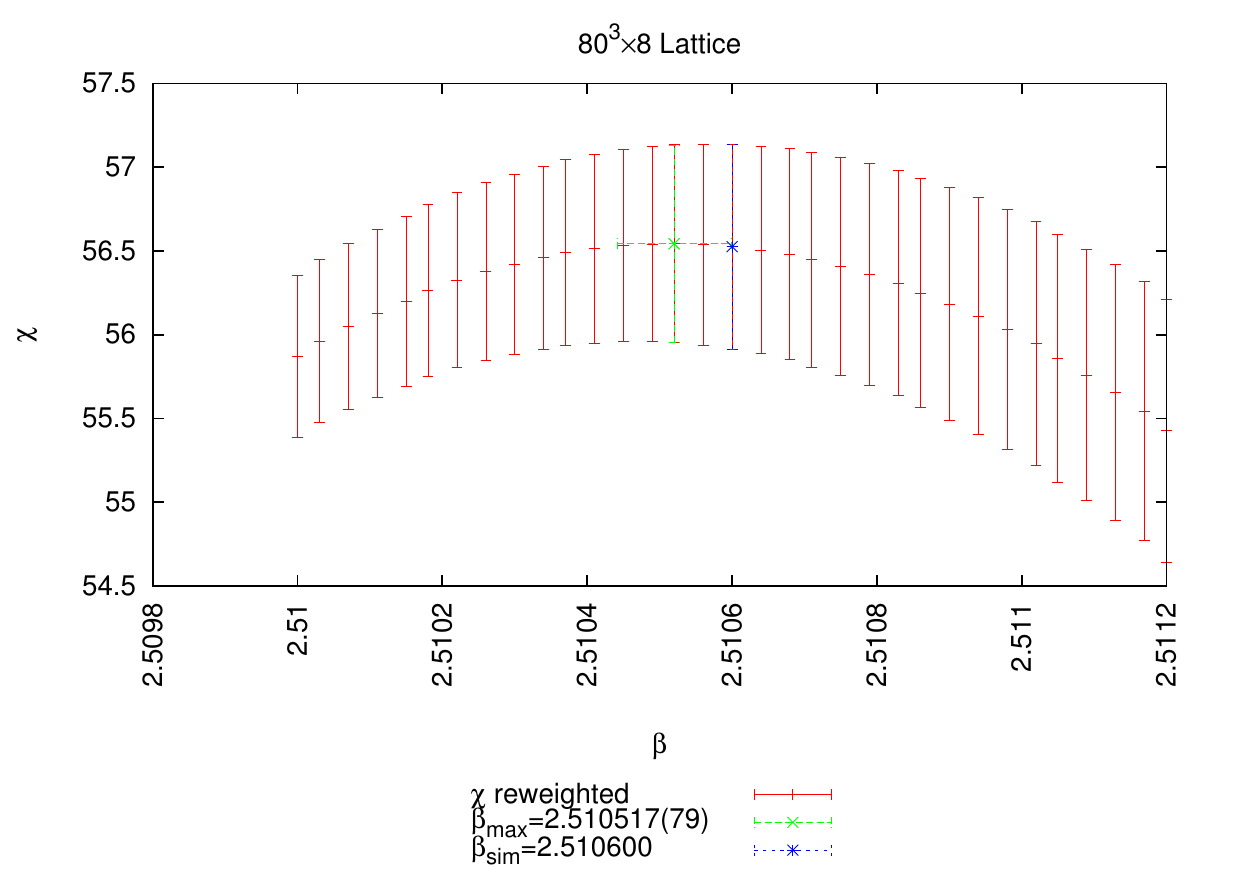}
  \caption[]{Continued.}
\end{figure}

% N_tau=10
\begin{figure}
  \centering
  \includegraphics[width=0.489\linewidth]{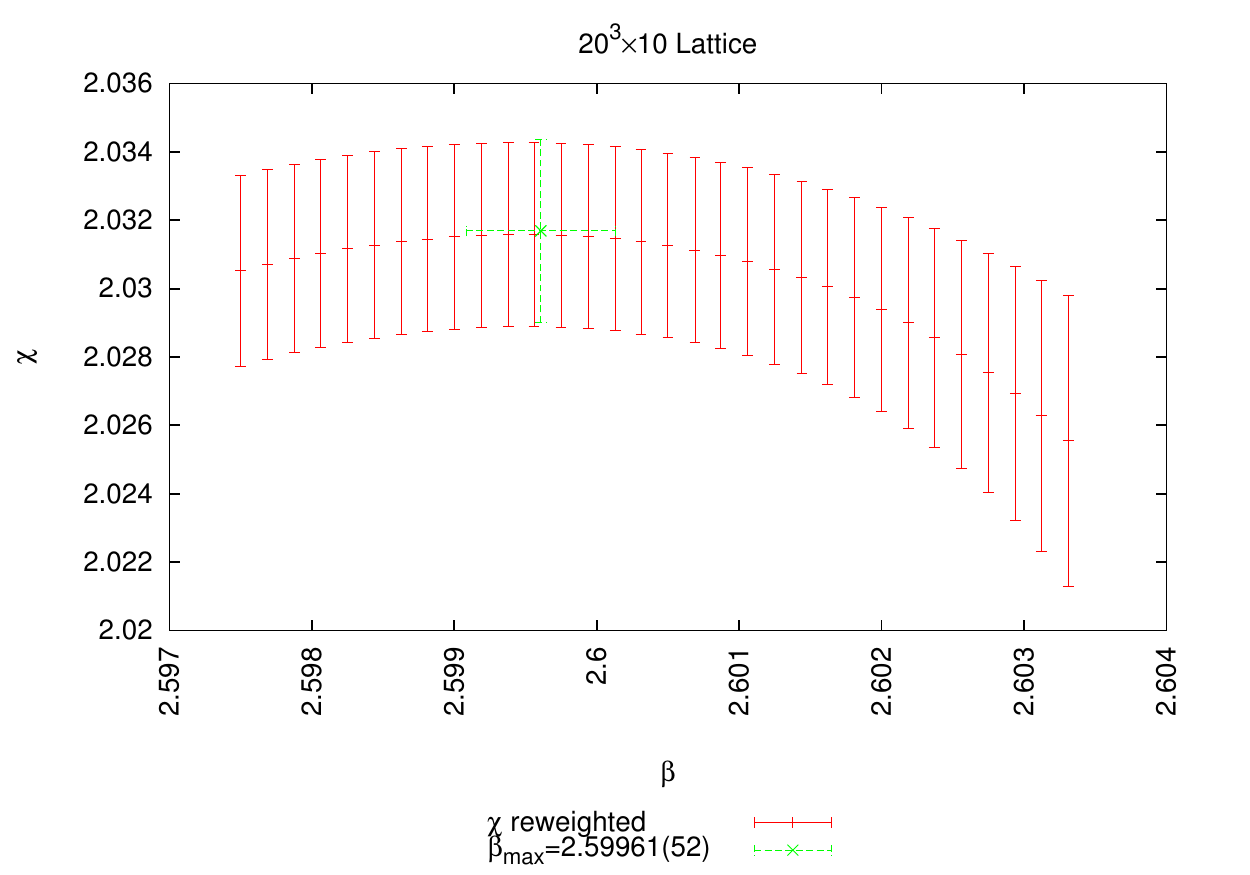}
  \includegraphics[width=0.489\linewidth]{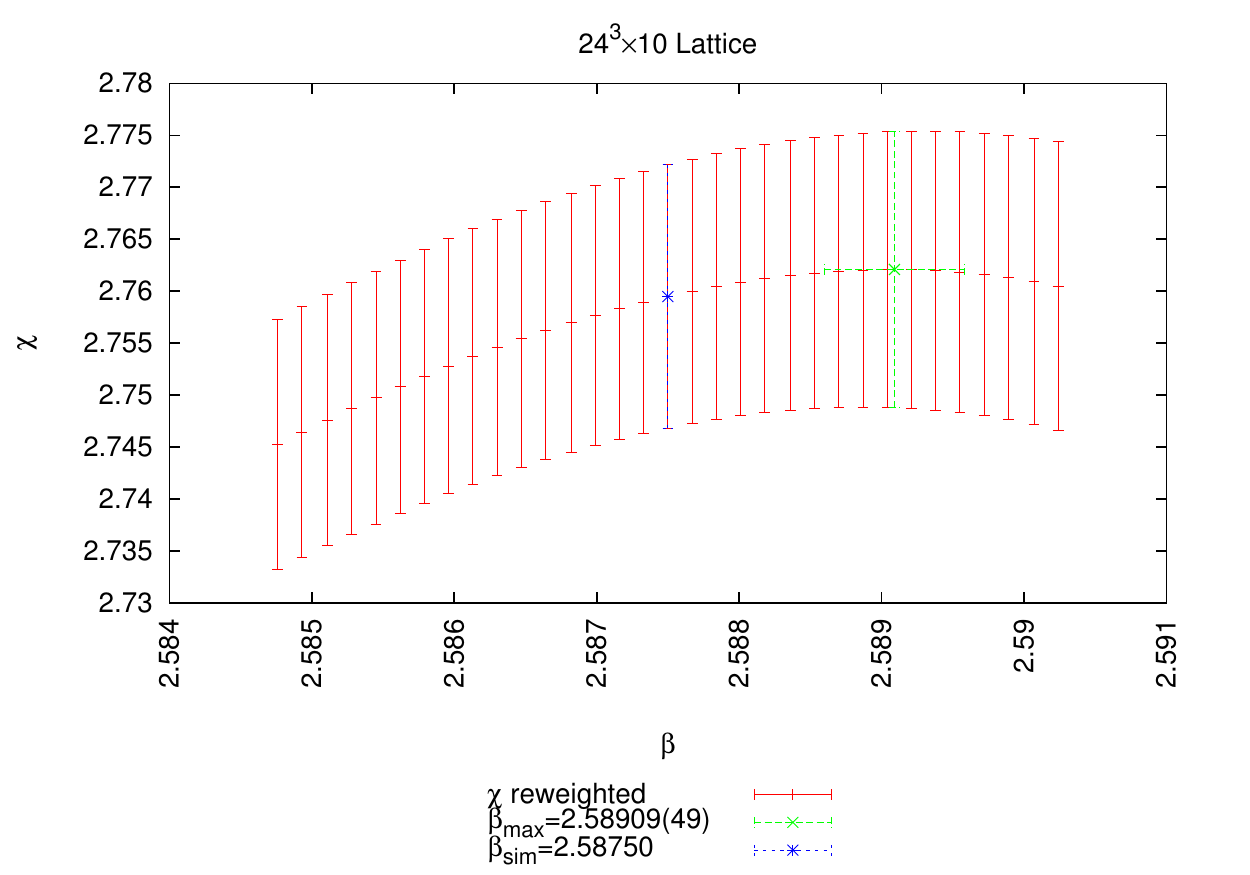}
  \includegraphics[width=0.489\linewidth]{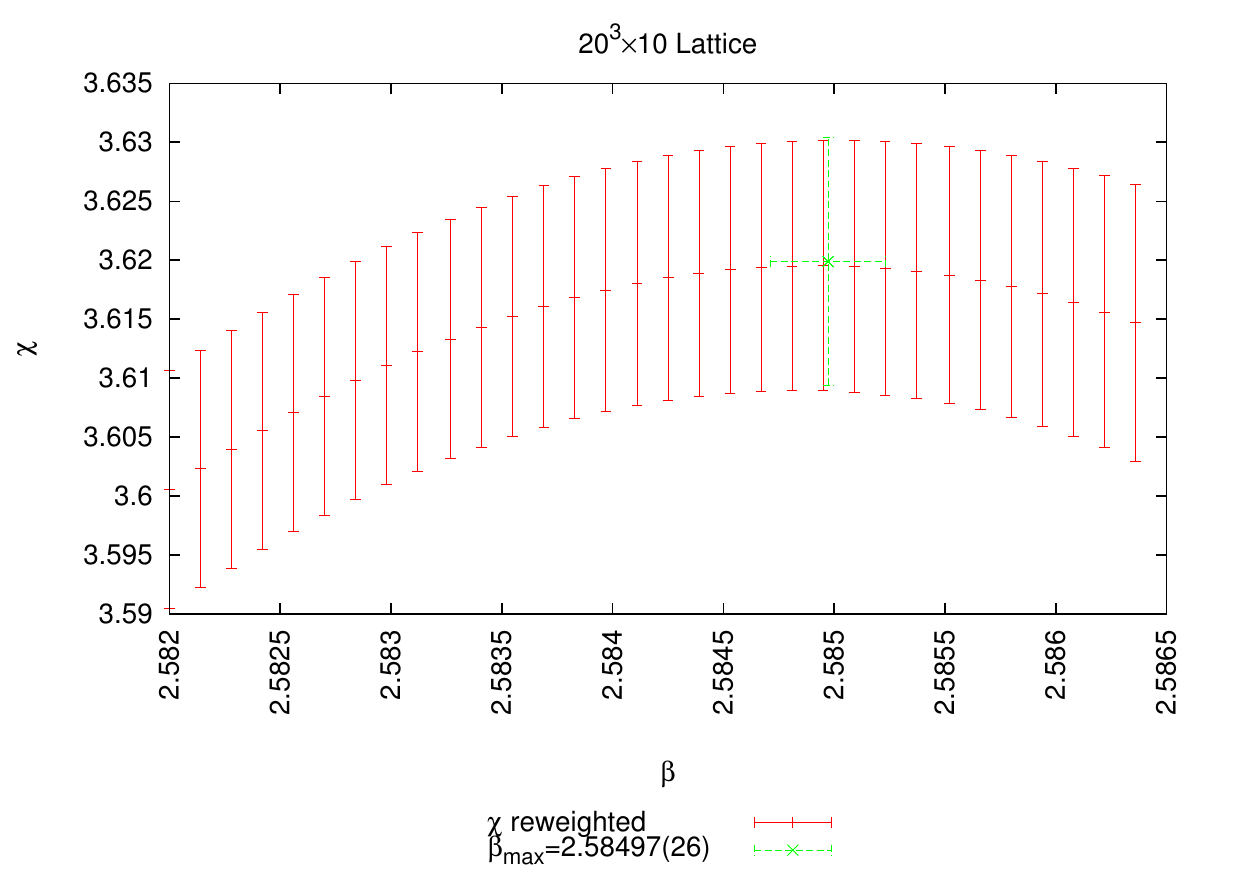}
  \includegraphics[width=0.489\linewidth]{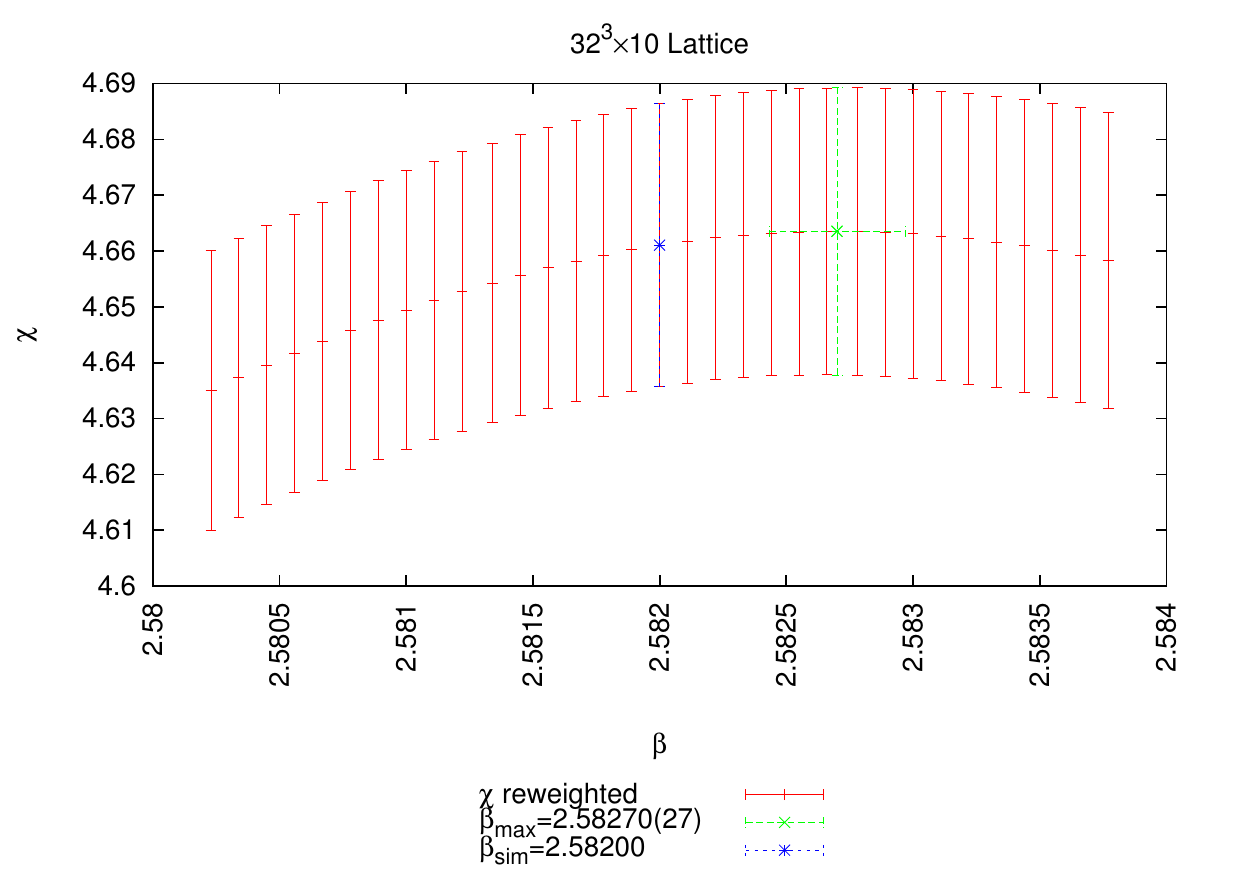}
  \includegraphics[width=0.489\linewidth]{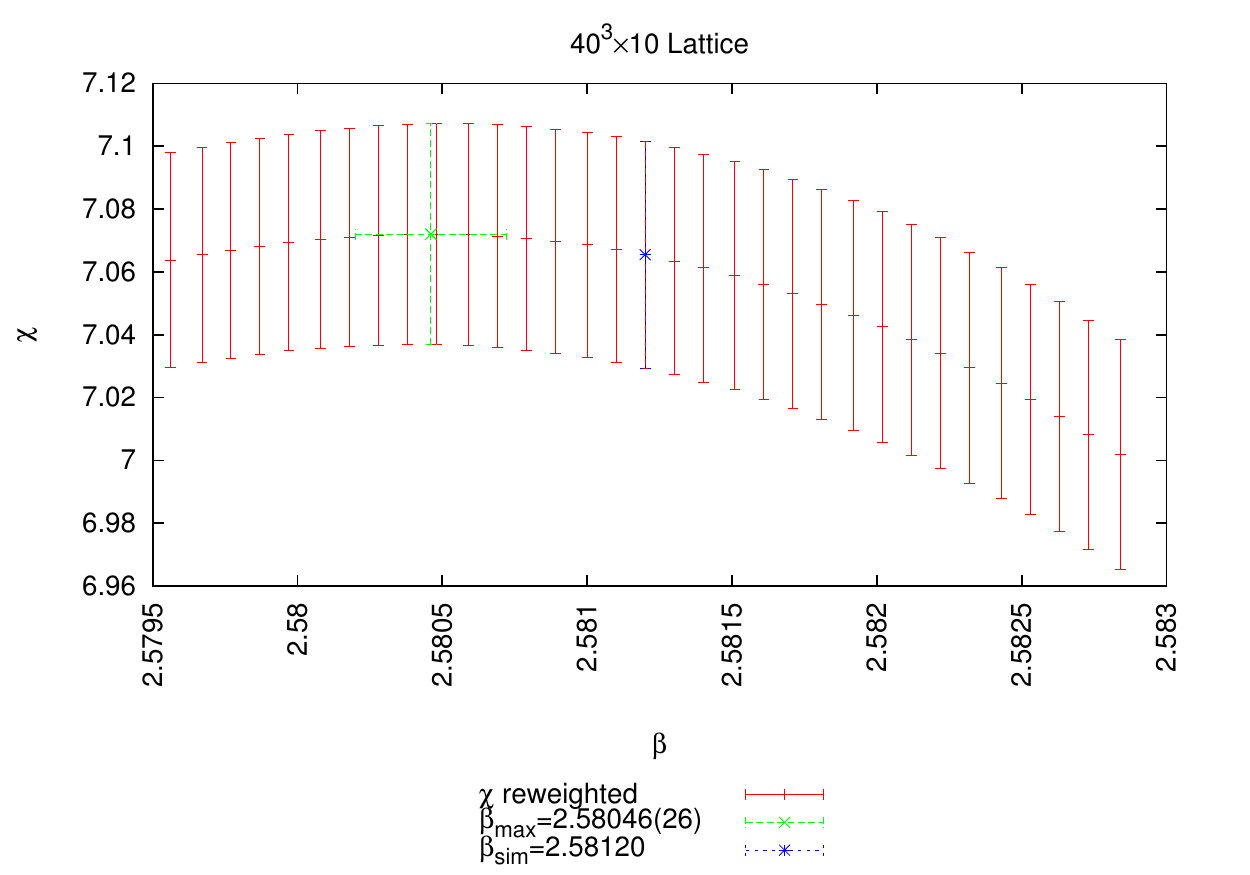}
  \includegraphics[width=0.489\linewidth]{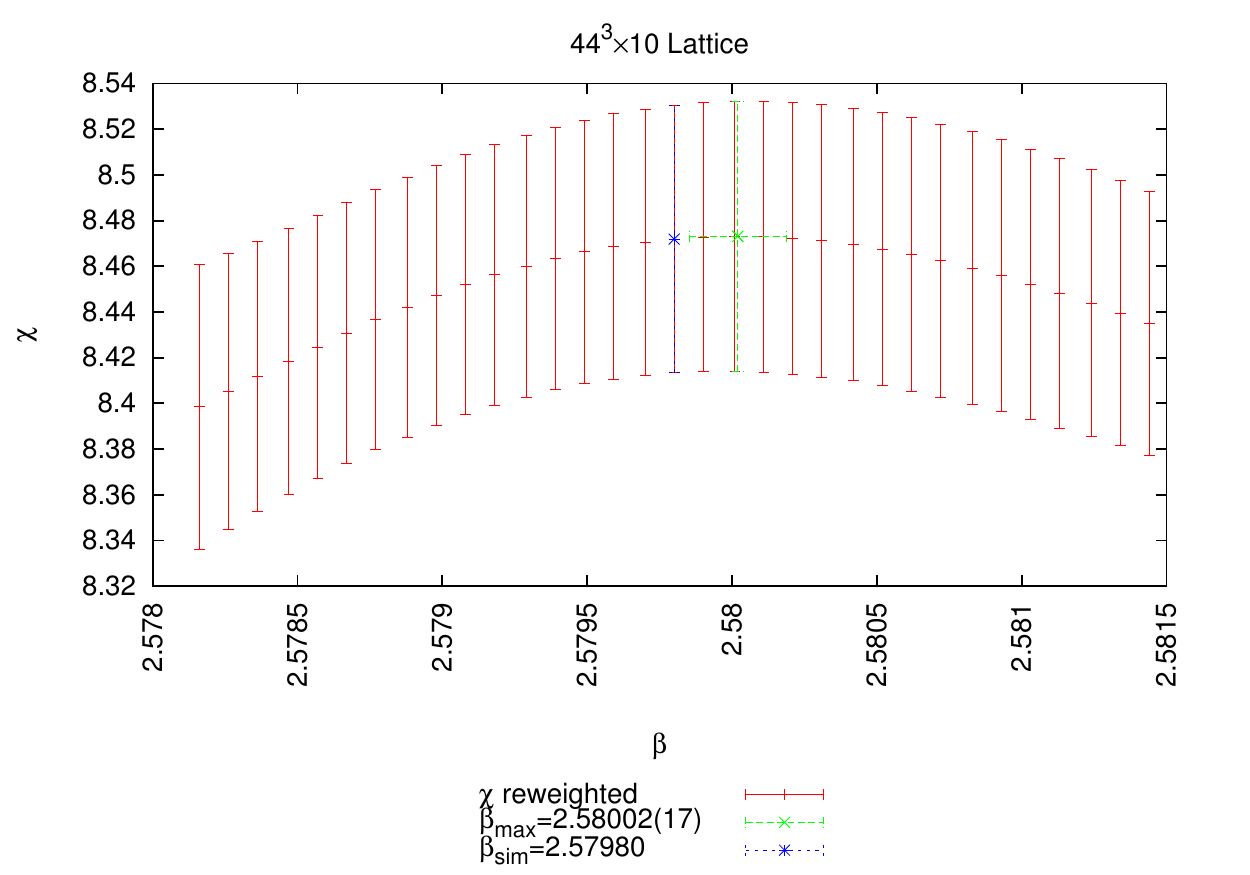}
  \caption{$N_\tau=10$ Polyakov loop reweighting.
           Figures without a simulation point have results combined from 
           data generated at multiple nearby simulation points.}
\end{figure}
\begin{figure}\ContinuedFloat
  \centering
  \includegraphics[width=0.489\linewidth]{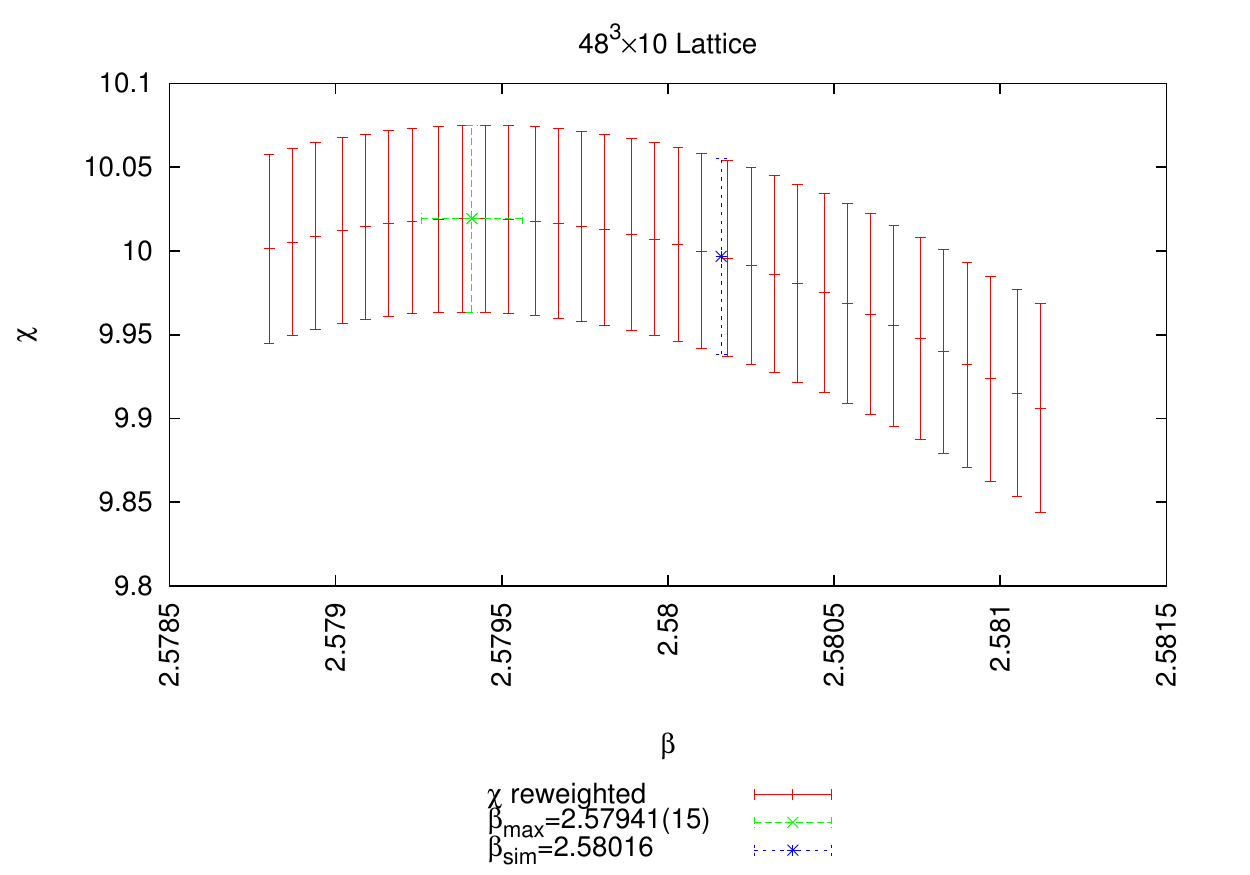}
  \includegraphics[width=0.489\linewidth]{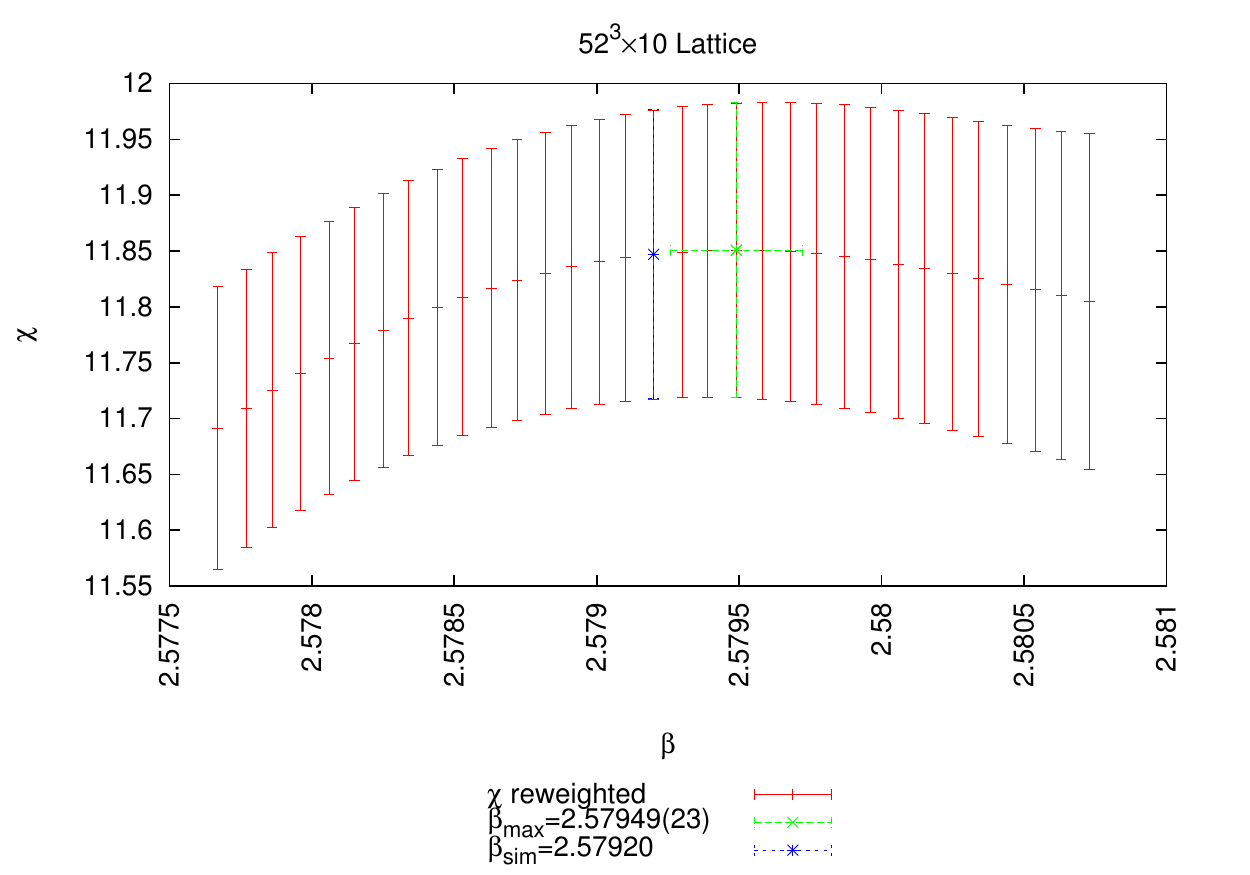}
  \includegraphics[width=0.489\linewidth]{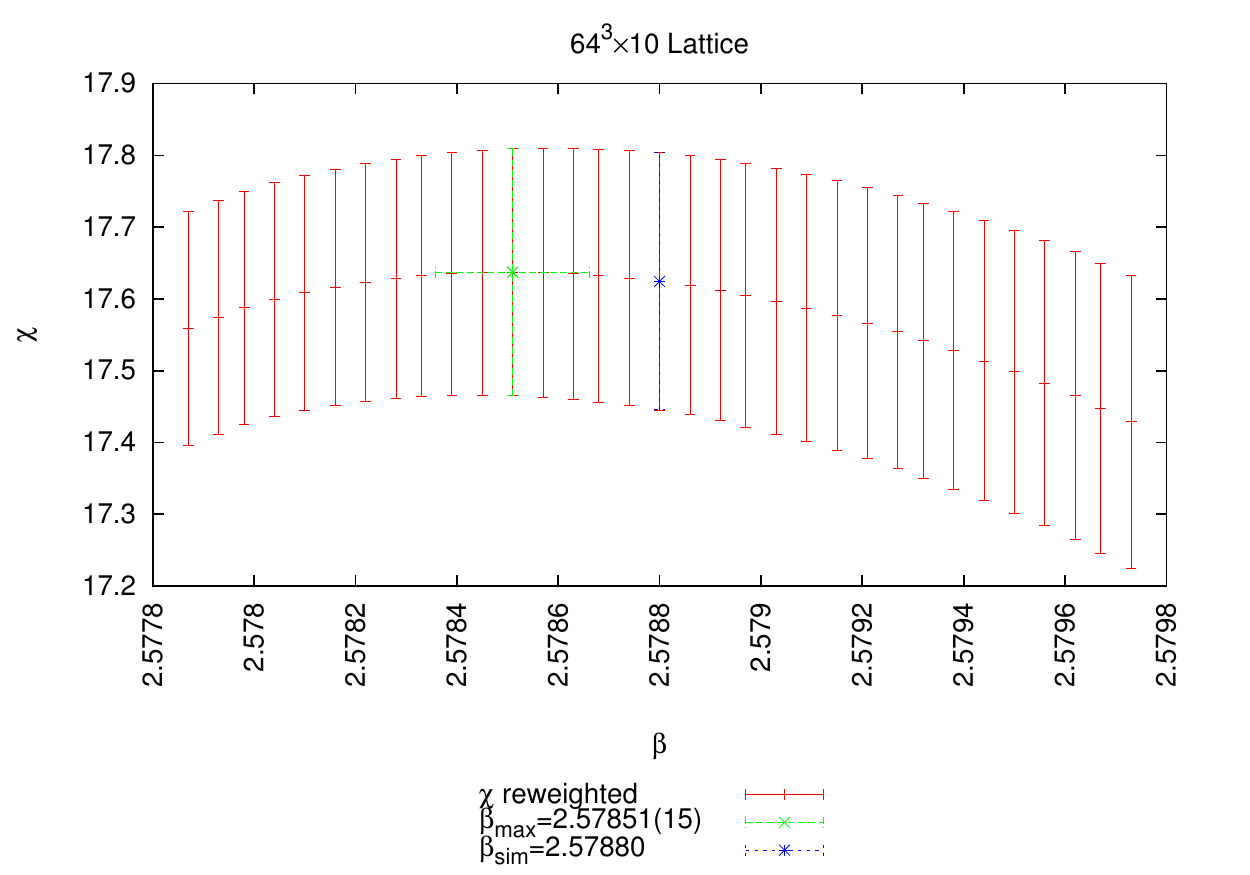}
  \caption[]{Continued.}
\end{figure}

% N_tau=12
\begin{figure}
  \centering
  \includegraphics[width=0.489\linewidth]{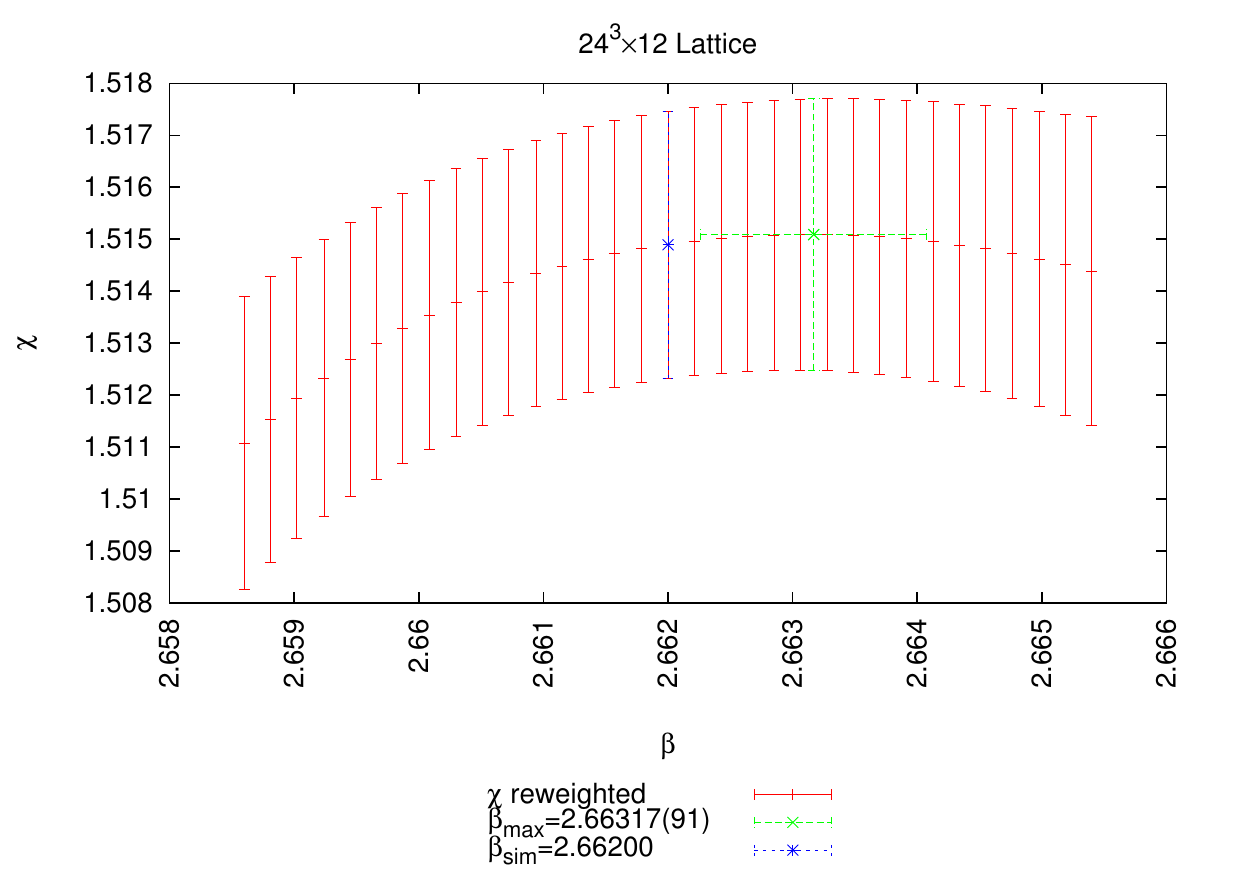}
  \includegraphics[width=0.489\linewidth]{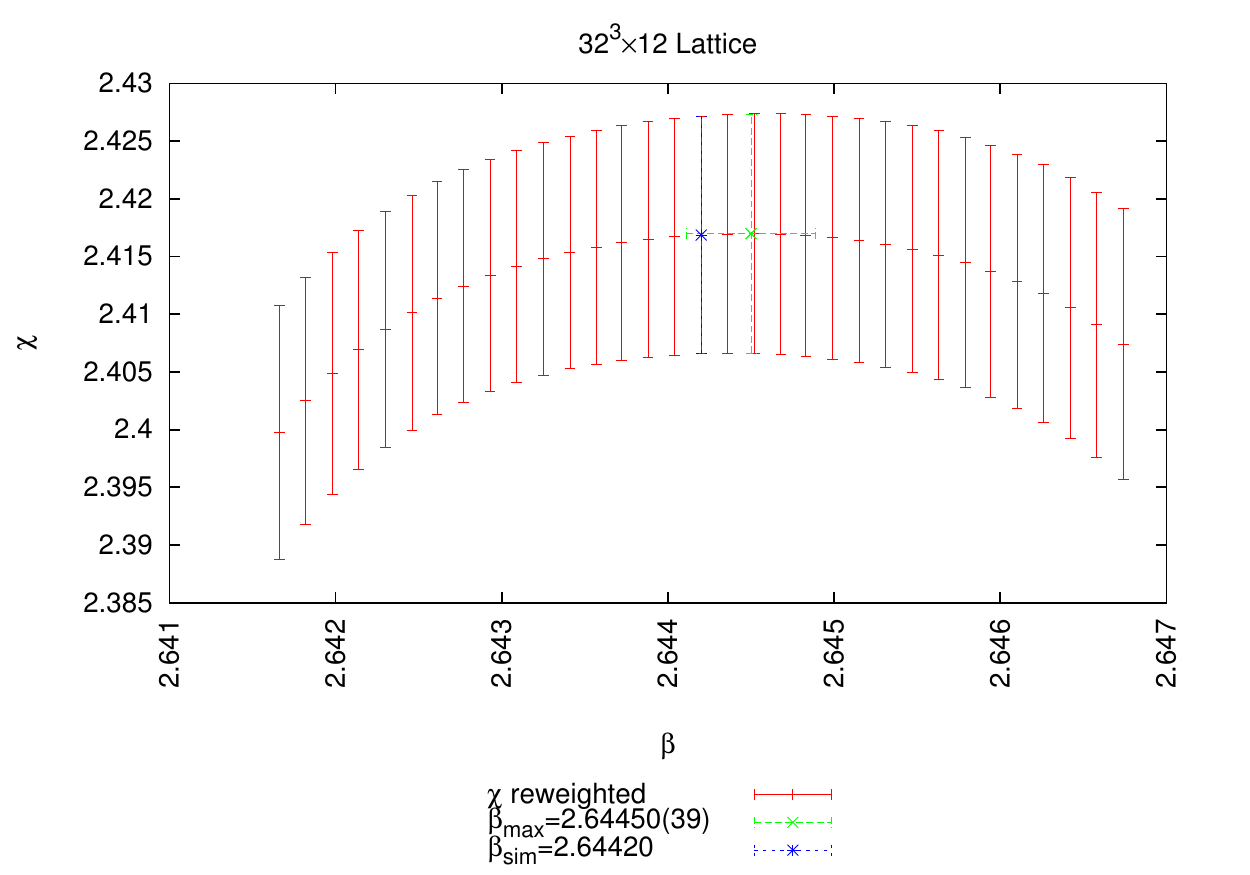}
  \caption{$N_\tau=12$ Polyakov loop reweighting.}
\end{figure}
\begin{figure}\ContinuedFloat
  \centering
  \includegraphics[width=0.489\linewidth]{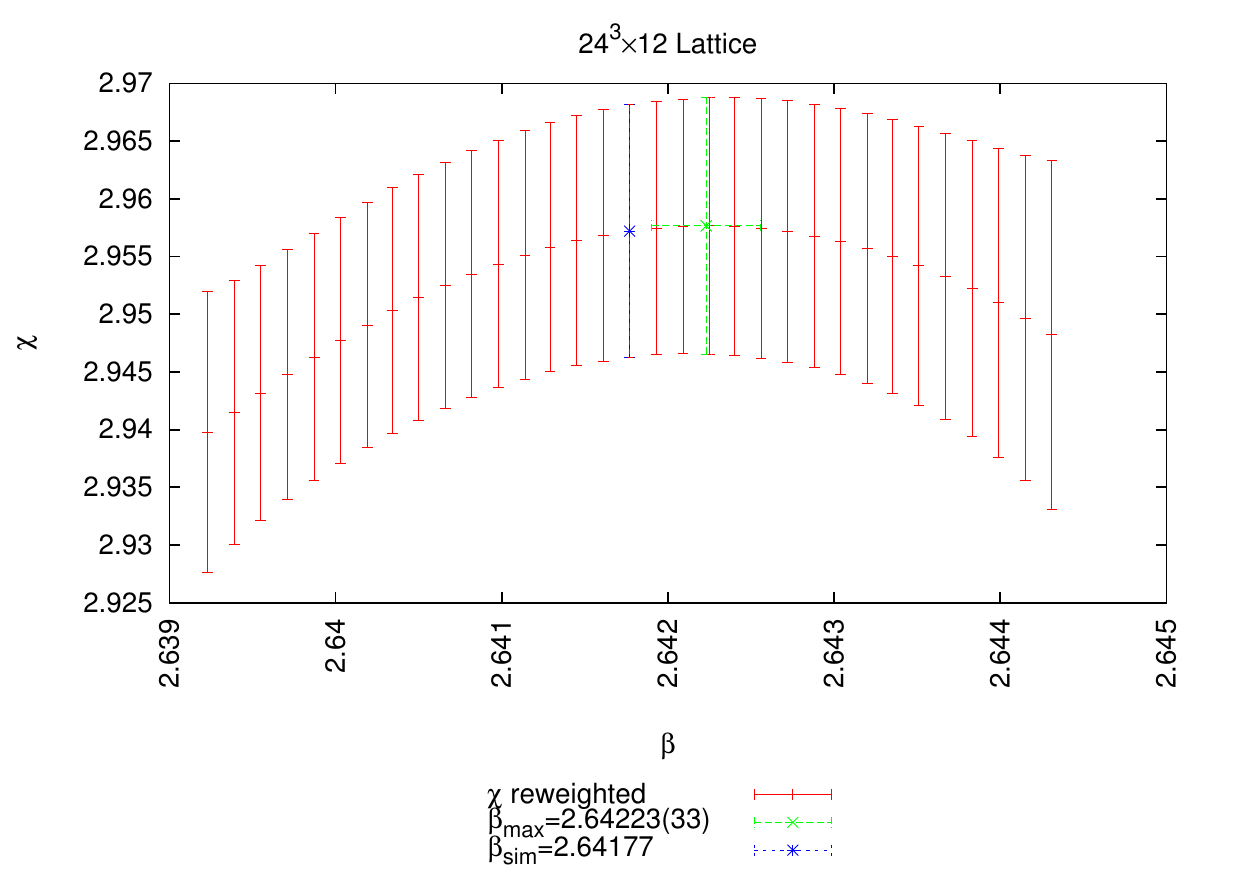}
  \includegraphics[width=0.489\linewidth]{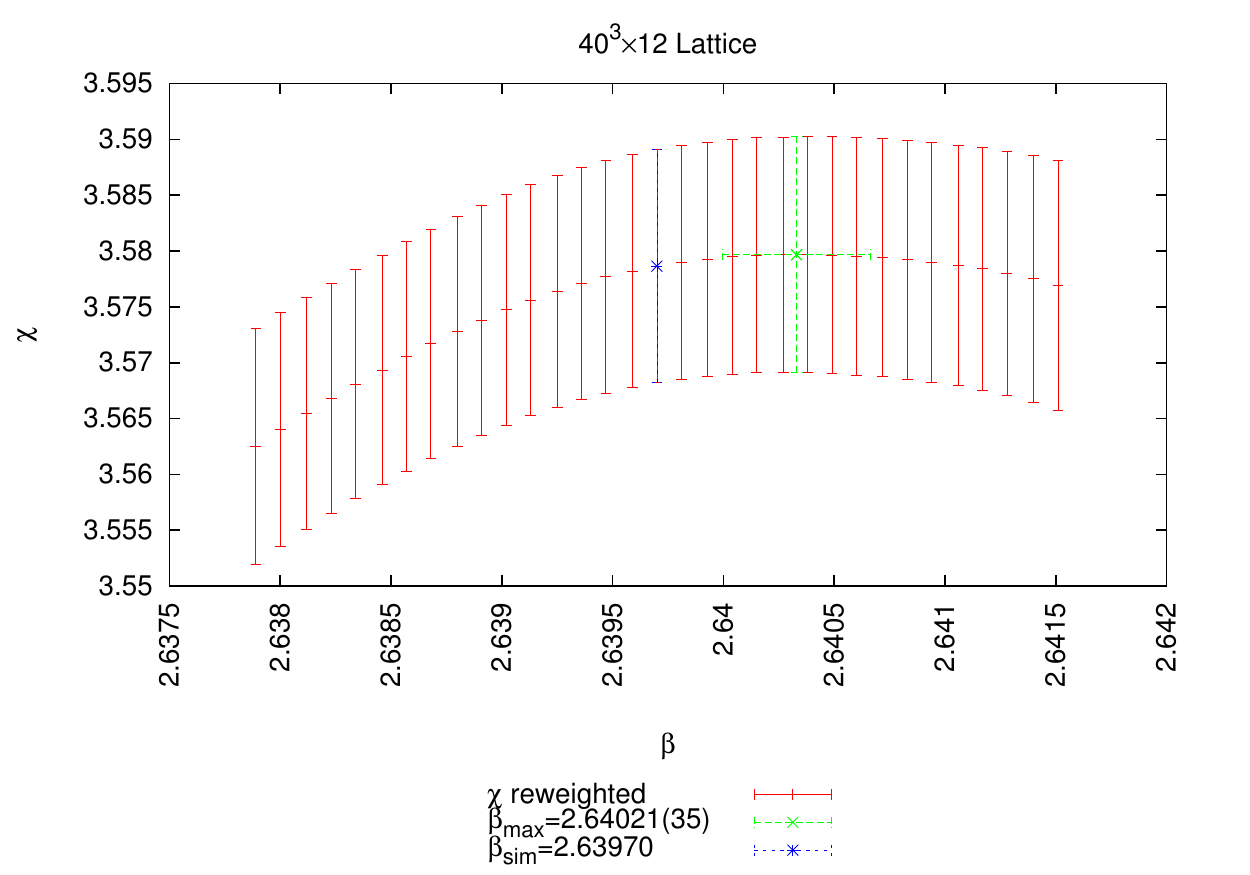}
  \includegraphics[width=0.489\linewidth]{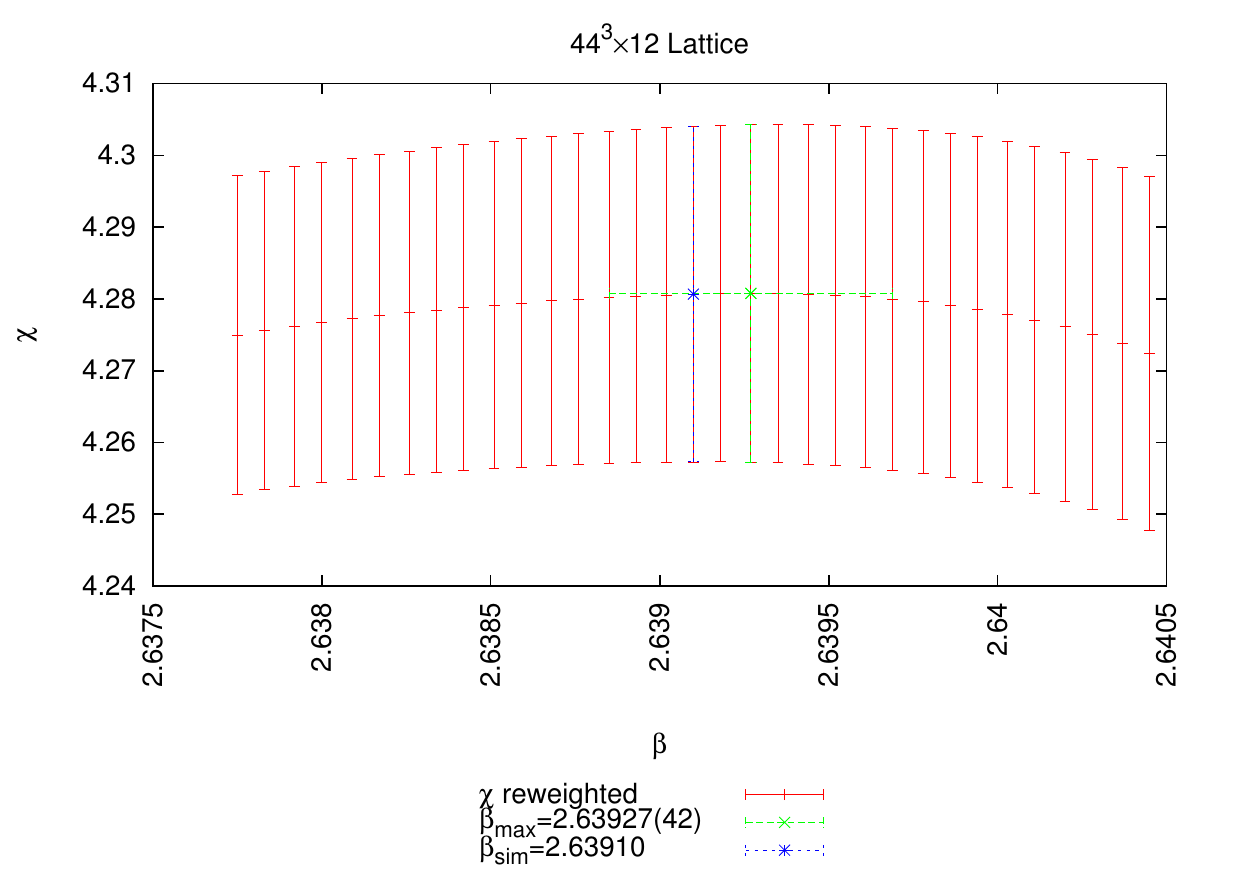}
  \includegraphics[width=0.489\linewidth]{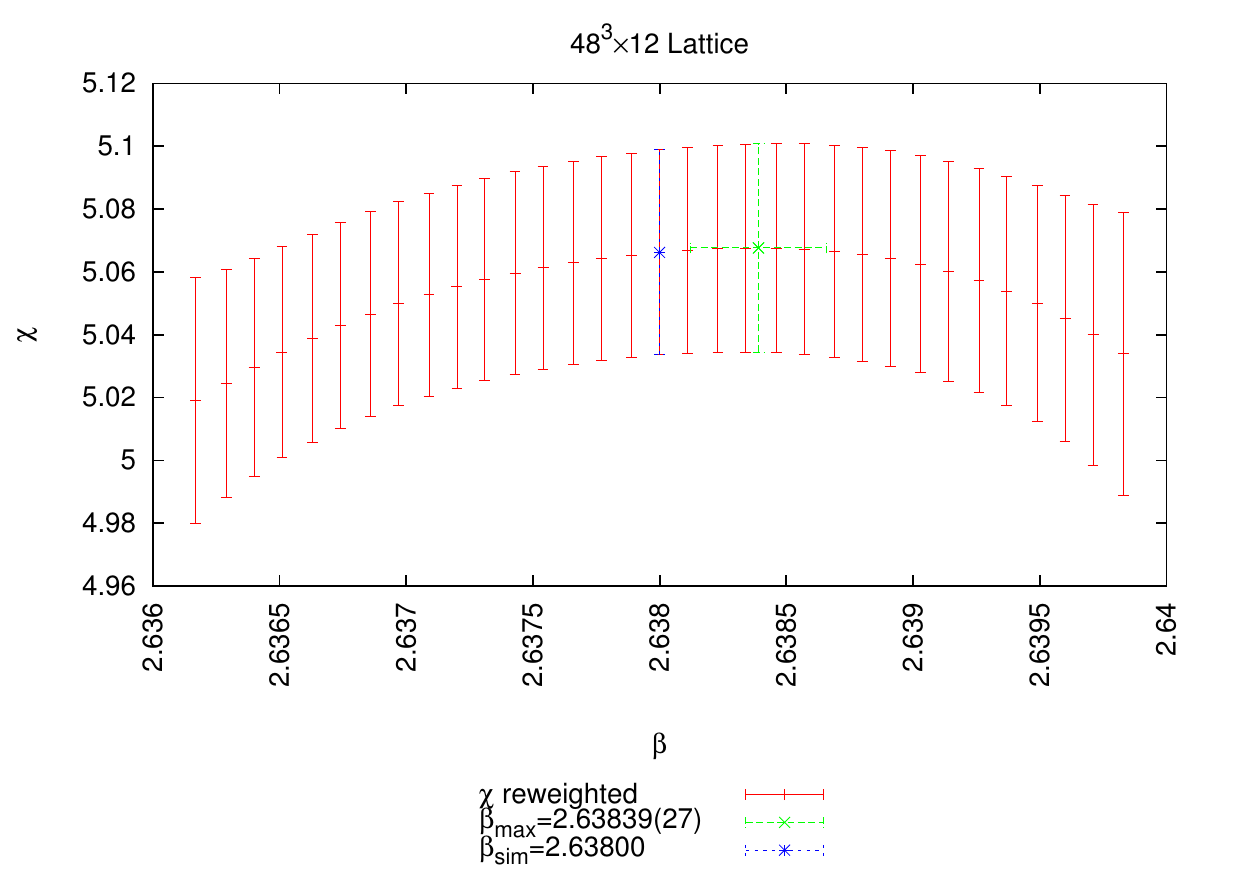}
  \caption[]{Continued.}
\end{figure}

% beta_c finite size fits
\begin{figure}
  \centering
  \includegraphics[width=0.489\linewidth]{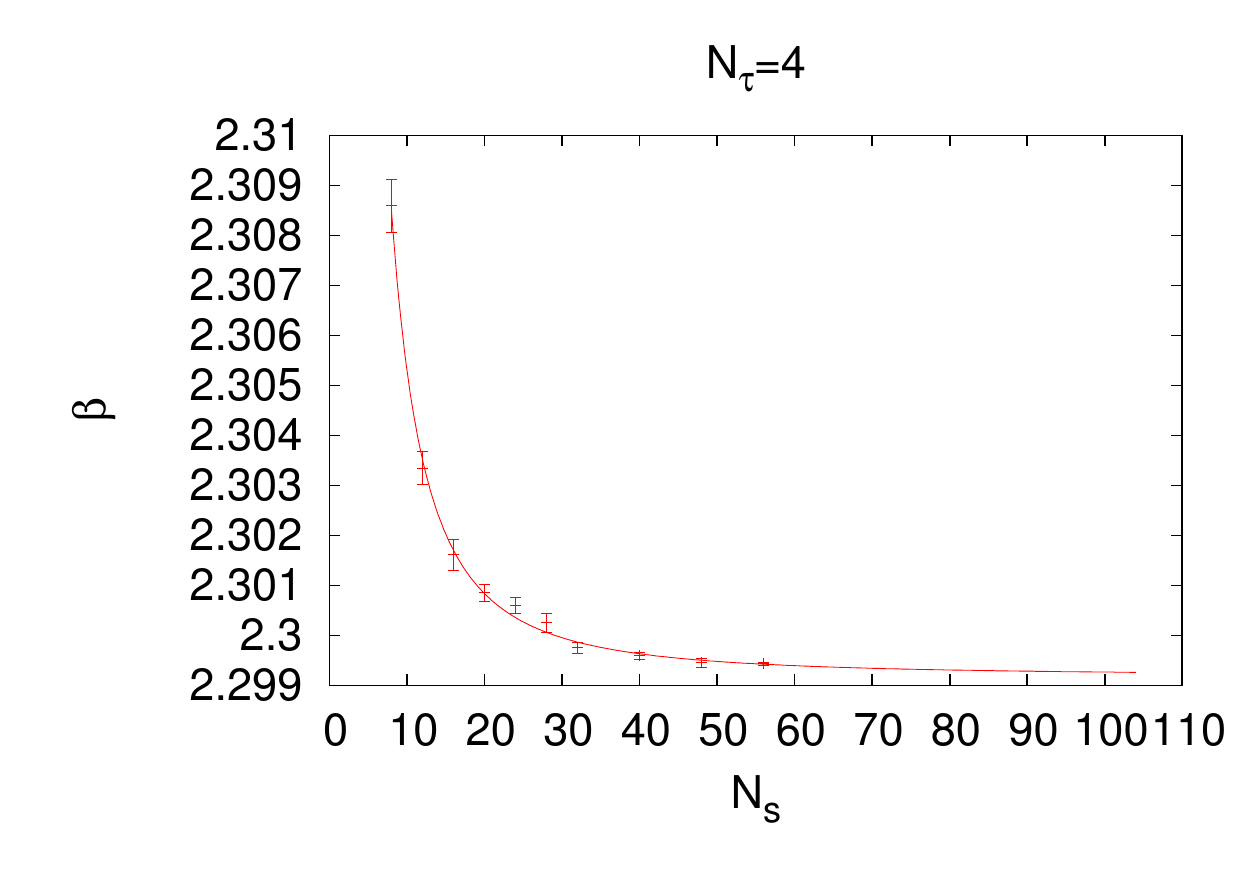}
  \includegraphics[width=0.489\linewidth]{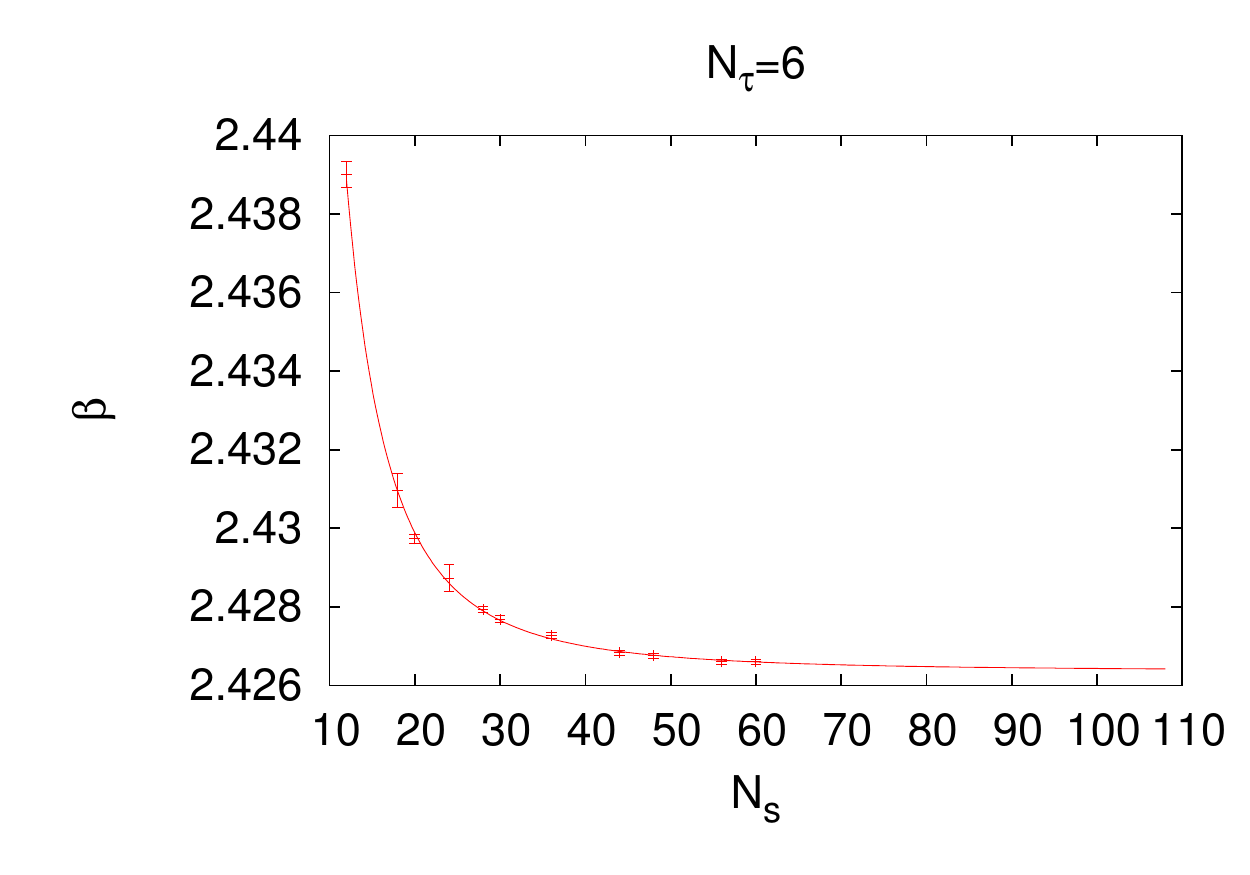}
  \includegraphics[width=0.489\linewidth]{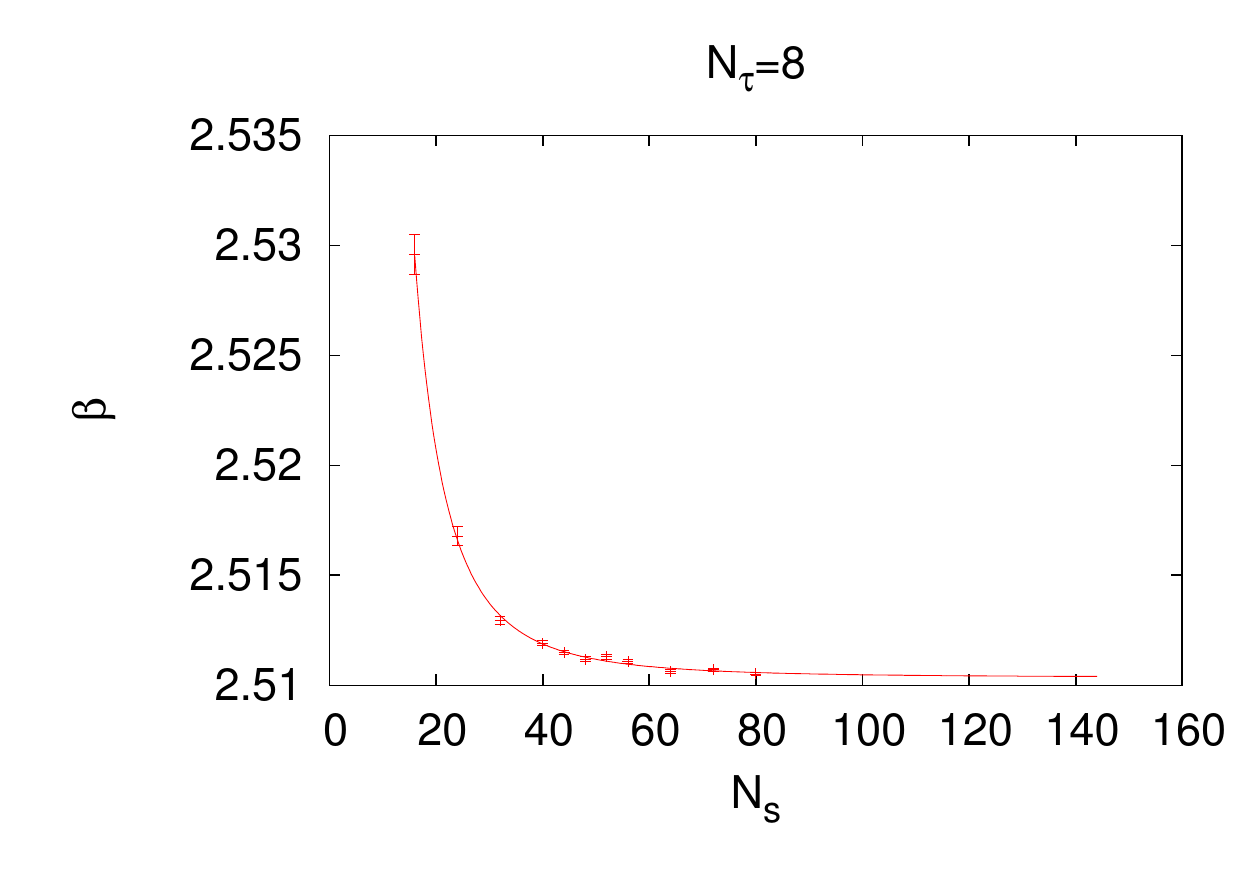}
  \includegraphics[width=0.489\linewidth]{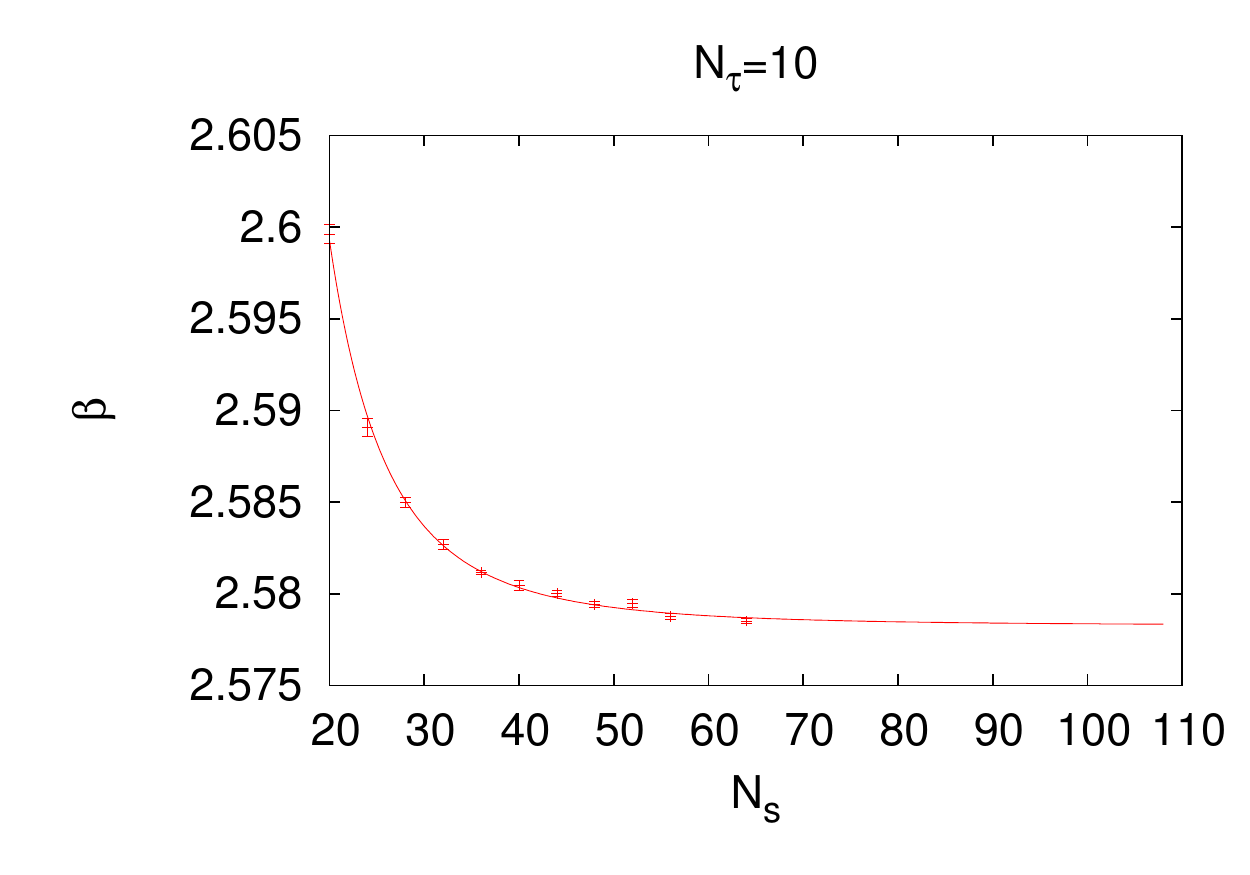}
  \caption{Finite size fits~\eqref{eq:3paramtc} for
           determining critical coupling constants $\beta_c(N_\tau)$.}
  \label{fig:supptcfss}
\end{figure}

%% file: appendix5.tex
\chapter{Probability and Statistics}\label{ap:prb}

This appendix is an introduction to the statistical
tools needed to analyze data, especially as generated by Markov Chain
Monte Carlo simulations. 
We will only be concerned with continuous random variables, 
and we will simply call them random variables.
We will denote random variables with capital letters.
Part of this presentation follows
Chapters~1 and~2 of Berg~\cite{berg_markov_2004}.

\section{Preliminaries}
%We start with a few definitions.
%The set of all possible outcomes $\Omega$ of an experiment is called the 
%{\it sample space}. {\it Events} $A$ are aggregates of points in the sample 
%space. If $\Omega$ is a measurable set, then a {\it random variable} is a 
%function $X:\Omega\to\mathbb{R}$. 

%Next we need a way to define probabilities. In the case that $|\Omega|<\infty$,
%we might say that the probability of event $A$ is $\pr{A}=|A|/|\Omega|$. In
%this instance, we say that the probability is {\it uniform} since every point 
%in the sample space is equally likely. However need not always be 
%the case; one outcome might be more likely than the other ones. 
For a random variable $X$ and an integrable function 
$f:\mathbb{R}\to\mathbb{R}$, we assign a 
probability that $X$ lies in the interval $[a,b]$ by
\begin{equation}\label{eq:cpr}
  \pr{X\in[a,b]}=\int_a^bdx\,f(x)~~~~
   \text{with}~~~~1=\int_{-\infty}^\infty dx\,f(x).
\end{equation}
%Often statisticians say $X$ is a {\it continuous} random variable, which is
%to be contrasted with {\it discrete} random variables. 
The function $f$ is called the {\it probability distribution function} (PDF).
The {\it cumulative distribution function} (CDF) is the function
$F(x)$ given by
\begin{equation}
  F(x)\equiv \pr{X<x}=\int_{-\infty}^xdt\,f(t).
\end{equation}
Two examples of important PDFs are the {\it Gaussian}
or {\it normal} distribution,
\begin{equation}
  \gau(x,\hat{x},\sigma)\equiv\frac{1}{\sigma\sqrt{2\pi}}
  \exp\Bigg(-\frac{(x-\hat{x})^2}{2\sigma^2}\Bigg)
\end{equation}
where $\sigma$ is the standard deviation
of the distribution and $\hat{x}$ is the mean, and the {\it Cauchy} 
distribution,
\begin{equation}
  \cau(x,\alpha)\equiv\frac{\alpha}{\pi\big(\alpha^2+x^2\big)}.
\end{equation}
We will refer to these PDFs later, particularly the normal 
distribution. We will call their CDFs $\Gau$ and $\Cau$, respectively.   

Now we present ways ways to characterize PDFs. We can get some information
from the mean and variance of a distribution. These are
both special cases of a more general concept.
Let $n\in\mathbb{N}$.
The {\it $n^{th}$ moment} of the distribution $f(x)$ is
\begin{equation}\label{dfn:mom}
  \ev{X^n}=\int_{-\infty}^\infty dx\,x^nf(x).
\end{equation}
The mean and variance are the special cases $\hat{x}=\ev{X}$ and 
$\sigma^2=\ev{(X-\hat{x})^2}$. Sometimes we call the mean the {\it expected
value} and sometimes we denote the variance $\variance$. Note that not all 
probability distributions have well-defined moments. The Cauchy distribution is
ill-behaved in this regard, since its $n^{th}$ moment diverges
$\Forall n\in\mathbb{N}$.

Generally in the lab, one draws random variables from distributions
about which one has no a priori knowledge, and therefore
does not know the true moments these distributions. The
definition \eqref{dfn:mom} suggests a way to estimate them. 
Suppose we draw a sample $X_1,...,X_N$.
  An {\it estimator} of the $n^{th}$ moment is
  \begin{equation}
    \bar{X}^n\equiv\frac{1}{N}\sum_{i=1}^N X_i^n.
  \end{equation}
In the case $n=1$ we obtain the ordinary arithmetic average.
We use the hat for true values and the bar for estimators.
For estimators of moments 
besides the mean, we must be more careful; this is discussed in 
Section~\ref{sec:bias}.

Consider two intervals $[a,b]$ and $[c,d]$ and two random variables
$X$ and $Y$ drawn from PDFs $f$ and $g$, respectively. Then $X$ and $Y$ are
said to be {\it independent} if
\begin{equation}\label{dfn:ind}
 \pr{X\in[a,b]\text{ and }Y\in[c,d]}=\int_a^b\int_c^d dx\,dy f(x)\,g(y)
\end{equation}
Hence the {\it joint PDF} of $X$ and $Y$ is $f(x)g(y)$.
We say $X$ and $Y$ are {\it uncorrelated} if
\begin{equation}
  \ev{XY}=\ev{X}\ev{Y}.
\end{equation}
The {\it covariance}
\begin{equation}\label{dfn:cov}
  \Cov[X,Y]\equiv\ev{XY}-\ev{X}\ev{Y}
\end{equation}
gives a measure of how correlated $X$ and $Y$ are. One can also
use the {\it correlation}
\begin{equation}\label{dfn:cor}
  \rho(X,Y)=\frac{\Cov[X,Y]}{\sqrt{\sigma^2_X\sigma^2_Y}}.
\end{equation}
So equivalently we say $X$ and $Y$ are uncorrelated if $\rho(X,Y)=0$.
It is worth emphasizing that if $X$ and $Y$ are independent,
it follows that they are uncorrelated. However if $X$ and $Y$ are
uncorrelated, {\it they can still be dependent}.
Here is an extreme example by Cosma Shalizi~\cite{cosma_indep}. Let $X$ be 
uniformly distributed
on [-1,1] and let $Y=|X|$. Then clearly $Y$ depends on $X$. However it 
is easy to see that $Y$ is uniform on [0,1] and $\ev{XY}=0=\ev{X}\ev{Y}$. 
Hence $X$ and $Y$ are not correlated.

The next two propositions show us how to add expectation values and
random variables. Let $X$ and $Y$ be independent random variables 
drawn from PDFs $f$ and $g$, respectively.
\begin{proposition}
  Let $a,b\in\mathbb{R}$ be constants. Then
  $$\ev{aX+bY}=a\ev{X}+b\ev{Y}.$$
  \begin{proof}
    Since $X$ and $Y$ are independent, their joint PDF is $fg$. Then
    \begin{equation*}
      \begin{aligned}
      \ev{aX+bY}&=\int dxdy\,(ax+by)f(x)g(y)\\
                &=a\int dxdy\,x\,f(x)g(y)+
                 b\int dxdy\,y\,f(x)g(y)\\
                &=a\int dx\,x\,f(x)+b\int dy\,y\,g(y)\\
                &=a\ev{X}+b\ev{Y}.
      \end{aligned}
    \end{equation*}
  \end{proof}
\end{proposition}

\begin{proposition}
  \label{prp:addvars}
  The PDF of the random variable $Z=X+Y$ is given by the convolution
  \begin{equation*}
    h(z)=\int_{-\infty}^\infty dx\,f(x)g(z-x)
  \end{equation*}
  \begin{proof}
    The CDF of $Y$ is, according to eq.~\eqref{dfn:ind},
    \begin{equation*}
      G(y)=\int_{x+y\leq z}dx\,dy\,f(x)g(y)
          =\int_{-\infty}^\infty dx\,f(x)\int_{-\infty}^{z-x}
            dy\,g(y).
    \end{equation*}
    The PDF $h$ follows from the Fundamental Theorem of Calculus:
    \begin{equation*}
      h(z)=\frac{dH}{dz}=\frac{dH}{d(z-x)}
          =\int_{-\infty}^\infty dx f(x)g(z-x).
    \end{equation*}
  \end{proof}
\end{proposition}

A sequence $\{X_N\}$ of random variables {\it converges in probability}
toward random variable $X$ if 
\begin{equation}
  \lim_{N\to\infty}\pr{|X_N-X|>\epsilon}=0,
\end{equation}
$\forall\epsilon>0$ 
If it does, we write
\begin{equation}
  X_N\xrightarrow{\text{P}}X.
\end{equation}
%The sequence converges to $X$ {\it almost surely} if
%\begin{equation}
%  \lim_{N\to\infty}\pr{X_N=X}=1,
%\end{equation}
%and in this case we write
%\begin{equation}
%  X_N\xrightarrow{\text{AS}}X.
%\end{equation}

\begin{theorem}[Chebyshev's Inequality]
  Let $X$ be drawn from a PDF with mean $\hat{x}$ and variance
  $\sigma^2$ and let $a>0$. Then
  \begin{equation*}
    \pr{|X-\hat{x}|>a\sigma}<a^{-2}.
  \end{equation*}
  \begin{proof}
    Let $T=(X-\hat{x})^2$ be a new random variable with PDF $g$. Then
    \begin{equation*}
      \pr{|X-\hat{x}|>a\sigma}=\pr{T>a^2\sigma^2}
                              =\int_{a^2\sigma^2}^\infty dt\,g(t)
    \end{equation*}
    But
    \begin{equation*}
      \begin{aligned}
        \sigma^2&=\int_{0}^\infty dt\,t\,g(t)
                =\Bigg(\int_0^{a^2\sigma^2}
                     +\int_{a^2\sigma^2}^\infty\Bigg)dt\,t\,g(t)\\
                &\geq\int_{a^2\sigma^2}^\infty dt\,t\,g(t)
                >a^2\sigma^2\int_{a^2\sigma^2}^\infty dt\,g(t)
                =a^2\sigma^2\pr{T>a^2\sigma^2}.
      \end{aligned}
    \end{equation*}
   Dividing through by $a^2\sigma^2$ completes the proof.
  \end{proof} 
\end{theorem}
Chebyshev's inequality tells us that large deviations from the mean are
unlikely. Intuitively one expects that as the number of measurements
increases, the sample average tends toward the true mean. This is
called the {\it Law of Large Numbers} (LLN). To prove it, we set up as 
follows: Let $X_1,...,X_N$ be a sequence of random variables drawn from a PDF
with mean $\hat{x}$ and variance $\sigma^2$.
\begin{theorem}[Weak LLN]
  $$
    \bar{X}\xrightarrow{\text{P}}\hat{x}.
  $$
  \begin{proof} Our proof will rely on Chebyshev's inequality, so we will
    first need to compute the mean and variance of the distribution of
    $\bar{X}$. All the $X_i$ are drawn from the same PDF, so
    $$
      \ev{\bar{X}}=\frac{1}{N}\sum_{i=1}^N\ev{X_i}
                  =\frac{N\hat{x}}{N}=\hat{x}.
    $$
    Meanwhile the variance of the distribution of $\bar{X}$ is
    $$
      \sigma^2_{\bar{X}}
      =\variance\sum_{i=1}^N \frac{X_i}{N}
      =\sum_{i=1}^N\frac{\sigma^2}{N^2}
      =\frac{\sigma^2}{N}.
    $$
    Now let $\epsilon>0$. Then $\exists\,a>0$ with 
    $\epsilon=a\,\sigma_{\bar{X}}$. Hence by Chebyshev's
    inequality we have
    $$
      \lim_{N\to\infty}\pr{|\bar{X}-\hat{x}|>\epsilon}
       \leq\lim_{N\to\infty}\frac{\sigma^2_{\bar{X}}}{\epsilon^2}
       =\lim_{N\to\infty}\frac{\sigma^2}{N\epsilon^2}=0.
    $$
    The probability can not be less than 0, so we are done.
  \end{proof}
\end{theorem}
The above proof relies on the PDF having a finite
variance. As it turns out, the Weak LLN is true
even when the variance is infinite! This can be proved
using characteristic functions. 
%But since we do not introduce characteristic
%functions until Section~\ref{sec:CLT}, and since we assume in practice
%that our data are drawn from PDFs with finite variance anyway,
%we direct the reader elsewhere. For example, a proof can be
%found on Wikipedia~\cite{Wiki_LLN}.
%
%For completeness we also list the Strong LLN, but without proof. 
%Like the Weak LLN, the Strong LLN is true even when the PDF variance
%is infinite.
%
%\begin{theorem}[Strong LLN]
%  $$
%    \bar{X}\xrightarrow{\text{AS}}\hat{x}.
%  $$
%\end{theorem}

\section{The normal distribution}
Now we focus on results about the normal distribution.
This first proposition will aid us in some of the calculations.

\begin{proposition}\label{prp:gauss}
  Let $\alpha>0$. Then
  \begin{equation*}
    \int_{-\infty}^\infty dx\,e^{-\alpha x^2}=\sqrt{\frac{\pi}{\alpha}}.
  \end{equation*}
  \begin{proof}
    Just square the LHS:
    \begin{equation*}
      \left(\int_{-\infty}^\infty dx\,e^{-\alpha x^2}\right)^2
      =\int_{-\infty}^\infty\int_{-\infty}^\infty dx\,dy\,
        e^{-\alpha(x^2+y^2)}
      =\int_0^\infty r\,dr\int_0^{2\pi}d\theta\,e^{-\alpha r^2}
      =\frac{\pi}{\alpha}.
    \end{equation*}
  \end{proof}
\end{proposition}

Let $X_1$ and $X_2$ be two independent random variables drawn from normal
distributions with respective means $\hat{x}_1$ and $\hat{x}_2$ and
standard deviations $\sigma_1$ and $\sigma_2$. 
\begin{proposition}
  \label{prp:addgauss}
  The random variable $Y=X_1+X_2$ is normally distributed with mean
  $\hat{x}_1+\hat{x}_2$ and variance $\sigma_1^2+\sigma_2^2$.
  \begin{proof}
    By Proposition \ref{prp:addvars}, the sum $Y$ has the distribution
    \begin{equation*}
      g(y)=\frac{1}{2\pi\sigma_1\sigma_2}
           \int_{-\infty}^\infty dx\,\exp\left[
             -\frac{(x-\hat{x}_1)^2}{2\sigma_1^2}
             -\frac{(y-x-\hat{x}_2)^2}{2\sigma_2^2}\right].
    \end{equation*}
    Pull everything out of the integral that does not depend on $x$,
    then complete the square with what remains.
    One obtains
    \begin{equation*}
      g(y)=\frac{1}{2\pi\sigma_1\sigma_2}
           \exp\left[-\frac{(y-\hat{x}_1-\hat{x}_2)^2}
                      {2(\sigma_1^2+\sigma_2^2)}\right]
           \int_{-\infty}^\infty dx\,
           \exp\left[-\left(\frac{\sigma_1^2+\sigma_2^2}
                     {2\sigma_1^2\sigma_2^2}\right)(x+C)^2\right]
    \end{equation*}
    where $C$ does not depend on $x$.
    Therefore one can make the substitution $u=x+C$ with $du=dx$ and
    carry out the new integral using Proposition \ref{prp:gauss}.
    The result is
    \begin{equation*}
      g(y)=\frac{1}{\sqrt{2\pi(\sigma_1^2+\sigma_2^2)}}
           \exp\left[-\frac{(y-\hat{x}_1-\hat{x}_2)^2}
                      {2(\sigma_1^2+\sigma_2^2)}\right].
    \end{equation*}
  \end{proof}
\end{proposition}

Since the normal distribution is so important, so must be its CDF.
The integral of the normal PDF is {\it non-elementary}; that is, 
it can not be expressed in terms of polynomials or standard
functions like $\sin$, $\cos$, or $\exp$. Therefore we give a name
to this special function. The {\it error function} is
\begin{equation}
  \erf(x)\equiv\frac{2}{\sqrt{\pi}}\int_0^xdt\,e^{-t^2}.
\end{equation}
Then we can write the Gaussian CDF with mean 0 as
\begin{equation}\label{eq:gaussCDF}
  \Gau(x,0,\sigma)=\frac{1}{\sqrt{2\pi}\sigma}
                   \int_{-\infty}^xdt\,e^{-t^2/2\sigma^2}
                  =\frac{1}{2}+\frac{1}{2}
                   \erf\left(\frac{x}{\sqrt{2}\sigma}\right).
\end{equation}
Now we can list some powerful applications of the normal distribution.
For instance one can compare two empirical estimates of some mean.
\begin{theorem}
  Suppose $\bar{X}$ and $\bar{Y}$ are normally distributed estimates 
  with the same mean, and call
  their respective standard deviations $\sigma_{\bar{X}}$ and
  $\sigma_{\bar{Y}}$. Then the probability that $\bar{X}$ 
  and $\bar{Y}$ differ by at least $D$ is
  \begin{equation*}
    \pr{\,|\bar{X}-\bar{Y}|>D}=1-\erf\left(\frac{D}
       {\sqrt{2\left(\sigma_{\bar{X}}^2+\sigma_{\bar{Y}}^2\right)}}\right).
  \end{equation*}
  \begin{proof}
    From Proposition \ref{prp:addgauss}, the random variable
    $\bar{X}-\bar{Y}$ is normally distributed with mean 0 
    and variance $\sigma_D^2=\sigma_{\bar{X}}^2+\sigma_{\bar{Y}}^2$. 
    Therefore by eq. \eqref{eq:gaussCDF}, the probability that $\bar{X}$ and 
    $\bar{Y}$ are at most $D$ apart is
    \begin{equation*}
      \begin{aligned}
        \pr{\,|\bar{X}-\bar{Y}|<D}
            &=\pr{-D<\bar{X}-\bar{Y}<D}\\
            &=\Gau(D,0,\sigma_D)-\Gau(-D,0,\sigma_D)\\
            &=1-2\Gau(-D,0,\sigma_D)\\
            &=\erf\left(\frac{D}{\sqrt{2}\sigma_D}\right).
      \end{aligned}
    \end{equation*}
    And of course, $\pr{\,|\bar{X}-\bar{Y}|>D}
     =1-\pr{\,|\bar{X}-\bar{Y}|<D}$.
  \end{proof}
\end{theorem}
The above theorem gives the probability that the
observed difference $|\bar{X}-\bar{Y}|$ is due to chance. This probability
is called the {\it q-value}. In practice one sets some threshold on $q$ 
below which one investigates further whether underlying distributions of the 
estimates are different.

\section{The central limit theorem}\label{sec:CLT}

Let $X$ and $Y$ be real random variables. Then we can construct a
complex random variable $F=X+iY$, and its expectation value will be
\begin{equation}
  \ev{F}=\ev{X}+i\ev{Y}.
\end{equation}
Let $X$ be drawn from the PDF $f$.
  The {\it characteristic function} of $X$ is
  \begin{equation}
    \phi(t)\equiv\ev{e^{itX}}=\int_{-\infty}^\infty dx\,e^{itx}f(x).
  \end{equation}
Knowing the characteristic function $X$ is equivalent to knowing its PDF,
because we can take the inverse Fourier transformation
\begin{equation}
  f(x)=\frac{1}{2\pi}\int_{-\infty}^\infty dt\,e^{-itx}\phi(t).
\end{equation}
The derivatives of the
characteristic function are easily calculated to be
\begin{equation}
  \phi^{(n)}(t)=i^n\int_{-\infty}^\infty dx\,x^ne^{itx}f(x);
\end{equation}
therefore
\begin{equation}
  \phi^{(n)}(0)=i^n\ev{X^n}.
\end{equation}
If $|f(x)|$ falls off faster than $x^m$ for any $m\in\mathbb{Z}$,
it follows from the above equation that all moments exist, and the
characteristic function is analytic in $t$ about $t=0$.

These are useful properties of characteristic functions.
Our main use for them is summarized in the next proposition.
\begin{proposition}\label{prp:addchars}
  The characteristic function of a sum of independent random variables equals 
  the product of their characteristic functions.
  \begin{proof}
    Let $X_1$,...,$X_N$ be drawn from PDFs $f_1$,...,$f_N$ with
    corresponding characteristic functions $\phi_1,...,\phi_N$, and
    let $Y=\sum_j X_j$. Then using the definition of the characteristic
    function we obtain
    $$
      \phi_Y(t)=\ev{e^{it\sum_j X_j}}=\ev{\prod_{j=1}^N e^{it X_j}}\\
               =\prod_{j=1}^N \ev{e^{it X_j}}=\prod_{j=1}^N\phi_j(t),
    $$
    where we used independence for the third equality.
  \end{proof}
\end{proposition}
Now suppose we are experimenters taking independent measurements of 
some observable. Furthermore suppose we do not know anything about the 
observable, except that it comes from some distribution with finite variance.
The central limit theorem (CLT) says that
the sample mean will become normally distributed about the true mean.
\begin{theorem}[Central limit theorem]
  Let $X_1,...,X_N$ be $N$ independent random variables drawn from PDF $f$.
  Suppose further that $f$ has mean $\hat{x}$ and variance $\sigma^2$. 
  Then the PDF of the estimator $\bar{X}$ converges to 
  $\gau(\bar{x},\hat{x},\sigma/\sqrt{N})$.
  \begin{proof}
    Our strategy is to look at the characteristic function 
    $\phi_S$ of the random variable
    $$
      S\equiv\bar{X}-\hat{x}=\frac{X_1+...+X_N-N\hat{x}}{N}.
    $$
    If we can show that $\phi_S$ converges to the characteristic function
    corresponding to $\gau(s,0,\sigma/\sqrt{N})$, then we are finished.
    In order to show this, we first need the characteristic function for
    the distribution $\gau(s,0,\sigma/\sqrt{N})$. By completing the
    square and using Proposition \ref{prp:gauss}, we find 
    \begin{equation*}
      \begin{aligned}
        \phi_{\text{gau}}
            &=\frac{1}{\sigma}\sqrt{\frac{N}{2\pi}}\int_{-\infty}^\infty ds\,
              e^{its}\exp\left[-\frac{s^2N}{2\sigma^2}\right]\\
            &=\frac{1}{\sigma}\sqrt{\frac{N}{2\pi}}
              \exp\left[-\frac{\sigma^2t^2}{2N}\right]
              \int_{-\infty}^\infty ds\,
              \exp\left[-\frac{N}{2\sigma^2}(s-C)^2\right]\\
            &=\exp\left[-\frac{\sigma^2t^2}{2N}\right],
      \end{aligned}
    \end{equation*}
    where $C$ is a number that does not depend on $s$. It remains to show 
    $\phi_S=\phi_{\text{gau}}$. By Proposition \ref{prp:addchars} we have
    $$
      \phi_S(t)=\phi_{\frac{1}{N}\sum X_i-\hat{x}}(t)
               =\left[\phi_{X-\hat{x}}\left(\frac{t}{N}\right)\right]^N,
    $$
    where $\phi_{X-\hat{x}}$ is the characteristic function corresponding
    to the random variable $X-\hat{x}$. Call its PDF $g$. From the
    properties of $f$, we know that $g$ has mean 0 and variance $\sigma^2$.
    Therefore by expanding $\phi_S$ about $t=0$ and using the 
    definition \eqref{dfn:mom}, we find
    $$
      \phi_S(t)=\left[1-\frac{\sigma^2t^2}{2N^2}
             +\mathcal{O}\left(\frac{t^3}{N^3}\right)\right]^N
               =\exp\left[-\frac{\sigma^2t^2}{2N}\right]
             +\mathcal{O}\left(\frac{t^3}{N^2}\right),
    $$
    as desired.
  \end{proof}
\end{theorem}
Since the variance of the estimator $\bar{X}$ tends to 0 for large $N$,
it follows that the sample mean converges to the true mean $\hat{x}$.
In particular for large $N$, we expect the true mean to be within
$\sigma/\sqrt{N}$ of the estimator roughly 68\% of the time.
Table \ref{tab:normal} gives the area under a Gaussian curve 
for different numbers of standard deviations away from the mean. 

\begin{table}
\centering
\caption{Table of areas under the curve for the normal distribution.
The last column gives the probability that a random variable 
drawn from the distribution falls at least the given number of error bars 
away from the mean.}
\vspace{2mm}
\begin{tabularx}{\linewidth}{LCR}
\hline\hline
Number of $\sigma$ from $\hat{x}$ & Area under curve & About 1 in ...\\
\hline
1 & 0.682 689 49 & 3\\
2 & 0.954 499 74 & 22\\
3 & 0.997 300 20 & 370\\
4 & 0.999 936 66 & 15 787\\
5 & 0.999 999 43 & 1 744 278\\
\hline\hline
\end{tabularx}
\label{tab:normal}
\end{table}

\section{Bias}\label{sec:bias}

For this section consider independent random variables $X_1,...,X_N$ 
drawn from a distribution with mean $\hat{x}$ and variance $\sigma^2$. 
Earlier we recovered the familiar estimator for the mean, which was
just the ordinary arithmetic average. But what about an estimator for 
the variance? Intuitively one might write
\begin{equation}\label{eq:bad}
  \bar{\sigma}^2_{\text{biased}}=\frac{1}{N}\sum_{i=1}^N(X_i-\bar{X})^2.
\end{equation} 
This estimator converges to the
exact result in the limit $N\to\infty$, but it disagrees
for small $N$. Most glaringly when $N=1$, the
estimator is zero, regardless of the exact result.
An estimator is said to be {\it biased} when its expectation value
does not agree with the exact result. The difference between the
expectation value of the estimator and the exact result is
correspondingly called the {\it bias}. When they agree, we say
the estimator is {\it unbiased}.
\begin{proposition}\label{prp:stdev}
  For $N\geq2$, an unbiased estimator of the variance is
  $$
    \bar{\sigma}^2=\frac{1}{N-1}\sum_{i=1}^N(X_i-\bar{X})^2.
  $$
  \begin{proof}
  To construct an unbiased estimator of the variance,
  we will determine the bias, then remove it. Note
  \begin{equation*}
    \ev{\bar{\sigma}^2_{\text{biased}}}=\frac{1}{N}\sum\limits_{i=1}^N
      \left(\ev{X_i^2}-2\ev{X_i\bar{X}}+\ev{\bar{X}^2}\right).
  \end{equation*}
  Let us analyze the above equation term by term. Since the random
  variables $X_i$ are drawn from the same distribution, the first term
  is an unbiased estimator of $\ev{X^2}$ for each $i$. Next the
  second term can be rewritten as
  \begin{equation*}
    \begin{aligned}
    \ev{X_i\bar{X}}&=\frac{1}{N}\left(\ev{X_i^2}+
                     \sum_{j,j\neq i}\ev{X_iX_j}\right)\\
                   &=\frac{1}{N}\left(\ev{X^2}+(N-1)\ev{X}^2\right)\\
                   &=\frac{1}{N}\left(\ev{X^2}-\ev{X}^2\right)+\ev{X}^2\\
                   &=\frac{\sigma^2}{N}+\hat{x}^2,
    \end{aligned}
  \end{equation*}
  where in the second line we used the independence of the $X_i$. Finally
  for the last term we have
  $$
    \ev{\bar{X}}=\ev{\frac{1}{N^2}\sum_{i,j}X_iX_j}
                =\frac{1}{N^2}\left(N\ev{X^2}+\sum_{i\neq j}\hat{x}^2\right)
                =\frac{\sigma^2}{N}+\hat{x}^2,
  $$
  where we again used independence in the second equality. Plugging
  everything into $\ev{\bar{\sigma}^2_{\text{biased}}}$ gives
  \begin{equation*}
      \ev{\bar{\sigma}^2_{\text{biased}}}
        =\frac{1}{N}\sum_{i=1}^N
          \left(\ev{X^2}-\frac{\sigma^2}{N}-\hat{x}^2\right)
        =\left(\frac{N-1}{N}\right)\sigma^2.
  \end{equation*}
  This equation shows us the bias is $-\sigma^2/N$. Therefore 
  an unbiased estimator of the variance is
  $$
    \bar{\sigma}^2=\left(\frac{N}{N-1}\right)\bar{\sigma}_\text{biased}^2
                  =\frac{1}{N-1}\sum_{i=1}^N(X_i-\bar{X})^2.
  $$
  \end{proof}
\end{proposition}

%In Excel there are two functions for calculating standard
%deviation, one called \texttt{stdev.s} and one called
%\texttt{stdev.p}. The difference is that the former assumes its numbers
%represent a sample, so it will be calculated according to Proposition
%\ref{prp:stdev}. By contrast the latter uses eq.~\eqref{eq:bad} to
%calculate the variance, which is fine if the numbers represent
%the entire population.

We saw that the bias of $\sigma^2_{\text{biased}}$ estimator goes like $1/N$. 
So one may wonder: How much bias does one typically expect to encounter? 
Bias problems appear whenever one wants to estimate some function 
of the mean $\hat{f}=f(\hat{x})$ that is not necessarily linear near the 
mean. One might be tempted to take the estimator
\begin{equation}
  \bar{f}_{\text{bad}}=\frac{1}{N}\sum_{i=1}^Nf_i,
\end{equation}
where $f_i\equiv f(X_i)$. However it turns out that
\begin{equation}
  \lim_{N\to\infty}\bar{f}_\text{bad}\neq\hat{f}.
\end{equation}
An estimator that never converges to its true value is called
{\it inconsistent}; otherwise it is {\it consistent}.
So this bad estimator is not a consistent estimator. 
A consistent estimator of $\hat{f}$ is
\begin{equation}
  \bar{f}=f(\bar{X}).
\end{equation} 
We can prove the consistency of $\bar{f}$ for a wide class of functions.
\begin{proposition}\label{prp:bias}
  Suppose $f:\mathbb{R}\to\mathbb{R}$ has a convergent Taylor series 
  in a region about $\hat{x}$. If $\bar{X}$ maps to this region, 
  then $\bar{f}$ has bias of order $1/N$.
  \begin{proof}
    If we consider $f$ as a function of the ordinary variable $x$, we can
    expand it about $\hat{x}$ as
    $$
      f(x)=f(\hat{x})+f'(\hat{x})(x-\hat{x})
           +\frac{1}{2}f''(\hat{x})(x-\hat{x})^2
           +\mathcal{O}\left((x-\hat{x})^3\right).
    $$
    Since $\bar{X}$ maps to the region in which this expansion is valid,
    we can plug it into the above formula and find its expected value.
    This gives
    $$
      \ev{\bar{f}}-\hat{f}=f'(\hat{x})\ev{\bar{X}-\hat{x}}
           +\frac{1}{2}f''(\hat{x})\ev{(\bar{X}-\hat{x})^2}
           +\mathcal{O}\left((\bar{X}-\hat{x})^3\right).
    $$
    The LHS of this equation is the bias of $\bar{f}$. To simplify the
    RHS, note that by the CLT $\ev{\bar{X}-\hat{x}}=0$ and
    $\ev{(\bar{X}-\hat{x})^2}=\sigma^2/N$. Therefore
    $$
      \ev{\bar{f}}-\hat{f}=\frac{1}{2}f''(\hat{x})\frac{\sigma^2}{N}
                           +\mathcal{O}\left(\frac{1}{N^2}\right).
    $$
  \end{proof}
\end{proposition}
According to the above proposition, the bias vanishes as $N\to\infty$, which
shows that $\bar{f}$ is consistent. For large $N$, $\bar{X}$ is very likely
to be close to $\hat{x}$ by the CLT, so Proposition \ref{prp:bias} will 
hold whenever $N$ is large and $f$ is a nice enough function.
There is another important consequence to this proposition: the bias
decreases faster than the statistical error bar.
Hence when $N$ becomes large enough, the bias
can be ignored.

\section{Jackknife resampling}
Let us consider a sample of independent measurements 
$X_1,...,X_N$ from some distribution with mean $\hat{x}$ and 
variance $\sigma^2$ and a function $f$ that has a Taylor series 
expansion near $\hat{x}$, but is not necessarily linear.
From Section \ref{sec:bias} we know that $\bar{f}=f(\bar{X})$ is
a consistent estimator of $\hat{f}=f(\hat{x})$. 
One could use error propagation to determine an error bar, however
for sufficiently complicated functions, the error propagation
formula is unwieldy. Moreover we can not use
\begin{equation}
  \bar{\sigma}^2_{\bar{f}}=\frac{\bar{\sigma}^2_{\bar{f}}}{N}
    =\frac{1}{N(N-1)}\sum_{i=1}^{N}
     \left(f(X_i)-\bar{f}\right)^2
\end{equation}
because $f(X_i)$ is not generally a valid sample point. (If it
were, then $\bar{f}_\text{bad}$ would have been a valid estimator.)
Finally, one may wish to estimate the bias.
Finding a simple method to estimate the error bar that also allows one
to estimate the bias motivates the jackknife.
Jackknife error bars agree with usual error bars when there is no bias,
so it makes sense to use the jackknife method generally. 

Here is how the jackknife method works: We throw away the first measurement 
from our sample, leaving a data set of $N-1$ resampled values. Statistical
analysis is done on this smaller sample. Then we resample again, this time
throwing out the second point, and so on.
  The {\it jackknife bins} are defined by
  \begin{equation}
    X_{J,i}\equiv\frac{1}{N-1}\sum_{j\neq i}X_j.
  \end{equation}
  They allow us to construct a {\it jackknife estimator} for the
  mean $\bar{f}_J$ by
  \begin{equation}\label{eq:jackmean}
    \bar{f}_J\equiv\frac{1}{N}\sum_{i=1}^N f_{J,i},
  \end{equation}
  where $f_{J,i}\equiv f(X_{J,i})$. The jackknife estimator for the 
  variance of $\bar{f}_J$ is
  \begin{equation}\label{eq:jackvar}
    \bar\sigma^2_{f_J}=\frac{N-1}{N}\sum_{i=1}^N(f_{J,i}-\bar{f}_J)^2.
  \end{equation}

  Consider the common problem of calculating the mean of the data and
  the variance of the mean. Using the unbiased estimator for the variance
  along with the CLT yields
  \begin{equation}
    \bar{X}=\frac{1}{N}\sum_{i=1}^N X_i~~~~\text{and}~~~~
     \bar{\sigma}^2_{\bar{X}}=\frac{1}{N(N-1)}\sum_{i=1}^N(X_i-\bar{X})^2.
  \end{equation}
  Meanwhile the jackknife estimator for the variance of $\bar{X}$ gives
  \begin{equation}
    \bar\sigma_{\bar{X}_J}^2=\frac{N-1}{N}\sum_{i=1}^N(X_{J,i}-\bar{X}_J)^2.
  \end{equation}
  Some simple algebra shows that $(N-1)(X_{J,i}-\bar{X}_J)=\bar{X}-X_i$.
  Therefore
  \begin{equation}
    \bar\sigma_{\bar{X}_J}^2=\bar{\sigma}^2_{\bar{X}}.
  \end{equation}

Next let us consider how the Jackknife lets us estimate bias.
From Proposition \ref{prp:bias} we know the bias of the estimator $\bar{f}$ 
is of order $1/N$, which we will write
\begin{equation}\label{eq:bias}
  \text{bias}\;\bar{f}=\frac{A}{N}+\mathcal{O}\left(\frac{1}{N^2}\right)
\end{equation}
for some constant $A$. Let us determine the bias of $\bar{f}_J$.
\begin{proposition}\label{prp:Jbias}
  If the measurements $X_i$ are distributed relatively close to $\hat{x}$, 
  then $\bar{f}_J$ has a bias of order $1/(N-1)$.
  \begin{proof}
    The assumption on the measurements is that they roughly fall within
    the series' radius of convergence. We rewrite
    $$
      X_{J,i}=\hat{x}+\frac{1}{N-1}\sum_{j\neq i}(X_j-\hat{x}).
    $$
    Then our strategy is the same as before: We expand $f$ in the same
    sense as before, and take the average value of $f_{J,i}$.
    We obtain
    \begin{equation*}
      \begin{aligned}
        \ev{f_{J,i}}&=\ev{f(X_{J,i})}\\
          &=\ev{f\left(\hat{x}
            +\frac{1}{N-1}\sum_{j\neq i}(X_j-\hat{x})\right)}\\
          &=\hat{f}+\frac{1}{2}f''(\hat{x})\frac{1}{(N-1)^2}
             \sum_{\substack{j\neq i\\k\neq i}}
              \ev{(X_j-\hat{x})(X_k-\hat{x})}
             +\mathcal{O}\left(\frac{1}{N^2}\right)\\
          &=\hat{f}+\frac{1}{2}f''(\hat{x})\frac{1}{(N-1)^2}\left(
             \sum_{j\neq i}\sigma^2+\sum_{j\neq k}\Cov(X_j,X_k)\right)
             +\mathcal{O}\left(\frac{1}{N^2}\right)\\
          &=\hat{f}+\frac{1}{2}f''(\hat{x})\frac{1}{N-1}\sigma^2
             +\mathcal{O}\left(\frac{1}{N^2}\right),
      \end{aligned}
    \end{equation*}
    where in third equality we used $\ev{X_j-\hat{x}}=0$ and in the
    last equality we used the independence of the measurements. Since
    the RHS is independent of $i$, it follows that
    $$
      \ev{\bar{f}_J}-\hat{f}=\frac{1}{2}f''(\hat{x})\frac{\sigma^2}{N-1}
       +\mathcal{O}\left(\frac{1}{N^2}\right).
    $$
  \end{proof}
\end{proposition}
Comparing the final steps of Propositions \ref{prp:bias} and \ref{prp:Jbias},
we see that they have the same lowest order contribution, except that $N$
is replaced by $N-1$. Therefore we can write
\begin{equation}
  \text{bias}\;\bar{f}_J=\frac{A}{N-1}+\mathcal{O}\left(\frac{1}{N^2}\right)
\end{equation}
with the same constant $A$ as with eq.~\eqref{eq:bias}. Combining both
of these equations, we conclude
\begin{equation}
  A=N(N-1)\left(\ev{\bar{f}}-\ev{\bar{f}_J}\right)
     +\mathcal{O}\left(\frac{1}{N}\right),
\end{equation}
which means that
\begin{equation}\label{eq:biasest}
  \overline{\text{bias}}=(N-1)(\bar{f}-\bar{f}_J)
\end{equation}
gives an estimator for the bias of $\bar{f}$, at least up to 
$\mathcal{O}(1/N^2)$.

\section{The $\chi^2$ distribution and fitting data}
Consider a sample of $N$ Gaussian, independent data points $(X_i,Y_i)$,
where the $Y_i$ have standard deviations $\sigma_i$. For now we will
assume the $X_i$ have no error. We will consider a situation where
we believe the $Y_i$ are measurements of some real function $y$ of $x$.
Abstractly we model these data with a fit that depends on some set
of $M$ parameters
\begin{equation}
  y=y(x;a),
\end{equation}
where $a=(a_1,...,a_M)$ is the vector of these parameters. Our goal
is to estimate the $a_j$ and their error bars, and then determine whether
this fit is consistent with the data.

Assuming that $y(x,a)$ is the exact law for the data, the joint PDF
of the measurements $Y_i$ is given by eq.~\eqref{dfn:ind} to be
\begin{equation}\label{eq:chi2NC}
  f(y_1,...,y_N)=\prod_{i=1}^N\frac{1}{\sqrt{2\pi}\sigma_i}
      \exp\left[\frac{-(y_i-y(x_i;a))^2}{2\sigma_i^2}\right].
\end{equation}
  The PDF given by eq.~\eqref{eq:chi2NC} is an example of the 
  {\it non-central $\chi^2$ distribution}. Generally this distribution
  has random variable
  \begin{equation}
    X^2=\sum_{i=1}^N\frac{(Y_i-\hat{y}_i)^2}{\sigma_i^2},
  \end{equation}
  where the random variables $Y_i$ are drawn from $\gau(y,\hat{y}_i,\sigma_i)$. 
  In the special case that the $Y_i$ are drawn from $\gau(y,0,1)$ we
  obtain the random variable
  \begin{equation}
    X^2=\sum\limits_{i=1}^NY_i^2.
  \end{equation}
  In this case the PDF of $X^2$ is called the {\it $\chi^2$ distribution}. 
  It simplifies to 
  \begin{equation}\label{eq:chi2dist}
    f(y_1,...,y_N)=\frac{1}{(2\pi)^{N/2}}
        \exp\left[-\frac{1}{2}\sum_{i=1}^Ny_i^2\right].
  \end{equation}

We will now think about a general, non-central $\chi^2$ PDF. The 
likelihood that the data fall within a region near what was observed is 
\begin{equation}
  \text{P}=\prod_{i=1}^N\frac{1}{\sqrt{2\pi}\sigma_i}
      \exp\left[\frac{-(y_i-y(x_i;a))^2}{2\sigma_i^2}\right]dy_i.
\end{equation}
Our strategy for determining the correct fit will be to find the vector $a$
that maximizes the above probability. The happens when the argument
of the exponential is closest to zero; i.e. when
\begin{equation}
  \chi^2\equiv\sum_{i=1}^N\frac{(y_i-y(x_i;a))^2}{2\sigma_i^2}
\end{equation}
is minimized. This is an example of a {\it maximum likelihood method}.
Once the parameters are found, one can then ask: What is the probability
that the discrepancy between the data and the fit is due to chance? 

To answer this question, we begin with the simpler case using
the $\chi^2$ CDF~\eqref{eq:chi2dist}. It is 
\begin{equation}
  F(\chi^2)=\pr{X^2\le\chi^2}
           =\frac{1}{(2\pi)^{N/2}}\int_{\sum y_i^2\le\chi^2}
            \prod dy_i\;e^{-y_i^2/2}.
\end{equation}
Switching to hyperspherical coordinates, this becomes
\begin{equation}
  F(\chi^2)=\frac{1}{(2\pi)^{N/2}}
              \int d\Omega\int_0^\chi dr\;r^{N-1}e^{-r^2/2}.
\end{equation}
The RHS looks similar to the gamma function. With this in mind,
we can make the substitution $t=r^2/2$ %and use Proposition~\ref{prp:nsolida}
to obtain
\begin{equation}\label{eq:chi2num}
  F(\chi^2)=\frac{1}{\Gamma(N/2)}\int_0^{\chi^2/2}dr\;t^{N/2-1}e^{-t}.
\end{equation}
  The integral
  \begin{equation}
    \Gamma(s,z)\equiv\frac{1}{\Gamma(s)}\int_0^zdt\;t^{s-1}e^{-t}
  \end{equation}
  with $\Re s>0$ is called the {\it incomplete gamma function}.
The CDF in the form \eqref{eq:chi2num} is well-suited for numerical
calculation because it is straightforward to compute the incomplete
gamma function.

\section{Statistical analysis of Markov chains}
Suppose we have computed using MCMC a time series of $N$ measurements
\{$X_1$, ..., $X_N$\}. In principle each element of this sample is drawn
from a PDF with mean $\ev{X_i}=\ev{X}=\hat{x}$ and variance 
$\sigma^2=\ev{(X_i-\hat{x})^2}$, i.e. they all have the same mean
and variance. Unbiased estimators for the mean and variance are
\begin{equation}\label{eq:umv}
  \bar{X}=\frac{1}{N}\sum_{i=1}^N X_i
  ~~~~\text{and}~~~~
  \bar{\sigma}^2=\frac{1}{N-1}\sum_{i=1}^N (X_i-\bar{X})^2.
\end{equation}
The variance of the random variable $\bar{X}$ is
\begin{equation}\label{eq:tsvar}
  \sigma_{\bar{X}}=\ev{(\bar{X}-\hat{x})^2}
                    =\frac{1}{N^2}\left(\sum_{i\neq j}\ev{X_iX_j}
                     +N\ev{X^2}\right)-\hat{x}^2.
\end{equation}
In the case that the measurements are uncorrelated, the expected values
factorize, and we obtain
\begin{equation}
  \sigma_{\bar{X}}=\sigma^2/N
\end{equation}
in agreement with the CLT. But in practice measurement
$i+1$ is often correlated with measurement $i+t$ because they are from
the same time series. To measure this we draw inspiration from 
definition \eqref{dfn:cor}. 
  The {\it autocovariance} between measurements $X_i$ and $X_{i+t}$ is
  \begin{equation}\label{dfn:acov}
    c(X_i,X_{i+t})\equiv\ev{(X_i-\hat{x})(X_{i+t}-\hat{x})}
     =\ev{X_iX_{i+t}}-\ev{X_i}\ev{X_{i+t}},
  \end{equation}
For a Markov
process in equilibrium, the autocorrelation depends only on the separation
$t$, so we define $c(t)\equiv c(X_i,X_{i+t})$. Finally note that
$c(0)=\sigma^2$, which motivates the definition of the {\it autocorrelation}
  \begin{equation}\label{dfn:acor}
    \gamma(t)\equiv\frac{c(t)}{\sigma^2}.
  \end{equation}
The autocorrelation decays in $t$ as a sum of exponentials, 
%I don't
%know why this is true, and I couldn't find a reference, but this is what
%everybody says. 
\begin{equation}
  \gamma(t)=A_\text{exp}\,e^{-t/\tau_\text{exp}}
            +\sum_{i=1}^\infty A_i\,e^{-t/\tau_i},
\end{equation}
where the $A$s are constants and we have picked out the leading
exponential behavior; i.e. for all $i$
\begin{equation}
  \tau_\text{exp}>\tau_i.
\end{equation}
  $\tau_\text{exp}$ is called the {\it exponential autocorrelation time}.

Plugging definition \eqref{dfn:acov} into eq.~\eqref{eq:tsvar} we have 
\begin{equation}
  \sigma^2_{\bar{X}}=\frac{1}{N^2}\sum_{i,j}c(X_i,X_j).
\end{equation}
In the last sum, $|i-j|=0$ occurs $N$ times, and $|i-j|=t$ occurs
$2(N-t)$ times. Note $1\leq t\leq N-1$. Therefore
\begin{equation}
  \sigma^2_{\bar{X}}=\frac{1}{N^2}
    \left(N\,c(0)+2\sum_{t=1}^{N-1}(N-t)c(t)\right).
\end{equation}
Finally we use $c(0)=\sigma^2$ to find
\begin{equation}\label{eq:IAC}
  \sigma^2_{\bar{X}}
    =\frac{\sigma^2}{N}\left(1+2\sum_{t=1}^{N-1}\left(1-\frac{t}{N}\right)
     \gamma(t)\right)
    \equiv\frac{\sigma^2}{N}\tau_\text{int}.
\end{equation}
  The quantity
  \begin{equation}\label{dfn:IAC}
    \tauint=\left(1+2\sum_{t=1}^{N-1}\left(1-\frac{t}{N}\right)
     \gamma(t)\right)
  \end{equation}
  is called the {\it integrated autocorrelation time}.
From eq.~\eqref{eq:IAC} we see that $\tauint$ is just the ratio between the
estimated variance of the sample mean and what this variance would have been 
if the data were uncorrelated. 

In practice, we often do not know the true mean $\hat{x}$ of the time series.
Therefore along the lines of eq.~\eqref{eq:umv}, we construct an unbiased
estimator of the autocovariance
\begin{equation}
  \bar{c}(t)=\frac{N}{(N-1)(N-t)}
    \sum_{i=1}^{N-t}(X_i-\bar{X})(X_{i+t}-\bar{X}),
\end{equation}
where it is the factor $N/(N-1)$ that removes the bias, just as with
the variance. 
Also in most situations we work in the limit where $N$ is large. In this
limit, we can construct an estimator for $\tau_\text{int}$ by
\begin{equation}\label{eq:IACest}
  \bar{\tau}_\text{int}(n)=1+2\sum_{t=1}^n\bar{\gamma}(t),
\end{equation}
where $n<N$. To understand the above estimator look at definition
\eqref{dfn:IAC}. When $t$ is small, $1-t/N\approx 1$. Large $t$ terms
are doubly suppressed by the exponential decay of $\gamma(t)$ and
by $1-t/N\approx 0$. Note that in the simplistic case where
$\gamma(t)$ has only one exponential term, one can prove
\begin{equation}
  \lim_{N\to\infty}\tau_\text{int}=1+2\sum_{t=1}^\infty\gamma(t),
\end{equation}
which parallels eq.~\eqref{eq:IACest} more closely. To construct a final
estimator for $\tauint$, one looks for a window in $n$ for which
eq.~\eqref{eq:IACest} becomes roughly independent of $n$. This serves
as the final $\bar{\tau}_\text{int}$.

%% file: appendix2.tex
\chapter{Calculational Details}\label{ap:calc}

This appendix includes proofs of some elementary facts that were
either stated without proof or used without proof earlier in the
dissertation. Unless stated otherwise,
$U\in\SU(N_c)$. We suppress space-time dependence when convenient. 

\begin{proposition}\label{prp:gaugecovar}
If the covariant derivative transforms as $D_\mu\to U D_\mu U^\dagger$
under a gauge transformation, then the vector potential must
transform as $A_\mu\to U A_\mu U^\dagger-(\partial_\mu U)U^\dagger.$
  \begin{proof} The transformed $D$ can be written 
  $U D_\mu U^\dagger=\partial_\mu'+A_\mu'$. Solving for $A_\mu'$ gives
    \begin{equation*}
    \begin{aligned}
         A_\mu'
         &=U(\partial_\mu+A_\mu)U^\dagger-\partial_\mu\\
         &=U(\partial_\mu U^\dagger)
               +U A_\mu U^\dagger-\partial_\mu\\
         &=\partial_\mu-(\partial_\mu U)U^\dagger
               +U A_\mu U^\dagger-\partial_\mu\\
         &=U A_\mu U^\dagger
                -(\partial_\mu U)U^\dagger.      
    \end{aligned}
    \end{equation*}
  \end{proof}
\end{proposition}

\begin{proposition}\label{prp:fieldtensor}
  $$F_{\mu\nu}=\left[D_\mu,D_\nu\right].$$
  \begin{proof}
    Use the definition of $D_\mu$ and apply the above commutator to some
    field $\psi$. We get 
    \begin{equation*}
    \begin{aligned}
      \left[D_\mu,D_\nu\right]\psi
         &=(\partial_\mu+A_\mu)(\partial_\nu\psi+A_\nu\psi)
                     -(\mu\leftrightarrow\nu)\\
         &=\partial_{\mu\nu}\psi+\partial_\mu A_\nu\psi
           +A_\nu\partial_\mu\psi+A_\mu\partial_\nu\psi
           +A_\mu A_\nu\psi-(\mu\leftrightarrow\nu)\\
         &=\partial_\mu A_\nu\psi-\partial_\nu A_\mu\psi
           +\left[A_\mu,A_\nu\right]\psi\\
         &=-ig\left(\partial_\mu A_\mu^a-\partial_\nu A_\mu^a\right)T^a\psi
           -ig^2A_\mu^bA_\nu^cf^{bca}T^a\psi\\
         &=-ig\left(\partial_\mu A_\mu^a-\partial_\nu A_\mu^a
                    +gf^{abc}A_\mu^bA_\nu^c\right)T^a\psi\\
         &=F_{\mu\nu}\psi.
    \end{aligned}
    \end{equation*} 
  \end{proof}
\end{proposition}

\begin{proposition}\label{prp:plaquette}
$$
U^\Box_{\mu\nu}(x)=\exp\left[-a^2F_{\mu\nu}(x)+\mathcal{O}\big(a^3\big)\right].
$$
  \begin{proof}
    Starting with the definition of the plaquette variable, we have
    \begin{equation*}\begin{aligned}
      U^\Box_{\mu\nu}(x)&=U(x,x+a\hat\nu)U(x+a\hat\nu,x+a\hat\nu+a\hat\mu)
               U(x+a\hat\mu+a\hat\nu,x+a\hat\mu)U(x+a\hat\mu,x)\\
            &=\exp\left[aA_\nu(x)\right]\exp\left[aA_\mu(x+a\hat\nu)\right]
                \exp\left[-aA_\nu(x+a\hat\mu)\right]
                \exp\left[-aA_\mu(x)\right]\\
            &=\exp\left[aA_\nu(x)\right]
               \exp\left[a\left( A_\mu(x)+a\Delta_\nu A_\mu(x) \right)
                                 +\mathcal{O}\left(a^3\right)\right]\\
               &\qquad\times
               \exp\left[-a\left(A_\nu(x)+a\Delta_\mu A_\nu(x) \right)
                                 +\mathcal{O}\left(a^3\right)\right]
               \exp\left[-aA_\mu(x)\right]\\
            &=\exp\left[aA_\nu+aA_\mu+a^2\Delta_\nu A_\mu
                 +\frac{1}{2}\left[aA_\nu,aA_\mu\right]
                 +\mathcal{O}\big(a^3\big)\right]\\
            &\qquad\times\exp\left[-aA_\nu-a^2\Delta_\mu A_\nu-aA_\mu
                 +\frac{1}{2}\left[-aA_\nu,-aA_\mu\right]
                 +\mathcal{O}\big(a^3\big)\right]\\
            &=\exp\left[a^2\Delta_\nu A_\mu+a^2\left[A_\mu,A_\nu\right]
                 -a^2\Delta_\mu A_\nu+\mathcal{O}\big(a^3\big)\right]\\
            &=\exp\left[-a^2F_{\mu\nu}+\mathcal{O}\big(a^3\big)\right].
    \end{aligned}\end{equation*}
    In the fourth step we applied the Campbell-Baker-Hausdorff formula and
    dropped the $x$ dependence for notational convenience, since at this
    step all the gauge fields depend on the same space-time point anyway.
    The fifth step uses another application of the Campbell-Baker-Hausdorff
    formula.
  \end{proof}
\end{proposition}

\begin{proposition} 
  $$S_W\approx-\frac{\beta}{4N_c}\sum_x a^4\tr F_{\mu\nu}(x)F_{\mu\nu}(x).$$
  \begin{proof} Using the definition~\eqref{eq:wilsonaction} and
    Proposition~\ref{prp:plaquette} we have
    \begin{equation*}
    \begin{aligned}
      S_W&=\beta\sum_{x,\mu<\nu}\left(1-\frac{1}{N_c}\Re\tr 
            U_{\mu\nu}^\Box(x)\right)\\
         &=\beta\sum_{x,\mu<\nu}\left(1-\frac{1}{2N_c}\tr 
            \left[U_{\mu\nu}^\Box(x)+
            U_{\mu\nu}^\Box(x)^\dagger\right]\right)\\
         &=\beta\sum_{x,\mu<\nu}\left(1-\frac{1}{2N_c}\tr
            \left[2\id+\frac{a^4}{2}F_{\mu\nu}(x)^2
                  +\mathcal{O}\left(a^5\right)\right]\right)\\
         &=\beta\sum_{x,\mu<\nu}\left(-\frac{a^4}{2N_c}\tr F_{\mu\nu}(x)^2
                  +\mathcal{O}\left(a^5\right)\right)\\
         &=-\frac{\beta}{4N}\sum_x a^4\tr F_{\mu\nu}(x)F_{\mu\nu}(x)
                  +\mathcal{O}\left(a^5\right).
    \end{aligned}
    \end{equation*}
    The cancellation of the $\mathcal{O}\big(a^2\big)$ term can be seen
    as follows: The role of the $\dagger$ in $\SU(N_c)$ is to take the inverse.
    For a path of link variables, this is the same as following the path
    in reverse, which is explained in Section~\ref{sec:latreg}.
    Following a plaquette in reverse just interchanges $\mu$ and $\nu$, 
    which flips the sign of the leading term in the exponential
    of Proposition~\ref{prp:plaquette} because $F_{\mu\nu}$
    is antisymmetric.
  \end{proof}
\end{proposition}

Next we prove some facts stated in Section~\ref{sec:topinvar}. 
We work at fixed $x_4$ and consider smooth maps
$U:\mathbb{R}^3\to\SU(2)$. The BC is $U(\infty)=U_0$, where $U_0$ is a 
constant matrix. $\delta U$ is a smooth deformation of $U$. 
Dependence on $\vec{x}$ is often suppressed for convenience.
%\begin{proposition} If $A_\mu$ is a gauge transformation of zero then
% $F_{\mu\nu}=0$.
%\begin{proof} Since $A_\mu$ is a gauge transformation of zero,
%  $A_\mu=U\partial_\mu U^\dagger$. Then
%  \begin{equation}\begin{aligned}
%    F_{\mu\nu}&=\partial_\mu A_\nu-\partial_\nu A_\mu+[A_\mu,A_\nu]\\
%    &= \partial_\mu\left(U\partial_\nu U^\dagger\right)
%      -\partial_\nu\left(U\partial_\mu U^\dagger\right)
%      +[U\partial_\mu U^\dagger,U\partial_\nu U^\dagger]\\
%    &=\partial_\mu U\partial_\nu U^\dagger+U\partial_{\mu\nu}U^\dagger
%      -\partial_\nu U\partial_\mu U^\dagger-U\partial_{\mu\nu} U^\dagger
%      -\partial_\mu UU^\dagger U\partial_\nu U^\dagger
%      +\partial_\nu UU^\dagger U\partial_\mu U^\dagger\\
%    &=0
%  \end{aligned}\end{equation}
%  since $U\partial_\mu U^\dagger=-\partial_\mu UU^\dagger$.
%\end{proof}
%\end{proposition}

\begin{lemma}\label{lem:variationrule}
  $\delta\left(U\partial_kU^\dagger\right)
   =-U\partial_k\left(U^\dagger\delta U\right)U^\dagger.$
\begin{proof} Note that $\delta U^\dagger=-U^\dagger\delta U\,U^\dagger$. Hence
  \begin{alignat*}{2}
   \delta\left(U\partial_kU^\dagger\right)
    &=\delta U\partial_kU^\dagger&&+U\partial_k\delta U^\dagger\\
    &=                           &&-U\partial_k\left(U^\dagger\delta
                                                     U\,U^\dagger\right)\\
    &=                           &&-U\left(\partial_kU^\dagger\delta UU^\dagger
                                        +U^\dagger\partial_k\delta U\,U^\dagger
                                        +U^\dagger\delta U\partial_k
                                         U^\dagger                     \right).
  \end{alignat*}
  Cancelling the first and last terms and using the product rule gives the
  result.
\end{proof}
\end{lemma}

\begin{theorem} The topological winding number is invariant under smooth
deformations of $U$.
\label{thm:wndeform}
\begin{proof} We integrate eq.~\eqref{eq:wn} over a time-slice of space-time,
which we call $\Omega$. Then
  \begin{align*}
   \delta n &=-\frac{1}{24\pi^2}~\delta\int_\Omega d^3x\,\epsilon_{ijk}\tr
                U\partial_iU^\dagger\,
                U\partial_jU^\dagger\,
                U\partial_kU^\dagger\\ 
            &=-\frac{1}{8\pi^2}\int_\Omega d^3x\,\epsilon_{ijk}\tr
                \delta\left(U\partial_iU^\dagger\right)
                            U\partial_jU^\dagger\,
                            U\partial_kU^\dagger\\ 
            &=+\frac{1}{8\pi^2}\int_\Omega d^3x\,\epsilon_{ijk}\tr
                \partial_i\left(U^\dagger\delta U\right)
                 U^\dagger\partial_jU\,
                 U^\dagger\partial_kU\\
            &=+\frac{1}{8\pi^2}\int_{\partial\Omega} dS_i\,\epsilon_{ijk}\tr
                U^\dagger\delta U\,
                U^\dagger\partial_j U\,
                U^\dagger\partial_k U
              -\frac{1}{8\pi^2}\int_\Omega d^3x\,\epsilon_{ijk}\tr
                U^\dagger\delta U\partial_i\left[
                U^\dagger\partial_j U\,
                U^\dagger\partial_k U\right].
  \end{align*}
  In the second step we used that the trace is cyclic.
  In the third step we used Lemma~\ref{lem:variationrule} as well as
  $U\partial_\mu U^\dagger=-\partial_\mu UU^\dagger$. In the last step
  we integrated by parts. The first integral is over the time-slice boundary
  evaluated at infinity. Since $\partial_jU=\partial_jU_0=0$ there, this term 
  vanishes. The integrand of the remaining integral is expanded as
  $$
    \epsilon_{ijk}\tr\Big[
     \partial_i U^\dagger\partial_j U U^\dagger\partial_k U
     +\partial_j U^\dagger\partial_i U U^\dagger\partial_k U
     +U^\dagger\partial_{ij}UU^\dagger\partial_k U
     +U^\dagger\partial_jUU^\dagger\partial_{ik}U\Big].
  $$
  Terms with double derivatives vanish, as they are symmetric with respect
  to exchange of indices, while $\epsilon$ is antisymmetric. The remaining
  terms are also shown to vanish using the antisymmetry of $\epsilon$
  in addition to cyclically permuting terms under the trace.
  This completes the proof.
\end{proof}
\end{theorem}

\begin{proposition}\label{prp:wnproof}
Consider the map $U:S^3\to\SU(2)$ given by
$$
U(\hat{x})=\left(\begin{array}{cc}
             \co_\chi+i\s_\chi\co_\psi     & i\s_\chi\s_\psi e^{-im\phi}\\
             i\s_\chi\s_\psi e^{im\phi}& c_\chi-i\s_\chi\co_\psi 
            \end{array}\right).
$$
Then $U$ has winding number $m$.
\begin{proof}
  Plugging this map into eq.~\eqref{eq:wn} we find 
  \begin{equation*}
  \begin{aligned}
    n&=-\frac{1}{24\pi^2}\int_{S^3} d^3x\,\epsilon_{ijk}\tr
        U\partial_i U^\dagger\,U\partial_j U^\dagger\,U\partial_k U^\dagger\\
     &=-\frac{1}{24\pi^2}\int_0^\pi d\chi\int_0^\pi d\psi\int_0^{2\pi} d\phi\,
        \epsilon_{\alpha\beta\gamma}\tr
        U\partial_\alpha U^\dagger\,U\partial_\beta U^\dagger\,
        U\partial_\gamma U^\dagger,\\
  \end{aligned}
  \end{equation*}
  where $\alpha,\,\beta,\,\gamma\in\{\chi,\,\psi,\,\phi\}$ and
  $\epsilon_{\chi\psi\phi}\equiv+1$.
  Since the trace is cyclic, all even permutations of $\chi,\,\psi,\,\phi$ give
  the same contribution to the integral, and similarly for all odd
  permutations. Hence
  \begin{equation*}
    n=-\frac{1}{8\pi^2}\int_0^\pi d\chi\int_0^\pi d\psi\int_0^{2\pi} d\phi\,
        \epsilon_{\chi\psi\phi}\tr\left(
        U\partial_\chi U^\dagger\,U\partial_\psi U^\dagger\,
        U\partial_\phi U^\dagger -
        U\partial_\chi U^\dagger\,U\partial_\phi U^\dagger\,
        U\partial_\psi U^\dagger\right).
  \end{equation*}
  Next we compute
  \begin{equation*}
  \begin{aligned}
    U^\dagger&=\left(\begin{array}{cc}
                  \co_\chi-i\s_\chi\co_\psi & -i\s_\chi\s_\psi e^{-im\phi}\\
                 -i\s_\chi\s_\psi e^{im\phi}& c_\chi+i\s_\chi\co_\psi 
                \end{array}\right)\\
    \partial_\chi U^\dagger
             &=\left(\begin{array}{cc}
                  -\s_\chi-i\co_\chi\co_\psi  & -i\co_\chi\s_\psi e^{-im\phi}\\
                 -i\co_\chi\s_\psi e^{im\phi} & -s_\chi+i\co_\chi\co_\psi 
                \end{array}\right)\\
    \partial_\psi U^\dagger
             &=\left(\begin{array}{cc}
                  +i\s_\chi\s_\psi & -i\s_\chi\co_\psi e^{-im\phi}\\
                 -i\s_\chi\co_\psi e^{im\phi} & -i\s_\chi\s_\psi 
                \end{array}\right)\\
    \partial_\phi U^\dagger
             &=\left(\begin{array}{cc}
                  0 & -m\s_\chi\s_\psi e^{-im\phi}\\
                 m\s_\chi\s_\psi e^{im\phi} & 0 
                \end{array}\right)
  \end{aligned}
  \end{equation*}
  and plug into the above equation. Plugging the integral into
  Mathematica,
  $$n=m.$$
\end{proof}
\end{proposition}

\begin{proposition} Let $U_n:S^3\to S^3$ have winding number $n$ and
  $U_k:S^3\to S^3$ have winding number $k$. Then the map
  $U_n U_k$ has winding number $n+k$.
  \begin{proof} 
    The total winding number for the map $U_n U_k$ can be written 
    $$
      w=\frac{1}{24\pi^2}\int_0^\pi d\chi\int_0^\pi d\psi
          \left(\int_0^\pi+\int_\pi^{2\pi}\right)d\phi\,
          \epsilon_{\alpha\beta\gamma}\tr
          U_nU_k\partial_\alpha(U_nU_k)^\dagger
          \left(\text{$\beta$ term}\right)
          \left(\text{$\gamma$ term}\right).
    $$
    From Theorem~\ref{thm:wndeform}, we know we can smoothly 
    deform $U_n$ to $\id$ for $x_3<0$ without changing $w$.
    Then for $0\leq\phi\leq\pi$, we have 
    $\partial_i U_k=0$ and $U_n U_k=U_k$, and we can clearly identify
    the first contribution to the above integral as $k$.
    Similarly, we smoothly deform $U_k$ to $\id$ for 
    $x_3>0$ and find the second contribution to be $n$. Thus,
    $$w=n+k.$$
  \end{proof}
\end{proposition}

%\begin{proposition} Let $H$ be a Hamiltonian satisfying
%$\bra{n}H\ket{n'}=f(|n-n'|)$ for some orthonormal vectors $\ket{n}$
%with $n\in\mathbb{N}$ and for some function $f$.
%Then the eigenvectors of $H$ are the theta vacua
%$$
%\ket{\theta}=\sum_{n=-\infty}^\infty e^{-in\theta}\ket{n}.
%$$
%  \begin{proof} It is enough to show that $H$ is diagonal in the
%    $\theta$ basis. We find
%    $$
%      \bra{\theta_1}H\ket{\theta_2}
%       =\sum_{n,n'}\bra{n}e^{in\theta_1}He^{-in'\theta_2}\ket{n'}
%       =\sum_{n,n'}e^{in\theta_1}e^{-in'\theta_2}f(|n-n'|).
%    $$
%    Consider the $n'$ sum at fixed $n$. We can replace the dummy index
%    $n'$ everywhere by $m=n'-n$ without changing the value
%    sum, because it ranges over all integers. Hence
%    $$
%      \bra{\theta_1}H\ket{\theta_2}
%      =\sum_n e^{in\theta_1}\sum_m e^{-i(n+m)\theta_2}f(|m|)
%      =\sum_n e^{in(\theta_1-\theta_2)}E_{\theta_2}
%      =2\pi\delta(\theta_1-\theta_2)E_{\theta_2},
%    $$
%    where we have defined
%    $$
%      E_\theta\equiv\sum_{m=-\infty}^\infty e^{-im\theta}f(|m|).
%    $$
%  \end{proof}
%\end{proposition}

Now in addition to the BC $U(\infty)=U_0$ for all $x_4$, we specify
$U(\vec{x})=U_+(\vec{x})$ at $x_4=\infty$ and $U(\vec{x})=U_-(\vec{x})$
at $x_4=-\infty$ with winding numbers $n_+$ and $n_-$, respectively.
As explained in Section~\ref{sec:topchargeandinstant}, the total
map $U$ then has winding number $n\equiv n_+-n_-$ on this surface.
Since the surface is homeomorphic to $S^3$, we can parameterize
points on the surface as we did in the map from Proposition~\ref{prp:wnproof}.

\begin{proposition}\label{prp:wnfsproof}
Consider the map of Proposition~\ref{prp:wnproof}. This map's winding 
number can be written in terms of the field strength as
$$
  n=\frac{1}{16\pi^2}\int d^4x\,\tr\dual{F_{\mu\nu}}F_{\mu\nu}.
$$
  \begin{proof} Starting from the definition of the winding number we have
    $$
      n=-\frac{1}{24\pi^2}\int d^3x\,
        \epsilon_{\nu\rho\sigma}\tr U\partial_\nu U^\dagger\,
        U\partial_\rho U^\dagger\,U\partial_\sigma U^\dagger.
    $$
    Recasting this integral as a 4D surface integral and noting
    that $\epsilon_{r\chi\psi\phi}=-1$, which by looking at the Jacobian 
    for this change of variables leads to an overall minus sign, we obtain
    $$
      n=\frac{1}{24\pi^2}\int dS_\mu\,
        \epsilon_{\mu\nu\rho\sigma}\tr U\partial_\nu U^\dagger\,
        U\partial_\rho U^\dagger\,U\partial_\sigma U^\dagger
       =\frac{1}{24\pi^2}\int dS_\mu\,
        \epsilon_{\mu\nu\rho\sigma}\tr A_\nu A_\rho A_\sigma.
    $$
    Next we recall the Chern-Simons current
    $$
      J^{CS}_\mu=2\epsilon_{\mu\nu\rho\sigma}\tr
        \left(A_\nu F_{\rho\sigma}+\frac{2}{3}A_\nu A_\rho A_\sigma\right).
    $$
    From the BCs we know that $F_{\rho\sigma}=0$ on this surface, so
    we are able to replace the integrand in the winding number
    with $J^{CS}$. We get
    $$
      n=\frac{1}{32\pi^2}\int dS_\mu\,J_\mu^{CS}
       =\frac{1}{32\pi^2}\int d^4x\,\partial_\mu J_\mu^{CS}
    $$
    by the divergence theorem. 

    It remains to compute $\partial_\mu J_\mu^{CS}$. The computation
    is somewhat tedious. We have
    \begin{equation*}
    \begin{aligned}
      \partial_\mu J_\mu^{CS}
       &=2\epsilon_{\mu\nu\rho\sigma}\tr\Big[\partial_\mu A_\nu F_{\rho\sigma}
           +A_\nu\partial_\mu F_{\rho\sigma}
           +\frac{2}{3}\left(\partial_\mu A_\nu A_\rho A_\sigma
              +A_\nu \partial_\mu A_\rho A_\sigma
              +A_\nu A_\rho \partial_\mu A_\sigma\right)\Big]\\
       &=2\epsilon_{\mu\nu\rho\sigma}\tr\Big[\partial_\mu A_\nu F_{\rho\sigma}
           +A_\nu\partial_\mu F_{\rho\sigma}
           +2\partial_\mu A_\nu A_\rho A_\sigma\Big]\\ 
       &=\epsilon_{\mu\nu\rho\sigma}\tr\Big[\partial_\mu A_\nu F_{\rho\sigma}
           -\partial_\nu A_\mu F_{\rho\sigma}
           +2A_\nu\partial_\mu[A_\rho,A_\sigma]
           +4\partial_\mu A_\nu A_\rho A_\sigma\Big]\\
       &=\epsilon_{\mu\nu\rho\sigma}\tr\Big[\partial_\mu A_\nu F_{\rho\sigma}
           -\partial_\nu A_\mu F_{\rho\sigma}
           +[A_\mu,A_\nu]\big(\partial_\rho A_\sigma-\partial_\sigma A_\rho
                          +[A_\rho,A_\sigma]\big)\Big]\\
       &=\epsilon_{\mu\nu\rho\sigma}\tr F_{\mu\nu}F_{\rho\sigma}\\
       &=2\tr \dual{F_{\mu\nu}}F_{\mu\nu}.
    \end{aligned}
    \end{equation*}
    To get to the second line, we used the fact that cyclic permutations
    of products under the trace leave the trace unchanged; the fact that
    $\epsilon$ is antisymmetric; and relabelled dummy indices. To get to
    the third line, we expanded the field strength tensor; and used the fact
    that terms with second-derivatives are symmetric and therefore vanish
    when contracted with $\epsilon$. Finally to get to the fourth line,
    one can use the same tricks as with the second line. In addition,
    note that $\epsilon\tr AAAA=0$ because cyclic permutations of
    four indices in $\epsilon$ flip the sign, while cyclic permutations
    of the $AAAA$ indices under the trace leave it unchanged; therefore
    we can add terms of this form inside the trace with impunity
    and obtain the $[A,A][A,A]$ term.
    Plugging this result back into our expression for the winding number
    completes the proof.
  \end{proof}
\end{proposition}

\begin{proposition} For configurations with topological charge $Q$, 
the action is bounded below by
$$
  S\geq\frac{8\pi^2|Q|}{g^2}.
$$
  \begin{proof} 
    Note that $\dual{F_{\mu\nu}}\dual{F_{\mu\nu}}=F_{\mu\nu}F_{\mu\nu}$, so
    $$
      \frac{1}{2}\tr\left(\dual{F_{\mu\nu}\pm F_{\mu\nu}}\right)^2
       =\tr F_{\mu\nu}F_{\mu\nu}\pm\tr\dual{F_{\mu\nu}}F_{\mu\nu}.
    $$
    The LHS of the above equation is non-negative, so
    $$
      \int d^4x\,\tr F_{\mu\nu}F_{\mu\nu}\geq
      \Bigg|\int d^4x\,\tr\dual{F_{\mu\nu}}F_{\mu\nu}\Bigg|.
    $$
    The LHS of the above equation is $2g^2S$ while the RHS
    is, according to Proposition~\ref{prp:wnfsproof}, $16\pi^2|Q|$.
    This completes the proof.
  \end{proof}
\end{proposition}

\begin{proposition} The equation
$$
\dual{F_{\mu\nu}}=F_{\mu\nu}
$$
is solved by
$$
  A_\mu(x)=\frac{r^2}{r^2+R^2}\,U(\hat{x})\partial_\mu U^\dagger(\hat{x}),
$$
where $\hat{x}=x/r.$
  \begin{proof} We start with the ansatz
    $$A_\mu(x)=f(r)U(\hat{x})\partial_\mu U^\dagger(\hat{x}),$$
    with $f(\infty)=1$ and $f(0)=0$. Plugging this ansatz into the
    field tensor, we get
    \begin{equation*}
    \begin{aligned}
      F_{\mu\nu}
        &=\partial_\mu f\,U\partial_\nu U^\dagger
          +f\partial_\mu U\partial_\nu U^\dagger
          +f^2U\partial_\mu U^\dagger U\partial_\nu U^\dagger
          -(\mu\leftrightarrow\nu)\\
        &=\partial_\mu f\,U\partial_\nu U^\dagger
          +f(1-f)\partial_\mu U\partial_\nu^\dagger
          -(\mu\leftrightarrow\nu).
    \end{aligned}
    \end{equation*}
    Terms symmetric in $\mu$ and $\nu$ vanished, and we utilized
    $\partial_\mu U^\dagger=-U^\dagger\partial_\mu UU^\dagger.$
    To proceed, we need to know the components of $\partial$.
    They are
    \begin{equation*}
     \partial=e_r\frac{\partial}{\partial r}
     +e_\chi\frac{1}{r}\frac{\partial}{\partial \chi}
     +e_\psi\frac{1}{rs_{\chi}}\frac{\partial}{\partial \psi}
     +e_\phi\frac{1}{rs_{\chi}s_{\psi}}\frac{\partial}{\partial \phi},
    \end{equation*}
    where $e_i$ is the unit vector in direction $i$. Since $f$ is
    a function of $r$ only and $U$ is a function of the angles
    only, this implies
    $$
      F_{r\chi}=\frac{1}{r}f'U\partial_\chi U^\dagger
    $$
    and
    $$
      F_{\psi\phi}=\frac{1}{r^2s^2_\chi s_\psi}
        f(1-f)\left(\partial_\psi U\partial_\phi U^\dagger
                    -\partial_\phi U\partial_\psi U^\dagger\right).
    $$
    From the definition of the dual tensor, we have
    $\dual{F_{r\chi}}=-F_{\psi\phi}$, since $\epsilon_{r\chi\psi\phi}=-1$.
    To satisfy the instanton equation $\dual{F_{\mu\nu}}=F_{\mu\nu}$
    we must therefore have $F_{r\chi}=-F_{\psi\phi}$. Because
    the variables are separated in $F$, we conclude
    $$
      kf'=kf(1-f)
    $$
    and
    $$
      U\partial_\chi U^\dagger=-\frac{1}{cs_\chi^2s_\psi}
               \left(\partial_\psi U\partial_\phi U^\dagger
                    -\partial_\phi U\partial_\psi U^\dagger\right) 
    $$
    for some constant $k$. Plugging the explicit mapping into the
    latter equation yields $k=2$. The former, ordinary differential
    equation is then easily solved. The result is
    $$
      f(r)=\frac{r^2}{r^2+R^2},
    $$
    where $R$ is a constant of integration.
\end{proof}
\end{proposition}

%\section{Chapter~\ref{ch:MCMC} details}
\begin{proposition}\label{prp:OR}
  Consider a lattice with underlying gauge group $\SU(2)$. Replacing
  a link variable $U$ of the configuration with 
  $$
    U'=\frac{1}{\det U^\sqcup}\left(U^\sqcup UU^\sqcup\right)^\dagger
  $$
  does not change the lattice's Wilson action.
  \begin{proof}
   Since $\det(kA)=k^n\det(A)$ for any constant $k$ and $n\times n$ matrix $A$, 
   one can show that the sum of two $\SU(2)$ matrices is proportional
   to an $\SU(2)$ matrix. Hence we can write
    $$
      U^\sqcup=u^\sqcup\sqrt{\det U^\sqcup}
    $$
    where $u^\sqcup\in\SU(2)$. After updating, the local contribution
    to the Wilson action becomes
    \begin{equation*}
      \tr U'U^\sqcup =\frac{1}{\det U^\sqcup}\tr
                       \left(U^\sqcup UU^\sqcup\right)^\dagger
                     =\tr U^\sqcup U\left(u^\sqcup\right)^\dagger u^\sqcup
                     =\tr U^\sqcup U,
    \end{equation*}
    which is what it was originally.
  \end{proof}
\end{proposition}

%% file: mythesis.bbl
\begin{thebibliography}{10}

\bibitem{adler_over-relaxation_1981}
S.~L. Adler.
\newblock Over-relaxation method for the {Monte} {Carlo} evaluation of the
  partition function for multiquadratic actions.
\newblock {\em Phys. Rev. D}, 23(12):2901--2904, 1981.

\bibitem{alles_topology_1997}
B.~All\'es, M.~D'Elia, and A.~Di~Giacomo.
\newblock Topology at zero and finite {T} in {SU}(2) {Yang}-{Mills} theory.
\newblock {\em Phys. Lett. B}, 412(1-2):119--124, 1997.

\bibitem{alle_three-loop_1997}
B.~All\'es, A.~Feo, and H.~Panagopoulos.
\newblock The three-loop $\beta$ function in {SU}({N}) lattice gauge theories.
\newblock {\em Nucl. Phys. B}, 491(1-2):498--512, 1997.

\bibitem{allton_lattice_1997}
C.~R. Allton.
\newblock Lattice {Monte} {Carlo} data versus perturbation theory.
\newblock {\em Nucl. Phys. B (Proc. Suppl.)}, 53(1-3):867--869, 1997.

\bibitem{aoki_finite_2007}
S.~Aoki, H.~Fukaya, S.~Hashimoto, and T.~Onogi.
\newblock Finite volume {QCD} at fixed topological charge.
\newblock {\em Phys. Rev. D}, 76(5), 2007.

\bibitem{aad_observation_2012}
{ATLAS Collaboration}.
\newblock Observation of a new particle in the search for the {Standard}
  {Model} {Higgs} boson with the {ATLAS} detector at the {LHC}.
\newblock {\em Phys. Lett. B}, 716(1):1--29, 2012.

\bibitem{barkai_can_1982}
D.~Barkai and K.~J.~M. Moriarty.
\newblock Can the {Monte} {Carlo} method for lattice gauge theory calculations
  be effectively vectorized?
\newblock {\em Comput. Phys. Commun.}, 27(2):105--111, 1982.

\bibitem{belavin_calculation_1974}
A.~A. Belavin and A.~A. Migdal.
\newblock Calculation of anomalous dimensionalities in non-{Abelian} gauge
  field theories.
\newblock {\em JETP Lett.}, 19(5):181--182, 1974.

\bibitem{belavin_pseudoparticle_1975}
A.~A. Belavin, A.~M. Polyakov, A.~S. Schwartz, and Y.~S. Tyupkin.
\newblock {Pseudoparticle} solutions of the {Yang-Mills} equations.
\newblock {\em Phys. Lett.}, 59(1):85--87, 1975.

\bibitem{berg_dislocations_1981}
B.~A. Berg.
\newblock Dislocations and topological background in the lattice {O}(3) sigma
  model.
\newblock {\em Phys. Lett. B}, 104(6):475--480, 1981.

\bibitem{berg_markov_2004}
B.~A. Berg.
\newblock {\em Markov {Chain} {Monte} {Carlo} {Simulations} and {Their}
  {Statistical} {Analysis}}.
\newblock World Scientific, Singapore, 2004.

\bibitem{berg_asymptotic_2015}
B.~A. Berg.
\newblock Asymptotic scaling and continuum limit of pure {SU}(3) lattice gauge
  theory.
\newblock {\em Phys. Rev. D}, 92(5):054501, 2015.

\bibitem{berg_deconfinement_2017}
B.~A. Berg and D.~A. Clarke.
\newblock Deconfinement, gradient, and cooling scales for pure {SU}(2) lattice
  gauge theory.
\newblock {\em Phys. Rev. D}, 95(9):094508, 2017.

\bibitem{berg_estimates_2018}
B.~A. Berg and D.~A. Clarke.
\newblock Estimates of scaling violations for pure {SU}(2) {LGT}.
\newblock {\em EPJ Web Conf.}, 175:10007, 2018.

\bibitem{berg_topological_2018}
B.~A. Berg and D.~A. Clarke.
\newblock Topological charge and cooling scales in pure {SU}(2) lattice gauge
  theory.
\newblock {\em Phys. Rev. D}, 97(5):054506, 2018.

\bibitem{bonati_comparison_2014}
C.~Bonati and M.~D'Elia.
\newblock Comparison of the gradient flow with cooling in {SU}(3) pure gauge
  theory.
\newblock {\em Phys. Rev. D}, 89(10):105005, 2014.

\bibitem{brower_qcd_2003}
R.~Brower, S.~Chandrasekharan, J.W. Negele, and U.-J. Wiese.
\newblock {QCD} at fixed topology.
\newblock {\em Phys. Lett. B}, 560(1-2):64--74, 2003.

\bibitem{cabibbo_new_1982}
N.~Cabibbo and E.~Marinari.
\newblock A new method for updating {SU}({N}) matrices in computer simulations
  of gauge theories.
\newblock {\em Phys. Lett. B}, 119(4-6):387--390, 1982.

\bibitem{callan_broken_1970}
C.~G. Callan.
\newblock Broken scale invariance in scalar field theory.
\newblock {\em Phys. Rev. D}, 2(8):1541--1547, 1970.

\bibitem{caswell_asymptotic_1974}
W.~E. Caswell.
\newblock Asymptotic behavior of non-abelian gauge theories to two-loop order.
\newblock {\em Phys. Rev. Lett.}, 33(4):244--246, 1974.

\bibitem{chatrchyan_observation_2012}
{CMS Collaboration}.
\newblock Observation of a new boson at a mass of 125 {GeV} with the {CMS}
  experiment at the {LHC}.
\newblock {\em Phys. Lett. B}, 716(1):30--61, 2012.

\bibitem{creutz_monte_1980}
M.~Creutz.
\newblock Monte {Carlo} study of quantized {SU}(2) gauge theory.
\newblock {\em Phys. Rev. D}, 21(8):2308--2315, 1980.

\bibitem{creutz_overrelaxation_1987}
M.~Creutz.
\newblock Overrelaxation and {Monte} {Carlo} simulation.
\newblock {\em Phys. Rev. D}, 36(2):515--519, 1987.

\bibitem{de_forcrand_topology_1997}
P.~De~Forcrand, M.~G. Perez, and I.~O. Stamatescu.
\newblock Topology of the {SU}(2) vacuum: a lattice study using improved
  cooling.
\newblock {\em Nucl. Phys. B}, 499(1-2):409--449, 1997.

\bibitem{debbio__2002}
L.~Del Debbio, H.~Panagopoulos, and E.~Vicari.
\newblock $\theta$ dependence of {SU}({N}) gauge theories.
\newblock {\em J. High Energy Phys.}, 2002(08):044, 2002.

\bibitem{degrand_topological_1997}
T.~DeGrand, A.~Hasenfratz, and T.~G. Kovacs.
\newblock Topological structure in the {SU}(2) vacuum.
\newblock {\em Nucl. Phys. B}, 505(1-2):417--441, 1997.

\bibitem{delia_phase_2013}
M.~D'Elia and F.~Negro.
\newblock Phase diagram of {Yang}-{Mills} theories in the presence of a
  $\theta$ term.
\newblock {\em Phys. Rev. D}, 88(3):034503, 2013.

\bibitem{engels_critical_1996}
J.~Engels, S.~Mashkevich, T.~Scheideler, and G.~Zinovjev.
\newblock Critical behaviour of {SU}(2) lattice gauge theory. {A} complete
  analysis with the $\chi^2$-method.
\newblock {\em Phys. Lett. B}, 365(1-4):219--224, 1996.

\bibitem{fabricius_heat_1984}
K.~Fabricius and O.~Haan.
\newblock Heat bath method for the twisted {Eguchi}-{Kawai} model.
\newblock {\em Phys. Lett. B}, 143(4-6):459--462, 1984.

\bibitem{ferrenberg_new_1989}
A.~M. Ferrenberg and R.~H. Swendsen.
\newblock New {Monte} {Carlo} technique for studying phase transitions.
\newblock {\em Phys. Rev. Lett.}, 63:1658, 1989.

\bibitem{fingberg_scaling_1993}
J.~Fingberg, U.~M. Heller, and F.~Karsch.
\newblock Scaling and asymptotic scaling in the {SU}(2) gauge theory.
\newblock {\em Nucl. Phys. B}, 392:493--517, 1993.

\bibitem{gattringer_quantum_2010}
C.~Gattringer and C.~B. Lang.
\newblock {\em Quantum {Chromodynamics} on the {Lattice}}.
\newblock Springer, Berlin, 2010.

\bibitem{gross_d.j._ultraviolet_1973}
D.~J. Gross and F.~Wilczek.
\newblock Ultraviolet behavior of non-abelian gauge theories.
\newblock {\em Phys. Rev. Lett.}, 30(26):1343--1346, 1973.

\bibitem{hirakida_thermodynamics_2018}
T.~Hirakida, E.~Itou, and H.~Kouno.
\newblock Thermodynamics for pure {SU(2)} gauge theory using gradient flow.
\newblock {\em arXiv preprint}, arXiv:1805.07106, 2018.

\bibitem{jones_two-loop_1974}
D.~R.~T. Jones.
\newblock Two-loop diagrams in {Yang}-{Mills} theory.
\newblock {\em Nucl. Phys. B}, 75(3):531--538, 1974.

\bibitem{kennedy_improved_1985}
A.~D. Kennedy and B.~J. Pendleton.
\newblock Improved heatbath method for {Monte} {Carlo} calculations in lattice
  gauge theories.
\newblock {\em Phys. Lett. B}, 156:393--399, 1985.

\bibitem{kronfeld_topological_1988}
A.~S. Kronfeld.
\newblock Topological aspects of lattice gauge theories.
\newblock {\em Nucl. Phys. B (Proc. Suppl.)}, 4:329--351, 1988.

\bibitem{levenberg_method_1944}
K.~Levenberg.
\newblock A method for the solution of certain non-linear problems in least
  squares.
\newblock {\em Quart. Appl. Math.}, 2(2):164--168, 1944.

\bibitem{lucini_su_2001}
B.~Lucini and M.~Teper.
\newblock {SU}({N}) gauge theories in four dimensions: {Exploring} the approach
  to {N}=$\infty$.
\newblock {\em J. High Energy Phys.}, 2001(6):050, 2001.

\bibitem{lucini_high_2004}
B.~Lucini, M.~Teper, and U.~Wenger.
\newblock The high temperature phase transition in {SU}({N}) gauge theories.
\newblock {\em J. High Energy Phys.}, 2004(1):061, 2004.

\bibitem{luscher_properties_2010}
M.~L\"uscher.
\newblock Properties and uses of the {Wilson} flow in lattice {QCD}.
\newblock {\em J. High Energy Phys.}, 2010(8):071, 2010.

\bibitem{luscher_stochastic_2018}
M.~L\"uscher.
\newblock Stochastic locality and master-field simulations of very large
  lattices.
\newblock {\em EPJ Web Conf.}, 175:01002, 2018.

\bibitem{luscher_lattice_2011}
M.~L\"uscher and S.~Schaefer.
\newblock Lattice {QCD} without topology barriers.
\newblock {\em J. High Energy Phys.}, 2011(7):036, 2011.

\bibitem{marquardt_algorithm_1963}
D.~W. Marquardt.
\newblock An algorithm for least-squares estimation of nonlinear parameters.
\newblock {\em SIAM J. Appl. Math.}, 11(2):431--441, 1963.

\bibitem{metropolis_equation_1953}
N.~Metropolis, A.~W. Rosenbluth, M.~N. Rosenbluth, A.~H. Teller, and E.~Teller.
\newblock Equation of state calculations by fast computing machines.
\newblock {\em J. Chem. Phys.}, 21(6):1087--1092, 1953.

\bibitem{montvay_quantum_1994}
I.~Montvay and G.~M\"unster.
\newblock {\em Quantum {Fields} on a {Lattice}}.
\newblock Cambridge, Cambridge, 1994.

\bibitem{peskin_introduction_1995}
M.~E. Peskin and D.~V. Schroeder.
\newblock {\em An {Introduction} to {Quantum} {Field} {Theory}}.
\newblock Westview, Boulder, 1995.

\bibitem{politzer_reliable_1973}
H.~D. Politzer.
\newblock Reliable perturbative results for strong interactions?
\newblock {\em Phys. Rev. Lett.}, 30(26):1346--1349, 1973.

\bibitem{rae_ground_2016}
T.~Rae, S.~Collins, S.~D\"urr, and S.~Hofmann.
\newblock Ground state charmed meson and baryon spectra for {$N_f$}=2+1+1
  {QCD}.
\newblock {\em PoS LATTICE 2016}, 367, 2016.

\bibitem{cosma_indep}
C.~Shalizi.
\newblock Reminder no. 1: Uncorrelated vs. independent, 2013.
\newblock [Online; accessed 25-May-2017].

\bibitem{sommer_scale_2014}
R.~Sommer.
\newblock Scale setting in lattice {QCD}.
\newblock {\em PoS LATTICE 2013}, 015, 2014.

\bibitem{srednicki_quantum_2007}
M.~Srednicki.
\newblock {\em Quantum {Field} {Theory}}.
\newblock Cambridge, Cambridge, 2007.

\bibitem{symanzik_small_1970}
K.~Symanzik.
\newblock Small distance behaviour in field theory and power counting.
\newblock {\em Commun. Math. Phys.}, 18(3):227--246, 1970.

\bibitem{symanzik_small-distance-behaviour_1971}
K.~Symanzik.
\newblock Small-distance-behaviour analysis and {Wilson} expansions.
\newblock {\em Commun. Math. Phys.}, 23(1):49--86, 1971.

\bibitem{t_hooft_computation_1976}
G.~'t~Hooft.
\newblock Computation of the quantum effects due to a four-dimensional
  pseudoparticle.
\newblock {\em Phys. Rev. D}, 14(12):3432--3450, 1976.

\bibitem{thomas_meson_2017}
C.~E. Thomas.
\newblock Meson spectroscopy from lattice {QCD}.
\newblock {\em Few-Body Systems}, 58(3), 2017.

\bibitem{wiki_SM}
{University of Zurich Physik-Institut}.
\newblock Standard model, 2018.
\newblock [Online; accessed 19-September-2018].

\bibitem{veneziano_u1_1979}
G.~Veneziano.
\newblock U(1) without instantons.
\newblock {\em Nucl. Phys. B}, 159(1-2):213--224, 1979.

\bibitem{vicari__2009}
E.~Vicari and H.~Panagopoulos.
\newblock $\theta$ dependence of {SU}({N}) gauge theories in the presence of a
  topological term.
\newblock {\em Phys. Rep.}, 470(3-4):93--150, 2009.

\bibitem{wilson_confinement_1974}
K.~G. Wilson.
\newblock Confinement of quarks.
\newblock {\em Phys. Rev. D}, 10(8):2445--2459, 1974.

\bibitem{witten_current_1979}
E.~Witten.
\newblock Current algebra theorems for the {U}(1) “{Goldstone} boson”.
\newblock {\em Nucl. Phys. B}, 156(2):269--283, 1979.

\end{thebibliography}
